\documentclass[usenglish, 11pt, a4paper]{article}
\usepackage[T1]{fontenc}
\usepackage{a4wide, url}
\usepackage[english]{babel}
\usepackage{graphicx}
\usepackage{natbib}
\usepackage{hyperref}
\usepackage[textfont=sc, labelfont=bf]{caption}
\usepackage[blocks]{authblk}
\usepackage{setspace}
\usepackage{multirow}
\usepackage{amsmath, amsthm, amssymb, amsfonts}
\usepackage{pgf}
\usepackage{bm}
\usepackage{paralist}
\usepackage{multirow}
\usepackage{color}
\usepackage{changepage}
\usepackage{enumitem}
\usepackage{titlesec}
\usepackage{amsfonts}
\usepackage{verbatim}
\usepackage{threeparttable}
\usepackage{xr}
\DeclareMathAlphabet\mathbfcal{OMS}{cmsy}{b}{n}
\newcommand{\mbf}{\mathbf}
\newcommand{\mc}{\mathcal}
\newcommand{\beq}{\begin{equation}}
\newcommand{\eeq}{\end{equation}}
\newcommand{\bea}{\begin{eqnarray}}
\newcommand{\eea}{\end{eqnarray}}
\newcommand{\beas}{\begin{eqnarray*}}
\newcommand{\eeas}{\end{eqnarray*}}
\newcommand{\bit}{\begin{itemize}}
\newcommand{\eit}{\end{itemize}}
\newcommand{\ben}{\begin{enumerate}}
\newcommand{\een}{\end{enumerate}}
\newcommand{\bpm}{\begin{pmatrix}}
\newcommand{\epm}{\end{pmatrix}}
\newcommand{\bbm}{\begin{bmatrix}}
\newcommand{\ebm}{\end{bmatrix}}

\renewcommand{\l}{\left}
\renewcommand{\r}{\right}
\def\wh{\widehat}
\def\wt{\widetilde}
\newcommand{\E}[0]{\mathsf{E}}

\newcommand{\p}{\mathsf{P}}

\newcommand{\nn}{\nonumber}

\theoremstyle{definition}
\newtheorem{thm}{Theorem}
\theoremstyle{definition}
\newtheorem{cor}{Corollary}
\theoremstyle{definition}
\newtheorem{lem}{Lemma}
\theoremstyle{definition}

\theoremstyle{definition}
\newtheorem{assum}{Assumption}
\theoremstyle{definition}

\theoremstyle{definition}

\theoremstyle{definition}

\usepackage[draft]{todonotes}   % notes showed

\begin{document}

\title{Modelling Large Dimensional Datasets with Markov Switching Factor Models}
\author{Matteo Barigozzi$^1$ \hskip 1cm Daniele Massacci$^{2}$}
\date{\small{\today}}

\maketitle

\begin{abstract}
	
\noindent We study a novel large dimensional approximate factor model with regime changes in the loadings driven by a latent first order Markov process. By exploiting the equivalent linear representation of the model, we first recover the latent factors by means of Principal Component Analysis. We then cast the model in state-space form, and we estimate loadings and transition probabilities through an EM algorithm based on a modified version of the Baum-Lindgren-Hamilton-Kim filter and smoother that makes use of the factors previously estimated. Our approach is appealing as it provides closed form expressions for all estimators. More importantly, it does not require knowledge of the true number of factors. We derive the theoretical properties of the proposed estimation procedure, and we show their good finite sample performance through a comprehensive set of Monte Carlo experiments. The empirical usefulness of our approach is illustrated through three applications to large U.S. datasets of stock returns,  macroeconomic variables, and inflation indexes.  \\ 


\end{abstract}

\thispagestyle{empty}

\footnotetext[1]{Universit\`a di Bologna, Department of Economics, \url{matteo.barigozzi@unibo.it}.} 

\footnotetext[2]{King's College London, King's Business School, \url{daniele.massacci@kcl.ac.uk}. Corresponding author. \smallskip

\noindent
The paper greatly benefited from comments from conference participants at the 10th Italian Congress of Econometrics and Empirical Economics, the 28th International Panel Data Conference, the EEA-ESEM Barcelona 2023, the 2023 NBER-NSF Time Series Conference, and the 16th Annual SoFiE Meeting. Any errors and omissions are the authors' own responsibility only.}

%%%%%%%%%%%%%%%%%%%%%%%%%%%%%%%%%%%%%%%%%%%%%%%%%%%%%%%%%%%%%%%%%%%%%%%%%%
%%%%%%%%%%%%%%%%%%%%%%%%%%%%%%%%%%%%%%%%%%%%%%%%%%%%%%%%%%%%%%%%%%%%%%%%%%

\section{Introduction}

%\begin{ext}
%Daniele: All comments for next version are in blue and can be deactivated by acting on lines 103-104 of the tex.
%\end{ext}

This paper develops a comprehensive approach for the analysis of large dimensional models exhibiting an approximate factor structure, in which the loadings are subject to regime shifts driven by a first order latent Markov process. We label these large dimensional Markov Switching factor models.

Since the works of \cite{hamilton1989new}, and \citet{diebold78measuring}, and inspired by the seminal paper of \cite{GQ1973}, Markov switching models have been widely used in the empirical analysis of macroeconomic and financial time series data: \cite{Hamilton_2016_Chapter} gives an overview from a macroeconomic perspective, and \cite{Doz_Ferrara_Pionnier_2020_WP} present recent evidence of their usefulness for turning-point detection and macroeconomic forecasting; \cite{Guidolin_2011_Chapter}, and \cite{AT2012}, provide a comprehensive survey in relation to financial markets; see also \cite{Qu_Zhuo_2021_ReStat} and references therein for more recent advances.
However, to the very best of our knowledge, the existing literature has focused on small dimensional Markov switching models, which are not applicable to high dimensional cross-sections. We aim at filling a gap in the literature by studying Markov switching models as applied to large panels.  

There now exists strong empirical evidence that macroecononomic and financial variables exhibit an approximate factor structure, as stressed in \cite{giannone_lenza_primiceri_2021}. This nature of the data naturally leads to approximate latent factor specifications as a tool to model time series comovement in large dimensional cross-sections. For example, following the seminal contribution of \cite{chamberlainrothschild83}, static approximate factor representations have been considered in \cite{Connor_Korajczyk_1986_JFE} to develop measures of portfolio performance, and in \citet{stockwatson02JASA,stockwatson02JBES} to forecast large macroeconomic panels and to build indexes of macroeconomic activity. The full inferential theory is developed by \cite{bai03}. Settings allowing for dynamic factor representations have been also extensively studied: see \cite{FHLZ17} and references therein. A broad overview of large factor models is provided in \cite{stockwatson16}. To the very best of our knowledge, the vast majority of existing contributions has looked at the linear setting. However, this may not be flexible enough to accommodate the discrete regimes typically observed in macroeconomic and financial series.

A number of contributions have extended linear static factor models to allow for discrete shifts in the loadings by assuming that these shifts are driven by an \textit{observable} state variable. A first and growing stream of literature assumes that this state variable is a deterministic time index, which leads to a factor model with structural instability in the loadings: see 
\citet{Breitung_Eickmeier_2011_JoE},
\citet{corradi2014testing},
\citet{baltagi2016estimation},
\citet{cheng2016shrinkage},
\citet{BCF18},
\cite{Barigozzi_Trapani_2020_SPA}, 
\citet{duan2022quasi},
among others, and \cite{Bai_Han_2016_FEC} for a survey of the literature.
The presence of structural breaks implies that regime changes are not recurrent and are related to events such as technological changes or shifts in monetary policy regimes. Alternatively, the states could be driven by the realisation of an observable stationary variable with respect to a reference value, in which case a threshold factor model would arise: see \cite{Massacci_2017_JoE,Massacci_2023_JFEcon}. Under this set up, regimes are recurrent and associated to cyclical events such as business and financial cycles. Smoothly varying loadings are considered in \citet{motta2011locally} and \citet{Pelger_Xiong_2022_JBES}. Finally, \cite{Chen_Chen_Chen_2023_WP} follow \cite{Su_Wang_2017_JoE} and propose a time-varying matrix factor model with smooth changes in the loadings driven by a time index.

In this paper, we are interested in large dimensional factor models in relation to recurrent regime changes. A major drawback of threshold factor models is that they require \textit{a priori} identification of the state variable. This may lead to model misspecification and unreliable empirical findings should the wrong state variable be employed to identify the regimes. In order to overcome this problem, we resort to the two-state Markov switching model of \cite{GQ1973} with a \textit{latent} state variable, and we extend it to allow for an underlying large dimensional factor structure. Within this setting, we make the following major methodological contributions: we propose an algorithm to estimate the conditional state probabilities, as well as the loadings and the factors; and we derive the asymptotic properties of the estimators for loadings and factors. Remarkably, our results do not require knowledge of the true number of factors in any regime, and they are robust to the number of factors being unknown and estimated. This is an important aspect of our paper. Estimating the number of factors is challenging in a linear setting, as evidenced by the high number of relevant contributions: \cite{baing02}, \cite{ABC10} and \cite{ahnhorenstein13}, develop model selection criteria; \cite{Kapetanios_2010_JBES}, \cite{onatski10}, and \cite{Trapani_2018_JASA}, propose inferential procedures. Dealing with an unknown number of factors clearly becomes even more engaging in the presence of regimes driven by a latent state variable and it therefore is an important contribution of our paper.  

To the very best of our knowledge, the literature on large dimensional Markov Switching factor models is still in its infancy. However, two existing contributions are important to discuss. First, \cite{Liu_Chen_2016_SS} study a model similar to ours, but their definition of common factors differs from ours in that they consider factors that are pervasive along the time dimension rather than along the cross-sectional dimension. As a consequence, their idiosyncratic components are assumed to be white noise.
Second, \cite{Urga_Wang_2022_WP} study a set up similar to ours, with some important differences: they assume \textit{a priori} knowledge of the number of factors; they consider a model with serially homoskedastic idiosyncratic components. In addition, the Maximum Likelihood estimation approach of \cite{Urga_Wang_2022_WP} adapts the EM algorithm by  \citet{RT82} and \citet{baili12} to the case of Gaussian mixtures, where the weights are given by the probability of the latent variables to be in a given regime. 
Furthermore, the fact that the proposed EM algorithm is just an approximation to Maximum Likelihood estimation is however not accounted for when deriving the asymptotic properties of the considered estimators, in other words no formal proof that such algorithm is a contraction towards the Maximum Likelihood estimator is given. 
%Last, the approach of  \cite{Urga_Wang_2022_WP} leads to estimators for the unknown parameters that do not have closed form solutions. This in an additional important difference with respect to our approach, which instead leads to closed form solutions. 

Our approach is as follows. We introduce an algorithm to estimate factors, loadings, and transition probabilities, which extends to high dimensional factor models the state-space approach advanced in \citet{hamilton1989new} and \cite{kim1994} to handle low dimensional Markov switching autoregressive models. In particular, we generalize the Baum-Lindgren-Hamilton-Kim filter and smoother, the original version of which was proposed to estimate Markov-switching VAR models: for example, see the reviews by \citet{Guidolin_2011_Chapter}, \citet{krolzig2013markov}, \citet{Hamilton_2016_Chapter}, and \citet{guidolin2018essentials}. An important feature of our approach is that it provides closed form expressions for all estimators. Even more remarkably, we not require \textit{a priori} knowledge of the number of factors in each regime, which is instead needed by \cite{Urga_Wang_2022_WP}.

We obtain our theoretical results by exploiting the well known property that a factor model with neglected discrete regime changes admits an equivalent representation with a higher number of factors: for example, see the discussions in \cite{Breitung_Eickmeier_2011_JoE}, \cite{BCF18}, and \citet{duan2022quasi}, in the case of structural breaks; and \cite{Massacci_2023_JFEcon} for threshold factor models. 
We use this property to estimate the latent factors by means of Principal Component Analysis (PCA) as applied to the linear representation. We then input these estimated factors into our algorithm, which allows us to recover the loadings and the transition probabilities. We then derive the asymptotic properties of the estimator for the loadings: we prove the asymptotic normality; we characterise the bias, which is induced both by the well known identification problem, and by the incomplete information related to the underlying data generated process. We also study the asymptotic properties of the estimated factors, which are obtained by projecting the data onto the estimated loadings. We corroborate our theoretical results through a comprehensive set of Monte Carlo experiments, which confirm the good finite sample properties of the estimation procedure we propose.

Finally, we assess the empirical validity of our model through three applications to large U.S. datasets of stock returns,  macroeconomic variables, and inflation indexes. Markov switching models have been widely used to capture the cyclical behaviour of small-dimensional portfolios of financial assets: see \cite{Guidolin_2011_Chapter}, and \cite{AT2012}, and references therein. We apply our Markov switching factor model to a large dimensional portfolio of financial assets: the results show that the regimes described by the model closely follow U.S. business cycle dynamics, and complement the findings in \cite{Massacci_Sarno_Trapani_2021_WP}, who identify the regimes based on an observable state variable. We then consider a large set of U.S. macroeconomic variables, and we use them to identify turning points in the U.S. business cycle in the spirit of \citet{Burns_Mitchell_1946_measuring}: through appropriate metrics, we show that our model performs very well also on this respect. Finally, building upon the recent contribution of \citet{AL}, we illustrate how our model may be employed to identify regimes in a large set of inflation indexes. Overall, these results confirm the usefulness of our theoretical framework to conduct empirical analysis.

The rest of the paper is organised as follows. Section \ref{section:MS_model} introduces the two-state model. Section \ref{section:Estimation} describes the estimation algorithm. Section \ref{section:Asymptotics} derives the asymptotic theory. Section \ref{section:num_regimes_factors} presents two further results related to estimation of the number of factors and to underspecification of the number of regimes.
Section \ref{section:Unobs_Hetero} deals with the issue of unobserved heterogeneity. Section \ref{section:Detecting_regime_changes} discusses the problem of testing for regime changes. Section \ref{section:Monte_Carlo} runs a comprehensive set of Monte Carlo experiments. Section \ref{section:Empirics}  presents the empirical applications. Finally, Section \ref{section:Conclusions} concludes. Details about the estimation algorithm are given in Appendix \ref{Appendix:Est}. Mathematical derivations are collected in Appendices \ref{section:Appendix} and \ref{Appendix:Under_model}. Additional Monte Carlo and empirical results are to be found in Appendices \ref{app:sim} and \ref{app:factors}, respectively.

%%%%%%%%%%%%%%%%%%%%%%%%%%%%%%%%%%%%%%%%%%%%%%%%%%%%%%%%%%%%%%%%%%%%%%%%%%
\subsection*{Notation}
We denote as $\otimes$ the Kronecker product, with $\odot$ the  element-wise (Hadamard) product, and with $\oslash$
the element-wise ratio.
For a vector $\bm v=(v_1\cdots v_m)'$ we denote its Euclidean norm as $\Vert \bm v\Vert=\sqrt{\sum_{i=1}^m v_i^2}$.
For a matrix $\mbf C$ we denote the spectral norm as $\Vert\mbf C\Vert=\sqrt{\mu_1(\mbf C\mbf C^\prime)}$, where $\mu_1(\mbf C\mbf C^\prime)$ indicates the largest eigenvalue of $\mbf C\mbf C^\prime$. If $\text{rk}(\mbf C)=r<\infty$, then, we sometimes use the same notation $\Vert\mbf C\Vert$ to denote also the Frobenius norm $\Vert\mbf C\Vert_F=\sqrt{\text{tr}(\mbf C\mbf C^\prime)}$. Indeed, $\Vert\mbf C\Vert_F\le \sqrt r \Vert\mbf C\Vert$ and since it is always true that $\Vert\mbf C\Vert\le \Vert\mbf C\Vert_F$, then, bounding the Frobenius or the spectral norm is asymptotically equivalent.

For a scalar discrete random variable $Z$,  the notation $\p( Z= z)$ is its probability mass function computed using the true value of the parameters. For random variables $\mbf Y$ and $\mbf W$ the notations $\E[\mbf Y]$ and  $\E[\mbf Y|\mbf W]$ are the expectation and conditional expectation given $\mbf W$, respectively, computed with respect to the true distributions $F_Y(\mbf y)$ and $F_{Y|W}(\mbf y|\mbf W)$ which in turn are computed using the true value of the parameters. If, in place of the true value of the parameters, we use an estimate of the parameters, say $\wh{\theta}$, then we adopt the notations $\p_{\wh{\theta}}( Z= z)$, $\E_{\wh{\theta}}[\mbf Y]$, and $\E_{\wh{\theta}}[\mbf Y|\mbf W]$, respectively.

Finally, we let $\mbf I_m$ be the identity matrix of dimension $m$, $\bm\iota_m$ an $m$-dimensional vector of ones, and $\mbf 0$ any matrix or vector of zeros whose dimensions depend on the context.

%%%%%%%%%%%%%%%%%%%%%%%%%%%%%%%%%%%%%%%%%%%%%%%%%%%%%%%%%%%%%%%%%%%%%%%%%%

\section{Markov switching factor model}
\label{section:MS_model}

%%%%%%%%%%%%%%%%%%%%%%%%%%%%%%%%%%%%%%%%%%%%%%%%%%%%%%%%%%%%%%%%%%%%%%%%%%

\subsection{Setup}
\label{section:MS_model_Setup}

We study a two-state large dimensional Markov switching factor model.  Formally,  we consider
%------------------------------------------------------------------------%
\begin{align}
\mathbf{x}_{t}&=\bm\Lambda _{1}\mathbf{f}_{1t}\mathbb{I}(s_{t}=1)+\bm\Lambda
_{2}\mathbf{f}_{2t}\mathbb{I}(s_{t}=2)+\mathbf{e}_{t},\qquad t\in\mathbb Z,
\label{eq:model}\\
\mathbf{e}_{t}&=\bm\Sigma _{e1}^{1/2}\mathbb{I}(s_{t}=1)\bm{\nu}_{t}+\bm\Sigma _{e2}^{1/2}\mathbb{I}(s_{t}=2)\bm{\nu}_{t} \label{eq:cov_mat}.
\end{align}
%------------------------------------------------------------------------%
We assume that the elements of the $N\times1$ vector process of observable dependent variables $\{\mathbf{x}_{t}\}$ have zero mean, and we consider the more general case in which they are allowed to have mean different from zero in Section \ref{section:Unobs_Hetero}; $\{\mathbf{f}_{jt}\}$ is the $r_{j}\times1$ vector process of latent factors such that $r_j$ is fixed and $r_{j}\ll N$, for $j=1,2$; $\bm\Lambda_j$ is the $N\times r_{j}$ matrix of factor loadings with rows equal to $\bm\lambda_{ji}^\prime$, for $i=1,\ldots, N$ and $j=1,2$; $\{\mathbf{e}_{t}\}$ is the $N\times1$ vector process of idiosyncratic components with innovations $\bm{\nu}_{t}\sim\left(\mathbf{0},\mathbf{I}_{N}\right)$. Note that we allow the elements of $\{\mbf e_t\}$ to be both serially and cross-sectionally weakly correlated, and we refer to Section \ref{section:Asymptotics} for the specific assumptions. It is also important to point out that the number of factors $r_j$ within each state is allowed to be unknown.

The model in \eqref{eq:model} and \eqref{eq:cov_mat} explicitly allows for two regimes: the case in which the number of states is actually underspecified is dealt with in Section \ref{section:num_regimes}. Also, the number of factors $r_1$ and $r_2$ is allowed to change between the regimes: in this, our approach is more general than in \cite{Liu_Chen_2016_SS}, who assume that $r_1=r_2$ and the dimension of the factor space is \textit{a priori} the same between the two regimes.

As it is standard in the literature, we assume that $s_{t}$ follows a discrete-state, homogeneous, irreducible and ergodic, first-order Markov chain such that
%------------------------------------------------------------------------%
\begin{equation*}
\begin{array}{lll}
\p \left( s_{t+1}=j\left\vert s_{t}=i\right. \right)=p_{ij} , & i,j=1,2, & 
\sum\limits_{j=1}^{2}p_{ij}=1,%
\end{array}%
\end{equation*}
%------------------------------------------------------------------------%
with matrix of transition probabilities
%------------------------------------------------------------------------%
\begin{equation}\label{eq:TransP}
\mathbf{P}=\left( 
\begin{array}{cc}
p_{11} & p_{12} \\ 
p_{21} & p_{22}%
\end{array}%
\right) =\left( 
\begin{array}{cc}
p_{11} & 1-p_{11} \\ 
1-p_{22} & p_{22}%
\end{array}%
\right).
\end{equation}
%------------------------------------------------------------------------%
Defining the $2\times1$ vector of state indicators
%------------------------------------------------------------------------%
\begin{equation}
\bm{\xi }_t=\left[
\begin{array}{c}
\mathbb I(s_t=1)\\
\mathbb I(s_t=2)
\end{array}
\right], \qquad t\in\mathbb Z,\label{eq:xidef}
\end{equation}
%------------------------------------------------------------------------%
allows us to write the transition equation
%------------------------------------------------------------------------%
\begin{equation}
\bm{\xi }_{t}=\mathbf{P}^{\prime }\bm{\xi }_{t-1}+\mathbf{v}_{t},\qquad t\in\mathbb Z,
\label{eq:trans}
\end{equation}%
%------------------------------------------------------------------------%
where $\{\mathbf{v}_{t}\}$ is a discrete-valued zero mean martingale difference sequence whose elements sum to zero. 
Because, $\Vert \mbf P\Vert <1$, $\{s_t\}$ follows an ergodic Markov chain, thus, there exists a stationary  vector of probabilities $\bar {\bm\xi}$ satisfying:
\[
\bar {\bm\xi}=\mbf P'\bar {\bm\xi}.
\]
Hence,  the elements of $\bar {\bm\xi}$ are long-run or unconditional state probabilities. In particular, we have  $\bar {\bm\xi}=\E[\bm\xi_t]$, such that
\begin{align}
\E[\bm\xi_t]=\E\l[\begin{array}{c}
\mathbb I(s_t=1)\\
\mathbb I(s_t=2)
\end{array}
\r] = 
\l[\begin{array}{c}
\p(s_t=1)\\
\p(s_t=2)
\end{array}
\r],\label{eq:UncondP}
\end{align}
where $0<\p(s_t=j)<1$, for $j=1,2$, by Assumption \ref{assum:F} in Section \ref{section:Asymptotics} below, which makes the Markov chain irreducible. In particular, \eqref{eq:TransP} and \eqref{eq:UncondP} are related by \citep[see, e.g.,][Chapter 9]{guidolin2018essentials}
\beq\label{eq:UncondP2}
\p(s_t=1) = \frac{1-p_{22}}{2-p_{11}-p_{22}}, \quad \p(s_t=2) = \frac{1-p_{11}}{2-p_{11}-p_{22}}.
\eeq
%%%%%%%%%%%%%%%%%%%%%%%%%%%%%%%%%%%%%%%%%%%%%%%%%%%%%%%%%%%%%%%%%%%%%%%%%%

Finally, unlike the low-dimensional model of \cite{diebold78measuring}, we do not specify the factor dynamics. In particular, \cite{diebold78measuring} allow for regime-specific factor mean, whereas the loadings do not vary: in this setting, the variance of the dependent variables remains constant over time. On the other hand, the large-dimensional model in \eqref{eq:model} and \eqref{eq:cov_mat} allows for regime-specific covariance matrix of $\mathbf{x}$: this is relevant for modelling both macroeconomic variables and financial returns, as stressed in \cite{McConnell_Perez-Quiros_2000_AER}, and \cite{Perez_Quiros_Timmermann_2000_JF,Perez_Quiros_Timmermann_2001_JoE}, respectively. We exploit this feature in the empirical analysis in Section \ref{section:Empirics}, where we use the model in \eqref{eq:model} and \eqref{eq:cov_mat} to study large U.S. datasets of stock returns,  macroeconomic variables, and inflation indexes. On the other hand, we explain in Section \ref{section:Unobs_Hetero} how we can deal with datasets displaying regime-specific individual effects.

\subsection{State space representation}
\label{section:ESSR}

Let the $\left(r_{1}+r_{2}\right)\times1$ vector process $\{\mathbf{g}_t\}$ be defined as
%------------------------------------------------------------------------%
\begin{equation}
\mathbf{g}_{t}=\left[ 
\begin{array}{c}
\mathbf{f}_{1t} \\ 
\mathbf{0}%
\end{array}%
\right] \mathbb{I}(s_{t}=1)+\left[ 
\begin{array}{c}
\mathbf{0} \\ 
\mathbf{f}_{2t}%
\end{array}%
\right] \mathbb{I}(s_{t}=2)=\l[\begin{array}{c}
\mbf f_{1t}\\
\mbf f_{2t}
\end{array}
\r]\odot\bm\xi_t,\qquad t\in\mathbb Z.
\label{eq:g_t}
\end{equation}
%------------------------------------------------------------------------%
Let $\mathbf{B}_{1}=[\bm\Lambda _{1}\ \mathbf{0}]$ and $\mathbf{B}_{2}=[\mathbf{0}\ \bm\Lambda_{2}]$, where $\mathbf{B}_{1}$ and $\mathbf{B}_{2}$ are $N\times \left(r_{1}+r_{2}\right)$ matrices.
The model in \eqref{eq:model}, \eqref{eq:cov_mat} and \eqref{eq:trans} admits the equivalent state space representation\footnote{Note that $\bm{\xi }_{t}\otimes \mathbf{g}_{t}= [\mbf f_{1t}'~\mbf 0_{}~\mbf f_{2t}'~\mbf 0_{}]'$.}
%------------------------------------------------------------------------%
\begin{align}
\label{eq:state_space_mes}
\mathbf{x}_{t}&=\left( \mathbf{B}_{1}~\mathbf{B}_{2}\right) \left( \bm{%
	\xi }_{t}\otimes \mathbf{g}_{t}\right) +\left( \mathbf{\Sigma }%
_{e1}^{1\left/ 2\right. }~\mathbf{\Sigma }_{e2}^{1\left/ 2\right. }\right)
\left( \bm{\xi }_{t}\otimes \mathbf{I}_{N}\right) \mathbf{e}_{t},\qquad t\in\mathbb Z,\\
\bm{\xi }_{t}&=\mathbf{P}^{\prime }\bm{\xi }_{t-1}+\mathbf{v}_{t}.\nn
\end{align}
Under standard assumptions, the term  $\left( \mathbf{B}_{1}~\mathbf{B}_{2}\right) \left( \bm{%
	\xi }_{t}\otimes \mathbf{g}_{t}\right) $ is identifiable up to a relabelling of the states. This means that the indices of the states can be permuted without changing the law governing the process for $\mathbf{x}_{t}$: on this, see Section 3 in \cite{LEROUX1992127}. Also note that, even for given $\bm{\xi }_{t}$, identification of  $\textbf{B}_{1}$ and $\textbf{B}_{2}$, and therefore of the elements of $\textbf{g}_{t}$, is in general possible only up to an invertible linear transformation (see \citealp{bai03}). 

\subsection{Linear representation}\label{sec:linearFM}
The model in \eqref{eq:state_space_mes} admits the same equivalent linear representation as a model with either one change point or a single threshold effect: see \citet{BCF18}, and \cite{Massacci_2017_JoE}, respectively. It can then be rewritten as the $r_1+r_2$ linear factor model
%------------------------------------------------------------------------%
\begin{equation}
\mathbf{x}_{t}=\mathbf{A}\mathbf{g}_{t}+\mathbf{e}_{t}, \qquad t\in\mathbb Z,
\label{eq:linear_model}
\end{equation}
%------------------------------------------------------------------------%
where $\mathbf{A}=\left[\mathbf{\Lambda}_{1} ~ \mathbf{\Lambda}_{2}\right]$. Therefore, large dimensional factor models with two discrete regimes, be them modelled through a permanent structural change, or through cyclical threshold or Markov switching dynamics, admit the same equivalent linear representation. Then $\mathbf{A}$ and $\mathbf{g}_{t}$  may be estimated by standard Principal Component Analysis (PCA) \citep{stockwatson02JASA,stockwatson02JBES,bai03}. Since PCA gives, as $N,T\to\infty$, consistent estimators of the factors up to premultiplication by an invertible matrix (see \citealp{bai03}), for ease of exposition we first consider estimation of the model in \eqref{eq:state_space_mes} by treating $\mbf g_t$ as known. We then briefly review the implementation of PCA and its effect on the estimation of the model in Section \ref{sec:factor_space}. 

\subsection{Log-likelihood}

Following the approaches by \citet{DGRqml}, \citet{BLqml}, and \citet{baili16}, all developed for QML estimation of linear factor models, we consider a misspecified Gaussian quasi-likelihood of an exact factor model with white noise idiosyncratic components. This implies that the idiosyncratic components are treated as if they were cross-sectionally and serially uncorrelated.  This approach is adopted also by \cite{Urga_Wang_2022_WP} in the case of Markov switching factor models. It is important to stress that we are not assuming that the idiosyncratic components are uncorrelated, as we are just considering likelihood estimation of a misspecified model. Furthermore, in the linear case, \citet{baili16} and \citet{BLqml}, show that such misspecifications are asymptotically negligible as $N,T\to\infty$.

The parameters of interest are then partitioned as
\beq
\bm\varphi = \left[ \mathrm{vec}\left( \mathbf{B}%
_{1}\right) ^{\prime },\mathrm{vec}\left( \mathbf{B}_{2}\right) ^{\prime },%
\text{diag}\left( \mathbf{\Sigma }_{e1}\right) ^{\prime },
\text{diag}\left( \mathbf{\Sigma }_{e2}\right) ^{\prime }
\right] ^{\prime }, \quad \bm\rho = \mathrm{vec}\left( \mathbf{P}\right),\nn
\eeq
so that the vector of parameters of interest, denoted as $\mathbf{q}$, is defined as
%------------------------------------------------------------------------%
\begin{equation}
\mathbf{q}=\left[ \bm\varphi^\prime,\bm\rho^\prime\right] ^{\prime }.\nn
\end{equation}
Notice that we estimate only the diagonal elements of $\bm\Sigma _{e1}$ and $\bm\Sigma _{e2}$ in \eqref{eq:cov_mat}.
%------------------------------------------------------------------------%
Let $\mathbf{X}=\left( \mathbf{x}%
_{1}^{\prime },\ldots ,\mathbf{x}_{T}^{\prime }\right) ^{\prime }$, $\bm{\mc G}=\left( \mathbf{g}_{1}^{\prime },\ldots ,\mathbf{g}_{T}^{\prime }\right)
^{\prime }$, where $\mathbf{X}$  is an $NT\times 1$ vector, $\bm{\mc G}$ is an $(r_1+r_2)T\times 1$ vector. These are $T$-dimensional realizations of the stochastic processes $\{\mbf x_t\}$ and $\{\mbf g_t\}$, respectively. Moreover, let $\bm{X}_{v}$ be the $\sigma$-algebra generated by the random variables $\{\mathbf{x}_{t}\}_{t=1}^v$, for $v=1,\ldots ,T$; in a similar way, define $\bm{G}_{v}$ as the $\sigma$-algebra generated by the random variables $\{\mathbf{g}_{t}\}_{t=1}^v$, for $v=1,\ldots ,T$. And for simplicity we write $\bm X\equiv \bm X_T$ and $\bm G\equiv \bm G_T$. 
		
The likelihood function, denoted by $f\left( \mathbf{X};\mathbf{q}\right)$, 
can be decomposed as%
%------------------------------------------------------------------------%
\begin{equation}
f\left( \mathbf{X};\mathbf{q}\right) =\frac{f\left( \mathbf{X},\bm{\mc G};\mathbf{q}\right) }{f\left( \bm{\mc G} \left\vert \bm{X};\mathbf{q}\right. \right)}
=\frac{f\left(\mathbf{X} \l\vert \bm{G};\mathbf{q}\right. \right) f\l(\bm{\mc G};\mathbf{q}\r)
}
{f\left( \bm{\mc G}\left\vert \bm{X};\mathbf{q}\right. \right)} =\frac{f\left(\mathbf{X} \l\vert \bm{G};\mathbf{q}\right. \right) f\l(\bm{\mc G}\r)
}
{f\left( \bm{\mc G}\left\vert \bm{X};\mathbf{q}\right. \right)}:
\label{eq:loglik1}
\end{equation}
in the last step we account for the fact that $f\l(\bm{\mc G};\mathbf{q}\r)\equiv f\l(\bm{\mc G}\r)$, since it does not depend on the parameters of our model, as we do not specify any dynamic model for the process $\{\mbf g_t\}$.

Furthermore, following \citet[Section 6.2]{krolzig2013markov}, we have
\beq
f\left(\mathbf{X} \l\vert \bm{G};\mathbf{q}\right. \right)=f\left(\mathbf{X} \l\vert \bm{G};\bm\varphi,\bm\rho\right. \right) =\sum_{\{\bm\xi_t\}_{t=1}^T\in \{0,1\}^{T}} f\left(\mathbf{X} \l\vert \bm{G},\{\bm\xi_t\}_{t=1}^T;\bm\varphi\r. \right) \p\l(\{\bm\xi_t\}_{t=1}^T |\bm G,\bm\rho\r).\label{eq:LLgiusta}
\eeq
Here, to avoid heavier notation, we use the same notation $\{\bm\xi_t\}_{t=1}^T$ both for a generic $T$ dimensional realization of the process $\{\bm\xi_t\}$ and for the $\sigma$-algebra generated by the random variables $\{\bm\xi_t\}_{t=1}^T$. 
Notice that the sum is over $2^T$ possible values since, given a realization for $\{\xi_{1t}\}_{t=1}^T$, the realizations of $\{\xi_{2t}\}_{t=1}^T$ are given by $\xi_{2t}=1-\xi_{1t}$ for all $t$. 

Given that we treat the idiosyncratic components as if they were uncorrelated, and using the Markov property of $\{\bm\xi_t\}$, up to omitted constant terms we have
\begin{align}
\log f&\left(\mathbf{X} \l\vert \bm{G},\{\bm\xi_t\}_{t=1}^T;\bm\varphi\r. \right)= \sum_{t=1}^T
\log f\left(\mathbf{x}_t \l\vert \mbf {g}_t, \bm\xi_t ;\bm\varphi\r. \right)\label{eq:LXGXIphi}\\
&\simeq
-\dfrac{1}{2}\sum\limits_{t=1}^{T}\log \det \mathbf{\Sigma} _{et} 
		 -\frac 12\sum_{t=1}^T \left\{ \mathbf{x}_{t}-\left( \mathbf{B}
_{1}~\mathbf{B}_{2}\right) \left( \bm{\xi }_{t} \otimes {\mathbf{g}}_{t}\right) \right\} ^{\prime }\left( \mathbf{\Sigma} _{et}\right) ^{-1}\left\{ \mathbf{x}_{t}-\left( \mathbf{B}_{1}~\mathbf{B}_{2}\right) \left( \bm{\xi }_{t}\otimes {\mathbf{g}}_{t}\right) \r\},\nn
\end{align}
where
$\mathbf{\Sigma} _{et}=\left(\text{diag}(\mathbf{\Sigma} _{e1})~\text{diag}(\mathbf{\Sigma} _{e2})\right) \left( \bm{\xi }%
_{t}\otimes \mathbf{I}_{N}\right)$. Note that in this case the likelihood \eqref{eq:LLgiusta} is not Gaussian; rather, it is a mixture of Gaussian distributions.
Finally, again by the Markov property of $\{\bm\xi_t\}$, we can write
\begin{align}\label{eq:XIrho}
\p\l(\{\bm\xi_t\}_{t=1}^T|\bm G;\bm\rho\r)&=\prod_{t=1}^T\p\l(\bm\xi_t|\bm\xi_{t-1},\bm G;\bm\rho\r) \p\l(\bm\xi_0\r).
\end{align}

\section{Estimation}
\label{section:Estimation}

In this section, we assume that the data generating process is characterised by two regimes as in the model in \eqref{eq:model} and \eqref{eq:cov_mat}. In Section \ref{section:num_regimes} we study the case in which the model is underspecified and the data generating process exhibits a higher number of regimes. We also assume that the dimension of the vector $\mathbf{g}_{t}$ in \eqref{eq:linear_model} is known. Should this not be the case, the dimension of $\mathbf{g}_{t}$ can be determined using information criteria such as those proposed in \cite{baing02}, \cite{ABC10}, and \cite{ahnhorenstein13}, or inferential techniques such as those developed in \cite{onatski10} and \cite{Trapani_2018_JASA}. This issue is discussed also in Section \ref{section:num_factors}. 
%For example, in the empirical analysis in Section \ref{section:Empirics}, we use the criteria developed in \cite{ahnhorenstein13}.

In what follows, Section \ref{section:E&M_steps} defines the steps of the proposed Expectation Maximization (EM) algorithm. Section \ref{section:BLHM} describes the Baum-Lindgren-Hamilton-Kim filter and smoother. Section \ref{sec:factor_space} details the estimator for the factor space. Section \ref{sec:EMparam} discusses the estimator for the parameters. Section \ref{sec:initialization} deals with initialization and convergence of the algorithm.

\subsection{EM algorithm}
\label{section:E&M_steps}

The algorithm outlined in this section is a generalization of the procedure described by \citet[Chapter 5]{krolzig2013markov}. 
The EM algorithm is made of two steps repeated at each iteration $k\ge 0$. The E step involves taking the expected value of  the log-likelihood derived from \eqref{eq:loglik1} conditional on $\bm{X}$ given an estimate of the parameters $\widehat{\mathbf{q}}^{\left(k\right)}$, namely
%------------------------------------------------------------------------%
\begin{equation}
\log f\left( \mathbf{X};\mathbf{q}\right) =\E_{\mathbf{\widehat{q}}^{\left(
		k\right) }}\left[ \log f\left( \mathbf{X}\left\vert \bm{G};\mathbf{q}\right. \right)\l\vert\bm X\r. \right] +\E_{\mathbf{\widehat{q}}^{\left(
		k\right) }}\left[ \log f \left( \bm{\mc G}\right)\l\vert\bm X\r. \right] -\E%
_{\mathbf{\widehat{q}}^{\left( k\right) }}\left[ \log f\left( \bm{\mc G}\left\vert 
\bm{X};\mathbf{q}\right. \right)\l\vert\bm X\r. \right].
\nn
\end{equation}
%------------------------------------------------------------------------%
The M step solves the constrained maximization problem with respect to $\mathbf{q}=\left[ \bm\varphi^\prime,\bm\rho^\prime\right] ^{\prime }$, that is
%------------------------------------------------------------------------%	
\begin{align}
\l({\widehat{\bm\varphi}}^{\left( k+1\right) },{\widehat{\bm\rho}}^{\left( k+1\right) }\r)&=\arg \max_{\bm\varphi,\bm\rho}\E_{\mathbf{\widehat{q}}^{\left(
		k\right) }}\left[ \log f\left( \mathbf{X}\left\vert \bm{G};\bm\varphi,\bm\rho\right. \right)\l\vert\bm X\r. \right]\nn\\
		&\text{s.t.}\quad  \mbf P\bm\iota_2=\bm\iota_2, 
		\label{eq:LLKgiusta}
\end{align}
where  the constraints ensure that probabilities add up to one. In principle, in the M step we should also account for the term $\E_{\mathbf{\widehat{q}}^{\left(	k\right) }}\left[ \log f \left( \bm{\mc G}\right)\l\vert\bm X\r. \right]$, which however in our context does not depend on any parameter.

It is well known that the iteration of these steps produces a series of increasing log-likelihoods. Indeed, $\E_{\mathbf{\widehat{q}}^{\left( k\right) }}\left[ \log f\left( \bm{\mc G}\left\vert \bm{X};\mathbf{q}\right. \right) \l\vert \bm X\r.\right]$ does not contribute to the convergence of the EM algorithm (see \citealp{DLR77}, and \citealp{wu83}). Moreover,  if the maximum is identified and unique, then the EM algorithm will eventually lead to the Maximum Likelihood estimator of $\mbf q$. As shown below,  the solution of the M step can be computed explicitly using the expressions given in \eqref{eq:LXGXIphi} and \eqref{eq:XIrho}. This solution is unique and in closed form. Therefore, no identification issue arises due to multiple maxima, or related to the existence of such maxima.

\subsection{Baum-Lindgren-Hamilton-Kim filter and smoother}
\label{section:BLHM}
 
From \eqref{eq:LXGXIphi} and \eqref{eq:XIrho}, in order to compute the expected likelihood in the E step
we need to compute $\E_{\widehat{\mathbf q}^{(k)}}[\bm\xi_t|\bm X]$, $\E_{\widehat{\mathbf q}^{(k)}}[\bm\xi_t\otimes \mathbf g_t|\bm X]$, and $\E_{\widehat{\mathbf q}^{(k)}}[(\bm\xi_t\otimes \mathbf g_t)(\bm\xi_t\otimes \mathbf g_t)'|\bm X]=\E_{\widehat{\mathbf q}^{(k)}}[(\mbf I_2\otimes \mbf g_t\mbf g_t')|\bm X]$. 

We start by considering the case in which both $\{\mbf g_t\}_{t=1}^T$ is observed and the true value of the parameters $\mbf q$ is known, while we postpone the discussion of the estimation of the factors to Section \ref{sec:factor_space}. Then, for the E step we just need to compute $\E[\bm\xi_t|\bm X]$, since in this case $\bm\xi_t$ and $\mathbf g_t$ are independent for all $t$. This is accomplished by means of a generalization the Baum-Lindgren-Hamilton-Kim filter and smoother explained in detail in Appendix \ref{Appendix:Est:filter}. 
It is an iterative procedure through which we first compute the sequences of conditional one-step-ahead predicted probabilities $\{\bm{\xi }_{t\left\vert t-1\right. }\}_{t=1}^{T}$, such that $\bm{\xi }_{t\left\vert t-1\right .}=\E \left[ \bm{\xi }_{t}\left\vert \bm{X}_{t-1}\right. \right]$,
and filtered probabilities $\{\bm{\xi }_{t\left\vert t\right. }\}_{t=1}^{T}$ such that $\bm\xi_{t|t}=\E[\bm\xi_t|\bm X_t]$. Second, by means of those sequences, we compute the sequence of smoothed probabilities $\{\bm{\xi }_{t\left\vert T\right. }\}_{t=1}^{T}$ such that $\bm\xi_{t|T}=\E[\bm\xi_t|\bm X]$. 

The final recursions for the filtered probabilities are given by (e.g., see \citealp[Chapter 5.1]{krolzig2013markov}, and \citealp{hamilton1989new})
\begin{align}
\bm{\xi }_{t\left\vert t-1\right. }&=\mathbf{P}^{\prime }\bm{\xi }%
_{t-1\left\vert t-1\right. }, \quad t=1,\ldots, T, \notag  \\ 
\bm{\xi }_{t\left\vert t\right. }&=\dfrac{\bm{\eta }_{t}\odot \bm{\xi }_{t\left\vert t-1\right. }}{\bm{\iota }_{2}^{\prime }\left( \bm{\eta 
	}_{t}\odot \bm{\xi }_{t\left\vert t-1\right. }\right) }, \quad t=1,\ldots, T,\label{eq:Hamiltonmain}
\end{align}
where 
\begin{align}
\bm{\eta }_{t}&=\left[ 
\begin{array}{c}
f\left( \mathbf{x}_{t}\left\vert \bm{\xi }_{t}=\left[ 1~0\right]^{\prime },\mathbf{{g}}_{t}\right. \right)  \\ 
f\left( \mathbf{x}_{t}\left\vert \bm{\xi }_{t}=\left[ 0~1\right]^{\prime },\mathbf{{g}}_{t}\right. \right) 
\end{array}\right].\nonumber
\end{align}
The filter can be started by setting either $\bm\xi_{0|0}=\left[ 1~0\right]^{\prime }$, or, equivalently, $\bm\xi_{0|0}=\left[ 0~1\right]^{\prime }$.

The final recursions for the smoothed probabilities are given by (e.g., see \citealp[Chapter 5.2]{krolzig2013markov}, and \citealp{kim1994}) 
\begin{equation}
\bm{\xi }_{t\left\vert T\right. }=\left[ \mathbf{P}\left( \bm{\xi }%
_{t+1\left\vert T\right. }\oslash \bm{\xi }_{t+1\left\vert t\right.
}\right) \right] \odot \bm{\xi }_{t\left\vert t\right. }, \quad t=1,\ldots, T.\label{eq:Kimmain}
\end{equation}
%------------------------------------------------------------------------%
This backward recursion is initiated at $\bm\xi_{T|T}$, which is the last iteration of the filter in \eqref{eq:Hamiltonmain}.

The above description of the Baum-Lindgren-Hamilton-Kim filter and smoother assumes that $\mbf q$ and $\mbf g_t$ are observed. However, in practice both need to be estimated. This is discussed in the next two Sections \ref{sec:factor_space} and \ref{sec:EMparam} below.

\subsection{Estimating the factor space}
\label{sec:factor_space}
In order to estimate the factors $\mbf g_t$, and their dimension $r_1+r_2$, we exploit the fact that the Markov switching factor model in \eqref{eq:model} is observationally equivalent to a linear factor model with $r_1+r_2$ common factors $\mbf g_t$ and factor loadings $\mbf A$: see Section \ref{sec:linearFM} and, in particular, equation \eqref{eq:linear_model}. The number of factors in \eqref{eq:linear_model} can be estimated using methods already available in the literature: for example, see \citet{baing02}, \citet{onatski10}, \citet{ahnhorenstein13}, and \citet{Trapani_2018_JASA}.
The factors  $\mbf g_t$ can be estimated by PCA as follows. First, the estimator $\wh{\mbf A}$ of the loadings matrix $\mbf A$ is obtained as $\sqrt N$ times the normalized eigenvectors corresponding to the $r_1+r_2$ largest eigenvalues of the sample $N\times N$ covariance matrix $T^{-1}\sum_{t=1}^T \mbf x_t\mbf x_t'$. Second, the factors are estimated by linear projection of the data $\mbf x_t$ onto the estimated loadings:
\beq
\wh{\mbf g}_t=\l(\wh{\mbf A}'\wh{\mbf A}\r)^{-1}\wh{\mbf A}'\mbf x_t=\frac{1}{N}\wh{\mbf A}'\mbf x_t, \quad t=1,\ldots, T.\label{eq:ghatPC}
\eeq
This is the same approach followed by \citet{stockwatson02JASA}. It is also the dual approach of the one adopted by \citet{bai03}. Consistency of $\wh{\mbf A}$ and $\wh{\mbf g}_t$ follow from Lemma \ref{Lemma:a_i_hat} and Lemma \ref{Lemma:avg}(a) in Appendix \ref{section:Appendix}, respectively. Note that the steps described in this section do not require knowing the latent state indicator $\bm\xi_t$, and they can be carried out independently. Because of these results, $\bm\xi_t$ and $\wh{\mathbf g}_t$ can also be treated as independent for all $t$. As a consequence, the Baum-Lindgren-Hamilton-Kim filter described in Section \ref{section:BLHM} can be implemented by just replacing the true factors $\mbf g_t$ with their estimator $\wh{\mathbf g}_t$ defined in \eqref{eq:ghatPC}.

\subsection{Estimating the parameters}\label{sec:EMparam}
At each iteration $k\ge 0$ of the EM algorithm, the filtered and smoothed probabilities, given in \eqref{eq:Hamiltonmain} and \eqref{eq:Kimmain}, respectively, and the smoothed cross-probabilities given in \eqref{eq:smooth4},
 are computed using an estimator $\wh{\mbf q}^{(k)}$ of the parameters and an estimator $\wh{\mbf g}_t$ of the factors. Hereafter, we denote as $\bm\xi_{t|t}^{(k)}$, $\bm\xi_{t|T}^{(k)}$, and $\bm\xi_{t,t-1|T}^{(k)}$ such estimators. This defines the E step.

In the M step we have to solve the constrained maximization problem in \eqref{eq:LLKgiusta}. Here we just give the final results, while we refer to Appendix \ref{Appendix:Mstep} for their derivation.
The estimates of the loadings $\mbf B_j$, $j=1,2$, are given by
\begin{equation}
\label{eq:B_j_hatmain}
\mathbf{\wh{B}}_{j}^{(k+1)}=\left( \sum_{t=1}^{T}{\xi }_{j,t\left\vert T\right. }^{(k)}\mathbf{x}_{t}\mathbf{\wh{g}}_{t}^{\prime }\right) \left(
\sum_{t=1}^{T}{\xi }_{j,t\left\vert T\right. }^{(k)}\wh{\mathbf{g}}_{t}\mathbf{\wh{g}}_{t}^{\prime }\right) ^{-1},\quad j=1,2,
\end{equation}
and, consistently with the fact that we use a mis-specified likelihood with uncorrelated idiosyncratic components, we set 
\begin{align}
[\wh{\mathbf{\Sigma }}_{ej}^{(k+1)}]_{ii}&=\l(\dfrac{\sum_{t=1}^{T}
\left( {x}_{it}-\mathbf{\wh{b}}_{ji}^{(k+1)\prime}\mathbf{\wh{g}}_{t}\right) ^2 
}
{
\sum_{t=1}^{T}{\xi }_{j,t\left\vert T\right. }^{(k)}
}\r),\quad  i=1,\ldots, N,\quad j=1,2,\label{eq:S_j_hatmain}\\
[\wh{\mathbf{\Sigma }}_{ej}^{(k+1)}]_{ik}&=0, \quad  i,k=1,\ldots, N,\quad i\ne k,\quad j=1,2,\nn
\end{align}
where $\mathbf{\wh{b}}_{ji}^{(k+1)\prime}$ is the $i$th row of $\mathbf{\wh{B}}_{j}^{(k+1)}$. 
Concerning the estimates of $\bm\rho$, which are subject to the adding up condition,
\beq
\wh{\bm\rho}^{(k+1)}=\l[\sum_{t=1}^T \bm\xi_{t,t-1|T}^{(k)}\r]\oslash \l[ \bm\iota_2\otimes \sum_{t=0}^{T-1}\bm\xi_{t|T}^{(k)}\r].\label{eq:rho_hatmain}
\eeq
By letting $k^*$ be the last iteration of the EM algorithm, we define our final estimator of the parameters as $\wh{\mbf q}\equiv \wh{\mbf q}^{(k^*+1)}$, as given by \eqref{eq:B_j_hatmain}, \eqref{eq:S_j_hatmain}, and \eqref{eq:rho_hatmain}.
The final estimator of $\bm\xi_t$ is defined as $\wh{\bm\xi}_{t|T}\equiv \bm\xi_{t|T}^{(k^*+1)}$, i.e., obtained by running one last time the Baum-Lindgren-Hamilton-Kim filter using the final estimates of the parameters.

\subsection{Initialization and convergence of the EM algorithm \label{sec:initialization}} 
To start the algorithm we need initial estimators $\wh{\mbf q}^{(0)}$ for the parameters.  Specifically, we set $\wh{\mbf B}_1^{(0)}=\wh{\mbf B}_2^{(0)}=\wh{\mbf A}$, as defined in Section \ref{sec:factor_space}. Then, given also $\wh{\mbf g}_t$ as in \eqref{eq:ghatPC}, let $\wh{\mbf e}_t=\mbf x_t- \wh{\mbf A}\wh{\mbf g}_t$, and we set $\wh{\bm\Sigma}_{e1}^{(0)}=\wh{\bm\Sigma}_{e2}^{(0)}=\text{diag}\l(T^{-1}\sum_{t=1}^T \wh{\mbf e}_t\wh{\mbf e}_t'\r)$. Finally, we set 
$$
\wh{\mbf P}^{(0)}= \l(\begin{array}{cc}
0.5+\omega_1 & 1-0.5-\omega_1\\
1-0.5-\omega_2&0.5+\omega_2
\end{array}
\r),
$$
where $\omega_1,\omega_2\in(0,0.5)$ and $\omega_1>\omega_2$. This initialization implicitly identifies state 1 as the most probable one, i.e., it is the state with largest unconditional probability as defined in \eqref{eq:UncondP2}.

We say that the EM algorithm converged at iterations $k^*$, where $k^*$ is the first value of $k$ such that:
\[
\frac{\l\vert\log f\left( \mathbf{X}\left\vert \bm{G};\wh{\bm\varphi}^{(k)},\wh{\bm\rho}^{(k)}\right. \right)-
\log f\left( \mathbf{X}\left\vert \bm{G};\wh{\bm\varphi}^{(k-1)},\wh{\bm\rho}^{(k-1)}\right. \right)
\r\vert}
{\frac 12 \l\{\vert\log f\left( \mathbf{X}\left\vert \bm{G};\wh{\bm\varphi}^{(k)},\wh{\bm\rho}^{(k)}\right. \right)+
\log f\left( \mathbf{X}\left\vert \bm{G};\wh{\bm\varphi}^{(k-1)},\wh{\bm\rho}^{(k-1)}\right. \right)
\r\}} < \epsilon,
\]
for some a priori chosen threshold $\epsilon>0$.

%%%%%%%%%%%%%%%%%%%%%%%%%%%%%%%%%%%%%%%%%%%%%%%%%%%%%%%%%%%%%%%%%%%%%%%%%%

\section{Asymptotic theory}
\label{section:Asymptotics}

In what follows, Section \ref{section:Asymptotics_Assumpions} states the assumptions, whereas Section \ref{section:Asymptotics_Results} presents the asymptotic properties of the estimators.

\subsection{Assumptions}
\label{section:Asymptotics_Assumpions}

For ease of reference, let us write \eqref{eq:model} and \eqref{eq:linear_model} in scalar notation as
%%-----------------------------------------------------------------------
\begin{equation*}
x_{it} =\sum_{j=1}^2\bm\lambda_{ji}^\prime \mbf f_{jt} \mathbb{I}(s_{t}=j)+ e_{it}=\mbf a_i^\prime\mbf g_t+ e_{it}, \quad i=1,\ldots, N, \; t\in\mathbb Z.
\end{equation*}
%%-----------------------------------------------------------------------
We consider the following set of assumptions, which generalizes to our framework the settings in \citet{bai03} and \citet{Massacci_2017_JoE}. 

\begin{assum}\label{assum:F}\textbf{ Factors.}
{\it $\,$ 
\begin{compactenum}[(a)]
\item For $j=1,2$, and all $t\in\mathbb Z$, $\E[\mathbf f_{jt}]=\mbf 0$ and $\E[\left\Vert \mathbf{f}_{jt}\right\Vert ^{4}]<\infty $.
\item For $j,k=1,2$, as $T\to\infty$, $T^{-1}\sum\nolimits_{t=1}^{T}\mathbb{I}\left( s_{t}=j\right) h_{kt} \mathbf{f}_{jt}\mathbf{f}_{jt}^{\prime }\overset{p}{\rightarrow }\mathbf{\Sigma }^{\left(k\right)}_{\mathbf{f}j}$, where $\mathbf{\Sigma }_{\mathbf{f}j}^{(k)}$ is $r_j\times r_j$ positive definite, and $\left\{h_{kt}\right\} _{t=1}^{T}$ is any sequence such that
\begin{inparaenum}[(i)]
\item $\p \left[0\leq h_{kt}\leq 1\right] =1$ and
\item $T^{-1}\sum\nolimits_{t=1}^{T}h_{kt} \overset{p}{\rightarrow }\bar{h}_{k}>0$.
\end{inparaenum}
\end{compactenum}}
\end{assum}

Assumption \ref{assum:F} restricts the factor processes $\left\{\mathbf{f}_{jt}\right\}$, for $j=1,2$, so that appropriate moments exist. The sequence $\{h_{kt}\}_{t=1}^T$ can be random or deterministic, and it is introduced to account for the fact that we estimate the \textit{expected} value of $\xi_{jt}$, and not its actual value. Assumption \ref{assum:F} implies that $0<\p\left[s_t=j\right]<1$, for $j=1,2$, thus ruling out the possibility that any of the states is absorbing, as discussed in Section \ref{section:MS_model}. It also implies that for $j=1,2$, as $T\rightarrow \infty $, 
\beq\label{eq:SFj}
\frac 1T\sum\limits_{t=1}^{T}\mathbb{I} \left( s_{t}=j\right) \mathbf{f}_{jt}\mathbf{f}_{jt}^{\prime }\overset{p}{\rightarrow }\mathbf{\Sigma }_{\mathbf{f}j},
\eeq 
where  $\mathbf{\Sigma }_{\mathbf{f}j}$ is positive definite and
\beq\label{eq:SG}
\frac 1T\sum_{t=1}^T\mbf g_t\mbf g_t^\prime\stackrel{p}{\to}\bm\Sigma_{\mbf g}=
\left( 
\begin{array}{cc}
\mathbf{\Sigma }_{\mathbf{f}_{1}} & \mathbf{0} \\ 
\mathbf{0} & \mathbf{\Sigma }_{\mathbf{f}_{2}}%
\end{array}
\right).
\eeq
In particular, note that \eqref{eq:SFj} allows the covariance matrix of $\mathbf{f}_j$ to be state-dependent, as advocated in \cite{Massacci_2023_JFEcon}. It is also easy to see that if $j\ne k$, then for all $T\in\mathbb N$
\beq\label{eq:SFjk}
\frac 1T\sum\limits_{t=1}^{T}\mathbb{I} \left( s_{t}=j\right) \mathbf{f}_{jt}\mathbf{f}_{kt}^{\prime }\mathbb{I} \left( s_{t}=k\right)=\mathbf{0}.
\eeq 

\begin{assum}\label{assum:FL}\textbf{Loadings.}
{\it $\,$
\begin{compactenum}[(a)]
\item For $j=1,2$, all $i=1,\ldots ,N$, and all $N\in\mathbb N$, $\left\Vert \bm{\lambda }_{ji}\right\Vert \leq \bar{\lambda}<\infty$, where $\bar{\lambda}$ is independent of $j$, $i$, and $N$.
\item For $j=1,2$, as $N\to\infty$, $ N^{-1}\mathbf{\Lambda }_{j}^{\prime }\mathbf{\Lambda }_{j}  \rightarrow \mathbf{\Sigma }_{\mathbf{\Lambda }_{j}}$, where $\mathbf{\Sigma }_{\mathbf{\Lambda }_{j}}$
is $r_{j}\times r_{j}$ positive definite.
\item  As $N\to\infty$,
$N^{-1} \mathbf{\Lambda }_{1}^{\prime }\mathbf{\Lambda }_{2}  \rightarrow \mathbf{\Sigma }_{\mathbf{\Lambda }_{12}}$, where $\mathbf{\Sigma }_{\mathbf{\Lambda }_{12}}$ is $r_{1} \times r_{2}$. 
\item For any $r_2 \times r_2$ full rank matrix $\mathbf{L}$, $\mathbf{\Lambda }_{1} \neq \mathbf{\Lambda }_{2}\mathbf{L}$.
\end{compactenum}
}
\end{assum}

According to Assumption \ref{assum:FL}, loadings are nonstochastic and factors have a nonnegligible effect on the variance of $\{\mathbf{x}_{t}\}$ within each regime. In particular, part (b) implies that at least one common factor is present within each regime. The condition in part (d) ensures that the regimes are identified and it is analogous to the alternative hypothesis in the test for change in loadings developed in \cite{Pelger_Xiong_2022_JBES}. This condition is trivially satisfied if $r_1 \neq r_2$, since the number of factors changes between regimes; if  instead $r_1 = r_2$, then part (d) rules out the possibility that the columns of $\mathbf{\Lambda }_{1}$ are a linear combination of the columns of $\mathbf{\Lambda }_{2}$, in which case the regimes cannot be separately identified. From Assumption \ref{assum:FL} it also follows that, as $N\to\infty$,
\beq\label{eq:SA}
\frac {\mbf A^\prime\mbf A}N\to \mathbf{\Sigma }_{\mathbf{A}}=\left( 
\begin{array}{cc}
\mathbf{\Sigma }_{\mathbf{\Lambda }_{1}} & \mathbf{\Sigma }_{\mathbf{\Lambda 
	}_{12}} \\ 
\mathbf{\Sigma }_{\mathbf{\Lambda }_{12}}^{\prime } & \mathbf{\Sigma }_{%
	\mathbf{\Lambda }_{2}}%
\end{array}%
\right),
\eeq
and
\begin{align}\label{eq:SB1B2}
&\frac {\mbf B_1^\prime\mbf B_1}N\to \mathbf{\Sigma }_{\mathbf{B}_1}=\left( 
\begin{array}{cc}
\mathbf{\Sigma }_{\mathbf{\Lambda }_{1}} & \mbf 0 \\ 
\mbf 0 & \mbf 0
\end{array}%
\right),\quad \frac {\mbf B_2^\prime\mbf B_2}N\to \mathbf{\Sigma }_{\mathbf{B}_2}=\left( 
\begin{array}{cc}
\mbf 0& \mbf 0 \\ 
\mbf 0 & \mathbf{\Sigma }_{\mathbf{\Lambda }_{2}} 
\end{array}%
\right),\quad \frac {\mbf B_j^\prime\mbf B_k}N\to \mathbf{0},\;\text{ if } j\ne k.
\end{align}

\begin{assum}\label{assum:TCSDH}\textbf{Idiosyncratic component.} {\it $\,$
\begin{compactenum}
	\item[(a)] For all $i=1,\ldots, N$, all $t\in\mathbb Z$, and all $N\in\mathbb N$, $\E\left[e_{it}\right]=0$ and $\E[e_{it}^8]\le M<\infty$, where $M$ is independent of $i$, $t$, and $N$.	
	\item[(b)] For $j,k=1,2$, for all $t\in\mathbb Z$, and $N\in\mathbb N$, 
	$$
	\frac 1N\sum_{i,l=1}^N 
	\l\vert\E[ \mathbb{I}\left(s_{t}=j\right) h_{kt} e_{it}e_{lt}]\r\vert\le M<\infty,
	$$ 
	where $\left\{h_{kt}\right\} _{t=1}^{T}$ is as in Assumption \ref{assum:F}(b), and $M$ is independent of $t$ and $N$.		
	\item[(c)] For $j,k=1,2$, all $i,l=1,\ldots, N$, all $N\in\mathbb N$, and all $T\in\mathbb N$,
	$$
	\E\left[ \left\vert \frac 1{\sqrt T}
	\sum\limits_{t=1}^{T}\left\{\mathbb{I}\left( s_{t}=j\right)h_{kt} e_{it}e_{lt}-\E\left[ \mathbb{I}\left( s_{t}=j\right)h_{kt} e_{it}e_{lt}%
	\right] \right\} \right\vert ^{4}\right] \leq M<\infty,
	$$ 
	where $\left\{h_{kt}\right\} _{t=1}^{T}$ is as in Assumption \ref{assum:F}(b), and $M$ is independent of $j$, $i$, $l$, $N$, and $T$.	
\end{compactenum}
}
\end{assum}

Part (b) of Assumption \ref{assum:TCSDH} controls the amount of cross-sectional correlation we can allow for. It implies the usual assumption for approximate factor models of nondiagonal idiosyncratic covariances $\bm\Sigma_{ej}$, $j=1,2$. Note that the sequence $\left\{h_{kt}\right\} _{t=1}^{T}$ has the same role as in Assumption \ref{assum:F}, which we refer to for further comments. Part (b) of Assumption \ref{assum:TCSDH} also implies
\begin{equation*}
\E\l[\l\vert \frac 1{\sqrt N}\sum_{i=1}^N \mathbb I(s_t=j)e_{it}\r\vert^2\r]\le M<\infty,
\end{equation*}
and hence $N^{-1/2}\Vert \mathbb I(s_t=j) \mbf e_t\Vert=O_p(1)$ for $j=1,2$, and for all $t\in\mathbb Z$. Part (c) of Assumption \ref{assum:TCSDH} limits time dependence, and it is guaranteed together with part (a) if we assume finite 8th order cumulants for the bivariate process $\{(e_{it},e_{lt})\}$. Notice that the constant $M$ in the three parts of the assumption does not have to be the same one.
 
\begin{assum}\label{assum:WD}\textbf{Weak dependence between common and idiosyncratic components.}
{\it  
For $j=1,2$, and all $N\in\mathbb N$, and all $T\in\mathbb N$,
$$
\E\left[ \frac 1N\sum\limits_{i=1}^{N}\left\Vert \frac 1{\sqrt T}\sum\limits_{t=1}^{T}\mathbb{I}\left( s_{t}=j\right)h_{kt} 
\mathbf{f}_{jt}e_{it}\right\Vert ^{2}\right]  \leq M<\infty,
$$
where $\left\{h_{kt}\right\} _{t=1}^{T}$ is as in Assumption \ref{assum:F}(b), and $M$ is independent of $N\in\mathbb N$ and $T\in\mathbb N$.
}
\end{assum}

Assumption \ref{assum:WD} limits the degree of dependence between factors, state variable $s_t$, and idiosyncratic components.

\begin{assum}\label{assum:eigenvalues}\textbf{ Eigenvalues.}
{\it The eigenvalues of the $\left(r_1 + r_2\right)\times \left(r_1 + r_2\right)$ matrix 
$\mathbf{\Sigma }_{\mathbf{A}}\mathbf{\Sigma }_{\mathbf{g}} $ are distinct, where $\bm\Sigma_{\mathbf{A}}$ is defined in \eqref{eq:SA} and $\bm\Sigma_{\mbf g}$ is defined in \eqref{eq:SG}.
 }
\end{assum}

Assumption \ref{assum:eigenvalues} guarantees a unique limit for $N^{-1}\mathbf{A}^{\prime }\mathbf{\widehat{A}}$, as stated in Lemma \ref{Lemma:var_hat} in Appendix \ref{section:Appendix}. By assuming distinct eigenvalues, we can  uniquely identify the space spanned by the eigenvectors, which are linear combinations of the columns of $\mbf A$. Notice that $\bm\Sigma_{\mbf g}$ is block diagonal because of \eqref{eq:SFjk}.

Assumptions \ref{assum:F} to \ref{assum:eigenvalues} are sufficient to prove the consistency of the estimators we propose. In order to derive their asymptotic distributions, we further introduce the following Assumptions \eqref{assum:MCLT} and \eqref{assum:rates}.

\begin{assum}\label{assum:MCLT}\textbf{Moments and Central Limit Theorems.} {\it 
$\,$
\begin{compactenum}
	\item[(a)] For $j=1,2$, all $i=1,\ldots,N$, all $N\in\mathbb N$ and all $T\in\mathbb N$,
	\begin{equation*}
	\E\l[\left\Vert \dfrac{1}{\sqrt{NT}}\sum\limits_{l=1}^{N}\sum%
	\limits_{t=1}^{T}\mathbf{a}_{l}\left\{ \mathbb{I}\left( s_{t}=j\right)e_{it}e_{lt}-%
	\E\left[ \mathbb{I}\left( s_{t}=j\right) e_{it}e_{lt}\right]
	\right\} \right\Vert ^{2}\r]\leq M<\infty,
	\end{equation*}
	where $M$ is independent of $j$, $i$, $N$, and $T$. 
	
	\item[(b)] For $j,k=1,2$, all $N\in\mathbb N$ and all $T\in\mathbb N$,
	\begin{equation*}
	\E\l[\left\Vert \dfrac{1}{\sqrt{NT}}\sum\limits_{i=1}^{N}\sum%
	\limits_{t=1}^{T}\mathbb{I}\left( s_{t}=j\right) \bm{\lambda }_{ki}%
	\mathbf{f}_{jt}^{\prime }e_{it}\right\Vert ^{2}\r]\leq M<\infty,
	\end{equation*}
	where $M$ is independent of $j$, $k$, $N$, and $T$. 
	
	\item[(c)] For $j,k=1,2$, all $i=1,\ldots, N$ and all $N\in\mathbb N$, as $T\to\infty$,
	\begin{equation*}
	\dfrac{1}{\sqrt{T}}\sum\limits_{t=1}^{T}\mathbb{I}\left( s_{t}=j\right)
	h_{kt}\mathbf{f}_{jt}e_{it}\overset{d}{\rightarrow }\mathcal{N}\left( \mathbf{%
		0},\mathbf{\Gamma }_{jki}\right), 
	\end{equation*}%
	where  $\left\{ h_{kt}\right\} _{t=1}^{T}$ is defined in Assumption \ref{assum:F}, and  
	$$
	\mathbf{\Gamma }%
	_{jki}=\lim_{T\rightarrow \infty }\frac 1 T
	\sum\limits_{t=1}^{T}\sum\limits_{v=1}^{T}\mathbb{I}\left(
	s_{t}=j\right) \mathbb{I}\left( s_{v}=j\right) h_{kt}h_{kv}\E[\mathbf{f}_{jt}%
	\mathbf{f}_{jv}^{\prime }e_{it}e_{iv}].
	$$
	
	\item[(d)] For all $t\in\mathbb Z$, as $N\to\infty$, 
	\begin{equation*}
	\dfrac{1}{\sqrt{N}}\sum_{i=1}^{N}
	\left[
	\begin{array}{c}
	\bm{\lambda }_{1i}\\
	\bm{\lambda }_{2i}\\
	\end{array}
	\right]
	e_{it}\overset{d}{\rightarrow } 
	\mathcal{N}\left( \mathbf{0},\l(
	\begin{array}{cc}
	\mathbf{\Phi }_{1t}&\mathbf{\Phi }_{12t}\\
	\mathbf{\Phi }_{12t}^\prime&\mathbf{\Phi }_{2t}
	\end{array}
	\r)	
	\right) ,
	\end{equation*}%
	where for $j,k=1,2$
	$$
	\bm\Phi_{jkt}=\lim_{N\to\infty} \frac 1N\sum_{i=1}^N\sum_{l=1}^N \bm\lambda_{ji}\bm\lambda_{kl}^\prime\E[e_{it}e_{lt}],
	$$ 
	and $\bm \Phi_{jt}=\bm \Phi_{jjt}$.
\end{compactenum}
}
\end{assum}

Parts (a) and (b) of Assumption \ref{assum:MCLT} are suitable moment bounds, whereas parts (c) and (d) are central limit theorems.

\begin{assum}
	\label{assum:rates}\textbf{Rates.} {\it As $N,T \rightarrow \infty$, $\sqrt{T}/N \rightarrow 0$ and $\sqrt{N}/T \rightarrow 0$.}
\end{assum}

Assumption \ref{assum:rates} imposes standard restrictions on the convergence rates.

Define the $\left(r_{1}+r_{2}\right) \times \left(r_{1}+r_{2}\right)$ matrix $\mathbf{\widehat{H}}$ as 
%------------------------------------------------------------------------%
\begin{equation}
\label{eq:hat_H}
\mathbf{\widehat{H}}=\dfrac{\mathbf{GG}^{\prime }}{T}\dfrac{\mathbf{A}%
	^{\prime }\mathbf{\widehat{A}}}{N}\mathbf{\widehat{V}} ^{-1},
\end{equation}%
%------------------------------------------------------------------------%
where $\mathbf{G}=\left( \mathbf{g}_{1},\ldots ,\mathbf{g}_{T}\right)$ and $\mathbf{\widehat{V}}$ is the $\left(r_{1}+r_{2}\right) \times \left(r_{1}+r_{2}\right)$ diagonal matrix containing the first $r_{1}+r_{2}$ eigenvalues of $\mathbf{\widehat{\Sigma}}_{\mathbf{x}}=\left( NT\right)
^{-1}\sum\nolimits_{t=1}^{T}\mathbf{x}_{t}\mathbf{x}_{t}^{\prime }$ sorted in decreasing order. In Lemma \ref{Lemma:var_hat} we prove that
%------------------------------------------------------------------------%
\beq\label{eq:QQQ}
p\lim_{N,T\rightarrow \infty }\frac{\mathbf{A}^{\prime }\mathbf{\widehat{A}}}N =\mbf Q, \text{ with } \mbf Q=\mathbf{\Sigma }_{\mathbf{g}}^{-1\left/ 2\right. }\mathbf{\Psi V}^{1\left/ 2\right. },
\eeq
%------------------------------------------------------------------------%
where  $\mathbf{V}$ is the $\left(r_1+r_2\right)\times\left(r_1+r_2\right)$ diagonal matrix of the first $\left(r_1+r_2\right)$ eigenvalues of $\mathbf{\Sigma }_{\mathbf{g}}^{1\left/ 2\right. }\mathbf{\Sigma }_{\mathbf{A}}\mathbf{\Sigma }_{\mathbf{g}}^{1\left/ 2\right. }$ in decreasing order, and $\mathbf{\Psi}$ is the corresponding matrix of eigenvectors such that $\mathbf{\Psi}^{\prime}\mathbf{\Psi}=\mathbf{I}_{r_1+r_2}$. Likewise define $\mathbf{Q}_{j}=p\lim_{N,T\rightarrow \infty }N^{-1}\mathbf{\Lambda}_{j}^{\prime }\mathbf{\widehat{A}}$, for $j=1,2$, which is an $r_j\times(r_1+r_2)$ matrix such that $\mathbf{Q}=\left[\mathbf{Q}_{1}^{\prime}\ \mathbf{Q}_{2}^{\prime}\right]^{\prime}$. Thus, by Lemma \ref{Lemma:Q_j} we have
%------------------------------------------------------------------------%
\beq\label{eq:Qj}
\mbf{Q}_{j}=\mathbf{\Sigma }_{\mathbf{f}j}^{-1\left/ 2\right. }\mathbf{%
	\Psi }_{j}\mathbf{V}^{1\left/ 2\right. }, \quad j=1,2,
\eeq
%------------------------------------------------------------------------%
where $\mathbf{\Psi}_{j}$ is the $r_j\times\left(r_1+r_2\right)$ matrix such that $\mathbf{\Psi}=\left[\mathbf{\Psi}_{1}^{\prime}\ \mathbf{\Psi}_{2}^{\prime}\right]^{\prime}$. Therefore, because of \eqref{eq:SG}, \eqref{eq:QQQ}, and by Lemma \ref{Lemma:Hat_V_NT} according to which $\wh{\mbf V}\overset{p}{\rightarrow } \mbf V$,
\beq\label{eq:Hlim}
{p}\lim\limits_{N,T\rightarrow \infty } \wh{\mbf H}=\mbf H, \text{ with }\mbf H=\bm\Sigma_{\mbf g}\mbf Q \mbf V^{-1}.
\eeq

\subsection{Asymptotic results}
\label{section:Asymptotics_Results}

For $j=1,2$, let $\mathbf{\wh{B}}_{j}=\mathbf{\wh{B}}_{j}^{(k^{*}+1)}$, where $k^{*}$ is the last iteration of the EM algorithm as defined in Section \ref{sec:EMparam}.  
For given $j=1,2$ and $i=1,\ldots,N$, let $\mathbf{\widehat{b}}_{ji}$ be the estimator for $\mathbf{b}_{ji}$ such that 
$\mathbf{\wh{B}}_{j}=[\mathbf{\widehat{b}}_{j1},\ldots,\mathbf{\widehat{b}}_{jN}]^{\prime}$ and $\mathbf{B}_{j}=[\mathbf{b}_{j1},\ldots,\mathbf{b}_{jN}]^{\prime}$. The following theorem states the asymptotic distribution of $\mathbf{\widehat{b}}_{ji}$.
%------------------------------------------------------------------------%
\begin{thm}
\label{th:asympt_dist}
{\it 
Let Assumptions \ref{assum:F} - \ref{assum:rates} hold. Then, for $k_{1},k_{2}=1,2$ with $k_{1} \neq k_{2}$, for any given $i=1,\ldots,N$, as $N,T\to\infty$,
%------------------------------------------------------------------------%
\begin{equation*}
%\begin{array}{cccc}
\sqrt{T}\left[\mathbf{\widehat{b}}_{k_{1}i}-\mathbf{\widehat{I}}^{\prime}_{\mathbf{\widehat{\xi}}k_{1}}
\mathbf{\widehat{H}}^{\prime }\mathbf{b}_{k_{1}i}-\left( \mathbf{I}_{r_1+r_2}-\mathbf{\widehat{I}}_{\mathbf{\widehat{\xi}}k_{1}}\right)^{\prime} \mathbf{\widehat{H}}^{\prime }\mathbf{b}_{k_{2}i}\right]\overset{d}{\rightarrow }\mathcal{N}\left( \mathbf{0},\mathbf{\Sigma }_{\widehat{\mathbf{b}}k_{1}i}\right),
%& k_{1},k_{2}=1,2,%
%& k_{1} \neq k_{2},%
%& i=1,\ldots,N,
%\end{array}
\end{equation*}
%------------------------------------------------------------------------%
where the $\left(r_1+r_2\right)\times\left(r_1+r_2\right)$ matrix $\mathbf{\widehat{I}}_{\mathbf{\widehat{\xi}}k_{1}}$ is defined as
%------------------------------------------------------------------------%
\begin{equation}
\label{eq:hat_I_xi}
\mathbf{\widehat{I}}_{\mathbf{\widehat{\xi}}k_{1}}=\left( \sum\limits_{t=1}^{T}\widehat{\xi}%
_{k_{1},t\left\vert T\right. }\mathbb{I}(s_t=k_1)\mathbf{\widehat{g}}_{t}\mathbf{\widehat{g}}%
_{t}^{\prime }\right) \left( \sum\limits_{t=1}^{T}\widehat{\xi}_{k_{1},t\left\vert
	T\right. }\mathbf{\widehat{g}}_{t}\mathbf{\widehat{g}}_{t}^{\prime }\right) ^{-1},
\end{equation}
%------------------------------------------------------------------------%
and where
%------------------------------------------------------------------------%
\begin{equation*}
\mathbf{\Sigma }_{\widehat{\mathbf{b}}k_{1}i}%
=\left(
\mathbf{Q}_{1}^{\prime}\mathbf{\Sigma}_{\mathbf{f}1}^{(k_{1})}\mathbf{Q}_{1}+
\mathbf{Q}_{2}^{\prime}\mathbf{\Sigma}_{\mathbf{f}2}^{(k_{1})}\mathbf{Q}_{2}
\right)^{-1}
\left(
\mathbf{Q}_{1}^{\prime}\mathbf{\Gamma}_{1k_{1}i}\mathbf{Q}_{1}+
\mathbf{Q}_{2}^{\prime}\mathbf{\Gamma}_{2k_{1}i}\mathbf{Q}_{2}
\right)
\left(
\mathbf{Q}_{1}^{\prime}\mathbf{\Sigma}_{\mathbf{f}1}^{(k_{1})}\mathbf{Q}_{1}+
\mathbf{Q}_{2}^{\prime}\mathbf{\Sigma}_{\mathbf{f}2}^{(k_{1})}\mathbf{Q}_{2}
\right)^{-1},
\end{equation*}
%------------------------------------------------------------------------%
with $\mbf Q_j$, $\mathbf{\Gamma}_{jk_{1}i}$, and $\mathbf{\Sigma}_{\mathbf{f}j}^{(k_{1})}$,
$j=1,2$, defined in \eqref{eq:Qj}, Assumption \ref{assum:MCLT}(c), 
and Assumption \ref{assum:F} when $h_{k_1}=\widehat{\xi}_{k_{1},t\left\vert T\right. }$, respectively.}
\end{thm}

Theorem \ref{th:asympt_dist} shows that the estimator $\mathbf{\widehat{b}}_{k_{1}i}$ for $\mathbf{b}_{k_{1}i}$ is subject to \textit{two sources of bias}. The first is standard and it is induced by the usual indeterminacy due to the latency of both factors and loadings, and it is captured by the invertible matrix $\mathbf{\widehat{H}}$ defined in (\ref{eq:hat_H}) (see \citealp{bai03}). If we assume $T^{-1}\sum_{t=1}^T \mbf g_t\mbf g_t^\prime=\mbf I_{r_1+r_2}$, then $\wh{\mbf H}$ becomes a rotation, namely an orthogonal matrix. However, additional restrictions on the loadings are necessary to reduce $\wh{\mbf H}$ to the identity: for a discussion on identification of factors see \textit{inter alia} \cite{baing13}. The second source of bias is induced by $\mathbf{\widehat{I}}_{\mathbf{\widehat{\xi}}k_{1}}$ defined in \eqref{eq:hat_I_xi}, which depends on the probability of the state being asymptotically correctly estimated. 
If the unconditional probability of being in state $k_1$ were correctly estimated with probability one, that is, if $\wh{\xi}_{k_1,t|T}\stackrel{p}{\to} \mathbb I(s_t=k_1)$, as $N,T\to\infty$, then $\mathbf{\widehat{I}}_{\mathbf{\widehat{\xi}}k_{1}}\stackrel{p}{\to}\mathbf{I}_{r_1+r_2}$ and $\mathbf{\widehat{b}}_{k_{1}i}$ would consistently estimate a linear transformation of $\mathbf{b}_{k_{1}i}$.

Therefore, $\mathbf{\widehat{b}}_{k_{1}i}$ estimates a linear transformations of $\mathbf{b}_{k_{1}i}$ and $\mathbf{b}_{k_{2}i}$, with weights determined by $\mathbf{\widehat{I}}_{\mathbf{\widehat{\xi}}k_{1}}$ and $( \mathbf{I}_{r_1+r_2}-\mathbf{\widehat{I}}_{\mathbf{\widehat{\xi}}k_{1}})$, respectively. This second source of bias is due to the fact that the process $s_t$ is latent, and it is specific to Markov switching models. As such, it does not affect threshold or structural break models, in which the state is identified with probability one.

Theorem \ref{th:asympt_dist} has implications for the estimation of the regime specific loadings $\bm\Lambda_j$, $j=1,2$. To see this, let $\wh{\mbf R}_{k}= \wh{\mbf H} \wh{\mbf I}_{\wh\xi k}$, for $k=1,2$, and consider the partition
%------------------------------------------------------------------------%
\begin{equation}
	\label{eq_R_k_H}
	\begin{array}{ccc}
		\wh{\mbf R}_{k}=
		\l[\begin{array}{cc}
			\wh{\mbf R}_{ k,11} &\wh{\mbf R}_{ k, 12}\\
			\wh{\mbf R}_{ k, 21}&\wh{\mbf R}_{ k, 22}
		\end{array}
		\r], 
		& &
		\wh{\mbf H} =
		\l[\begin{array}{cc}
			\wh{\mbf H}_{ 11} &\wh{\mbf H}_{ 12}\\
			\wh{\mbf H}_{ 21}&\wh{\mbf H}_{  22}
		\end{array}
		\r],
	\end{array}
\end{equation}
%------------------------------------------------------------------------%
where $\wh{\mbf R}_{ k, j\ell}$, $k,j,\ell=1,2$ and $\wh{\mbf H}_{ j\ell}$, $j,\ell=1,2$,  are $r_j\times r_\ell$. Then, from Theorem \ref{th:asympt_dist}, for any given $i=1,\ldots,N$, as $N,T\to\infty$, we obtain
%------------------------------------------------------------------------%
\begin{align}
\label{eq:b_1i_hat}
&\sqrt T\l\{\wh{\mbf b}_{1i}^\prime- [\bm\lambda_{1i}^\prime~\mbf 0]\,\wh{\mbf R}_{1} -
[\mbf 0~\bm\lambda_{2i}^\prime]\, \l(\wh{\mbf H}-\wh{\mbf R}_1\r)\r\}\nn\\
&=\sqrt T\l\{\wh{\mbf b}_{1i}^\prime-\bm\lambda_{1i}^\prime[\wh{\mbf R}_{ 1,11}\, \wh{\mbf R}_{ 1,12}]-\bm\lambda_{2i}^\prime
\l[ \l(\wh{\mbf H}_{21}-\wh{\mbf R}_{1,21}\r)\, \l(\wh{\mbf H}_{22}-\wh{\mbf R}_{1,22}\r)
\r]\r\}\stackrel{d}{\to}\mathcal N\l(\mbf 0,\mathbf{\Sigma }_{\widehat{\mathbf{b}}{1}i}\r),
\end{align}
%------------------------------------------------------------------------%
and
%------------------------------------------------------------------------%
\begin{align}
\label{eq:b_2i_hat}
&\sqrt T\l\{\wh{\mbf b}_{2i}^\prime- [\mbf 0~\bm\lambda_{2i}^\prime]\,\wh{\mbf R}_{2} -
[\bm\lambda_{1i}^\prime~\mbf 0]\, \l(\wh{\mbf H}-\wh{\mbf R}_2\r)\r\}\nn\\
&=\sqrt T\l\{\wh{\mbf b}_{2i}^\prime-\bm\lambda_{2i}^\prime[\wh{\mbf R}_{ 2,21}\, \wh{\mbf R}_{ 2,22}]-\bm\lambda_{1i}^\prime
\l[ \l(\wh{\mbf H}_{11}-\wh{\mbf R}_{2,11}\r)\, \l(\wh{\mbf H}_{12}-\wh{\mbf R}_{2,12}\r)
\r]\r\}\stackrel{d}{\to}\mathcal N\l(\mbf 0,\mathbf{\Sigma }_{\widehat{\mathbf{b}}{2}i}\r).
\end{align}
%------------------------------------------------------------------------%
This means that $r_1+r_2$ columns of $\wh{\mbf B}_j$, $j=1,2$,  estimate two different linear transformations of the columns of $[\bm\Lambda_1 \,\bm\Lambda_2]$. 
We can distinguish two cases.
%From (\ref{eq:b_1i_hat}) and (\ref{eq:b_2i_hat}), 
%if we consistently estimate the unconditional probability of being in a given state $j=1,2$, then $\wh{\mbf R}_k\stackrel{p}{\to} \wh{\mbf H}$ as $N,T\to\infty$. \textcolor{red}{LA FRASE PRECEDENTE NON HA SENSO, CHE C'ENTRANO (\ref{eq:b_1i_hat}) and (\ref{eq:b_2i_hat})?}
%
On the one hand, if $r_1=r_2=r$, as assumed for example in \citet{Liu_Chen_2016_SS},  
%all submatrices $\wh{\mbf R}_{ k, j\ell}$ and $\wh{\mbf H}_{ k, j\ell}$ are invertible and 
there is no need to know the true values of $r_1$ and $r_2$ to get consistent estimates of the space spanned by the true loadings in the two different regimes. Indeed, in this case $\mbf B_1$ and $\mbf B_2$ have an even number of columns, equal to $2r$, and from the first line of (\ref{eq:b_1i_hat}) and (\ref{eq:b_2i_hat}) we see that we can consider the first half of the columns of either $\wh{\mbf B}_1$ or $\wh{\mbf B}_2$ as an estimator of a linear transformation of $\bm\Lambda_1$ and the second half of the columns of either $\wh{\mbf B}_1$ or $\wh{\mbf B}_2$ as an estimator of a linear transformation of $\bm\Lambda_2$. Hence, we can define the following estimators of the loadings:
\begin{align}\label{eq:Ljhat}
\wh{\bm\lambda}_{1i} = \wh{\mbf b}_{1i,1:r},\;\; \wh{\bm\lambda}_{2i} = \wh{\mbf b}_{2i,r+1:2r}, \quad i=1,\ldots,N,
\end{align}
or
\begin{align}\label{eq:Ljtilde}
\wt{\bm\lambda}_{1i} = \wh{\mbf b}_{2i,1:r},\;\; \wt{\bm\lambda}_{2i} = \wh{\mbf b}_{1i,r+1:2r} \quad i=1,\ldots,N,
\end{align}
where $\wh{\mbf b}_{ji,1:r}$ denotes the first $r$ elements of $\wh{\mbf b}_{ji}$, and $\wh{\mbf b}_{ji,r+1:2r}$ denotes the second $r$ elements of $\wh{\mbf b}_{ji}$, for $j=1,2$ and $i=1,\ldots,N$. The property of these estimators are formalized in the following corollary, which is a direct consequence of Theorem \ref{th:asympt_dist}, and of (\ref{eq:b_1i_hat}) and (\ref{eq:b_2i_hat}).

\begin{cor}
\label{cor:loadings}
{\it 
Let Assumptions \ref{assum:F} - \ref{assum:rates} hold and assume $r_1=r_2=r$. Then, for any given $i=1,\ldots,N$, as $N,T\to\infty$,
\begin{align}
\sqrt{T}\left[ \widehat{\bm{\lambda }}_{1i}^{\prime }-\bm{\lambda }_{1i}^{\prime }\widehat{\mathbf{R}}_{1,11}-\bm{\lambda }_{2i}^{\prime
}\left( \widehat{\mathbf{H}}_{21}-\widehat{\mathbf{R}}_{1,21}\right) \right] 
\overset{d}{\rightarrow }\mathcal{N}\left( \mathbf{0},\mathbf{\Sigma }_{\widehat{\bm{\lambda }}1i}\right),\nn\\
%\end{equation*}
%and
%\begin{equation*}
\sqrt{T}\left[ \widehat{\bm{\lambda }}_{2i}^{\prime }-\bm{\lambda }_{2i}^{\prime }\widehat{\mathbf{R}}_{2,22}-\bm{\lambda }_{1i}^{\prime
}\left( \widehat{\mathbf{H}}_{12}-\widehat{\mathbf{R}}_{2,12}\right) \right] 
\overset{d}{\rightarrow }\mathcal{N}\left( \mathbf{0},\mathbf{\Sigma }_{\widehat{\bm{\lambda }}2i}\right) ,\nn
\end{align}
and
\begin{align}
\sqrt{T}\left[ \wt{\bm{\lambda }}_{1i}^{\prime }-\bm{\lambda }_{2i}^{\prime }\widehat{\mathbf{R}}_{2,21}-\bm{\lambda }_{1i}^{\prime
}\left( \widehat{\mathbf{H}}_{11}-\widehat{\mathbf{R}}_{2,11}\right) \right] 
\overset{d}{\rightarrow }\mathcal{N}\left( \mathbf{0},\mathbf{\Sigma }_{\wt{\bm{\lambda }}1i}\right),\nn\\
%\end{equation*}
%and
%\begin{equation*}
\sqrt{T}\left[ \wt{\bm{\lambda }}_{2i}^{\prime }-\bm{\lambda }_{1i}^{\prime }\widehat{\mathbf{R}}_{1,12}-\bm{\lambda }_{2i}^{\prime
}\left( \widehat{\mathbf{H}}_{22}-\widehat{\mathbf{R}}_{1,22}\right) \right] 
\overset{d}{\rightarrow }\mathcal{N}\left( \mathbf{0},\mathbf{\Sigma }_{\wt{\bm{\lambda }}2i}\right) ,\nn
\end{align}
where $\mathbf{\Sigma }_{\widehat{\bm{\lambda }}1i}$, $\mathbf{\Sigma 
}_{\widehat{\bm{\lambda }}2i}$, $\mathbf{\Sigma }_{\wt{\bm{\lambda }}1i}$, and $\mathbf{\Sigma 
}_{\wt{\bm{\lambda }}2i}$ are the suitable $r\times r$ blocks of $\mathbf{\Sigma }_{\widehat{\mathbf{b}}1i}$ and $\mathbf{\Sigma }_{\widehat{\mathbf{b}}2i}$, respectively.
%\begin{align}
%&\sqrt T\l\{\wh{\bm\lambda}_{1i}^\prime-\bm\lambda_{1i}^\prime\wh{\mbf R}_{1}\r\}\stackrel{d}{\to}\mathcal N\l(\mbf 0,\mathbf{\Sigma }_{\widehat{\mathbf{b}}{1}i}\r),\;\; \sqrt T\l\{\wh{\bm\lambda}_{2i}^\prime-\bm\lambda_{2i}^\prime\wh{\mbf R}_{2}\r\}\stackrel{d}{\to}\mathcal N\l(\mbf 0,\mathbf{\Sigma }_{\widehat{\mathbf{b}}{2}i}\r),\nn\\
%&
%\sqrt T\l\{\wt{\bm \lambda}_{1i}^\prime-\bm\lambda_{1i}^\prime\l(\wh {\mbf H}-\wh{\mbf R}_{2}\r)\r\}\stackrel{d}{\to}\mathcal N\l(\mbf 0,\mathbf{\Sigma }_{\widehat{\mathbf{b}}{2}i}\r),\;\;\sqrt T\l\{\wt{\bm\lambda}_{2i}^\prime-\bm\lambda_{2i}^\prime\l(\wh{\mbf H}-\wh{\mbf R}_{1}\r)\r\}\stackrel{d}{\to}\mathcal N\l(\mbf 0,\mathbf{\Sigma }_{\widehat{\mathbf{b}}{1}i}\r).\nn
%\end{align}
}
\end{cor}

This corollary has some interesting implications. If we strengthen Assumption \ref{assum:FL}(c) to add the identification constraint $\bm\Sigma_{\bm\Lambda_{12}}=\mbf 0$, which is natural given Asssumption \ref{assum:FL}(d), then it is immediate to see that $\wh{\mbf H}_{12}\stackrel{p}{\to} \mbf 0$ and $\wh{\mbf H}_{21}\stackrel{p}{\to} \mbf 0$, as $N,T\to\infty$, in other words $\wh{\mbf H}\stackrel{p}{\to} \mbf H$ which is now a block-diagonal matrix (see \eqref{eq:Hlim} and recall that $\bm\Sigma_{\mbf g}$ is block-diagonal by construction). 
It follows that if the unconditional probability of being in a given state were correctly estimated with probability one, so that, as $N,T\to\infty$, we had $\mathbf{\widehat{I}}_{\mathbf{\widehat{\xi}}k_{1}}\stackrel{p}{\to}\mathbf{I}_{r_1+r_2}$, then, as $N,T\to\infty$, for $k=1,2$ we have $\wh{\mbf R}_{k}\stackrel{p}{\to} \mbf H$, which implies $\widehat{\bm{\lambda }}_{ki}^{\prime }\stackrel{p}{\to}\bm{\lambda }_{ki}^{\prime }\widehat{\mathbf{H}}_{kk}$, while $\wt{\bm{\lambda }}_{ki}^{\prime }\stackrel{p}{\to}\mbf 0$. These results, which allow for a clear separation of $\bm\Lambda_1$ and $\bm
\Lambda_2$, hold only under the restrictive assumption $\bm\Sigma_{\bm\Lambda_{12}}=\mbf 0$. However, in general it is not possible to verify such condition and the two sets of estimators 
$\wh{\bm{\lambda }}_{1i}^{\prime }$ and $\wh{\bm{\lambda }}_{2i}^{\prime }$
or $\wt{\bm{\lambda }}_{1i}^{\prime }$ and $\wt{\bm{\lambda }}_{2i}^{\prime }$ will estimate consistently only a linear combination of the true loadings in both regimes.

% is not imposed in the empirical exercise of Section \ref{section:Empirics}.

On the other hand, if $r_1\ne r_2$, 
%then the first $r_1$ columns of $\wh{\mbf B}_1$, and the last $r_2$ columns of $\wh{\mbf B}_2$, estimate an invertible linear transformation of  the columns of $\bm\Lambda_1$ and $\bm\Lambda_2$, respectively. However, the last $r_2$ columns of $\wh{\mbf B}_1$, and the first $r_1$ columns of $\wh{\mbf B}_2$, estimate $\bm\Lambda_1\wh{\mbf R}_{1,12}$ and $\bm\Lambda_2\wh{\mbf R}_{2,21}$, respectively, none of which is invertible \textcolor{red}{NON CAPISCO COSA IMPORTI L'INVERTIBILITA' SECONDO ME L'UNCO PROBLEMA QUI E' CHE NON SAI COME SPLITTARE LE COLONNE PERCHE' NON SAI QUANTI FATTORI PER REGIME HAI}. In this case, 
we need consistent estimators of $r_1$ and $r_2$ in order to be able to isolate the first $r_1$ columns of $\wh{\mbf B}_1$ and the last $r_2$ columns of $\wh{\mbf B}_2$, respectively. Therefore, if we only know that $r_1\ne r_2$ without knowing their true values, then we can consistently estimate a linear transformation of the columns of $\mbf B_j$, but nothing can be said about $\bm\Lambda_j$, $j=1,2$.

Theorem \ref{th:asympt_dist} describes the asymptotic properties of the estimator for the factor loadings $\mathbf{\wh{B}}_{1}$ and $\mathbf{\wh{B}}_{2}$. Complementary results can be obtained with respect to the estimated factors associated to the loading matrices $\mathbf{\wh{B}}_{1}$ and $\mathbf{\wh{B}}_{2}$. Formally, the true factors that correspond to $\mathbf{B}_{1}$ and $\mathbf{B}_{2}$ are $\xi _{1t}\mathbf{g}_{t}$  and $\xi _{2t}\mathbf{g}_{t}$, respectively, and their estimators are $\widehat{\xi}_{1,t\left\vert T\right. }\mathbf{\wh{g}}_{t}$ and $\widehat{\xi}_{2,t\left\vert T\right. }\mathbf{\wh{g}}_{t}$, respectively. The following theorem states the asymptotic distribution of these estimators.

\begin{thm}
\label{th:asympt_dist_factors} {\it Let Assumptions \ref{assum:F} - \ref{assum:rates} hold. Then, for any given $t=1,\dots,T$, as $N,T\to\infty$,
%------------------------------------------------------------------------%
\begin{equation*}
\sqrt{N}\left\{ \left( 
\begin{array}{c}
\widehat{\xi}_{1,t\left\vert T\right. }\mathbf{\wh{g}}_{t} \\ 
\widehat{\xi}_{2,t\left\vert T\right. }\mathbf{\wh{g}}_{t}
\end{array}%
\right) -\mathbf{\widehat{H}}_{\mathbf{\xi }}^{-1}\left( 
\begin{array}{c}
\xi _{1t}\mathbf{g}_{t} \\ 
\xi _{2t}\mathbf{g}_{t}
\end{array}%
\right) \right\} \overset{d}{\rightarrow }\mathcal{N}\left( \mathbf{0},%
\mathbf{\Sigma }_{\mathbf{\wh{\xi}\otimes\widehat{g}},t}\right),
\end{equation*}%
%------------------------------------------------------------------------%
where
%------------------------------------------------------------------------%
\begin{equation*}
\mathbf{\widehat{H}}_{\mathbf{\xi }}=\left[ 
\begin{array}{cc}
\mathbf{\widehat{H}\widehat{I}}_{\wh{\mathbf{\xi }}1} & \mathbf{\widehat{H}}\left( 
\mathbf{I}_{r_1+r_2}-\mathbf{\widehat{I}}_{\wh{\mathbf{\xi }}2}\right) \\ 
\mathbf{\widehat{H}}\left( \mathbf{I}_{r_1+r_2}-\mathbf{\widehat{I}}_{\wh{\mathbf{\xi }}1}\right)
 & \mathbf{\widehat{H}\widehat{I}}_{\wh{\mathbf{\xi }}2}%
\end{array}%
\right],
\end{equation*}
with $\mathbf{\widehat{H}}$ and $\mathbf{\widehat{I}}_{\wh{\mathbf{\xi }}j}$ defined in \eqref{eq:hat_H} and \eqref{eq:hat_I_xi}, respectively, and where
%------------------------------------------------------------------------%
\begin{equation*}
\mathbf{\Sigma }_{\mathbf{\wh{\xi}\otimes\widehat{g}},t}=\left\{ \mathbf{H}_{\mathbf{\xi }}\left( 
\begin{array}{cc}
\mathbf{\Sigma }_{\mathbf{B}1} & \mathbf 0 \\ 
\mbf 0 & \mathbf{\Sigma }_{\mathbf{B}2}%
\end{array}%
\right) \mathbf{H}_{\mathbf{\xi }}^{\prime }\right\} ^{-1}\left( \mathbf{H}_{%
	\mathbf{\xi }}\mathbf{\Sigma }_{\mathbf{Be}t}\mathbf{H}_{\mathbf{\xi }%
}^{\prime }\right) \left\{ \mathbf{H}_{\mathbf{\xi }}\left( 
\begin{array}{cc}
\mathbf{\Sigma }_{\mathbf{B}1} & \mbf 0 \\ 
\mbf 0 & \mathbf{\Sigma }_{\mathbf{B}2}%
\end{array}%
\right) \mathbf{H}_{\mathbf{\xi }}^{\prime }\right\} ^{-1},
\end{equation*}
%------------------------------------------------------------------------%
where $\mathbf{\Sigma }_{\mathbf{B}j}$, $j=1,2$, is defined in \eqref{eq:SB1B2},
\begin{equation*}
\mathbf{\Sigma }_{\mathbf{Be}t}
=\l(\begin{array}{cccc}
\bm\Phi_{1t}&\mbf 0&\mbf 0&\bm\Phi_{12t}\\
\mbf 0&\mbf 0&\mbf 0&\mbf 0\\
\mbf 0&\mbf 0&\mbf 0&\mbf 0\\
\bm\Phi_{12t}^\prime&\mbf 0&\mbf 0&\bm\Phi_{2t}
\end{array}
\r),
\end{equation*}
with $\bm\Phi_{jt}$ and $\bm\Phi_{jkt}$, $j,k=1,2$, defined in Assumption \ref{assum:MCLT}(d),
and where
%----------------------------------------------------------------------------%
\begin{equation*}
\mathbf{H}_{\mathbf{\xi }}=\left[ 
\begin{array}{cc}
\mathbf{HI}_{\mathbf{\xi }1} & \mathbf{H}\left( \mathbf{I}_{r_1+r_2}-\mathbf{I}_{\mathbf{\xi }2}\right) \\ 
\mathbf{H}\left( \mathbf{I}_{r_1+r_2}-\mathbf{I}_{\mathbf{\xi }1}\right) & 
\mathbf{HI}_{\mathbf{\xi }2}%
\end{array}%
\right],
\end{equation*}
with $\mbf H$ defined in \eqref{eq:Hlim} and
%----------------------------------------------------------------------------%
\begin{equation*}
	\mathbf{I}_{\mathbf{\xi }j}=\textrm{p}\lim\nolimits_{N,T\rightarrow \infty }\mathbf{\widehat{I}}_{\wh{\mathbf{\xi }}j}=\mathbf{H}^{-1}\left[ 
	\begin{array}{cc}
		\mathbb{I}\left( j=1\right) \mathbf{I}_{r_{1}} & \mathbf{0} \\ 
		\mathbf{0} & \mathbb{I}\left( j=2\right) \mathbf{I}_{r_{2}}%
	\end{array}%
	\right] \mathbf{H,}
\end{equation*}
as defined in Lemma \ref{Lemma:Hat_I_xi} in Appendix \ref{section:Appendix} 
%\textcolor{red}{COSA VOLE DIRE BY LEMMA...? QUESTA NON E' UNA PROOF CI SERVE LA DEFINIZIONE DI $\mathbf{I}_{\mathbf{\xi }j}$ CHE VIENE DAL LEMMA? IN QUEL CASO SE LA SOSTITUIAMO .
%}
%----------------------------------------------------------------------------%
}

\end{thm}

In general, $\wh{\mathbf{I}}_{\wh{\mathbf{\xi }}j}\ne \mbf I_{r_1+r_2}$ and so also $\mathbf{I}_{\mathbf{\xi }j}\ne \mbf I_{r_1+r_2}$. Then, because of Theorem \ref{th:asympt_dist}, the estimator $\wh{\mbf b}_{ji}$ is biased and it is straightforward to see that the asymptotic covariance in Theorem \ref{th:asympt_dist_factors} is positive definite. Note that if we know that $r_1=r_2=r$ holds, then we can build consistent estimators for linear combinations of ${\mbf f}_{jt}$, $j=1,2$, by simply regressing $\mbf x_t$ onto the estimators $\wh{\bm\Lambda}_j$ or $\wt{\bm\Lambda}_j$ which are defined in \eqref{eq:Ljhat} and \eqref{eq:Ljtilde}, respectively, and, as shown in Corollary \ref{cor:loadings}, are consistent for linear transformation of ${\bm\Lambda}_j$. Formally, this means we can build the sequence of factor estimators by running the cross-sectional regressions
%----------------------------------------------------------------------------%
%\begin{align}\label{eq:fjhat}
%\wh{\mbf f}_{jt} =\frac 1N \wh{\xi}_{j,t|T}\wh{\bm\Lambda}_j^\prime\mbf x_t, \qquad j=1,2.
%\end{align}
\begin{equation}
	\label{eq:fjhat}
	\begin{array}{ccc}
		\widehat{\mathbf{f}}_{jt}=\widehat{\xi }_{j,t\left\vert T\right. }\left( 
		\widehat{\mathbf{\Lambda }}_{j}^{\prime }\widehat{\mathbf{\Lambda }}%
		_{j}\right) ^{-1}\left( \widehat{\mathbf{\Lambda }}_{j}^{\prime }\mathbf{x}%
		_{t}\right) , & j=1,2, & t=1,\ldots,T,%
	\end{array}%
\end{equation}
or
\begin{equation}
	\label{eq:fjtilde}
	\begin{array}{ccc}
		\wt{\mathbf{f}}_{jt}=\widehat{\xi }_{j,t\left\vert T\right. }\left( 
		\wt{\mathbf{\Lambda }}_{j}^{\prime }\wt{\mathbf{\Lambda }}%
		_{j}\right) ^{-1}\left( \wt{\mathbf{\Lambda }}_{j}^{\prime }\mathbf{x}%
		_{t}\right) , & j=1,2, & t=1,\ldots,T.%
	\end{array}%
\end{equation}
%----------------------------------------------------------------------------%
If the unconditional probability of being in a given state is correctly estimated then $\wh{\mathbf{I}}_{\wh{\mathbf{\xi }}j}\stackrel{p}{\to} \mbf I_{r_1+r_2}$  as $N,T\to\infty$, and Theorem \ref{th:asympt_dist_factors} is redundant: in this case, asymptotic normality of \eqref{eq:fjhat} and of \eqref{eq:fjtilde} follows from arguments analogous to those in \citet{bai03}. In the more general case we are considering, the asymptotic distribution of $\wh{\mbf f}_{jt}$ is stated in the following theorem (an analogous result holds for $\wt{\mbf f}_{jt}$ and it is omitted for brevity).

\begin{thm}
\label{th:asympt_dist_fjt}
{\it Let Assumptions \ref{assum:F} - \ref{assum:rates} hold and $r_1=r_2$. Then, for $j,k=1,2$ with $j \neq k$, and for any given $t=1,\dots,T$, as $N,T\to\infty$,
%----------------------------------------------------------------------------%
\begin{equation*}
	\begin{array}{cl}
		& \sqrt{N}\left\{ \widehat{\mathbf{f}}_{jt}-\left\{ 
		\begin{array}{c}
			\left[ \dfrac{\left( \mathbf{\Lambda }_{j}\widehat{\mathbf{H}}_{jj}+\mathbf{%
					\Lambda }_{k}\widehat{\mathbf{H}}_{kj}\right) ^{\prime }\left( \mathbf{%
					\Lambda }_{j}\widehat{\mathbf{H}}_{jj}+\mathbf{\Lambda }_{k}\widehat{\mathbf{%
						H}}_{kj}\right) }{N}\right] ^{-1} \\ 
			\times \dfrac{\left( \mathbf{\Lambda }_{j}\widehat{\mathbf{H}}_{jj}+\mathbf{%
					\Lambda }_{k}\widehat{\mathbf{H}}_{kj}\right) ^{\prime }\widehat{\xi }%
				_{j,t\left\vert T\right. }\left( \mathbb I(s_t=j)\mathbf{\Lambda }_{j}\mathbf{%
					f}_{jt}+\mathbb I(s_t=k)\mathbf{\Lambda }_{k}\mathbf{f}_{kt}\right) }{N}%
		\end{array}%
		\right\} \right\}   
		\overset{d}{\rightarrow }  \mathcal{N}\left( \mathbf{0},\mathbf{\Sigma}_{\widehat{\mathbf{f}}%
			_{jt}}\right),
	\end{array}%
\end{equation*}
%----------------------------------------------------------------------------%
where 
%----------------------------------------------------------------------------%
\begin{equation*}
	\mathbf{\Sigma}_{\widehat{\mathbf{f}}_{jt}}=\left( \xi _{j,t}^{\ast }\right)
	^{2}\left( \mathbf{H}_{11}^{\prime }\mathbf{\Phi }_{1t}\mathbf{H}_{11}+%
	\mathbf{H}_{jj}^{\prime }\mathbf{\Phi }_{jkt}\mathbf{H}_{kj}+\mathbf{H}%
	_{kj}^{\prime }\mathbf{\Phi }_{jkt}^{\prime }\mathbf{H}_{jj}+\mathbf{H}%
	_{22}^{\prime }\mathbf{\Phi }_{2t}\mathbf{H}_{22}\right), 
\end{equation*}
%----------------------------------------------------------------------------%
with $\xi _{j,t}^{\ast }=p\lim_{N,T\rightarrow \infty }\widehat{\xi }_{j,t\left\vert T\right. }$ and 
$\mathbf{\Phi }_{1t}$, $\mathbf{\Phi }_{2t}$, and
$\mathbf{\Phi }_{jkt}$, defined in Assumption \ref{assum:MCLT}(d).}
\end{thm}

According to Theorem \ref{th:asympt_dist_fjt}, $\widehat{\mathbf{f}}_{jt}$ estimates the space spanned by either $\mathbf{f}_{jt}$ or $\mathbf{f}_{kt}$, for $j,k=1,2$, with $j \neq k$, depending on which the true underlying regime is in period $t$.

%%%%%%%%%%%%%%%%%%%%%%%%%%%%%%%%%%%%%%%%%%%%%%%%%%%%%%%%%%%%%%%%%%%%%%%%%%
%%%%%%%%%%%%%%%%%%%%%%%%%%%%%%%%%%%%%%%%%%%%%%%%%%%%%%%%%%%%%%%%%%%%%%%%%%

\section{On the number of factors and regimes}
\label{section:num_regimes_factors}

This section deals with two further issues related to the model in \eqref{eq:model} and \eqref{eq:cov_mat}. Section \ref{section:num_factors} studies estimation of the number of factors within each regime. Section \ref{section:num_regimes} discusses the consequences of an underspecified model.

\subsection{Estimating the number of factors within each regime \label{section:num_factors}}

Theorems \ref{th:asympt_dist} and \ref{th:asympt_dist_factors} rely on the factor estimator $\mathbf{\wh{g}}_{t}$ obtained from the equivalent linear representation in \eqref{eq:linear_model}. This estimator does not embed any information related to the likelihood of observing a regime $j$ at a given point in time $t$, for $j=1,2$ and $t\in\mathbb Z$. We now study the property of the estimator for the dimension of the factor space that is obtained when such information is accounted for. In particular, we are interested in separately identifying the number of factors within each regime, namely $r_{1}$ and $r_{2}$, given the dimension $r_1+r_2$ of the factor space of the equivalent linear representation in \eqref{eq:linear_model}. Note that under Assumption \ref{assum:FL}(b), at least one factor is present in each regime, which means that $r_1\geq1$ and $r_2\geq1$. Our framework is then more general than \cite{Liu_Chen_2016_SS} and \cite{Urga_Wang_2022_WP}: in the former $r_1=r_2$, and the two regimes have the same number of factors; the latter assumes that $r_1$ and $r_2$ are both known and do no have to be estimated. We do not impose any restriction on $r_1$ and $r_2$, except that $r_1\geq1$ and $r_2\geq1$, as required in Assumption \ref{assum:FL}(b). This is the natural extension of the linear set up, and it is aligned to Assumption B in \cite{baing02}.

Formally, for $j=1,2$, we consider the regime-specific covariance matrix
%----------------------------------------------------------------------------%
\begin{equation}
\label{eq:sigma_hat_x_j}
	\mathbf{\widehat{\Sigma}}_{\widehat{\xi},\mathbf{x}j}=\dfrac{\sum\nolimits_{t=1}^{T}\widehat{\xi}%
		_{jt\left\vert T\right. }\mathbf{x}_{t}\mathbf{x}_{t}^{\prime }}{%
		N\sum\nolimits_{t=1}^{T}\widehat{\xi}_{jt\left\vert T\right. }},
\end{equation}%
%----------------------------------------------------------------------------%
where $0 < \sum\nolimits_{t=1}^{T}\widehat{\xi}_{jt\left\vert T\right. } < T$. The matrix $\mathbf{\widehat{\Sigma}}_{\widehat{\xi},\mathbf{x}j}$ includes information about the regimes through the estimated sequence $\{\widehat{\xi}_{jt\left\vert T\right. }\}_{t=1}^{T}$. Define the $r_{j}\times 1$ vectors
%----------------------------------------------------------------------------%
\begin{equation*}
	\begin{array}{ccc}
		\mathbf{f}_{jjt}=\mathbb{I}_{jt}\mathbf{f}_{jt}, & \mathbf{f}_{\widehat{\xi}%
			,kjt}=\widehat{\xi}_{kt\left\vert T\right. }\mathbf{f}_{jt}, & j,k=1,2,
	\end{array}%
\end{equation*}
%----------------------------------------------------------------------------%
and the $r_{j}\times T$ matrices
%----------------------------------------------------------------------------%
\begin{equation*}
	\begin{array}{ccc}
		\mathbf{F}_{jj}=\left( \mathbb{I}_{j1}\mathbf{f}_{j1},\ldots ,\mathbb{I}_{jT}%
		\mathbf{f}_{jT}\right) , & \mathbf{F}_{\widehat{\xi},kj}=\left( \widehat{\xi}%
		_{k1\left\vert T\right. }\mathbf{f}_{j1},\ldots ,\widehat{\xi}_{kT\left\vert
			T\right. }\mathbf{f}_{jT}\right) , & j,k=1,2.%
	\end{array}%
\end{equation*}
%----------------------------------------------------------------------------%
For $1 \leq p \leq \bar{p}$, with $\bar{p} < \infty$, let $\mathbf{\widehat{V}}_{\widehat{\xi},j}^{\left( p\right) }$ be the $p \times p$ diagonal matrix containing the first $p$ eigenvalues of $\mathbf{\widehat{\Sigma}}_{\widehat{\xi},\mathbf{x}j}$ in decreasing order. Finally, let $\mathbf{\widehat{\Lambda}}_{\widehat{\xi},j}^{\left(p\right) }=[\bm{\widehat{\lambda}}_{\widehat{\xi},j1}^{\left( p\right)
},\ldots,\bm{\widehat{\lambda}}_{\widehat{\xi},jN}^{\left( p\right)
}]^{\prime}$ be the $N \times p$ matrix estimator for $\bm{\Lambda}_{j}$, which is obtained as $\sqrt N$ times the normalized eigenvectors corresponding to the $p$ largest eigenvalues of the $N\times N$ sample covariance matrix $\mathbf{\widehat{\Sigma}}_{\widehat{\xi},\mathbf{x}j}$ in \eqref{eq:sigma_hat_x_j}. The following theorem characterises the mean square convergence of $\bm{\widehat{\lambda}}_{\widehat{\xi},ji}^{\left( p\right)
}$ for a given value of $p$.

\begin{thm}
\label{th:n_factors_regimes}
{\it Let Assumptions \ref{assum:F} - \ref{assum:WD} hold. Then, for any fixed $1 \leq p \leq \bar{p}$ with $\bar{p} < \infty$, and for $j,k=1,2$ with $j \neq k$, there exists $r_{j}\times p$ matrices $\mathbf{\widehat{H}}_{\widehat{\xi},kj}^{\left( p\right) }$ such that
%----------------------------------------------------------------------------%
\begin{equation}
\label{eq:H_hat_p}
	\mathbf{\widehat{V}}_{\widehat{\xi},j}^{\left( p\right) }\mathbf{\widehat{H}}_{\widehat{\xi}%
		,kj}^{\left(p\right)}=\dfrac{\mathbf{F}_{\widehat{\xi},kj}\mathbf{F}%
		_{jj}^{\prime }}{\sum\nolimits_{t=1}^{T}\widehat{\xi}_{jt\left\vert T\right. }}\dfrac{\mathbf{\Lambda }_{j}^{\prime }\mathbf{\widehat{\Lambda}}_{\widehat{\xi},j}^{\left(p\right) }}{N}
\end{equation}
%----------------------------------------------------------------------------%
with $\mathrm{rank}\left( \mathbf{\widehat{H}}_{\widehat{\xi},kj}^{\left( p\right)
}\right) =\min \left\{ r_{j},p\right\} $, 
%and $C_{NT}=\min \left\{ \sqrt{N},%
%\sqrt{T}\right\} $, 
which satisfy
%----------------------------------------------------------------------------%
\begin{equation*}
	\min \left\{ {N},%
{T}\right\} \left\{ \dfrac{1}{N}\sum\limits_{i=1}^{N}\left\Vert \left[ \bm{\widehat{\lambda}}_{\widehat{\xi},ji}^{\left( p\right)
	}-\left( \mathbf{\widehat{H}}_{\widehat{\xi},jj}^{\left( p\right) \prime }\bm{\lambda }_{ji}+\mathbf{\widehat{H}}_{\widehat{\xi},kj}^{\left( p\right) \prime }\bm{\lambda }_{ki}\right) \right] \right\Vert ^{2}\right\} =O_{p}\left(1\right) .
\end{equation*}
}
\end{thm}

Theorem \ref{th:n_factors_regimes} extends Theorem 1 in \cite{baing02} and Theorem 3.4 in \cite{Massacci_2017_JoE} to the case of the Markov switching factor model in \eqref{eq:model} and \eqref{eq:cov_mat}. For $j,k=1,2$ with $j\neq k$, the theorem shows that $\bm{\widehat{\lambda}}_{\widehat{\xi},ji}^{\left( p\right)
}$ estimates a linear combination of the vector $\left(\bm{\lambda }^{\prime}_{ji},\bm{\lambda }^{\prime}_{ki}\right)^{\prime}$ and not just of $\bm{\lambda }_{ji}$. It implies that the dimension of the estimated underlying factor space is $r_{1}+r_{2}$ even when the available information about the regimes is accounted for. Imperfect knowledge of the regimes therefore leads to an enlarged factor space: this makes our setting analogous to large dimensional change point factor models, as previously discussed in Section \ref{sec:linearFM}. This complements what proved in \cite{Breitung_Eickmeier_2011_JoE}, and \cite{corradi2014testing}, who show that model misspecification in the form of omitted discrete regime shifts leads to an inflated number of factors. More generally, Theorem \ref{th:n_factors_regimes} implies that, without further assumptions on the number of factors within each regime, it is not possible to separately estimate $r_1$ and $r_2$ even when the dimension $r_1+r_2$ of the equivalent linear representation in \eqref{eq:linear_model} has been accurately estimated.

%The result in Theorem \ref{th:n_factors_regimes} has practical implications.

As in \cite{Liu_Chen_2016_SS}, we now make the additional assumption that $r_1=r_2$, which means that the number of factors is equal across regimes. If the estimated number of factors in the equivalent linear representation in \eqref{eq:linear_model} is an even number, we can recover the number of factors within each regime, as this is equal to $r_1=r_2=\left(r_1+r_2\right)/2$. On the other hand, if the estimated number of factors in the linear representation in \eqref{eq:linear_model} is an odd number, an additional third regime might actually be neglected, as discussed in Section \ref{section:num_regimes} below.

%Obviously if we find evidence of just one common factor, then only the idiosyncratic covariances can be regime specific, while the loadings remain constant.

Finally, under the assumption that both $r_1$ and $r_2$ are known as in \cite{Urga_Wang_2022_WP}, the number of factors is known in both regimes and does not have to be estimated.

%Therefore, our approach is general enough to handle the situation in which the number of factors is unknown in both regimes. By Assumption 2(b), it only requires that $r_1\geq1$ and $r_2\geq1$, which means that at least one factor is present in each regime. 

\subsection{The case of an underspecified number of regimes\label{section:num_regimes}}

Up to know we have \textit{a priori} assumed that the data are generated according to the model with two regimes in \eqref{eq:model} and \eqref{eq:cov_mat}. This is consistent with existing empirical studies employing Markov switching models: for example, see \cite{diebold78measuring}. However, in some cases the underlying data generating process of the dependent variables of interest displays a higher number of regimes: for example, \cite{Guidolin_Timmermann_JAE_2006} show that the joint distribution of stock and bond returns requires a four-state model. Therefore, the two-regime specification in \eqref{eq:model} and \eqref{eq:cov_mat} leads to model misspecification in case the joint distribution of the dependent variables $\mathbf{x}_{t}$ is characterised by a higher number of regimes. 

We now study the case in which the model is underspecified and the data are generated by a process with a number of regimes that is finite and greater than two. 
%In what follows, we present the main arguments and result, and we provide details in Appendix \ref{Appendix:Under_model}.

Since the number of regimes is finite, without loss of generality we consider the model with three regimes
%----------------------------------------------------------------------------%
\begin{align}
%	\begin{array}{rcl}
		\mathbf{x}_{t}  =&\,  \mathbb{I}\left( s_{t}=1\right) \left( \mathbf{\Lambda }%
		_{1}\mathbf{f}_{1t}+\mathbf\Sigma _{e1}^{1\left/ 2\right. }\mathbf{e}_{t}\right) +%
		\mathbb{I}\left( s_{t}=2\right) \left( \mathbf{\Lambda }_{2}\mathbf{f}%
		_{2t}+\mathbf\Sigma _{e2}^{1\left/ 2\right. }\mathbf{e}_{t}\right)\nonumber\\  
		&+\mathbb{I}\left( s_{t}=3\right) \left( \mathbf{\Lambda }_{3}\mathbf{f}%
		_{3t}+\mathbf\Sigma _{e3}^{1\left/ 2\right. }\mathbf{e}_{t}\right), \quad t\in\mathbb Z, 
%	\end{array}
	\label{eq: reg_0}
\end{align}
%----------------------------------------------------------------------------%
and let
\begin{equation*}
	\begin{array}{cc}
		\mathbf{g}_{t}=\left[ 
		\begin{array}{c}
			\mathbf{f}_{1t} \\ 
			\mathbf{0} \\ 
			\mathbf{0}%
		\end{array}%
		\right] \mathbb{I}\left( s_{t}=1\right) +\left[ 
		\begin{array}{c}
			\mathbf{0} \\ 
			\mathbf{f}_{2t} \\ 
			\mathbf{0}%
		\end{array}%
		\right] \mathbb{I}\left( s_{t}=2\right) +\left[ 
		\begin{array}{c}
			\mathbf{0} \\ 
			\mathbf{0} \\ 
			\mathbf{f}_{3t}%
		\end{array}%
		\right] \mathbb{I}\left( s_{t}=3\right) , & t\in \mathbb{Z}.%
	\end{array}%
\end{equation*}
%----------------------------------------------------------------------------%
Suppose that only two regimes are accounted for. Given a natural ordering of the regimes, this means that we have to consider two cases, namely: $\left(a\right)$ $s_t=1$ and $s_t \neq 1$; $\left(b\right)$ $s_t=3$ and $s_t \neq 3$. The model in $\left( \ref{eq: reg_0}\right)$ admits the following two equivalent two-regime representations
%----------------------------------------------------------------------------%
\begin{align}	
\mathbf{x}_{t} & =  \left( \mathbf{B}_{1}^{\left( j\right) }~\mathbf{B}_{2}^{\left( j\right) }\right) \left( \bm{\xi }_{t}^{\left( j\right)
			}\otimes \mathbf{g}_{t}\right) +\left( \bm{\Sigma}_{e1}^{\left( j\right)
				,1\left/ 2\right. }~\bm{\Sigma}_{e2}^{\left( j\right) ,1\left/ 2\right. }\right)
			\left( \bm{\xi }_{t}^{\left( j\right) }\otimes \bm{\xi }_{t}\otimes 
			\mathbf{I}_{N}\right) \mathbf{e}_{t}, \;\; t\in\mathbb Z, 	\label{eq: reg_1}\\ 
			\bm{\xi }_{t}^{\left( j\right) } & =  \mathbf{P}^{\left( j\right)
				\prime }\bm{\xi }_{t-1}^{\left( j\right) }+\mathbf{v}_{t}^{\left(
				j\right) } ,\qquad\qquad\qquad\qquad\qquad\qquad\qquad\qquad\qquad\qquad\qquad  j=1,3,\nonumber
\end{align}
%----------------------------------------------------------------------------%
where the loadings are defined as
%----------------------------------------------------------------------------%
$\mathbf{B}_{1}^{\left( 1\right) }=\left( \mathbf{\Lambda }_{1}~\mathbf{0}~%
		\mathbf{0}\right)$, $\mathbf{B}_{2}^{\left( 1\right) }=\left( \mathbf{0}~%
		\mathbf{\Lambda }_{2}~\mathbf{\Lambda }_{3}\right)$, $\mathbf{B}_{1}^{\left( 3\right) }=\left( \mathbf{\Lambda }_{1}~\mathbf{%
			\Lambda }_{2}~\mathbf{0}\right)$,  $\mathbf{B}_{2}^{\left( 3\right) }=\left( 
		\mathbf{0}~\mathbf{0}~\mathbf{\Lambda }_{3}\right)$,
%----------------------------------------------------------------------------%
the latent state process is defined as
%----------------------------------------------------------------------------%
\begin{equation*}
	\begin{array}{cc}
		\bm{\xi }_{t}^{\left( 1\right) }=\left[ 
		\begin{array}{c}
			\mathbb{I}\left( s_{t}=1\right)  \\ 
			\mathbb{I}\left( s_{t}=2\right) +\mathbb{I}\left( s_{t}=3\right)
		\end{array}%
		\right], &
		\bm{\xi }_{t}^{\left( 3\right) }=\left[ 
		\begin{array}{c}
			\mathbb{I}\left( s_{t}=1\right) +\mathbb{I}\left( s_{t}=2\right)  \\ 
			\mathbb{I}\left( s_{t}=3\right) 
		\end{array}%
		\right],
	\end{array}
\end{equation*}
%----------------------------------------------------------------------------%
%----------------------------------------------------------------------------%
the idiosyncratic covariance matrices are defined as
%----------------------------------------------------------------------------%
$\bm{\Sigma}_{e1}^{\left( 1\right) }=( \bm{\Sigma}
		_{e1}~\mathbf{0}~\mathbf{0})$, $\bm{\Sigma}
		_{e2}^{\left( 1\right) }=( \mathbf{0}~\bm{\Sigma}
		_{e2}~\bm{\Sigma} _{e3})$, $\bm{\Sigma}_{e1}^{\left( 3\right)}=( \bm{\Sigma}
		_{e1}~\bm{\Sigma} _{e2}~\mathbf{0})$, $\bm{\Sigma} _{e2}^{\left( 3\right) }=( \mathbf{0}~%
		\mathbf{0~}\bm{\Sigma} _{e3})$,
%----------------------------------------------------------------------------%
and the transition probabilities are equal to
%----------------------------------------------------------------------------%
\begin{equation*}
	\begin{array}{c}
		\mathbf{P}^{\left( 1\right)}=\left( 
		\begin{array}{cc}
			p_{11} & p_{1, \neq 1} \\ 
			p_{\neq 1, 1} & p_{\neq 1, \neq 1}%
		\end{array}%
		\right) =\left( 
		\begin{array}{cc}
			p_{11} & 1-p_{11} \\ 
			1-p_{\neq 1, \neq 1} & p_{\neq 1, \neq 1}
		\end{array}
		\right),
		\\
		\\
		\mathbf{P}^{\left( 3\right)}=\left( 
		\begin{array}{cc}
			p_{\neq 3, \neq 3} & p_{\neq 3, 3} \\ 
			p_{3, \neq 3} & p_{3, 3}%
		\end{array}%
		\right) =\left( 
		\begin{array}{cc}
			p_{\neq 3, \neq 3} & 1-p_{\neq 3, \neq 3} \\
			1-p_{3, 3} & p_{3, 3}%
		\end{array}%
		\right).
	\end{array}
\end{equation*}
%----------------------------------------------------------------------------%

For $j=1,3$, define the vector of parameters $\mathbf{q}^{\left(j\right)}=\left[ \bm\varphi^{\left(j\right) \prime},\bm\rho^{\left(j\right)\prime}\right] ^{\prime }$, where
%----------------------------------------------------------------------------%
\begin{equation*}
	\begin{array}{cc}
		\bm{\varphi }^{\left( j\right) }=\left[ \mathrm{vec}\left( \mathbf{B}%
		_{1}^{\left( j\right) }\right) ^{\prime },\mathrm{vec}\left( \mathbf{B}%
		_{2}^{\left( j\right) }\right) ^{\prime },\mathrm{diag}\left( \bm{\Sigma}_{e1}^{\left(j\right)}\right) ^{\prime },\mathrm{diag}\left( \bm{\Sigma}_{e2}^{\left(j\right)}\right)
		^{\prime }\right] , & \bm{\rho }^{\left( j\right) }=\mathrm{vec}\left( 
		\mathbf{P}^{\left( j\right) }\right).
	\end{array}%
\end{equation*}%
%----------------------------------------------------------------------------%
Let $\left(NT\right)^{-1}\log f\left( \mathbf{X};\mathbf{q }^{\left( j\right) } \right) $ be the normalised log-likelihood function of $%
\left( \ref{eq: reg_1}\right) $. Assume that 
%----------------------------------------------------------------------------%
\begin{equation}
	\E\left[ \dfrac{1}{NT}%
	\log f\left( \mathbf{X};\mathbf{q }^{\left(
		1\right) } \right)  \right] >\E
	\left[ \dfrac{1}{NT}\log f\left( \mathbf{X};\mathbf{%
		q }^{\left( 3\right) }
	\right)  \right].
	\label{eq: lik_in}
\end{equation}
%----------------------------------------------------------------------------%
In a likelihood sense, the condition in (\ref{eq: reg_1}) captures a larger regime shift for $j=1$ than for $j=3$. Further, let $\widehat{\mathbf{q}}$ be the generic maximum likelihood estimator for the parameter of an underspecified model that allows for only two regimes when in fact the data generating process is given by \eqref{eq: reg_0}.

We proceed by contradiction, see also Appendix \ref{Appendix:Under_model} for more details. If $\widehat{\mathbf{q}}$ were an estimator for $\mathbf{q}^{\left( 3\right) }$, then
%----------------------------------------------------------------------------%
\begin{equation}
	\label{eq:contradiction}
	\E\left[ \dfrac{1}{NT}\log
	f\left( \mathbf{X};\widehat{\mathbf{q}}\right)
	\right] -\E\left[ \dfrac{1}{NT}\log f\left( \mathbf{X};\mathbf{q}^{\left( 1\right) }\right) \right] = - C + o_{p}\left(1\right),
\end{equation}
%----------------------------------------------------------------------------%
which leads to a contradiction since $\left(NT\right)^{-1}\log f\left( \mathbf{X};\widehat{\mathbf{q}}\right)$ is the estimated log-likelihood function. On the other hand, if  $\widehat{\mathbf{q}}$ were an estimator for $\mathbf{q}^{\left( 1\right) }$, then
%----------------------------------------------------------------------------%
\begin{equation*}
\E\left[ \dfrac{1}{NT}\log
f\left( \mathbf{X};\widehat{\mathbf{q}}\right)
\right] -\E\left[ \dfrac{1}{NT}\log f\left( \mathbf{X};\mathbf{q}^{\left( 1\right) }\right) \right] = o_{p}\left(1\right).
\end{equation*}%
%----------------------------------------------------------------------------%

Therefore, when one regime is neglected, the maximum likelihood estimator estimates the regimes that maximise the likelihood according to the inequality in (\ref{eq: lik_in}). Provided that a sufficient number of iterations is done, the EM algorithm proposed in Section \ref{section:Estimation} delivers an estimator that is close enough to the maximum likelihood estimator, such that the inequality in (\ref{eq: lik_in}) is preserved: see \cite{MR93,MR94}. Therefore, the EM algorithm delivers the estimator for the underspecified representation that is associated to the highest likelihood. This also implies that when running the filter with just two regimes the estimated state $\wh{\xi}_{1,t|T}$ is still correctly estimating the conditional expectation of the indicator related to the most likely regime, i.e., $\mathsf E[\mathbb I(s_t=1)|\bm X]$.

This result is consistent with the homologous finding in \cite{bai_1997}, and \cite{Bai_Perron_1998_Econometrica}, in relation to regression models with structural instability. Therefore, our result is the potential starting point for an inferential procedure on the number of regimes in large dimensional Markov switching factor models. It is also important to note that any neglected regime will be accounted for by an enlarged factor space, as discussed in Section \ref{sec:linearFM}.

\section{Unobserved heterogeneity}
\label{section:Unobs_Hetero}

The model in $\left(\ref{eq:model}\right)$ assumes no individual effects. However, these may be important when modelling macroeconomic series as in \cite{diebold78measuring}. In our set up, individual effects can be introduced by extending \cite{baili12,baili16} and considering
%------------------------------------------------------------------------%
\begin{equation}
\label{eq:model_unobs_hetero}
\mathbf{x}_{t}%
=\left(\bm{\alpha}_{1}+\bm\Lambda_{1}\mathbf{f}_{1t}\right)\mathbb{I}(s_{t}=1)%
+\left(\bm{\alpha}_{2}+\bm\Lambda_{2}\mathbf{f}_{2t}\right)\mathbb{I}(s_{t}=2)%
+\mathbf{e}_{t},
\end{equation}
%------------------------------------------------------------------------%
where $\bm{\alpha}_{j}=\left(\alpha_{j1},\ldots,\alpha_{jN}\right)^{\prime}$, for $j=1,2$, and $\alpha_{ji}$ captures the individual effect of cross-sectional unit $i$ within regime $j$. The vectors $\bm{\alpha}_{1}$ and $\bm{\alpha}_{2}$ introduce unobserved heterogeneity. If the state variable driving the regimes were observable, the resulting identification problem could be solved by expressing the model in terms of deviations of $\mathbf{x}_{t}$ from the \textit{conditional} means within each regime: on this, see \cite{Massacci_Sarno_Trapani_2021_WP}. However, since the state variable $s_{t}$ in $\left(\ref{eq:model_unobs_hetero}\right)$ is latent, this strategy no longer is applicable since the state is not observable with probability one. For this reason, we express the model in terms of the deviation of $\mathbf{x}_{t}$ from the \textit{unconditional} mean.

Formally, consider the $N\times 1$ vector of centred variables $\mathbf{y}_{t}$ defined as
%------------------------------------------------------------------------%
\begin{equation*}
%\begin{array}{rcl}
\mathbf{y}_{t}  =  \mathbf{x}_{t}-\E\left( \mathbf{x}_{t}\right) 
%& = & \left( \bm{\alpha }_{1}+\mathbf{\Lambda }_{1}\mathbf{f}%
%_{1t}\right) \mathbb{I}\left( s_{t}=1\right) +\left( \bm{\alpha }_{2}+%
%\mathbf{\Lambda }_{2}\mathbf{f}_{2t}\right) \mathbb{I}\left( s_{t}=2\right) +%
%\mathbf{e}_{t} \\ 
%&  & -\E\left[ \left( \bm{\alpha }_{1}+\mathbf{\Lambda }_{1}%
%\mathbf{f}_{1t}\right) \mathbb{I}\left( s_{t}=1\right) \right] +\E%
%\left[ \left( \bm{\alpha }_{2}+\mathbf{\Lambda }_{2}\mathbf{f}%
%_{2t}\right) \mathbb{I}\left( s_{t}=2\right) \right]  \\ 
 =  \bm{\alpha }_{1}d_{1t}+\mathbf{\Lambda }_{1}\mathbf{f}_{1t}%
\mathbb{I}\left( s_{t}=1\right) +\bm{\alpha }_{2}d_{2t}+\mathbf{%
	\Lambda }_{2}\mathbf{f}_{2t}\mathbb{I}\left( s_{t}=2\right) +\mathbf{e}_{t},
%\end{array}%
\end{equation*}
%------------------------------------------------------------------------%
where $d_{jt}=\mathbb{I}\left( s_{t}=j\right) -\E\left[ \mathbb{I}%
\left( s_{t}=j\right) \right]$, $j=1,2$. If $\bm{\alpha }_{1}=\bm{\alpha }_{2}$, $\mathbf{x}_{t}$ has the same expected value in both regimes, and $\mathbf{y}_{t}=\mathbf{\Lambda }_{1}\mathbf{f}_{1t}\mathbb{I}\left( s_{t}=1\right) +\mathbf{\Lambda }_{2}\mathbf{f}_{2t}\mathbb{I}\left( s_{t}=2\right) +\mathbf{e}_{t}$. In the more general case in which $\bm{\alpha }_{1}\neq \bm{\alpha }_{2}$, unconditional demeaning leads to a larger factor space of dimension $r_{1}+r_{2}+2$. The additional two factors $d_{1t}$ and $d_{2t}$ take only two values, namely $d_{jt}= -\E\left[ \mathbb{I}\left( s_{t}=j\right) \right]$ or $d_{jt}= 1-\E\left[ \mathbb{I}\left( s_{t}=j\right) \right]$, depending on whether $\mathbb{I}\left( s_{t}=j\right)=0$ or $\mathbb{I}\left( s_{t}=j\right)=1$, respectively, for $j=1,2$. In this case, the equivalent linear representation in $\left(\ref{eq:linear_model}\right)$ holds with $\mathbf{g}_{t}=\left[ d_{1t},\mathbb{I}\left( s_{t}=1\right) \mathbf{f%
}_{1t}^{\prime },d_{2t},\mathbb{I}\left( s_{t}=2\right) \mathbf{f}%
_{2t}^{\prime }\right]$ and $\mathbf{A}=\left[ \bm{\alpha }_{1},\mathbf{\Lambda }_{1},\bm{\alpha }_{2},\mathbf{\Lambda }_{2}\right]$. The measurement equation in $\left(\ref{eq:state_space_mes}\right)$ of the state space representation remains valid with $\mathbf{B}_{1}=\left[ \bm{\alpha }_{1},\mathbf{\Lambda }_{1},\bm{\alpha }_{2},\mathbf{0}\right]$ and $\mathbf{B}_{2}=\left[ \bm{\alpha }_{1},\mathbf{0},\bm{\alpha }_{2},\mathbf{\Lambda }_{2}\right]$. Therefore, the tools developed in this paper can be applied to the sample counterpart of $\mathbf{y}_{t}$, namely to $\mathbf{\widehat{y}}_{t}=\mathbf{x}_{t}-\left(T^{-1}\sum_{t=1}^{T}\mathbf{x}_{t}\right)$, which consistently estimates $\mathbf{y}_{t}$ as $T\rightarrow \infty $. Corollary \ref{cor:loadings} holds accordingly with respect to $\left(\alpha_{1i},\bm{\lambda}^{\prime}_{1i}\right)^{\prime}$ and $\left(\alpha_{2i},\bm{\lambda}^{\prime}_{2i}\right)^{\prime}$ instead of with respect to $\bm{\lambda}_{1i}$ and $\bm{\lambda}_{2i}$ only, respectively, for $i=1,\ldots,N$.
%------------------------------------------------------------------------%

%%%%%%%%%%%%%%%%%%%%%%%%%%%%%%%%%%%%%%%%%%%%%%%%%%%%%%%%%%%%%%%%%%%%%%%%%%
%%%%%%%%%%%%%%%%%%%%%%%%%%%%%%%%%%%%%%%%%%%%%%%%%%%%%%%%%%%%%%%%%%%%%%%%%%

\section{Detecting regime changes}
\label{section:Detecting_regime_changes}

The model in \eqref{eq:model} and \eqref{eq:cov_mat} \textit{a priori} assumes the existence of two regimes. However, in practice Markov switching dynamics should be detected with suitable statistical tools. The development of rigorous inference goes beyond the purpose of this paper. In what follows, we give an overview of the relevant literature, which we use to discuss a possible starting point to run inference on the number of regimes in large dimensional Markov switching factor models. 

First of all, it is however important to note that the Monte Carlo experiments in Section \ref{section:Monte_Carlo} show that, when we fit the model in \eqref{eq:model} and \eqref{eq:cov_mat} to a linear factor model with just one regime (which means a model with no regime change), the algorithm detailed in Section \eqref{section:Estimation} assigns probability almost equal to unity to one state and therefore does not require any inferential procedure on the number of regimes. We refer to Appendix \ref{app:sim} and the related Tables \ref{tab:no_change1} and \ref{tab:no_change2} for all relevant details.

As discussed in \cite{Qu_Zhuo_2021_ReStat}, there exist three approaches to detect Markov regime switching in low dimensional models. A first one involves testing parameter homogeneity against heterogeneity: this is done in \cite{Carrasco_Hu_Ploberger_2014_Econometrica}, who develop a class of tests for parameter constancy in random coefficient models; the power of these tests may however be limited, as they detect parameter heterogeneity of general form and are not specific to Markov switching models. A second approach, put forward in \cite{Hamilton_1996_JoE}, proposes specification tests in Markov switching models: if the null hypothesis of correct model specification is rejected, as a solution one may include additional regimes; however, also this approach may suffer from low power, as it detects model misspecification of unknown form. Finally, a third approach proposes likelihood ratio based tests for the null hypothesis of a given number of regimes against the alternative of a higher number of regimes: this is followed in \cite{Hansen1992} and \cite{Qu_Zhuo_2021_ReStat}, and it needs to account for the problem highlighted in \cite{Davies1977,Davies1987} as the additional transition probabilities are identified only under the alternative.

The above mentioned contributions are valid for low dimensional models. They are not directly applicable to large dimensional factor models, as these require imposing a number of restrictions on the loadings that goes to infinity as $N\rightarrow\infty$. This problem has been addressed when the variable driving the state is observable. \cite{Chen_Dolado_Gonzalo_2014_JoE}, and \cite{Han_Inoue_2015_ET}, test for a break in the loadings by testing for a change in the covariance matrix of the estimated factors. This approach, also used in \cite{Massacci_2017_JoE} in threshold factor models, is valid provided that the covariance matrix of the true factors is stable over time. However, this may not be realistic in practice, as discussed in \cite{Chen_Dolado_Gonzalo_2014_JoE}. \cite{Massacci_2023_JFEcon} develops an inferential procedure for threshold factor models that is robust to factor heteroskedasticity. However, these solutions are not directly applicable to large dimensional Markov switching factor models, since the state variable is latent rather than observable.

Given the above discussion, a possible strategy to conduct inference on the number of regimes in large dimensional Markov switching factor models is to merge the tests available for low dimensional models with those in use for large dimensional factor models with observable state variable. This is a complex problem that goes beyond the purpose of this paper and will be addressed in future research.

%%%%%%%%%%%%%%%%%%%%%%%%%%%%%%%%%%%%%%%%%%%%%%%%%%%%%%%%%%%%%%%%%%%%%%%%%%
%%%%%%%%%%%%%%%%%%%%%%%%%%%%%%%%%%%%%%%%%%%%%%%%%%%%%%%%%%%%%%%%%%%%%%%%%%

\section{Monte Carlo}
\label{section:Monte_Carlo}

We set $N=\{100,200\}$ and $T=\{250, 500, 750, 1000\}$. At each time period $t=1,\ldots, T$, we simulate the $N\times 1$ vector of data $\mbf x_t$ according to \eqref{eq:model} and \eqref{eq:cov_mat}. This requires to simulate the latent state $\bm \xi_t$, the loadings $\bm\Lambda_1$ and $\bm\Lambda_2$, the factors $\mbf f_{1t}$ and $\mbf f_{2t}$, and the idiosyncratic components $\mbf e_t$.

We simulate the latent state $\bm \xi_t$ according to \eqref{eq:trans}, with $\mbf P$ having entries $p_{11}=0.9$ and $p_{22}=0.7$, so that $p_{12}=0.1$ and $p_{21}=0.3$. This configuration corresponds to the unconditional probabilities to be equal to $\p(s_t=1)=\E[\xi_{1t}]=\frac{1-p_{22}}{2-p_{11}-p_{22}}=0.75$ and $\p(s_t=2)=\E[\xi_{2t}]=\frac{1-p_{11}}{2-p_{11}-p_{22}}=0.25$. Then, we generate the innovations $\mbf v_t$ of the VAR in \eqref{eq:trans} as follows: at each given $t$ we generate $u_t\sim \mathcal U[0,1]$ and
\begin{inparaenum}
\item [(i)] if $\xi_{1,t-1}=1$ and $u_t\le p_{11}$ then $\mbf v_t=[1\; 0]^\prime-\mbf P^\prime\bm\xi_{t-1}$;
\item [(ii)] if $\xi_{1,t-1}=1$ and $u_t> p_{11}$ then $\mbf v_t=[0\; 1]^\prime-\mbf P^\prime\bm\xi_{t-1}$;
\item [(iii)] if $\xi_{1,t-1}=0$ and $u_t\le p_{21}$ then $\mbf v_t=[1\; 0]^\prime-\mbf P^\prime\bm\xi_{t-1}$;
\item [(iv)] if $\xi_{1,t-1}=0$ and $u_t> p_{21}$ then $\mbf v_t=[0\; 1]^\prime-\mbf P^\prime\bm\xi_{t-1}$.
\end{inparaenum}

We set the number of factors in each state to $r_j=r=\{1,2\}$,  $j=1,2$. The common component is generated according to model \eqref{eq:model}. Let $\chi_{it}=\bm\lambda_{1i}^\prime \mathbf{f}_{1t}\mathbb I(s_t=1)+\bm\lambda_{2i}^\prime \mathbf{f}_{2t}\mathbb I(s_t=2)$, $i=1,\ldots, N$, $t=1,\ldots, T$. The $r$ entries of $\bm\lambda_{1i}$ and $\bm\lambda_{2i}$ are generated from a $\mathcal N(1,1)$ distribution. The matrices $\bm\Lambda_1$ and $\bm\Lambda_2$ are then transformed in such a way that $\bm\Lambda_1^\prime \bm\Lambda_1$ and $\bm\Lambda_2^\prime \bm\Lambda_2$ are diagonal matrices. The factors are such that $\mbf f_{jt}=\mbf f_t$, $j=1,2$, and satisfy $T^{-1}\sum_{t=1}^T\mbf f_t\mbf f_t^\prime =\mbf I_r$, where each component of $\mbf f_t$ is such that $f_{kt}=\rho_f f_{k,t-1}+z_{kt}$, $k=1,\ldots,r$, with $\rho_f=\{0,0.7\}$ and $z_{kt}\sim \mathcal N(0,1)$.

The idiosyncratic components are generated according to  \eqref{eq:cov_mat}, where $\mbf \Sigma_{je}=\mbf \Sigma_{je,a}+\mbf \Sigma_{je,b}$, $j=1,2$, with $\mbf \Sigma_{je,a}$ diagonal and $\mbf \Sigma_{je,b}$ banded. Specifically, the entries of $\mbf \Sigma_{1e,a}$ are generated from a $\mathcal U[0.25,1.25]$ and those of $\mbf \Sigma_{2e,a}$ are generated from a $\mathcal U[0.75,1.75]$, while $\mbf \Sigma_{1e,b}$ is a Toeplitz matrix with $\tau^{k}$ on the $k$th diagonal for $k=1,2$ and zero elsewhere, and, finally $\mbf \Sigma_{2e,b}$ is a Toeplitz matrix with $\tau^{k-1}$ on the $k$th diagonal for $k=1,2,3$ and zero elsewhere. We set $\tau=\{0,0.5\}$. Moreover, each component of $\bm\nu_t$ is such that $\nu_{it}=\rho_{i}\nu_{i,t-1}+\omega_{it}$, $i=1,\ldots, N$, $t=1,\ldots,T$, with $\rho_i=\{0,\rho\}$ and $\rho\sim\mathcal U[0,0.5]$. Finally, we set the average noise-to-signal ratio across all $N$ simulated time series to be $N^{-1}\sum_{i=1}^N\frac{\sum_{t=1}^T e_{it}^2}{\sum_{t=1}^T \chi_{it}^2}=0.5$.

We simulate the model above 100 times for different values of $r$, $\rho_f$, $\tau$, and $\rho$. 
The EM is run allowing for at most 100 iterations and using a convergence threshold equal to $10^{-6}$. We initialize the algorithm using PCA as described in Section \ref{sec:initialization}. Since the states are identified only up to a permutation at each iteration of the algorithm we assign label 1 to the state with the highest estimated unconditional probability.\footnote{Note that the initialization such that $\omega_1=\omega_2=0.5$ is not empirically feasible, as it leads to no convergence of the EM algorithm. We conjecture that this has to do with the relabelling issue discussed in Section \ref{section:ESSR}, since for $\omega_1=\omega_2=0.5$ both states are equally likely.} 

Results are collected in Tables \ref{tab:Phat1}-\ref{tab:Phat4} and are organised as follows: $\left(i\right)$ $r=1$, $\rho_f=0$, $\tau=0$, $\rho=0$ in Table \ref{tab:Phat1}; $\left(ii\right)$ $r=1$, $\rho_f=0.7$, $\tau=0.5$, $\rho=0.5$ in Table \ref{tab:Phat2}; $\left(iii\right)$ $r=2$, $\rho_f=0$, $\tau=0$, $\rho=0$ in Table \ref{tab:Phat3}; $\left(iv\right)$ $r=2$, $\rho_f=0.7$, $\tau=0.5$, $\rho=0.5$ in Table \ref{tab:Phat4}.

The first four columns of Tables \ref{tab:Phat1}-\ref{tab:Phat4} report the mean and, between brackets, the corresponding standard deviation over all replications of the estimated diagonal entries of the transition matrix $\wh p_{jj}$, $j=1,2$, of the unconditional probabilities $\p(s_t=j)$, estimated as  $\bar{\widehat{\xi}}_{j,t|T}=T^{-1}\sum_{t=1}^T \widehat{\xi}_{j,t|T}$, $j=1,2$.

\begin{table}[t!]
\caption{Simulation results - $r=1$, $\rho_f=0$, $\tau=0$, $\rho=0$.}\label{tab:Phat1}
\vskip .2cm
\small
\centering
\begin{tabular}{ll c c c c | c c | c}
\hline
\hline
\\[-10pt]
$T$ & $N$	&	$\wh p_{11}$ &	$\wh p_{22}$ &  $\bar{\widehat{\xi}}_{1,t|T}$ &  $\bar{\widehat{\xi}}_{2,t|T}$ & $R^2_{B^*}$ & MSE($\chi$) & avg. iter \\
\hline
250	&	100	&			0.89		&			0.64		&			0.76		&			0.24		&	0.97	&	0.02	&	13.78	\\
	&		&	\footnotesize	$(	0.03	)$	&	\footnotesize	$(	0.13	)$	&	\footnotesize	$(	0.06	)$	&	\footnotesize	$(	0.06	)$	&		&		&		\\
500	&	100	&			0.90		&			0.68		&			0.76		&			0.24		&	0.98	&	0.01	&	12.55	\\
	&		&	\footnotesize	$(	0.01	)$	&	\footnotesize	$(	0.04	)$	&	\footnotesize	$(	0.03	)$	&	\footnotesize	$(	0.03	)$	&		&		&		\\
750	&	100	&			0.90		&			0.69		&			0.75		&			0.25		&	0.98	&	0.01	&	12.71	\\
	&		&	\footnotesize	$(	0.01	)$	&	\footnotesize	$(	0.03	)$	&	\footnotesize	$(	0.03	)$	&	\footnotesize	$(	0.03	)$	&		&		&		\\
1000	&	100	&			0.90		&			0.69		&			0.75		&			0.25		&	0.98	&	0.01	&	12.05	\\
	&		&	\footnotesize	$(	0.01	)$	&	\footnotesize	$(	0.03	)$	&	\footnotesize	$(	0.03	)$	&	\footnotesize	$(	0.03	)$	&		&		&		\\
250	&	200	&			0.89		&			0.64		&			0.76		&			0.24		&	0.97	&	0.01	&	11.98	\\
	&		&	\footnotesize	$(	0.02	)$	&	\footnotesize	$(	0.11	)$	&	\footnotesize	$(	0.06	)$	&	\footnotesize	$(	0.06	)$	&		&		&		\\
500	&	200	&			0.89		&			0.68		&			0.75		&			0.25		&	0.97	&	0.01	&	21.23	\\
	&		&	\footnotesize	$(	0.02	)$	&	\footnotesize	$(	0.04	)$	&	\footnotesize	$(	0.03	)$	&	\footnotesize	$(	0.03	)$	&		&		&		\\
750	&	200	&			0.89		&			0.68		&			0.75		&			0.25		&	0.97	&	0.02	&	37.37	\\
	&		&	\footnotesize	$(	0.02	)$	&	\footnotesize	$(	0.04	)$	&	\footnotesize	$(	0.03	)$	&	\footnotesize	$(	0.03	)$	&		&		&		\\
1000	&	200	&			0.90		&			0.69		&			0.75		&			0.25		&	0.98	&	0.02	&	36.22	\\
	&		&	\footnotesize	$(	0.01	)$	&	\footnotesize	$(	0.03	)$	&	\footnotesize	$(	0.03	)$	&	\footnotesize	$(	0.03	)$	&		&		&		\\
\hline
\hline
\end{tabular}
\end{table}

\begin{table}[t!]
\caption{Simulation results - $r=1$, $\rho_f=0.7$, $\tau=0.5$, $\rho=0.5$.}\label{tab:Phat2}
\vskip .2cm
\small
\centering
\begin{tabular}{ll c c c c | c c | c}
\hline
\hline
\\[-10pt]
$T$ & $N$	&	$\wh p_{11}$ &	$\wh p_{22}$ &  $\bar{\widehat{\xi}}_{1,t|T}$ &  $\bar{\widehat{\xi}}_{2,t|T}$ & $R^2_{B^*}$ & MSE($\chi$) & avg. iter \\
\hline
250	&	100	&			0.89		&			0.62		&			0.77		&			0.23		&	0.97	&	0.02	&	20.14	\\
	&		&	\footnotesize	$(	0.03	)$	&	\footnotesize	$(	0.17	)$	&	\footnotesize	$(	0.07	)$	&	\footnotesize	$(	0.07	)$	&		&		&		\\
500	&	100	&			0.90		&			0.68		&			0.76		&			0.24		&	0.98	&	0.02	&	15.28	\\
	&		&	\footnotesize	$(	0.02	)$	&	\footnotesize	$(	0.05	)$	&	\footnotesize	$(	0.04	)$	&	\footnotesize	$(	0.04	)$	&		&		&		\\
750	&	100	&			0.90		&			0.69		&			0.76		&			0.24		&	0.98	&	0.02	&	14.43	\\
	&		&	\footnotesize	$(	0.01	)$	&	\footnotesize	$(	0.03	)$	&	\footnotesize	$(	0.03	)$	&	\footnotesize	$(	0.03	)$	&		&		&		\\
1000	&	100	&			0.90		&			0.66		&			0.77		&			0.23		&	0.98	&	0.01	&	14.07	\\
	&		&	\footnotesize	$(	0.02	)$	&	\footnotesize	$(	0.14	)$	&	\footnotesize	$(	0.05	)$	&	\footnotesize	$(	0.05	)$	&		&		&		\\
250	&	200	&			0.89		&			0.62		&			0.77		&			0.23		&	0.98	&	0.02	&	11.95	\\
	&		&	\footnotesize	$(	0.03	)$	&	\footnotesize	$(	0.14	)$	&	\footnotesize	$(	0.07	)$	&	\footnotesize	$(	0.07	)$	&		&		&		\\
500	&	200	&			0.89		&			0.67		&			0.75		&			0.25		&	0.98	&	0.01	&	20.21	\\
	&		&	\footnotesize	$(	0.02	)$	&	\footnotesize	$(	0.04	)$	&	\footnotesize	$(	0.04	)$	&	\footnotesize	$(	0.04	)$	&		&		&		\\
750	&	200	&			0.89		&			0.69		&			0.75		&			0.25		&	0.98	&	0.01	&	19.17	\\
	&		&	\footnotesize	$(	0.01	)$	&	\footnotesize	$(	0.04	)$	&	\footnotesize	$(	0.02	)$	&	\footnotesize	$(	0.02	)$	&		&		&		\\
1000	&	200	&			0.90		&			0.69		&			0.75		&			0.25		&	0.98	&	0.01	&	21.82	\\
	&		&	\footnotesize	$(	0.01	)$	&	\footnotesize	$(	0.03	)$	&	\footnotesize	$(	0.03	)$	&	\footnotesize	$(	0.03	)$	&		&		&		\\
\hline
\hline
\end{tabular}
\end{table}

\begin{table}[t!]
\caption{Simulation results - $r=2$, $\rho_f=0$, $\tau=0$, $\rho=0$.}\label{tab:Phat3}
\vskip .2cm
\small
\centering
\begin{tabular}{ll c c c c | c c | c}
\hline
\hline\\[-10pt]
$T$ & $N$	&	$\wh p_{11}$ &	$\wh p_{22}$ &  $\bar{\widehat{\xi}}_{t|T,1}$ &  $\bar{\widehat{\xi}}_{t|T,2}$ & $R^2_{B^*}$ & MSE($\chi$) & avg. iter \\
\hline
250	&	100	&			0.88		&			0.46		&			0.81		&			0.19		&	0.97	&	0.04	&	19.32	\\
	&		&	\footnotesize	$(	0.04	)$	&	\footnotesize	$(	0.22	)$	&	\footnotesize	$(	0.08	)$	&	\footnotesize	$(	0.08	)$	&		&		&		\\
500	&	100	&			0.89		&			0.65		&			0.76		&			0.24		&	0.97	&	0.03	&	14.63	\\
	&		&	\footnotesize	$(	0.02	)$	&	\footnotesize	$(	0.04	)$	&	\footnotesize	$(	0.03	)$	&	\footnotesize	$(	0.03	)$	&		&		&		\\
750	&	100	&			0.90		&			0.67		&			0.76		&			0.24		&	0.97	&	0.03	&	14.46	\\
	&		&	\footnotesize	$(	0.01	)$	&	\footnotesize	$(	0.04	)$	&	\footnotesize	$(	0.03	)$	&	\footnotesize	$(	0.03	)$	&		&		&		\\
1000	&	100	&			0.90		&			0.68		&			0.76		&			0.24		&	0.97	&	0.03	&	13.83	\\
	&		&	\footnotesize	$(	0.01	)$	&	\footnotesize	$(	0.03	)$	&	\footnotesize	$(	0.02	)$	&	\footnotesize	$(	0.02	)$	&		&		&		\\
250	&	200	&			0.87		&			0.48		&			0.78		&			0.22		&	0.97	&	0.03	&	13.72	\\
	&		&	\footnotesize	$(	0.04	)$	&	\footnotesize	$(	0.22	)$	&	\footnotesize	$(	0.08	)$	&	\footnotesize	$(	0.08	)$	&		&		&		\\
500	&	200	&			0.89		&			0.65		&			0.75		&			0.25		&	0.97	&	0.02	&	10.40	\\
	&		&	\footnotesize	$(	0.02	)$	&	\footnotesize	$(	0.05	)$	&	\footnotesize	$(	0.04	)$	&	\footnotesize	$(	0.04	)$	&		&		&		\\
750	&	200	&			0.89		&			0.67		&			0.75		&			0.25		&	0.97	&	0.02	&	10.86	\\
	&		&	\footnotesize	$(	0.01	)$	&	\footnotesize	$(	0.04	)$	&	\footnotesize	$(	0.03	)$	&	\footnotesize	$(	0.03	)$	&		&		&		\\
1000	&	200	&			0.90		&			0.68		&			0.75		&			0.25		&	0.97	&	0.01	&	10.81	\\
	&		&	\footnotesize	$(	0.01	)$	&	\footnotesize	$(	0.03	)$	&	\footnotesize	$(	0.02	)$	&	\footnotesize	$(	0.02	)$	&		&		&		\\
\hline
\hline
\end{tabular}
\end{table}

\begin{table}[t!]
\caption{Simulation results - $r=2$, $\rho_f=0.7$, $\tau=0.5$, $\rho=0.5$.}\label{tab:Phat4}
\vskip .2cm
\small
\centering
\begin{tabular}{ll c c c c | c c | c}
\hline
\hline\\[-10pt]
$T$ & $N$	&	$\wh p_{11}$ &	$\wh p_{22}$ &  $\bar{\widehat{\xi}}_{t|T,1}$ &  $\bar{\widehat{\xi}}_{t|T,2}$ & $R^2_{B^*}$ & MSE($\chi$) & avg. iter \\
\hline
250	&	100	&			0.91		&			0.38		&			0.86		&			0.14		&	0.98	&	0.04	&	17.40	\\
	&		&	\footnotesize	$(	0.03	)$	&	\footnotesize	$(	0.20	)$	&	\footnotesize	$(	0.07	)$	&	\footnotesize	$(	0.07	)$	&		&		&		\\
500	&	100	&			0.90		&			0.65		&			0.77		&			0.23		&	0.97	&	0.03	&	20.36	\\
	&		&	\footnotesize	$(	0.02	)$	&	\footnotesize	$(	0.04	)$	&	\footnotesize	$(	0.04	)$	&	\footnotesize	$(	0.04	)$	&		&		&		\\
750	&	100	&			0.90		&			0.67		&			0.76		&			0.24		&	0.97	&	0.03	&	17.20	\\
	&		&	\footnotesize	$(	0.01	)$	&	\footnotesize	$(	0.04	)$	&	\footnotesize	$(	0.03	)$	&	\footnotesize	$(	0.03	)$	&		&		&		\\
1000	&	100	&			0.90		&			0.68		&			0.76		&			0.24		&	0.98	&	0.03	&	16.61	\\
	&		&	\footnotesize	$(	0.01	)$	&	\footnotesize	$(	0.03	)$	&	\footnotesize	$(	0.03	)$	&	\footnotesize	$(	0.03	)$	&		&		&		\\
250	&	200	&			0.89		&			0.41		&			0.83		&			0.17		&	0.97	&	0.03	&	14.55	\\
	&		&	\footnotesize	$(	0.04	)$	&	\footnotesize	$(	0.21	)$	&	\footnotesize	$(	0.09	)$	&	\footnotesize	$(	0.09	)$	&		&		&		\\
500	&	200	&			0.89		&			0.66		&			0.76		&			0.24		&	0.97	&	0.02	&	13.41	\\
	&		&	\footnotesize	$(	0.01	)$	&	\footnotesize	$(	0.06	)$	&	\footnotesize	$(	0.04	)$	&	\footnotesize	$(	0.04	)$	&		&		&		\\
750	&	200	&			0.90		&			0.67		&			0.76		&			0.24		&	0.97	&	0.02	&	14.56	\\
	&		&	\footnotesize	$(	0.01	)$	&	\footnotesize	$(	0.03	)$	&	\footnotesize	$(	0.03	)$	&	\footnotesize	$(	0.03	)$	&		&		&		\\
1000	&	200	&			0.90		&			0.68		&			0.76		&			0.24		&	0.98	&	0.02	&	11.96	\\
	&		&	\footnotesize	$(	0.01	)$	&	\footnotesize	$(	0.03	)$	&	\footnotesize	$(	0.02	)$	&	\footnotesize	$(	0.02	)$	&		&		&		\\
\hline
\hline
\end{tabular}
\end{table}

Since the loadings are not identified, in the fifth column of Tables \ref{tab:Phat1}-\ref{tab:Phat4} we report the multiple $R^2$ coefficient obtained from regressing the columns of $\wh{\mbf B}_1$ onto the columns of $\mbf B_1^*=\mbf B_1 \mathbf{\widehat{I}}_{\mathbf{\widehat{\xi}}1}+ \mbf B_2 (\mbf I_{2r}-\mathbf{\widehat{I}}_{\mathbf{\widehat{\xi}}1})$, thus correcting for the bias described in Theorem \ref{th:asympt_dist}. Namely, we compute 
\[
R^2_{B^*} = \frac{\text{tr}\l\{
\l(\mbf B_1^{*\prime} \wh{\mbf B}_1\r)
\l(\wh{\mbf B}_1^{\prime} \wh{\mbf B}_1\r)^{-1}
\l(\wh{\mbf B}_1^\prime\mbf B_1^{*}\r)
\r\}}
{\text{tr}\l({\mbf B}_1^{*\prime}{\mbf B}_1^{*}\r)}.
\]
The closer this number is to one, the closer is the space spanned by the columns of $\wh{\mbf B}_1$ to the space spanned by the columns of ${\mbf B}_1^*$ (see \citealp{DGRqml}). 

In the sixth column of Tables  \ref{tab:Phat1}-\ref{tab:Phat4} we report the MSE of the estimated common components defined as
\[
\text{MSE}(\chi) = \frac{\sum_{i=1}^N\sum_{t=1}^T (\wh \chi_{it}-\chi_{it})^2}{\sum_{i=1}^N\sum_{t=1}^T \chi_{it}^2},
\]
where $\wh \chi_{it} = \left( \wh{\mathbf{b}}_{1i}~\wh{\mathbf{b}}_{2i}\right)^\prime \left( \wh{\bm{%
	\xi }}_{t}\otimes \wh{\mathbf{g}}_{t}\right)$.

In the last column of Tables \ref{tab:Phat1}-\ref{tab:Phat4} we report the average number of iterations needed for the EM algorithm to converge.

The results in Tables \ref{tab:Phat1}-\ref{tab:Phat4} confirm the empirical validity of the estimation procedure detailed in Section \ref{section:Estimation}. In all four scenarios, as $N$ and $T$ increase the estimators $\wh p_{11}$, $\wh p_{22}$, $\bar{\widehat{\xi}}_{t|T,1}$ and $\bar{\widehat{\xi}}_{t|T,2}$ all converge to the true values of the corresponding parameters. In addition, $R^2_{B^*}$ and MSE($\chi$) are very to $1.00$ and $0.00$, respectively. Finally, note that the average number of iterations declines almost monotonically as $N$ and $T$ increase.

So far, the considered data generating process studies the performance of the proposed EM algorithm when in the model in \eqref{eq:model}-\eqref{eq:cov_mat} the loadings and idiosyncratic covariances are regime specific but the factors and their number do not change. We then consider three more scenarios which we briefly describe here while we refer to Appendix \ref{app:sim} for details on the data generating process and simulation results. 
%(1) 
%(2) the case of constant loadings and idiosyncratic covariances but with factors having regime specific autocorrelation, and (3) the mis-specified case of no Markov switching. .} 

First, we consider the same data generating process as the one considered in this section, but when setting a different number of factors in each regime, specifically, we set $r_1=3$ and $r_2=1$. We the run our EM algorithm initialized by means of PCA using $r_1+r_2=4$ factors. Results show that we correctly estimate the conditional and unconditional probabilities, as well as we correctly retrieve the loadings space (see Tables \ref{tab:numf_change1} and \ref{tab:numf_change2}).

Second, we set $r=r_j=1$, $j=1$, and we let only the autocorrelation of the factors be regime specific, while the loadings and idiosyncratic covariances are constant. In this case the EM algorithm wrongly overestimates the probability of being in the regime with highest simulated probability, thus it does not find evidence of a Markov switching dynamics, but it correctly retrieves the constant loadings space as the PCA estimator would do. Indeed, PCA is known to deliver consistent estimates of the loadings space even when the factors dynamics is piecewise constant \citep{BCF18,duan2022quasi} (see Tables \ref{tab:acf_change1} and \ref{tab:acf_change2}). 

Last, we simulate data from a linear factor model with $r=2$ factors, i.e., when no change is present, but then we fit on the same data our Markov switching model as if there were two regimes. The EM algorithm correctly assigns 97\% probability to one regime at all time periods, i.e., as if there were just one regime (see Tables \ref{tab:no_change1} and \ref{tab:no_change2}).

Overall, our Monte Carlo findings provide evidence in support of the estimation algorithm proposed in Section \ref{section:Estimation}.

%%%%%%%%%%%%%%%%%%%%%%%%%%%%%%%%%%%%%%%%%%%%%%%%%%%%%%%%%%%%%%%%%%%%%%%%%%
%%%%%%%%%%%%%%%%%%%%%%%%%%%%%%%%%%%%%%%%%%%%%%%%%%%%%%%%%%%%%%%%%%%%%%%%%%

\section{Empirical analysis}
\label{section:Empirics}

In this section we show how the methodological framework we propose can be used to model three different large U.S. datasets involving stock returns, macroeconomic time series, and inflation indexes. This is done in Sections \ref{section:Empirics_SR}, \ref{section:Empirics_macro}, and \ref{section:Empirics_inflation}, respectively. For each application, the estimated factors $\wh{\mbf f}_{jt}$, as defined in \eqref{eq:fjhat} for $j=1,2$, are shown in Appendix \ref{app:factors}.

\subsection{Stock returns}
\label{section:Empirics_SR}

This application relates to a vast literature that models stock return dynamics using Markov switching specifications. \cite{Perez_Quiros_Timmermann_2000_JF,Perez_Quiros_Timmermann_2001_JoE} document business cycle asymmetries in U.S. stock returns using decile-sorted portfolios. \cite{Ang_Bekaert_2002_RFS}, and \cite{Guidolin_Timmermann_RFS_2008}, study portfolio allocation in international equity markets under regime switching. In a multi asset setting, \cite{Guidolin_Timmermann_JAE_2006} describe the joint distribution of equity and bonds under regime switching. \cite{Guidolin_2011_Chapter}, and \cite{AT2012}, provide a review of the literature. We contribute to this literature by characterizing stock return dynamics using a Markov switching model in a large dimensional setting. To the very best of our knowledge, we are the first to do so. 

The vector of observable dependent variables $\mathbf{x}_{t}$ in $\left(\ref{eq:model}\right)$ is made of monthly value weighted returns in excess of the risk-free rate from the $N=49$ industry portfolios kindly made publicly available on Kenneth French website.\footnote{See \url{https://mba.tuck.dartmouth.edu/pages/faculty/ken.french/data_library.html}.} Consistently with the discussion in Section \ref{section:Unobs_Hetero}, the unconditional mean of $\mathbf{x}_t$ is equal to $\mathbf{0}$, which means that the returns have been demeaned along the time series dimension over the whole sample period. To obtain a balanced panel, the sample runs from July 1969 through December 2021, a total of $T=630$ time periods.

Using the eigenvalue ratio criterion of \cite{ahnhorenstein13} as applied to the equivalent linear representation in (\ref{eq:linear_model}), we find that the dimension of the vector $\mathbf{g}_t$ is equal to $r_1+r_2=2$ common factors. As commonly assumed in the related literature (see \citealp{AT2012}), we let the number of regimes be equal to two. Therefore, there is one common factor in each regime, so $r_1=r_2=r=1$. Based on this result, we apply the algorithm detailed in Section \ref{section:Estimation}. We stress that, in this case, it is crucial to allow for heteroskedastic idiosyncratic components, namely $\bm\Sigma_{e1}\ne \bm\Sigma_{e2}$ as assumed in the general model specification in \eqref{eq:cov_mat}, since the idiosyncratic components on average account for about 35\% of the total variation in the data. 
Given this set up, the EM algorithm converges in  22 iterations. 

The realisation of the estimator $\wh{\mbf P}$ for the matrix of conditional probabilities $\mbf P$ in (\ref{eq:TransP}) is
\begin{equation*}
\wh{\mbf P}=\l(\begin{array}{cc}
0.9194&	0.0806\\
0.3395&	0.6605\\
\end{array}
\r).
\end{equation*}
The estimated unconditional probability for regime $j$ is equal to the sample average $\bar{\widehat{\xi}}_{j|T}=T^{-1}\sum_{t=1}^{T}\widehat{\xi}_{j,t|T}$, for $j=1,2$. It follows that $\bar{\widehat{\xi}}_{1|T}=0.8044$ and $\bar{\widehat{\xi}}_{2|T}=0.1956$.\footnote{The analytical formulas of the unconditional probabilities in \eqref{eq:UncondP2} give $\bar{\widehat{\xi}}_{1|T}=0.8081$ and $\bar{\widehat{\xi}}_{2|T}=0.1919$.} Therefore, regime $j=1$ is approximately four times more frequent than regime $j=2$. This lead us to label $\wh{\xi}_{2,t|T}$ as the probability of a recession, since expansions occur more often than recessions. 

\begin{figure}[t!]
\caption{Estimated conditional probabilities $\wh{\bm\xi}_{t|T}$ - Stock returns. }\label{fig:xi2FF}
\centering
\begin{tabular}{cc}
\includegraphics[width=.5\textwidth]{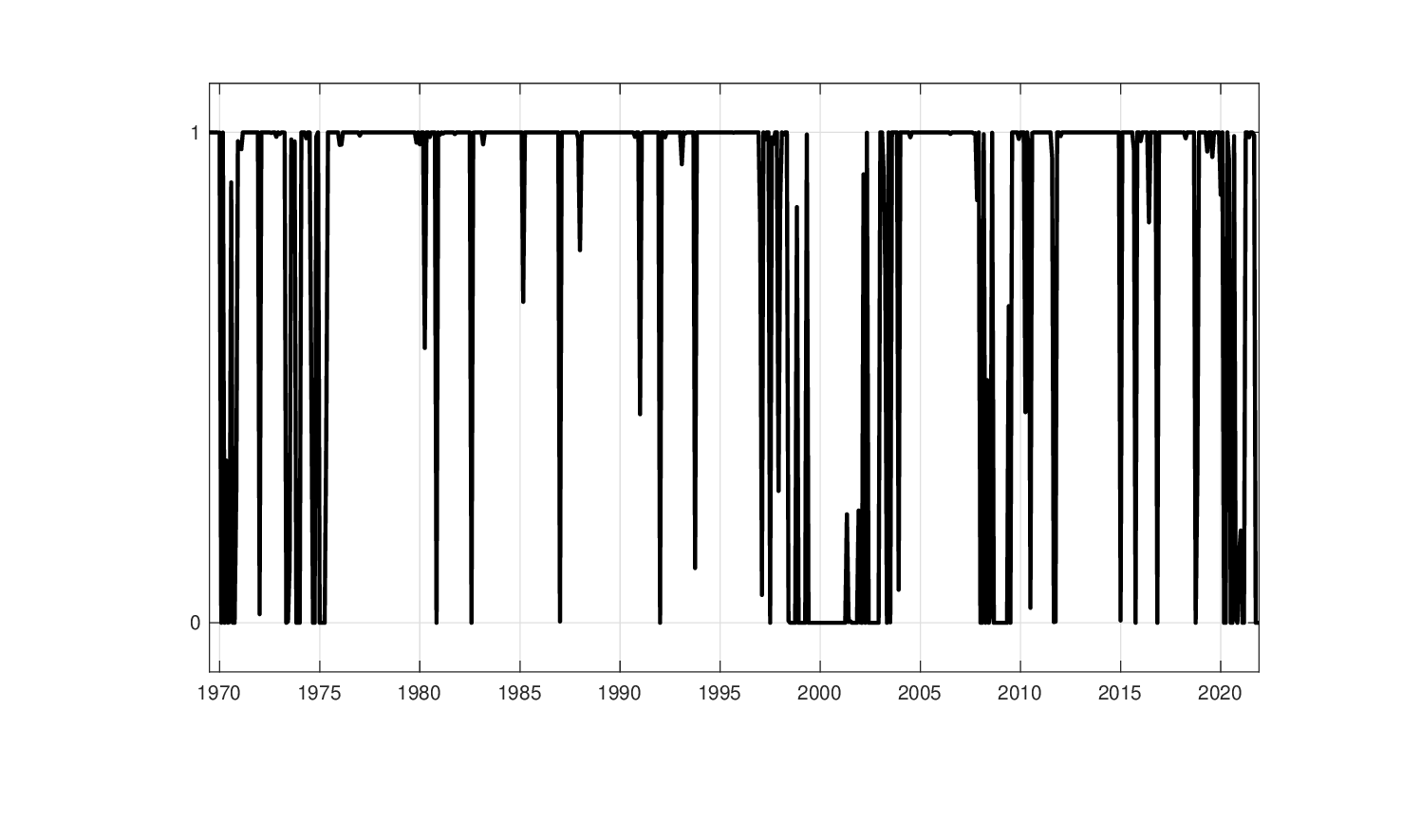}&\includegraphics[width=.5\textwidth]{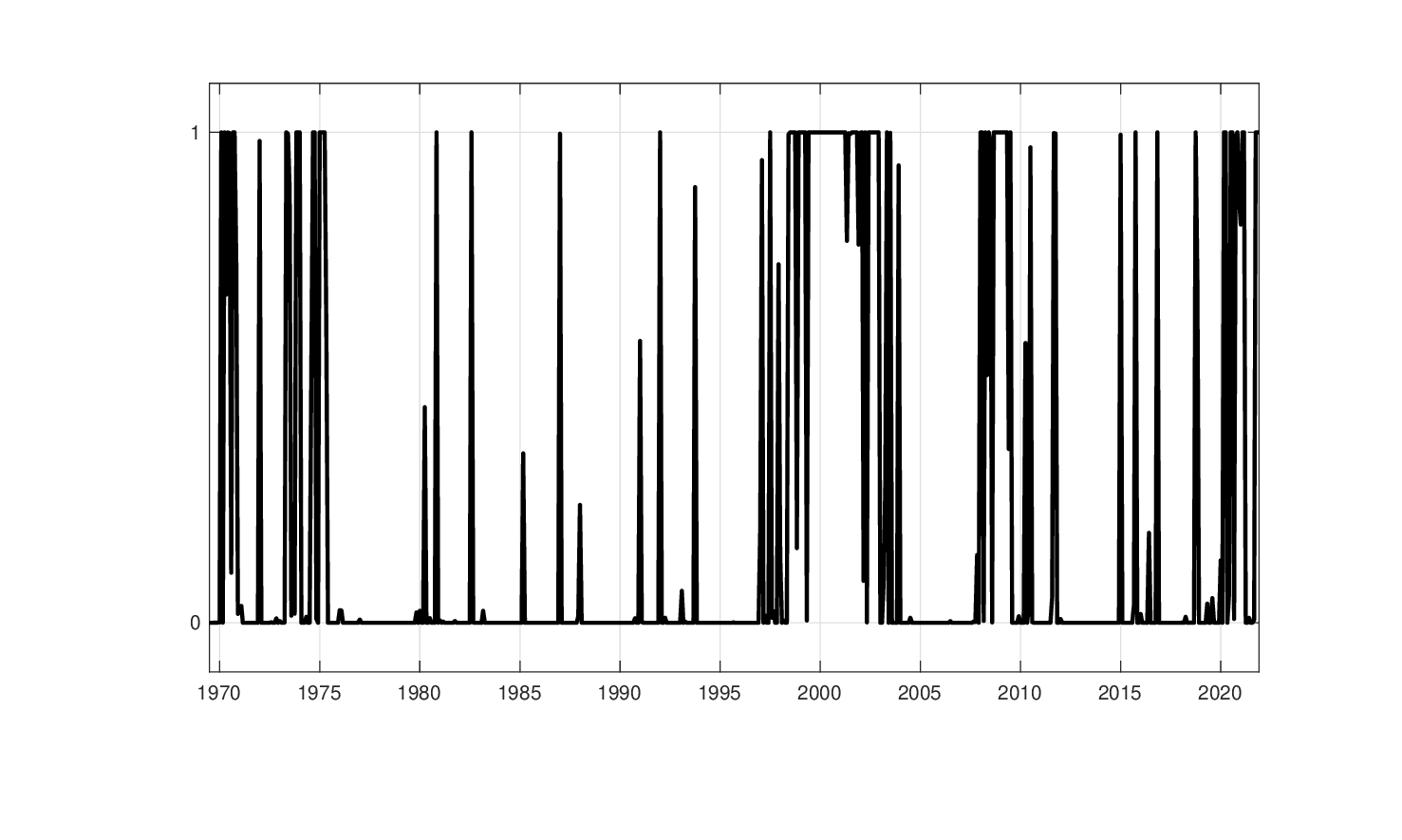}\\[-20pt]
(a): $\wh{\xi}_{1,t|T}$ & (b): $\wh{\xi}_{2,t|T}$
\end{tabular}
\begin{tablenotes}
		\small
		\item This figure plots the series of the estimated conditional probabilities $\wh{\xi}_{1,t|T}$ (panel (a)) and $\wh{\xi}_{2,t|T}$ (panel (b)), for $t=1,\ldots,T$, estimated from the Markov switching factor model in \eqref{eq:state_space_mes} for the stock returns dataset.
	\end{tablenotes}
\end{figure}

Figure \ref{fig:xi2FF} plots the sequences of estimates $\wh{\xi}_{1,t|T}$ and $\wh{\xi}_{2,t|T}$, for $t=1,\ldots,T$. 
In order to provide economic understanding of the regimes described by the model, we define the estimated recession indicator $\wh{REC}_t$ as being equal to one if $\widehat{\xi}_{2,t|T}\geq 0.5$ and to zero otherwise. Formally, this means that $\wh{REC}_t=\mathbb{I}\left(\widehat{\xi}_{2,t|T}\geq 0.5\right)$. Note that $\wh{REC}_t$ has correlation equal to $0.99$ with $\wh{\xi}_{2,t|T}$, which suggests that the underlying states are precisely estimated.  We then follow \citet{harding2006synchronization} and compute the degree of concordance between the estimated recession indicator and the NBER recession indicator, denoted as $REC_t$.\footnote{The NBER recession indicator is publicly available at \url{https://fred.stlouisfed.org/series/USREC}.} The degree of concordance is given by
\beq\label{eq:DOC}
DoC=T^{-1}\sum_{t=1}^T \l\{\wh{REC}_t\, REC_t+ (1-\wh{REC}_t)\, (1-REC_t)\r\}.
\eeq 
For the dataset of stock returns we consider, we have $DoC=0.8048$. We also compute the probabilities of misclassification, which are given by $FP=T^{-1}\sum_{t=1}^T \wh{REC}_t\, (1-REC_t)$ (namely, the frequency of false positives) and $FN=T^{-1}\sum_{t=1}^T (1-\wh{REC}_t)\, REC_t$ (namely, the frequency of false negatives). We obtain $FP=0.1286$ and $FN=0.0667$. Therefore, the state $j=1$ is related to periods of economic expansions, whereas the state $j=2$ is more likely to occur during recessionary phases. Our model therefore captures regime changes in equity markets related to business cycle dynamics.  

We then turn to the estimated factors. Since $r_1=r_2$, the estimators for $\bm\Lambda_j$, for $j=1,2$, are readily available from \eqref{eq:Ljhat} or \eqref{eq:Ljtilde}. Next, by projecting the data onto the estimated loadings weighted by the probability of being in a given state, we obtain the estimated scalar factors $\widehat{{f}}_{jt}$ and $\wt{{f}}_{jt}$, for $j=1,2$ and $t=1,\ldots,T$, as given in \eqref{eq:fjhat} and \eqref{eq:fjtilde}, respectively. 

Table \ref{tab:Corr2} displays the correlations between the estimated latent factors and the six observable factors considered in \cite{FF2016RFS}, namely: the value-weighted return on the market portfolio in excess of the one-month Treasury bill rate ($RM_t$); size ($SMB_t$); value ($HML_t$); profitability ($RMW_t$); investment ($CMA_t$); momentum ($MOM_t$).  These correlations are computed both over the whole sample period, as well as within regimes. These in turn are defined in two ways: through the NBER recession indicator $REC_{t}$ (Panel A); through the predicted NBER recession indicator $\wh{REC}_t$ previously defined (Panel B). The results in Table \ref{tab:Corr2} show that, over the whole sample period, $\widehat{{f}}_{1t}$ is strongly correlated with $RM_t$, and reasonably correlated with $SMB_t$, $HML_t$ and $CMA_t$. The estimate $\widehat{{f}}_{2t}$  is correlated with $MOM_t$. A similar picture comes from $\wt{{f}}_{1t}$ and $\wt{{f}}_{2t}$. When we compute the correlations during NBER expansions and recessions, additional findings arise (Panel A). On one hand, in expansionary periods, the correlations between $\widehat{{f}}_{1t}$ and $\wt{{f}}_{1t}$, and $RM_t$, $SMB_t$, $HML_t$ and $CMA_t$, are similar to those computed over the whole sample period. On the other hand, $\widehat{{f}}_{2t}$ and $\wt{{f}}_{2t}$ display sizeable correlations in recession with $SMB_{t}$ and $HML_{t}$, as well as with $MOM_{t}$. The homologous correlations calculated for the regime $j=2$ identified by the model are generally of lower magnitude, with the exception of those related to $MOM_{t}$ (Panel B). This confirms that $f_{2t}$ is a factor that drives the cross-section of equity returns during macroeconomic recessionary periods. Whereas a linear factor model would not be able to uncover this feature, our model can detect these asymmetric dynamics. This shows the empirical usefulness of our framework to model large dimensional portfolios of financial assets.

\begin{table}[t!]
	\caption{Factor correlations - Stock returns.}\label{tab:Corr2}
	\vskip .2cm
	{\small
	\centering
	\begin{tabular}{lrr|rr|rr|rr|rr|rr}
		\hline\hline
		\multicolumn{13}{c}{Panel A: NBER Regimes} \\ \hline
		& \multicolumn{4}{c|}{Whole Sample} & \multicolumn{4}{c|}{Expansions} & 
		\multicolumn{4}{c}{Recessions} \\ 
		& $\widehat{f}_{1t}$ & $\widehat{f}_{2t}$ & $\widetilde{f}_{1t}$ & $%
		\widetilde{f}_{2t}$ & $\widehat{f}_{1t}$ & $\widehat{f}_{2t}$ & $\widetilde{f%
		}_{1t}$ & $\widetilde{f}_{2t}$ & $\widehat{f}_{1t}$ & $\widehat{f}_{2t}$ & $%
		\widetilde{f}_{1t}$ & $\widetilde{f}_{2t}$ \\ \hline
		$RM_{t}$ & 0.74 & 0.01 & 0.74 & -0.02 & 0.80 & 0.04 & 0.80 & 0.01 & 0.55 & 
		-0.06 & 0.55 & -0.08 \\ 
		$SMB_{t}$ & 0.32 & 0.07 & 0.32 & -0.03 & 0.32 & -0.08 & 0.32 & -0.16 & 0.34
		& 0.38 & 0.34 & 0.27 \\ 
		$HML_{t}$ & -0.17 & 0.06 & -0.17 & 0.06 & -0.14 & -0.09 & -0.14 & -0.03 & 
		-0.30 & 0.33 & -0.30 & 0.26 \\ 
		$RMW_{t}$ & -0.06 & 0.02 & -0.06 & 0.10 & -0.09 & 0.06 & -0.10 & 0.16 & 0.14
		& -0.07 & 0.14 & -0.03 \\ 
		$CMA_{t}$ & -0.22 & -0.06 & -0.13 & -0.01 & -0.17 & -0.11 & -0.17 & -0.04 & 
		-0.41 & 0.05 & -0.40 & 0.04 \\ 
		$MOM_{t}$ & -0.01 & -0.23 & -0.01 & -0.17 & 0.01 & -0.18 & 0.01 & -0.15 & 
		-0.09 & -0.32 & -0.09 & -0.21 \\ 
		\hline
		\hline
		\multicolumn{13}{c}{} \\ \hline\hline
		\multicolumn{13}{c}{Panel B: Model Regimes} \\ \hline
		& \multicolumn{4}{c|}{Whole Sample} & \multicolumn{4}{c|}{$j=1$} & 
		\multicolumn{4}{c}{$j=2$} \\ 
		& $\widehat{f}_{1t}$ & $\widehat{f}_{2t}$ & $\widetilde{f}_{1t}$ & $%
		\widetilde{f}_{2t}$ & $\widehat{f}_{1t}$ & $\widehat{f}_{2t}$ & $\widetilde{f%
		}_{1t}$ & $\widetilde{f}_{2t}$ & $\widehat{f}_{1t}$ & $\widehat{f}_{2t}$ & $%
		\widetilde{f}_{1t}$ & $\widetilde{f}_{2t}$ \\ \hline
		$RM_{t}$ & 0.74 & 0.01 & 0.74 & -0.02 & 0.97 & 0.04 & 0.97 & 0.05 & 0.21 & 
		0.01 & 0.21 & -0.03 \\ 
		$SMB_{t}$ & 0.32 & 0.07 & 0.32 & -0.03 & 0.44 & 0.00 & 0.44 & 0.00 & 0.14 & 
		0.11 & 0.14 & -0.03 \\ 
		$HML_{t}$ & -0.17 & 0.06 & -0.17 & 0.06 & -0.23 & -0.08 & -0.22 & -0.07 & 
		-0.10 & 0.10 & -0.10 & 0.11 \\ 
		$RMW_{t}$ & -0.06 & 0.02 & -0.06 & 0.10 & -0.10 & -0.07 & -0.10 & -0.08 & 
		0.02 & 0.04 & 0.02 & 0.15 \\ 
		$CMA_{t}$ & -0.22 & -0.06 & -0.13 & -0.01 & -0.29 & -0.08 & -0.29 & -0.07 & 
		-0.16 & -0.05 & -0.16 & 0.01 \\ 
		$MOM_{t}$ & -0.01 & -0.23 & -0.01 & -0.17 & -0.01 & 0.00 & -0.02 & -0.01 & 
		-0.05 & -0.31 & -0.05 & -0.23 \\ 
		\hline
		\hline
	\end{tabular}%
	\vskip .1cm
	
	\begin{tablenotes}
		\small
		\item This table reports the correlation coefficients between the estimated factors $\wh f_{1t}$, $\wh f_{2t}$, $\wt f_{1t}$, and $\wt f_{2t}$ obtained from the Markov switching factor model in (\ref{eq:model}) according to \eqref{eq:fjhat} and \eqref{eq:fjtilde}, and the following six observable factors from \cite{FF2016RFS}: the value-weighted return on the market portfolio in excess of the one-month Treasury bill rate ($RM_t$); size ($SMB_t$); value ($HML_t$); profitability ($RMW_t$); investment ($CMA_t$); momentum ($MOM_t$). Correlations are computed over the whole sample period, as well as during: $(i)$ expansions and recessions as identified through the NBER recession indicator (Panel A); $(ii)$ regimes $j=1$ and $j=2$, where regime $j$ occurs at time $t$ if $\wh{\xi}_{j,t|T} \geq 0.5$ (Panel B).
	\end{tablenotes}
	}
\end{table}

\subsection{Macroeconomic time series}
\label{section:Empirics_macro}

We now apply our methodology to a large set of macroeconomic variables to measure the probability of recessions and expansions in the U.S. economy. This relates our work to a large literature on business cycle dating, which goes back to the pioneering work of \citet{Burns_Mitchell_1946_measuring}: see \citet{Romer_Romer_2020_WP} for a recent discussion of the topic. We follow \cite{hamilton1989new}, \citet{diebold78measuring}, and \citet{Chauvet_1998_IER}, in employing a Markov switching approach. In the spirit of \citet{Stock_Watson_2014_JoE}, we use a large set of time series data to estimate recession and expansion probabilities. Finally, we study the ability of our model in dating turning points both using the full-sample and in real-time in a spirit similar to  \citet{Chauvet_Piger_2008}.

Formally, the vector of observable dependent variables $\mathbf{x}_{t}$ in $\left(\ref{eq:model}\right)$ is made of the monthly macroeconomic dataset FRED-MD described by \citet{mccracken2016fred}
formed of $N=126$ times series covering both the real and nominal sectors of the U.S. economy and including also labor market indicators, and financial variables.\footnote{See \url{https://research.stlouisfed.org/econ/mccracken/fred-databases/}.} The data is transformed to stationarity and missing values are imputed by means of the routines made available by \citet{mccracken2016fred}, which
produce a balanced panel, with a sample running from April 1959 through March 2024, for a total of $T=780$ time periods.

Using the information criterion of \cite{baing02} as applied to the equivalent linear representation in (\ref{eq:linear_model}), we find that the dimension of the vector $\mathbf{g}_t$ is equal to $r_1+r_2=8$ common factors. As commonly assumed in the literature  \citep{Romer_Romer_2020_WP}, we consider two regimes. Therefore, under the assumption that the number of factor is the same across states, there are four common factors in each regime, namely $r_1=r_2=r=4$. We then apply the algorithm detailed in Section \ref{section:Estimation}. We further impose homoskedastic idiosyncratic components, namely $\bm\Sigma_{e1}=\bm\Sigma_{e2}$. This is because, in the dataset in use, idiosyncratic components are often negligible, explaining on average less than 10\% of the total variation of real variables \citep{BN06}.\footnote{Results with heteroskedastic idiosyncratic components are similar and available upon request.} In this set up, the EM algorithm converges in  12 iterations. 

The estimate of the matrix of conditional probabilities $\mbf P$ in (\ref{eq:TransP}) is equal to
\begin{equation*}
\wh{\mbf P}=\l(\begin{array}{cc}
0.9576&	0.0424\\
0.1399&	0.8601\\
\end{array}
\r).
\end{equation*}
The estimated unconditional probabilities are $\bar{\widehat{\xi}}_{1|T}=0.8354$ and $\bar{\widehat{\xi}}_{2|T}=0.1646$.\footnote{The analytical formulas in \eqref{eq:UncondP2} give unconditional probabilities equal to $\bar{\widehat{\xi}}_{1|T}=0.8362$ and $\bar{\widehat{\xi}}_{2|T}=0.1638$.} In this sample, the unconditional probability of a recession, as measured by the NBER recession indicator, is 0.1218. Therefore, we can identify regime $j=2$ as the recession regime.

\begin{figure}[t!]
\caption{Estimated conditional probabilities $\wh{\bm\xi}_{t|T}$ - Macroeconomic time series. }\label{fig:xi2FRED}
\centering
\begin{tabular}{cc}
\includegraphics[width=.5\textwidth]{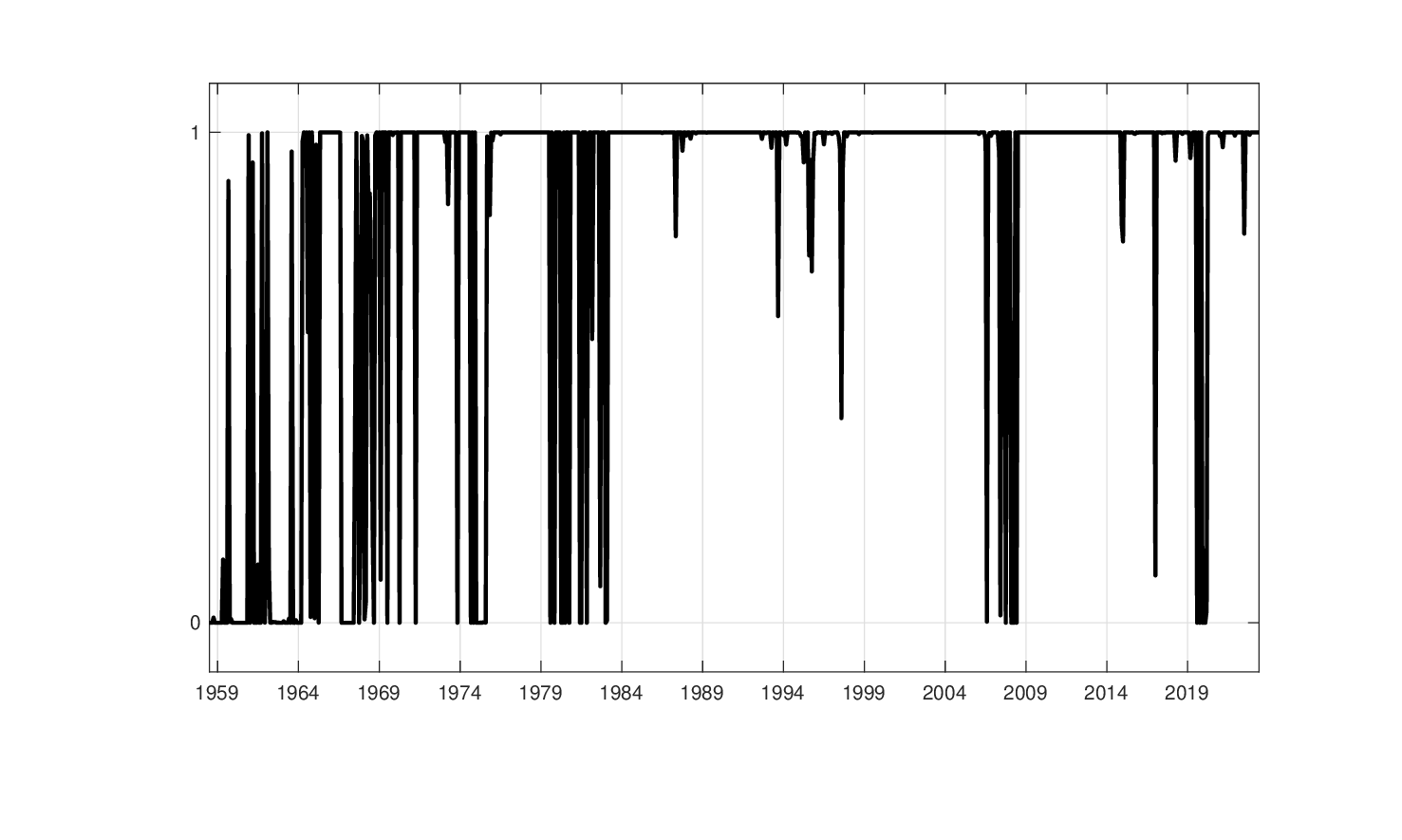}&\includegraphics[width=.5\textwidth]{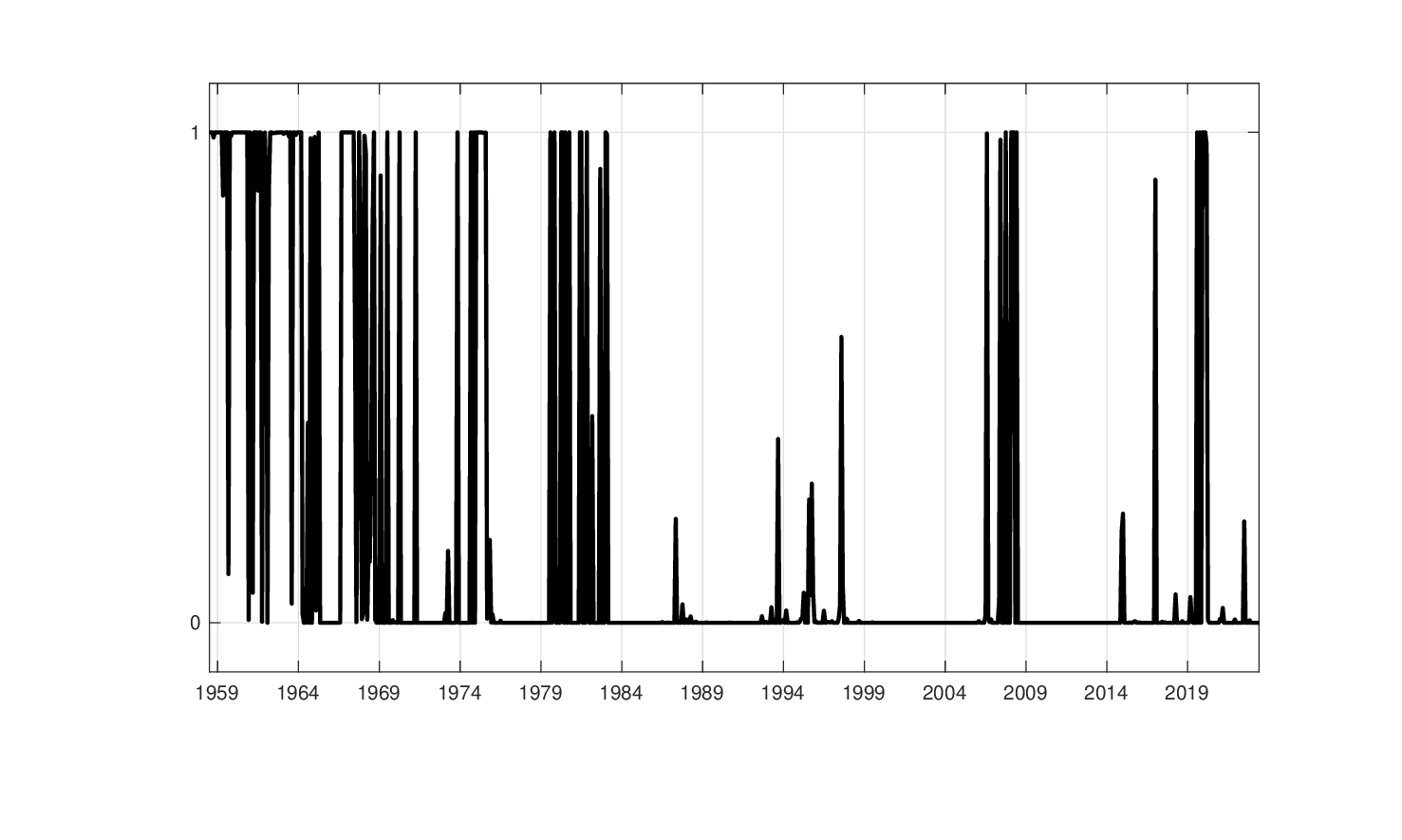}\\[-20pt]
(a): $\wh{\xi}_{1,t|T}$ & (b): $\wh{\xi}_{2,t|T}$
\end{tabular}
\begin{tablenotes}
		\small
		\item This figure plots the series of the estimated conditional probabilities $\wh{\xi}_{1,t|T}$ (panel (a)) and $\wh{\xi}_{2,t|T}$ (panel (b)), for $t=1,\ldots,T$, estimated from the Markov switching factor model in \eqref{eq:state_space_mes} for the macroeconomic time series dataset.
	\end{tablenotes}
\end{figure}

Figure \ref{fig:xi2FRED} plots the sequences of estimates $\wh{\xi}_{1,t|T}$ and $\wh{\xi}_{2,t|T}$, for $t=1,\ldots,T$. The two most recent main recessions, which are due to the Great Financial Crisis (2007-2009) and the Covid19 pandemic (2020-2021), are well captured. To quantify the performance of our model, we once again follow \citet{harding2006synchronization} and compute the degree of concordance $DoC$ in \eqref{eq:DOC} between the estimated recession indicator $\widehat{REC}_t$ defined as in Section \ref{section:Empirics_SR}, and the NBER recession indicator. We obtain $DoC=0.7718$, with frequency of false positives and false negatives equal to $FP=0.1333$ and $FN=0.0949$, respectively. All these measures show the goodness of our method to ex-post dating business cycle turning points.

Turning to real-time dating of turning points, for each month, starting from February 1980 up to March 2024, we re-estimate our model from April 1959 up to that month and compute the filtered probability of recession, $\widehat{\xi}_{1,t|t}$ as given in \eqref{eq:Hamiltonmain}, for the last observation in the considered sample. So our first prediction is for February 1980. This is the same approach as \citet{Urga_Wang_2022_WP} with two main differences. First, our indicator of recessions is very stable meaning that  most of the times our indicator is equal either 0 or 1 and a thresholding procedure is seldom needed. Second, we do not use a sub-set of the $N$ series but include all of them.
In Table \ref{tab:tpdate}, we report the time delay of our method in detecting turning points as defined by the NBER recession indicator $REC_t$. We compare our results with those reported by \citet{Urga_Wang_2022_WP}. A negative delay means that we anticipate the turning point. Our method predicts well the starting of recessions sometimes with a smaller delay than its competitors, while it tends to underestimate their duration, thus anticipating the end of recessions and resulting in a negative delay in predicting expansions.

\begin{table}[t!]
	\caption{Out of sample turning points detection.}\label{tab:tpdate}
	\vskip .2cm
	{\small
	\centering
	\begin{tabular}{l | c c c c c c}
	\hline
	\hline
	& Recession & Expansion& Recession & Expansion & Recession & Expansion\\
	&Feb-80
	&Aug-80	
	&Aug-81
	&Nov-82
	&Aug-90
	&Apr-91\\
	\hline
	\citet{Chauvet_Piger_2008} 	&6&5&7&6&7&6\\
	\citet{Urga_Wang_2022_WP}	&3&2&3&7&NA&1\\
	This paper				&1&-1&4&-7&NA&NA\\
	\hline
	&Recession & Expansion & Recession & Expansion& Recession & Expansion\\
&Apr-01
&Dec-01
&Jan-08
&Jul-09
&Mar-20
&May-20\\
	\hline
	\citet{Chauvet_Piger_2008}	&10&7&13&7&0&-1\\
	\citet{Urga_Wang_2022_WP}	&8&7&11&10&0&4\\
	This paper				&6&2&9&-4&1&2\\
	\hline
	\end{tabular}
		\vskip .1cm
		
		\begin{tablenotes}
		\small
		\item This table reports the delay in detecting turning points for the methods proposed by \citet{Chauvet_Piger_2008},
	\citet{Urga_Wang_2022_WP}, and this paper. Negative delays mean the date of the turning point is predicted earlier than the true one. Delays for the method by \citet{Chauvet_Piger_2008} are taken from Table 2 in \citet{Urga_Wang_2022_WP} with the exception of the last recession and expansion turning  points for which the delay is computed using the smoothed recession probability indicator available at \url{https://fred.stlouisfed.org/series/RECPROUSM156N}.
	\end{tablenotes}
	}
\end{table}

\subsection{Inflation indexes}
\label{section:Empirics_inflation}

In the last application, we consider a panel of $N=142 $   U.S. disaggregated Personal Consumption Expenditure (PCE) price monthly inflation rates from February 1959 to December 2023, for a total of $T=779$ time periods. The dataset is built as described in \citet{AL}, who analyze the same data by means of a time-varying linear dynamic factor model allowing for both short and long memory dynamics. They show evidence of a structural change in the mid/end-1980s or even mid-1990s, depending on the size of the moving window considered; using the \citet{hallinliska07} information criterion, they find evidence of one factor before and after the change-point.

In Section \ref{sec:linearFM} we discussed that the model in \eqref{eq:state_space_mes} admits the same equivalent linear representation as a model with one change point. We then apply the algorithm detailed in Section \ref{section:Estimation} with two regimes and one common factor in each regime, namely $r_1=r_2=r=1$. Note that, in this application, it is crucial to allow for heteroskedastic idiosyncratic components, namely with $\bm\Sigma_{e1}\ne \bm\Sigma_{e2}$, as assumed in the general specification of our model in \eqref{eq:cov_mat}: in this case, idiosyncratic components on average account for about 80\% of the total variation in the data. The EM algorithm converges in 10 iterations.

\begin{figure}[t!]
\caption{Estimated conditional probabilities $\wh{\bm\xi}_{t|T}$ - Inflation indexes. }\label{fig:xi2PCE}
\centering
\begin{tabular}{cc}
\includegraphics[width=.5\textwidth]{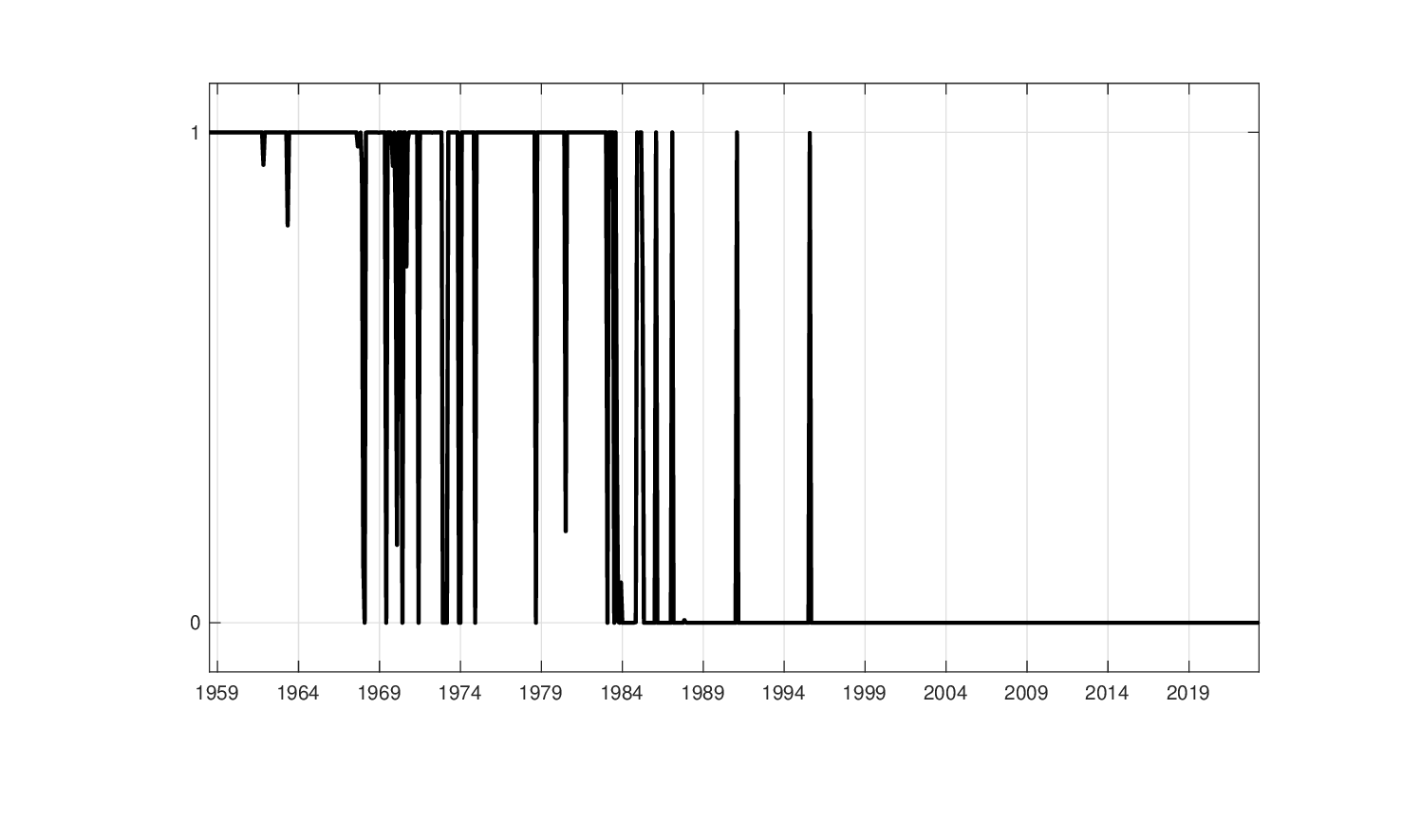}&\includegraphics[width=.5\textwidth]{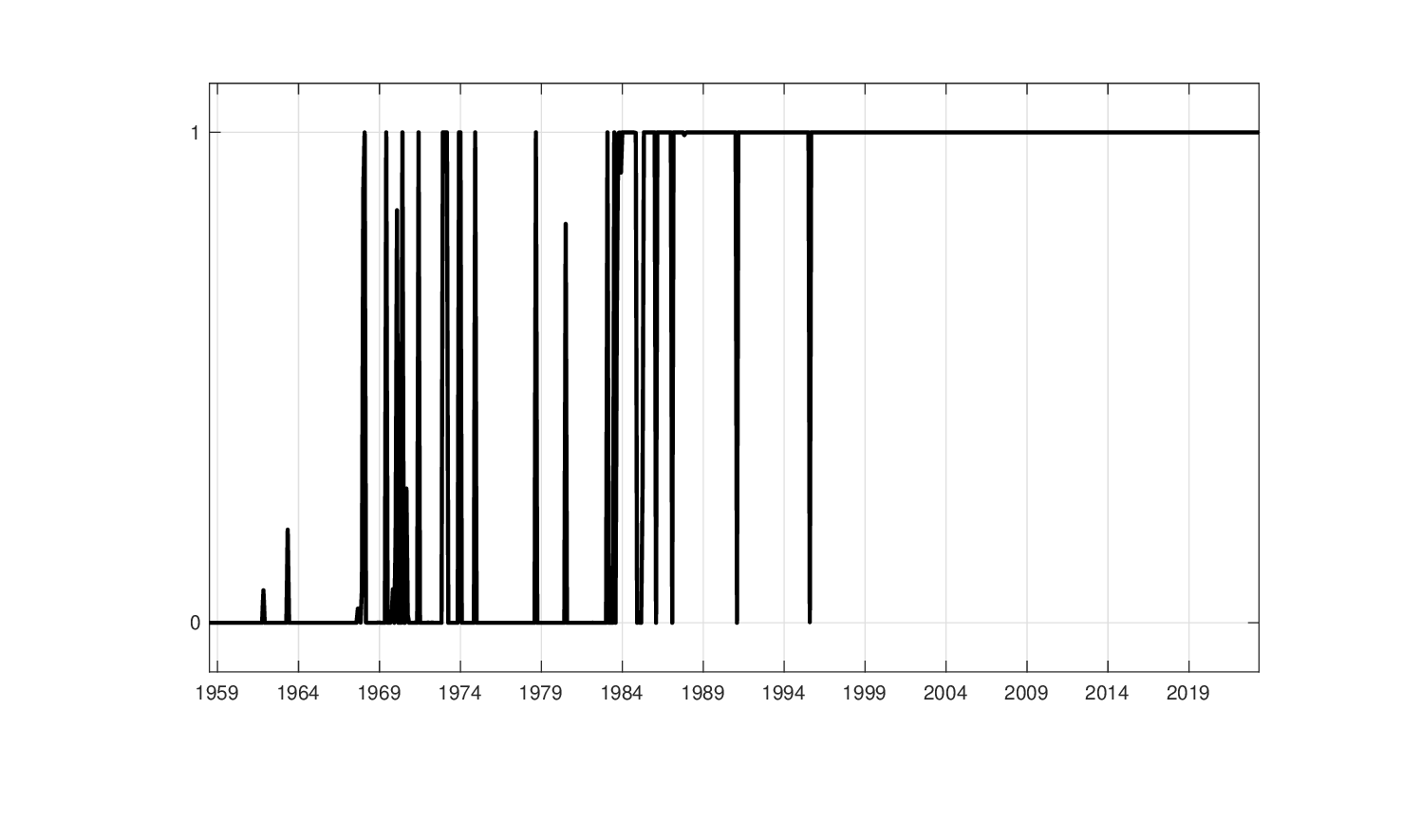}\\[-20pt]
(a): $\wh{\xi}_{1,t|T}$ & (b): $\wh{\xi}_{2,t|T}$
\end{tabular}
\begin{tablenotes}
		\small
		\item This figure plots the series of the estimated conditional probabilities $\wh{\xi}_{1,t|T}$ (panel (a)) and $\wh{\xi}_{2,t|T}$ (panel (b)), for $t=1,\ldots,T$, estimated from the Markov switching factor model in \eqref{eq:state_space_mes} for the inflation indexes dataset.
	\end{tablenotes}
\end{figure}

The estimate of the matrix of conditional probabilities $\mbf P$ in (\ref{eq:TransP}) is equal to
\begin{equation*}
%\beq
\wh{\mbf P}=\l(\begin{array}{cc}
0.9368&	0.0632\\
0.0449&	0.9551\\
\end{array}
\r).
%\eeq
\end{equation*}
The estimated unconditional probabilities are $\bar{\widehat{\xi}}_{1|T}=0.3770$ and $\bar{\widehat{\xi}}_{2|T}=0.6230$.\footnote{The analytical formulas in \eqref{eq:UncondP2} give $\bar{\widehat{\xi}}_{1|T}=0.4154$ and $\bar{\widehat{\xi}}_{2|T}=0.5846$.} By just looking at these numbers, it may seem hard to interpret the two regimes. However, by plotting  $\wh{\xi}_{1,t|T}$ and $\wh{\xi}_{2,t|T}$ as in Figure \ref{fig:xi2PCE}, we immediately see that, from March 1996 onwards, regime $j=2$ occurs with probability one in all time periods. Therefore, this regime can be identified with the most recent part of the sample. On the other hand, in the first part of the sample regime $j=1$ is often the most likely to occur. This finding is consistent with the results in \citet{AL}: they show that the first part of the sample, in which regime $j=1$ is more likely to happen, is characterized by periods of high volatility and long memory, namely by persistent dynamics; conversely, the second part of the sample, which corresponds to regime $j=2$, is characterized by low volatility and short memory, namely by fast mean reversion. More generally, this shows that our model can also be used as a starting point to model stochastic breaks in large dimensional factor models, in the spirit of \citet{Chib_1998_JoE}. 
%\textcolor{red}{This shows that our methodology is able to capture not only changes in volatility due to changes in the loadings and the idiosyncratic variances, but, indirectly also changes due to time-varying autocorrelation in the factors, as the latter also induces changes in the overall volatility of the observed data. NON SONO SICURO CHE CONVENGA METTERE QUESTA PARTE. SI RICOLLEGA ALLA SIMULAZIONE NUMERO 4 RICHIESTA DAL REFEREE. IN BASE ALLE NOSTRE IPOTESI (IN PARTICOLARE, ASSUMPTION 1), LE DINAMICHE DEI FATTORI NON DOVREBBERO INFLUENZARE I REGIMI. EMPIRICAMENTE, FORSE QUELLO CHE SUCCEDE E' CHE CI SONO DEI CAMBIAMENTI NEL LOADINGS E AL TEMPO STESSO CAMBIANO LE DINAMICHE DEI FATTORI, MA SOLO I PRIMI PERMETTONO DI IDENTIFICARE I REGIMI}

%%%%%%%%%%%%%%%%%%%%%%%%%%%%%%%%%%%%%%%%%%%%%%%%%%%%%%%%%%%%%%%%%%%%%%%%%%
%%%%%%%%%%%%%%%%%%%%%%%%%%%%%%%%%%%%%%%%%%%%%%%%%%%%%%%%%%%%%%%%%%%%%%%%%%

\section{Concluding remarks}
\label{section:Conclusions}

This paper develops estimation and inferential theory for high dimensional factor models with discrete regime changes in the loadings driven by a latent first order Markov process. Our estimator employs a EM algorithm based on a modified version of the Baum-Lindgren-Hamilton-Kim filter and smoother. Remarkably, the estimator does not need knowledge of the number of factors in either states. It only requires the true number of factors in the equivalent linear representation, which can be estimated using existing techniques. We derive convergence rates and asymptotic distributions of the estimators for factors and loadings, and we show their good finite sample performance through an extensive set of Monte Carlo experiments. Finally, we empirically validate our methodology through three applications to large U.S. datasets of stock returns,  macroeconomic variables, and inflation indexes.

Our work can be extended along several dimensions. Two are worth mentioning. Our model allows for two regimes and the case of multiple states to capture richer dynamics is worth exploring. The challenging task of making inference on the number of regimes is also worth considering. These extensions are part of our ongoing research agenda and will be studied in future work.

%%%%%%%%%%%%%%%%%%%%%%%%%%%%%%%%%%%%%%%%%%%%%%%%%%%%%%%%%%%%%%%%%%%%%%%%%%
%%%%%%%%%%%%%%%%%%%%%%%%%%%%%%%%%%%%%%%%%%%%%%%%%%%%%%%%%%%%%%%%%%%%%%%%%%
{\small {\setlength{\bibsep}{.2cm} 
		\bibliographystyle{chicago}
		\bibliography{BM_Biblio}
}}

%%%%%%%%%%%%%%%%%%%%%%%%%%%%%%%%%%%%%%%%%%%%%%%%%%%%%%%%%%%%%%%%%%%%%%%%%%
\appendix
\small
\setcounter{equation}{0}
\numberwithin{equation}{section}
\numberwithin{figure}{section}
\numberwithin{table}{section}

\section{Details of estimation\label{Appendix:Est}}

\subsection{Baum-Lindgren-Hamilton-Kim filter}\label{Appendix:Est:filter}

For simplicity of notation, in this appendix we will consider both the factors $\{\mbf g_t\}_{t=1}^T$ and the true values of the parameters $\mbf q$ to be known. To simplify notation, let $\bm{\varepsilon }_{1}=\left[ 1~0\right]^{\prime }$ and $\bm{\varepsilon }_{2}=\left[ 0~1\right]^{\prime }$, so that $\p(s_t=j)\equiv \p(\bm\xi_t=\bm\varepsilon_j)$, $j=1,2$, and therefore, in the following, we can just use  $\bm\xi_t$ as defined in \eqref{eq:xidef}, without the need of referring also to $s_t$.  Then, for any $v=1,\ldots, T$, we use the notation
\beq\label{eq:notationfilter}
\bm{\xi }_{t\left\vert v\right. }= \E\left[ \bm{\xi }_{t}\left\vert \bm{X}_{v}\right. \right]=\l[\begin{array}{c}
\p\left( \bm{\xi }_{t}=\bm{\varepsilon }_{1}\left\vert \bm{X}_{v}\right. \right)\\
\p\left( \bm{\xi }_{t}=\bm{\varepsilon }_{2}\left\vert \bm{X}_{v}\right. \right)
\end{array}
\r].
\eeq
Notice also that, since $\{\bm\xi_t\}_{t=1}^u$ is independent of $\bm G_v$ for all $u,v=1,\ldots, T$, because we consider the factors as observed, we can always write $\bm{\xi }_{t\left\vert v\right. }=\E\left[ \bm{\xi }_{t}\left\vert \bm{X}_{v}\right. \right]=\E\left[ \bm{\xi }_{t}\left\vert \mathbf{X}_{v},\bm G_v\right. \right]$.	
	
The one-step-ahead predictions and the filtered probabilities are computed by means of the following steps which are similar to the Hamilton filter, see, e.g., \citet[Chapter 5.1]{krolzig2013markov} and \citet{hamilton1989new}.

Then, the one-step-ahead predicted probabilities are obtained through the prior probability 
\begin{align}
\p\left( \bm{\xi }_{t}=\bm{\varepsilon }_{i}\left\vert \bm{X}_{t-1},\bm G_{t-1}\right. \right) &=\sum\limits_{j=1}^{2}\p \left( \bm{\xi }_{t}=\bm{\varepsilon }_{i}\left\vert \bm{\xi }_{t-1}=\bm{\varepsilon }_{j}\right. \right) \p \left( \bm{\xi }_{t-1}=\bm{\varepsilon }_{j}\left\vert \bm{X}_{t-1},\bm G_{t-1}\right. \right)\nn\\
&=\sum\limits_{j=1}^{2}\p \left( \bm{\xi }_{t}=\bm{\varepsilon }_{i}\left\vert \bm{\xi }_{t-1}=\bm{\varepsilon }_{j}\right. \right) \p \left( \bm{\xi }_{t-1}=\bm{\varepsilon }_{j}\left\vert \bm{X}_{t-1}\right. \right), \qquad i=1,2.
\label{eq:prediction}
\end{align}
So that, because of \eqref{eq:notationfilter}, we have 
\beq
\bm{\xi }_{t\left\vert t-1\right. }=\mathbf{P}^{\prime }\bm{\xi }%
_{t-1\left\vert t-1\right. }, \quad t=1,\ldots, T.\label{eq:predictionvec}
\eeq
The update involves the posterior probability:
\begin{align}
\p\left( \bm{\xi }_{t}=\bm{\varepsilon }_{i}\left\vert \bm{X}_{t}\right. \right)&=\p\left( \bm{\xi }_{t}=\bm{\varepsilon }_{i}\left\vert \bm{X}_{t},\bm G_t\right. \right) =\p\left( \bm{\xi }_{t}=\bm{\varepsilon }_{i}\left\vert \mathbf x_{t},\bm{X}_{t-1},\bm G_t\right. \right)\nn\\
&=
\dfrac{f \left( 
	\mathbf{x}_{t},\bm{\xi }_{t}=\bm{\varepsilon }_{i}\left\vert \bm{X}_{t-1},\bm G_t\right. \right) }{f\left( \mathbf{x}_{t}\left\vert 
	\bm{X}_{t-1},\bm G_t\right. \right) }\nn\\
	&=\dfrac{f \left( \mathbf{x}_{t}\left\vert \bm{\xi }_{t}=\bm{\varepsilon }_{i},\bm{X}_{t-1},\bm G_t\right. \right) \p \left( 
	\bm{\xi }_{t}=\bm{\varepsilon }_{i}\left\vert \bm{X}_{t-1},\bm G_t\right.
	\right) }{f \left( \mathbf{x}_{t}\left\vert \bm{X}_{t-1},\bm G_t\right. \right) }, \quad i=1,2.\label{eq:update}
\end{align}
Then, since $\mathbf{x}_{t}$ depends on $\bm{X}_{t-1}$ only through $\bm{\xi }_{t-1}$ and it depends on $\bm G_t$ only through $\mbf g_t$
\begin{align}\label{eq:numfilter1}
f\left( \mathbf{x}_{t}\left\vert \bm{\xi }_{t}=\bm{\varepsilon }_{i},\bm{X}_{t-1},\bm G_t\right. \right)   =  f \left( \mathbf{x}_{t}\left\vert \bm{\xi }_{t}=\bm{\varepsilon }_{i},\mbf g_t\right. \right), \quad i=1,2.
\end{align}
Let, 
%$\bm \eta_t =\E\left[ \mathbf{x}_{t}\left\vert \bm{\xi }_{t},\mathbf{g}_{t}\right. \right]  $, therefore, because of \eqref{eq:notationfilter}, 
\begin{align}
\bm{\eta }_{t}&=\left[ 
\begin{array}{c}
f\left( \mathbf{x}_{t}\left\vert \bm{\xi }_{t}=\bm{\varepsilon }_{1},\mathbf{{g}}_{t}\right. \right)  \\ 
f\left( \mathbf{x}_{t}\left\vert \bm{\xi }_{t}=\bm{\varepsilon }_{2},\mathbf{{g}}_{t}\right. \right) 
\end{array}\right]\nn\\
& =\dfrac{1}{\left( 2\pi \right) ^{N\left/ 2\right. }}\left\{ 
\begin{array}{c}
\left\vert \text{diag}(\bm\Sigma _{e1})\right\vert ^{-1\left/ 2\right. }\exp \left[ -\dfrac{1}{2}\left( \mathbf{x}_{t}-\mathbf{B}_{1}\mathbf{{g}}_{t}\right) ^{\prime}\left( \text{diag}(\bm\Sigma _{e1})\right) ^{-1}\left( \mathbf{x}_{t}-\mathbf{B}_{1}\mathbf{{g}}_{t}\right) \right]  \\ 
	\\
\left\vert \text{diag}(\bm\Sigma _{e2})\right\vert ^{-1\left/ 2\right. }\exp \left[ -\dfrac{1}{2}\left( \mathbf{x}_{t}-\mathbf{B}_{2}\mathbf{{g}}_{t}\right) ^{\prime}\left( \text{diag}(\bm\Sigma _{e2})\right) ^{-1}\left( \mathbf{x}_{t}-\mathbf{B}_{2}\mathbf{{g}}_{t}\right) \right] 
\end{array}
\label{eq:eta}
\right\} .
\end{align}
Further, notice that,  from \eqref{eq:notationfilter} and \eqref{eq:eta}, the denominator of \eqref{eq:update} be written as:
\begin{align}
f\left( \mathbf{x}_{t}\left\vert \bm{X}_{t-1},\bm G_t\right. \right) &=\sum\limits_{j=1}^{2}f\left( \mathbf{x}_{t}\left\vert \bm{\xi }_{t}=\bm{\varepsilon }_{j},\bm{X}_{t-1},\bm G_t\right. \right) 
\p\left( \bm{\xi }_{t}=\bm{\varepsilon }_{j},\left\vert \bm{X}_{t-1},\bm G_t\right. \right)\nn\\
&=\sum\limits_{j=1}^{2}f\left( \mathbf{x}_{t}\left\vert \bm{\xi }_{t}=\bm{\varepsilon }_{j},\mbf g_t\right. \right) 
\p\left( \bm{\xi }_{t}=\bm{\varepsilon }_{j},\left\vert \bm{X}_{t-1}\right. \right)=\bm{\eta }_{t}^{\prime }\bm{\xi 
}_{t\left\vert t-1\right. }.\label{eq:denomfilter}
\end{align}
%------------------------------------------------------------------------%
%%------------------------------------------------------------------------%
%\begin{equation}
%\p\left( \mathbf{x}_{t}\left\vert \mathbf{X}_{t-1}, \mbf G_t\right. \right) =\bm{\eta }_{t}^{\prime }\bm{\xi 
%}_{t\left\vert t-1\right. }.\label{eq:denomfilter}
%\end{equation}
%------------------------------------------------------------------------%
Taking into account \eqref{eq:notationfilter}, \eqref{eq:prediction}, \eqref{eq:numfilter1}, and \eqref{eq:denomfilter}, the filtered probabilities are obtained from \eqref{eq:update} as
%------------------------------------------------------------------------%
\begin{equation}
%\begin{array}{c}
\bm{\xi }_{t\left\vert t\right. }=\dfrac{\bm{\eta }_{t}\odot \bm{\xi }_{t\left\vert t-1\right. }}{\bm{\eta }_{t}^{\prime }\bm{\xi }_{t\left\vert t-1\right. }}=\dfrac{\bm{\eta }_{t}\odot \bm{\xi }_{t\left\vert t-1\right. }}{\bm{\iota }_{2}^{\prime }\left( \bm{\eta 
	}_{t}\odot \bm{\xi }_{t\left\vert t-1\right. }\right) }, \quad t=1,\ldots, T,\label{eq:Hamilton}
\end{equation}
where $\bm\eta_t$ is computed as in \eqref{eq:eta}. The filter can started by setting either $\bm\xi_{0|0}=\bm\varepsilon_1$, or, equivalently, $\bm\xi_{0|0}=\bm\varepsilon_2$.
%------------------------------------------------------------------------%
%The filter may be started at a given $\bm{\xi }_{1\left\vert 0\right. }$. See Section \ref{} for a possible choice.
%, e.g. $\bm{\xi }_{1\left\vert 0\right. }=\E\left( \bm{\xi }_{t}\right) =\mathbf{%
%	\bar{\xi}}$. 
%------------------------------------------------------------------------%

We then run the Kim smoother, see e.g., \citet[Chapter 5.2]{krolzig2013markov}  and \cite{kim1994}. Notice that (recall that $\bm X\equiv\bm{X}_{T}$ and $\bm G\equiv\bm{{G}}_{T}$):%------------------------------------------------------------------------%
\begin{align}
\p&\left( \bm{\xi }_{t}=\bm\varepsilon_i\left\vert \bm{X},\bm{{G}}\right. \right)=
\sum_{j=1}^{2}
\p\left( \bm{\xi }_{t}=\bm\varepsilon_i\left\vert \bm{\xi }_{t+1}=\bm{\varepsilon }_{j},\bm{X},\bm{{G}}\right. \right) 
\p\left( \bm{\xi }_{t+1}=\bm{\varepsilon }_{j}\left\vert \bm{X},\bm{{G}}\right. \right) \nn\\
& =\sum_{j=1}^{2}
\dfrac{
	\p \left( \bm{\xi }_{t}=\bm{\varepsilon }_i\left\vert \bm{\xi}_{t+1}=\bm{\varepsilon }_{j},\bm{X}_{t},\bm{{G}}_{t}\right.\right) 
	f\left( \left\{ \mathbf{x}_{s},\mathbf{{g}}_{s}\right\} _{s=t+1}^{T}\left\vert \bm{\xi }_{t}=\bm\varepsilon_i,\bm{\xi }_{t+1}=\bm{\varepsilon }_{j},\bm{X}_{t},\bm G_t\right. \right) 
	}
	{
	f\left( \left\{ \mathbf{x}_{s},\mathbf{{g}}_{s}\right\}_{s=t+1}^{T}\left\vert \bm{\xi }_{t+1}=\bm{\varepsilon }_{j},\bm{X}_{t},\bm{{G}}_{t}\right. \right) 
	} 
	\p \left( \bm{\xi }_{t+1}=\bm{\varepsilon}_{j}\left\vert \bm{X},\bm{{G}}\right. \right) \nn\\
& =  \sum_{j=1}^{2}\p \left( \bm{\xi }_{t}=\bm{\varepsilon}_{i}\left\vert \bm{\xi }_{t+1}=\bm{\varepsilon }_{j},\bm{X}_{t},
\bm{{G}}_{t}\right. \right) 
\p\left( \bm{\xi }_{t+1}=\bm{\varepsilon }_{j}\left\vert \bm{X},\bm{{G}}\right. \right)\nn  \\
& =  \sum_{j=1}^{2}
\dfrac{
	\p\left( \bm{\xi}_{t}=\bm{\varepsilon}_{i}\left\vert \bm{X}_{t},\bm{{G}}_{t}\right. \right) 
	\p\left( \bm{\xi }_{t+1}=\bm{\varepsilon }_{j}\left\vert \bm{\xi }_{t}=\bm{\varepsilon}_{i},\bm{X}_{t},\bm{{G}}_{t}\right. \right) 
	}
	{
	\p \left( \bm{\xi }_{t+1}=\bm{\varepsilon }_{j}\left\vert \bm{X}_{t},\bm{{G}}_{t}\right. \right) 
	}
	\p\left( \bm{\xi }_{t+1}=\bm{\varepsilon }_{j}\left\vert \bm{X},\bm{{G}}\right. \right), \quad i=1,2, \nn
\end{align}
which by \eqref{eq:notationfilter} implies that the sequence of smoothed probabilities is given by
%------------------------------------------------------------------------%
\begin{equation}
\bm{\xi }_{t\left\vert T\right. }=\left[ \mathbf{P}\left( \bm{\xi }%
_{t+1\left\vert T\right. }\oslash \bm{\xi }_{t+1\left\vert t\right.
}\right) \right] \odot \bm{\xi }_{t\left\vert t\right. }, \quad t=1,\ldots, T.\label{eq:Kim}
\end{equation}
%------------------------------------------------------------------------%
This backward recursion is initiated at $\bm\xi_{T|T}$ which is the last iteration of the filter in \eqref{eq:Hamilton}.

Finally, for the implementation of the EM algorithm we need to compute also the smoothed cross-probabilities, see \citet[Chapter 5.A.2]{krolzig2013markov},
\beq\label{eq:smooth4}
\bm\xi_{t,t-1|T}= \l[\begin{array}{c}
\p(\bm\xi_t=\bm\varepsilon_1,\bm\xi_{t-1}=\bm\varepsilon_1|\bm X)\\
\p(\bm\xi_t=\bm\varepsilon_2,\bm\xi_{t-1}=\bm\varepsilon_1|\bm X)\\
\p(\bm\xi_t=\bm\varepsilon_1,\bm\xi_{t-1}=\bm\varepsilon_2|\bm X)\\
\p(\bm\xi_t=\bm\varepsilon_2,\bm\xi_{t-1}=\bm\varepsilon_2|\bm X)
\end{array}\r]=\bm\rho\, \odot\l[\l(\bm\xi_{t|T}\oslash\bm\xi_{t|t-1}
\r)\otimes\bm\xi_{t-1|t-1}
\r], \quad t=1,\ldots, T.
\eeq
%%%%%%%%%%%%%%%%%%%%%%%%%%%%%%%%%%%%%%%%%%%%%%%%%%%%%%%%%%%%%%%%%%%%%%%%%%
\subsection{M-step \label{Appendix:Mstep}}

In the M step we have to solve the constrained maximization problem in \eqref{eq:LLKgiusta}. Let us start with estimation of $\bm\varphi$. From \eqref{eq:LLgiusta}, we have:
\begin{align}
&\frac{\partial\log f\left( \mathbf{X}\left\vert \bm{G};\bm\varphi,\bm\rho\right. \right) }{\partial\bm\varphi^\prime }=\frac 1{f\l(\mathbf{X}\left\vert \bm{G};\bm\varphi,\bm\rho\right. \right)}\sum_{\{\bm\xi_t\}_{t=1}^T}\frac{\partial f\left(\mathbf{X} \l\vert \bm{G},\{\bm\xi_t\}_{t=1}^T;\bm\varphi\r. \right)}{\partial \bm\varphi^\prime} \p\l(\{\bm\xi_t\}_{t=1}^T |\bm G,\bm\rho\r)\nn\\
&= \frac 1{f\l(\mathbf{X}\left\vert \bm{G};\bm\varphi,\bm\rho\right. \right)}\sum_{\{\bm\xi_t\}_{t=1}^T}\frac{\partial \log f\left(\mathbf{X} \l\vert \bm{G},\{\bm\xi_t\}_{t=1}^T;\bm\varphi\r. \right)}{\partial \bm\varphi^\prime}
f\left(\mathbf{X} \l\vert \bm{G},\{\bm\xi_t\}_{t=1}^T;\bm\varphi\r. \right) \p\l(\{\bm\xi_t\}_{t=1}^T |\bm G;\bm\rho\r)\nn\\
&=\mathcal C \sum_{\{\bm\xi_t\}_{t=1}^T}\frac{\partial \log f\left(\mathbf{X} \l\vert \bm{G},\{\bm\xi_t\}_{t=1}^T;\bm\varphi\r. \right)}{\partial \bm\varphi^\prime}
\p\l(\{\bm\xi_t\}_{t=1}^T |\bm X,\bm G;\bm\varphi,\bm\rho\r),\label{eq:focphi0}
\end{align}
where $\mathcal C$ is a positive normalization constant.\footnote{Specifically, we have:
\[
\p\l(\{\bm\xi_t\}_{t=1}^T |\bm X,\bm G;\bm\varphi,\bm\rho\r) = \frac{f\left(\mathbf{X} \l\vert \bm{G},\{\bm\xi_t\}_{t=1}^T;\bm\varphi\r. \right) \p\l(\{\bm\xi_t\}_{t=1}^T |\bm G;\bm\rho\r)}{\sum_{\{\bm\xi_t\}_{t=1}^T}f\left(\mathbf{X} \l\vert \bm{G},\{\bm\xi_t\}_{t=1}^T;\bm\varphi\r. \right) \p\l(\{\bm\xi_t\}_{t=1}^T |\bm G;\bm\rho\r)},
\]
so $\mathcal C =\frac{\sum_{\{\bm\xi_t\}_{t=1}^T}f\left(\mathbf{X} \l\vert \bm{G},\{\bm\xi_t\}_{t=1}^T;\bm\varphi\r. \right) \p\l(\{\bm\xi_t\}_{t=1}^T |\bm G;\bm\rho\r)}{f\l(\mathbf{X}\left\vert \bm{G};\bm\varphi,\bm\rho\right. \right) }$.
}
Therefore, from \eqref{eq:LXGXIphi}, \eqref{eq:LLKgiusta}, and \eqref{eq:focphi0}, if we observed $\bm G$, the first order conditions would be:
\begin{align}
\mbf 0&=\l. \frac{\partial \E_{\wh{\mbf q}^{(k)}} \l[\log f\left( \mathbf{X}\left\vert \bm{G};\bm\varphi,\bm\rho\right. \right)\l\vert\bm X\r.\r]}{\partial \bm\varphi^\prime}
\r\vert_{\bm\varphi=\wh{\bm\varphi}^{(k+1)}}  \nn\\
%&=\sum_{t=1}^T
%\sum_{j=1}^2 \l.\frac{\partial \E_{\wh{\mbf q}^{(k)}} \l[\log f\left(\mathbf{x}_t \l\vert \mbf {g}_t, \bm\xi_t=\bm\varepsilon_j ;\bm\varphi\r. \right)\l\vert \bm X\r.\r]}{\partial \bm\varphi^\prime}\r\vert_{\bm\varphi=\wh{\bm\varphi}^{(k+1)}} \p\l(\bm\xi_t=\bm\varepsilon_j |\bm X,\bm G;\wh{\bm\varphi}^{(k)},\wh{\bm\rho}^{(k)}\r)\nn\\
&= \sum_{t=1}^T 
\sum_{j=1}^2 \l.\frac{\partial \E_{\wh{\mbf q}^{(k)}} \l[\log f\left(\mathbf{x}_t \l\vert \mbf {g}_t, \bm\xi_t=\bm\varepsilon_j ;\bm\varphi\r. \right)\l\vert \bm X\r.\r]}{\partial \bm\varphi^\prime}\r\vert_{\bm\varphi=\wh{\bm\varphi}^{(k+1)}} \p\l(\bm\xi_t=\bm\varepsilon_j |\bm X;\wh{\bm\varphi}^{(k)},\wh{\bm\rho}^{(k)}\r)\nn\\
&= \sum_{t=1}^T\sum_{j=1}^2\l.\frac{\partial \E_{\wh{\mbf q}^{(k)}} \l[\log f\left(\mathbf{x}_t \l\vert \mbf {g}_t, \bm\xi_t=\bm\varepsilon_j ;\bm\varphi\r. \right)\l\vert \bm X\r.\r]}{\partial \bm\varphi^\prime}\r\vert_{\bm\varphi=\wh{\bm\varphi}^{(k+1)}}\xi_{j,t|T}^{(k)},\label{eq:focphi}
\end{align}
where ${\xi }_{j,t\left\vert T\right. }^{(k)}=\E_{\wh{\mbf q}^{(k)}}[\xi_{jt}|\mbf X]=\p(\bm\xi_t=\bm\varepsilon_j\l\vert \bm X;\wh{\bm\varphi}^{(k)},\wh{\bm\rho}^{(k)}\r.)$ is the $j$th component of $\bm{\xi }_{t\left\vert T\right. }^{(k)}$.

Then, by substituting \eqref{eq:LXGXIphi} into \eqref{eq:focphi}, and by replacing true factors with estimated ones, we get
\begin{equation}
\label{eq:B_j_hat}
\mathbf{\wh{B}}_{j}^{(k+1)}=\left( \sum_{t=1}^{T}{\xi }_{j,t\left\vert T\right. }^{(k)}\mathbf{x}_{t}\mathbf{\wh{g}}_{t}^{\prime }\right) \left(
\sum_{t=1}^{T}{\xi }_{j,t\left\vert T\right. }^{(k)}\wh{\mathbf{g}}_{t}\mathbf{\wh{g}}_{t}^{\prime }\right) ^{-1},\quad j=1,2,
\end{equation}
and, consistently with the fact that we use a mis-specified likelihood with uncorrelated idiosyncratic components, we set 
\begin{align}
[\wh{\mathbf{\Sigma }}_{ej}^{(k+1)}]_{ii}&=\l(\dfrac{\sum_{t=1}^{T}
\left( {x}_{it}-\mathbf{\wh{b}}_{ji}^{(k+1)\prime}\mathbf{\wh{g}}_{t}\right) ^2
%\left( {x}_{it}-\mathbf{\wh{b}}_{ji}^{(k+1)}^\prime\mathbf{\wh{g}}_{t}\right) 
}
{
\sum_{t=1}^{T}{\xi }_{j,t\left\vert T\right. }^{(k)}
}\r),\quad  i=1,\ldots, N,\quad j=1,2,\label{eq:S_j_hat}\\
[\wh{\mathbf{\Sigma }}_{ej}^{(k+1)}]_{ik}&=0, \quad  i,k=1,\ldots, N,\quad i\ne k,\quad j=1,2,\nn
\end{align}
where $\mathbf{\wh{b}}_{ji}^{(k+1)\prime}$ is the $i$th row of $\mathbf{\wh{B}}_{j}^{(k+1)}$. 

%\matt{riscrivere come $[\wh{\mathbf{\Sigma }}_{ej}^{(k+1)}]_{ii}$, $i=1,\ldots, N$}
Moving to estimation of $\bm\rho$, from \eqref{eq:LLgiusta}, we have:
\begin{align}
&\frac{\partial\log f\left( \mathbf{X}\left\vert \bm{G};\bm\varphi,\bm\rho\right. \right) }{\partial\bm\rho^\prime }=
\frac 1{f\left( \mathbf{X}\left\vert \bm{G};\bm\varphi,\bm\rho\right. \right)}
\sum_{\{\bm\xi_t\}_{t=1}^T} f\left(\mathbf{X} \l\vert \bm{G},\{\bm\xi_t\}_{t=1}^T;\bm\varphi\r. \right)\frac{\partial \p\l(\{\bm\xi_t\}_{t=1}^T |\bm G;\bm\rho\r)}{\partial\bm\rho^\prime}\nn\\
&=\frac 1{f\left( \mathbf{X}\left\vert \bm{G};\bm\varphi,\bm\rho\right. \right)}
\sum_{\{\bm\xi_t\}_{t=1}^T}
\frac{\partial \log\p\l(\{\bm\xi_t\}_{t=1}^T |\bm G;\bm\rho\r)}{\partial\bm\rho^\prime}
f\left(\mathbf{X} \l\vert \bm{G},\{\bm\xi_t\}_{t=1}^T;\bm\varphi\r. \right)\p\l(\{\bm\xi_t\}_{t=1}^T |\bm G;\bm\rho\r)\nn\\
&=\mathcal C\sum_{\{\bm\xi_t\}_{t=1}^T}\frac{\partial \log\p\l(\{\bm\xi_t\}_{t=1}^T |\bm G;\bm\rho\r)}{\partial\bm\rho^\prime}
\p\l(\{\bm\xi_t\}_{t=1}^T |\bm X,\bm G;\bm\varphi,\bm\rho\r),\label{eq:focrho0}
\end{align}
where $\mathcal C$ is the same positive normalization constant as in \eqref{eq:focphi0}. And, because of \eqref{eq:XIrho} and \eqref{eq:focrho0}, if we observed $\bm G$ the derivatives with respect to the generic $(i,j)$th element of $\bm\rho$, i.e, $p_{ij}$, $i,j=1,2$,  would be (treating $\bm\xi_0$ as known)
\begin{align}
&\frac{\partial\log f\left( \mathbf{X}\left\vert \bm{G};\bm\varphi,\bm\rho\right. \right) }{\partial p_{ij} }\nn\\
&=\sum_{t=1}^T \sum_{h=1}^2\sum_{\ell=1}^2
\frac{\partial \log\p\l(\bm\xi_t =\bm\varepsilon_h|\bm\xi_{t-1}=\bm\varepsilon_\ell;\bm\rho\r)}{\partial p_{ij}}
\p\l(\bm\xi_t=\bm\varepsilon_h,\bm\xi_{t-1}=\bm\varepsilon_\ell |\bm X;{\bm\varphi},{\bm\rho}\r)\nn\\
%&=\sum_{t=1}^T \sum_{h=1}^2\sum_{\ell=1}^2
%\frac{\partial \log\p\l(\bm\xi_t =\bm\varepsilon_h|\bm\xi_{t-1}=\bm\varepsilon_\ell;\bm\rho\r)}{\partial p_{ij}}\p\l(\bm\xi_t=\bm\varepsilon_h,\bm\xi_{t-1}=\bm\varepsilon_\ell |\bm X;{\bm\varphi},{\bm\rho}\r)\nn\\
&=\sum_{t=1}^T \sum_{h=1}^2\sum_{\ell=1}^2
\frac 1{\p\l(\bm\xi_t =\bm\varepsilon_h|\bm\xi_{t-1}=\bm\varepsilon_\ell;\bm\rho\r)}\frac{\partial \p\l(\bm\xi_t =\bm\varepsilon_h|\bm\xi_{t-1}=\bm\varepsilon_\ell;\bm\rho\r)}{\partial p_{ij}}\p\l(\bm\xi_t=\bm\varepsilon_h,\bm\xi_{t-1}=\bm\varepsilon_\ell |\bm X;{\bm\varphi},{\bm\rho}\r)\nn\\
&=\sum_{t=1}^T \sum_{h=1}^2\sum_{\ell=1}^2
\frac{\mathbb I\l(\bm\xi_t =\bm\varepsilon_j,\bm\xi_{t-1}=\bm\varepsilon_i\r)}{
\p\l(\bm\xi_t =\bm\varepsilon_h|\bm\xi_{t-1}=\bm\varepsilon_\ell;\bm\rho\r)}\p\l(\bm\xi_t=\bm\varepsilon_h,\bm\xi_{t-1}=\bm\varepsilon_\ell |\bm X;{\bm\varphi},{\bm\rho}\r)\nn\\
&=\sum_{t=1}^T 
\frac{\p\l(\bm\xi_t=\bm\varepsilon_j,\bm\xi_{t-1}=\bm\varepsilon_i |\bm X;{\bm\varphi},{\bm\rho}\r)}{
\p\l(\bm\xi_t =\bm\varepsilon_j|\bm\xi_{t-1}=\bm\varepsilon_i;\bm\rho\r)}
=\sum_{t=1}^T 
\frac{\p\l(\bm\xi_t=\bm\varepsilon_j,\bm\xi_{t-1}=\bm\varepsilon_i |\bm X;{\bm\varphi},{\bm\rho}\r)}{
p_{ij}}.\label{eq:focpij0}
\end{align}

Now, from  \eqref{eq:LLKgiusta} and \eqref{eq:focrho0}, the first order conditions are:
\begin{align}
\mbf 0=&\l.\l\{\frac{\partial\E_{\wh{\mbf q}^{(k)}}\l[\log f\left( \mathbf{X}\left\vert \bm{G};\bm\varphi,\bm\rho\right. \right)\l\vert\bm X\r.\r] }{\partial(\text{vec}(\mbf P))^\prime }
-\bm\kappa^\prime\l(\bm\iota_2^\prime\otimes \mathbf I_2\r)\r\}\r\vert_{\text{vec}(\mbf P)=\text{vec}(\wh{\mbf P}^{(k+1)})},
\label{eq:focPhat}
\end{align}
where $\bm\kappa$ is the $2$-dimensional vector of Lagrange multipliers, thus it has positive entries. Then, from \eqref{eq:focpij0} 
 \begin{align}
\frac{\partial\E_{\wh{\mbf q}^{(k)}}\l[\log f\left( \mathbf{X}\left\vert \bm{G};\bm\varphi,\bm\rho\right. \right) \l\vert \bm X\r.\r]
}{\partial p_{ij} }&=\sum_{t=1}^T 
\frac{\p\l(\bm\xi_t=\bm\varepsilon_j,\bm\xi_{t-1}=\bm\varepsilon_i |\bm X;\wh{\bm\varphi}^{(k)},\wh{\bm\rho}^{(k)}\r)}{p_{ij}}.\label{eq:focpij}
\end{align}
By collecting all 4 terms deriving from \eqref{eq:focpij} into a vector, we have
 \begin{align}
\frac{\partial\E_{\wh{\mbf q}^{(k)}}\l[\log f\left( \mathbf{X}\left\vert \bm{G};\bm\varphi,\bm\rho\right. \right) \l\vert \bm X\r.\r]
}{\partial \bm\rho^\prime }&=\sum_{t=1}^T \bm\xi_{t,t-1|T}^{(k)\prime} \oslash \bm\rho^\prime,\label{eq:focP}
\end{align}
where $ \bm\xi_{t,t-1|T}^{(k)}$ is defined in \eqref{eq:smooth4}. Finally, from the first order conditions \eqref{eq:focPhat}, we must have:
\beq
\mbf 0 =\l. \l\{ \sum_{t=1}^T \bm\xi_{t,t-1|T}^{(k)\prime} \oslash \bm\rho^\prime-\bm\kappa^\prime\l(\bm\iota_2^\prime\otimes \mathbf I_2\r)
\r\}\r\vert_{\bm\rho=\wh{\bm\rho}^{(k+1)}}.\label{eq:uffa}
\eeq
Let $\bm\kappa=(\kappa_1,\kappa_2)^\prime$, and let $\wt{\bm\kappa}=\l(\bm\iota_2\otimes\bm\kappa\r)=(
\kappa_1, \kappa_2, \kappa_1,\kappa_2)^\prime$. Then, \eqref{eq:uffa} gives
\beq\label{eq:rhoklag}
\wh{\bm\rho}^{(k+1)}=\sum_{t=1}^T \bm\xi_{t,t-1|T}^{(k)}\oslash
\wt{\bm\kappa}.
\eeq
By applying the adding up condition to \eqref{eq:rhoklag}:
\begin{align}
\bm\iota_2&=\l(\bm\iota_2^\prime\otimes\mbf I_2\r)\wh{\bm\rho}^{(k+1)} =
\l(\bm\iota_2^\prime\otimes\mbf I_2\r)\l(\sum_{t=1}^T \bm\xi_{t,t-1|T}^{(k)}\oslash\wt{\bm\kappa}\r)
=\l(\bm\iota_2^\prime\otimes\mbf I_2\r)\sum_{t=1}^T\l(\begin{array}{c}
 \frac{\xi_{11,t,t-1|T}^{(k)}}{\kappa_1}\\
 \frac{\xi_{21,t,t-1|T}^{(k)}}{\kappa_2}\\
  \frac{\xi_{12,t,t-1|T}^{(k)}}{\kappa_1}\\
   \frac{\xi_{22,t,t-1|T}^{(k)}}{\kappa_2}
\end{array}
\r)
\nn\\
&=\sum_{t=1}^T \sum_{j=1}^2\l(\begin{array}{c}
 \frac{\xi_{1j,t,t-1|T}^{(k)}}{\kappa_1} \\
\frac {\xi_{2j,t,t-1|T}^{(k)}} {\kappa_2}
\end{array}
\r)=\sum_{t=1}^{T} \l(\begin{array}{c}
 \frac{\xi_{1,t-1|T}^{(k)}}{\kappa_1} \\
\frac {\xi_{2,t-1|T}^{(k)}} {\kappa_2}
\end{array}
\r)
=\sum_{t=0}^{T-1} \l(\begin{array}{c}
 \frac{\xi_{1,t|T}^{(k)}}{\kappa_1} \\
\frac {\xi_{2,t|T}^{(k)}} {\kappa_2}
\end{array}
\r)=\sum_{t=0}^{T-1} \bm\xi_{t|T}^{(k)}\oslash {\bm\kappa},\nn
%=\bm\iota_2.\label{eq:uffakappa}
%&= \l[\l(\bm\iota_2^\prime\otimes\mbf I_2\r) \sum_{t=1}^T \bm\xi_{t,t-1|T}^{(k)}\r]\oslash \wt{\bm\kappa}
\end{align}
which implies ${\bm\kappa} =  \sum_{t=0}^{T-1} \bm\xi_{t|T}^{(k)}$. Therefore, from \eqref{eq:rhoklag},
\beq
\wh{\bm\rho}^{(k+1)}=\l[\sum_{t=1}^T \bm\xi_{t,t-1|T}^{(k)}\r]\oslash \l[ \bm\iota_2\otimes \sum_{t=0}^{T-1}\bm\xi_{t|T}^{(k)}\r].\label{eq:rho_hat}
\eeq
%%%%%%%%%%%%%%%%%%%%%%%%%%%%%%%%%%%%%%%%%%%%%%%%%%%%%%%%%%%%%%%%%%%%%%%%%%
%%%%%%%%%%%%%%%%%%%%%%%%%%%%%%%%%%%%%%%%%%%%%%%%%%%%%%%%%%%%%%%%%%%%%%%%%%
\setcounter{equation}{0}
\numberwithin{equation}{section}

\section{Mathematical proofs\label{section:Appendix}}

Define $C_{NT}=\min\left\{\sqrt N,\sqrt T\right\}$ Let $\mathbb{I}_{1t}=\mathbb{I}(s_{t}=1)$ and $\mathbb{I}_{2t}=\mathbb{I}(s_{t}=2)$. For $j=1,2$, and $i,l=1,\ldots,N$, define
%------------------------------------------------------------------------%
\begin{equation}
\label{eq:aux_quant}
\begin{array}{rl}
\sigma _{jil}=\E\left( \dfrac{1}{T}\sum\limits_{t=1}^{T}\mathbb{I}%
_{jt}e_{it}e_{lt}\right) , & \chi _{jil}=\dfrac{1}{T}\sum\limits_{t=1}^{T}%
\mathbb{I}_{jt}e_{it}e_{lt}-\E\left( \dfrac{1}{T}\sum%
\limits_{t=1}^{T}\mathbb{I}_{jt}e_{it}e_{lt}\right) , \\ 
\varphi _{jil}=\dfrac{1}{T}\sum\limits_{t=1}^{T}\mathbb{I}_{jt}\bm{%
	\lambda }_{ji}^{\prime }\mathbf{f}_{jt}e_{lt}, & \varphi _{jli}=\dfrac{1}{T}%
\sum\limits_{t=1}^{T}\mathbb{I}_{jt}\bm{\lambda }_{jl}^{\prime
}\mathbf{f}_{jt}e_{it}.%
\end{array}%
\end{equation}%
%------------------------------------------------------------------------%
\subsection{Lemmas}
%%%%%%%%%%%%%%%%%%%%%%%%%%%%%%%%%%%%%%%%%%%%%%%%%%%%%%%%%%%%%%%%%%%%%%%%%%
\begin{lem}
{\it
\label{Lemma:a_i_hat} Under Assumptions \ref{assum:F} - \ref{assum:WD}, and given $\mathbf{\widehat{H}}$ defined in $\left(\ref{eq:hat_H}\right)$, we have
%----------------------------------------------------------------------------%
\begin{equation*}
	\dfrac{1}{N}\sum\limits_{i=1}^{N}\left\Vert \mathbf{\widehat{a}}_{i}-\mathbf{\widehat{H}} ^{\prime }\mathbf{a}_{i}\right\Vert
	^{2}=O_{p}\left( \dfrac{1}{C_{NT}^{2}}\right) .
\end{equation*}
%----------------------------------------------------------------------------%
}
\end{lem}
%%%%%%%%%%%%%%%%%%%%%%%%%%%%%%%%%%%%%%%%%%%%%%%%%%%%%%%%%%%%%%%%%%%%%%%%%%
\begin{lem}
	\label{Lemma:a_i_hat_components}
	{\it
	Let Assumptions \ref{assum:F} - \ref{assum:MCLT} hold. Then:
	\begin{description}
		\item[(a)] $N^{-1}\sum\nolimits_{l=1}^{N}\mathbf{\widehat{a}}_{l}\sigma
		_{jil}=O_{p}\left( \tfrac{1}{\sqrt{N}C_{NT}}\right) $;
		
		\item[(b)] $N^{-1}\sum\nolimits_{l=1}^{N}\mathbf{\widehat{a}}_{l}\chi
		_{jil}=O_{p}\left( \tfrac{1}{\sqrt{T}C_{NT}}\right) $;
		
		\item[(c)] $N^{-1}\sum\nolimits_{l=1}^{N}\mathbf{\widehat{a}}_{l}\varphi
		_{jil}=O_{p}\left( \tfrac{1}{\sqrt{T}C_{NT}}\right) $;
		
		\item[(d)] $N^{-1}\sum\nolimits_{l=1}^{N}\mathbf{\widehat{a}}_{l}\varphi
		_{jli}=O_{p}\left( \tfrac{1}{\sqrt{T}}\right) $.
	\end{description}
	}
\end{lem}
%%%%%%%%%%%%%%%%%%%%%%%%%%%%%%%%%%%%%%%%%%%%%%%%%%%%%%%%%%%%%%%%%%%%%%%%%%
\begin{lem}
\label{Lemma:3}
{\it
Under Assumptions \ref{assum:F} - \ref{assum:MCLT},
%----------------------------------------------------------------------------%
\begin{equation*}
	N^{-1}\left( \mathbf{\widehat{A}}-\mathbf{A\widehat{H}}\right) ^{\prime }\mathbf{%
		\widehat{A}}=O_{p}\left( \dfrac{1}{C_{NT}^{2}}\right) .
\end{equation*}
%----------------------------------------------------------------------------%
}
\end{lem}
%%%%%%%%%%%%%%%%%%%%%%%%%%%%%%%%%%%%%%%%%%%%%%%%%%%%%%%%%%%%%%%%%%%%%%%%%%
\begin{lem}
\label{Lemma:4}
{\it 
Under Assumptions \ref{assum:F} - \ref{assum:MCLT},%
%----------------------------------------------------------------------------%
\begin{equation*}
	N^{-1}\left( \mathbf{\widehat{A}}-\mathbf{A\widehat{H}}\right) ^{\prime }\mathbf{e}%
	_{t}=O_{p}\left( \dfrac{1}{C^{2}_{NT}}\right) .
\end{equation*}
%----------------------------------------------------------------------------%
}
\end{lem}
%%%%%%%%%%%%%%%%%%%%%%%%%%%%%%%%%%%%%%%%%%%%%%%%%%%%%%%%%%%%%%%%%%%%%%%%%%
\begin{lem}
\label{Lemma:avg}{\it Let Assumptions \ref{assum:F} - \ref{assum:MCLT} hold. Then:
	\begin{description}
		\item[(a)] $\mathbf{\widehat{g}}_{t}-\mathbf{\widehat{H}}^{-1}\mathbf{g}_{t}=O_{p}\left( \tfrac{1}{\sqrt{N}}\right)+O_{p}\left( \tfrac{1}{C_{NT}^{2}}\right)$, for $t=1,\dots,T$;
		
		\item[(b)] $\tfrac{1}{T}\sum\nolimits_{t=1}^{T}\left( \mathbf{\widehat{g}}_{t}-\widehat{%
			\mathbf{H}}^{-1}\mathbf{g}_{t}\right) \mathbf{\wh{g}}_{t}^{\prime }=O_{p}\left( \tfrac{1}{C_{NT}^{2}}\right).$
\end{description}}
\end{lem}
%%%%%%%%%%%%%%%%%%%%%%%%%%%%%%%%%%%%%%%%%%%%%%%%%%%%%%%%%%%%%%%%%%%%%%%%%%
\begin{lem}
\label{Lemma:var_hat}
{\it Under Assumptions \ref{assum:F} - \ref{assum:eigenvalues}, and given $\mathbf{Q}$ defined in \eqref{eq:QQQ},
	%----------------------------------------------------------------------------%
	\begin{equation*}
		p\lim_{N,T\rightarrow \infty }\dfrac{\mathbf{A}^{\prime }\mathbf{\widehat{A}}}{N}%
		= \mathbf{Q}.
	\end{equation*}
	%----------------------------------------------------------------------------%
}
\end{lem}
%%%%%%%%%%%%%%%%%%%%%%%%%%%%%%%%%%%%%%%%%%%%%%%%%%%%%%%%%%%%%%%%%%%%%%%%%%
\begin{lem}
\label{Lemma:Q_j}
{\it
Let Assumptions \ref{assum:F} - \ref{assum:eigenvalues} hold, and consider the matrix $\mathbf{Q}$ defined in \eqref{eq:QQQ}. Then, for $j=1,2$, the $r_j\times \left(r_1+r_2\right)$ matrix $\mathbf{Q}_j$ satisfying $\mathbf{Q}=\left[\mathbf{Q}_{1}^{\prime}\ \mathbf{Q}_{2}^{\prime}\right]^{\prime}$ is such that
%----------------------------------------------------------------------------%
\begin{equation*}
	\mathbf{Q}_{j}=\mathbf{\Sigma }_{\mathbf{f}j}^{-1\left/ 2\right. }\mathbf{%
		\Psi }_{j}\mathbf{V}^{1\left/ 2\right. },
\end{equation*}
%----------------------------------------------------------------------------%
where $\mathbf{\Sigma }_{\mathbf{f}j}$ is defined in (\ref{eq:SFj}), and $\mathbf{\Psi }_{j}$ is the $r_j\times \left(r_1+r_2\right)$ matrix such that $\mathbf{\Psi}=\left[\mathbf{\Psi}_{1}^{\prime}\ \mathbf{\Psi}_{2}^{\prime}\right]^{\prime}$, with $\mathbf{\Psi}$ as in \eqref{eq:QQQ}.
}
\end{lem}
%%%%%%%%%%%%%%%%%%%%%%%%%%%%%%%%%%%%%%%%%%%%%%%%%%%%%%%%%%%%%%%%%%%%%%%%%%
\begin{lem}
\label{Lemma:Hat_V_NT}
{\it
Let $\mathbf{\widehat{V}}$ be the $\left(r_{1}+r_{2}\right) \times \left(r_{1}+r_{2}\right)$ diagonal matrix containing the first $r_{1}+r_{2}$ eigenvalues of $\mathbf{\widehat{\Sigma}}_{\mathbf{x}}=\left( NT\right)
^{-1}\sum\nolimits_{t=1}^{T}\mathbf{x}_{t}\mathbf{x}_{t}^{\prime }$ in decreasing order. Define $\mathbf{V}$ as the $\left(r_1+r_2\right)\times\left(r_1+r_2\right)$ diagonal matrix of the first $r_{1}+r_{2}$ eigenvalues of $\mathbf{\Sigma }_{\mathbf{g}}^{1\left/ 2\right. }\mathbf{\Sigma }_{\mathbf{A}}\mathbf{\Sigma }_{\mathbf{g}}^{1\left/ 2\right. }$ in decreasing order, where $\mathbf{\Sigma }_{\mathbf{g}}$ and $\mathbf{\Sigma }_{\mathbf{A}}$ are defined in (\ref{eq:SG}) and (\ref{eq:SA}), respectively. Then, under Assumptions \ref{assum:F} - \ref{assum:WD},
\begin{equation*}
	\wh{\mbf V}\overset{p}{\rightarrow } \mbf V.
\end{equation*}
}
\end{lem}

\begin{lem}
\label{Lemma:Hat_I_xi}
{\it
Let Assumptions \ref{assum:F} - \ref{assum:MCLT} hold. Then, as $N,T\rightarrow \infty$,
%----------------------------------------------------------------------------%
\begin{equation*}
	\begin{array}{cc}
		\mathbf{\widehat{I}}_{\mathbf{\widehat{\xi}}j}\overset{p}{\rightarrow }\mathbf{I}_{\mathbf{\xi }j}=\mathbf{H}^{-1}\left[ 
		\begin{array}{cc}
			\mathbb{I}\left( j=1\right) \mathbf{I}_{r_{1}} & \mathbf{0} \\ 
			\mathbf{0} & \mathbb{I}\left( j=2\right) \mathbf{I}_{r_{2}}%
		\end{array}%
		\right] \mathbf{H,} & j=1,2,
	\end{array}%
\end{equation*}
%----------------------------------------------------------------------------%
where $\mathbf{H}$ is defined in (\ref{eq:Hlim}).
}
\end{lem}

\begin{lem}
\label{Lemma:V_hat_j}
{\it
Let Assumptions \ref{assum:F} - \ref{assum:WD} hold. Then, for any fixed $1 \leq p \leq \bar{p}$ with $\bar{p} < \infty$, and for $j=1,2$, $\left\Vert \mathbf{\widehat{V}}_{\widehat{\xi},j}^{\left( p\right) }\right\Vert =O_{p}\left( 1\right) $, where $\mathbf{\widehat{V}}_{\widehat{\xi},j}^{\left( p\right) }$ is the $p \times p$ diagonal matrix containing the first $p$ eigenvalues of $\mathbf{\widehat{\Sigma}}_{\widehat{\xi},\mathbf{x}j}$ defined in \eqref{eq:sigma_hat_x_j} in decreasing order.
}
\end{lem}

\begin{lem}
\label{Lemma:O_p_1}{\it Let Assumption \ref{assum:TCSDH} hold. For $j,k=1,2$, and $i,l=1,\ldots,N$, all $N\in\mathbb N$, consider
	%----------------------------------------------------------------------------%
	\begin{equation*}
		\sigma _{\widehat{\xi},jkil}=\dfrac{1}{T}\sum\limits_{t=1}^{T}\E\left( 
		\mathbb{I}_{jt}\widehat{\xi}_{kt\left\vert T\right. }e_{it}e_{lt}\right).
	\end{equation*}
	%----------------------------------------------------------------------------%
	Then
	%----------------------------------------------------------------------------%
	\begin{equation*}
		\dfrac{1}{N}\sum\limits_{i=1}^{N}%
		\sum\limits_{l=1}^{N}\sigma _{\widehat{\xi},jkil}^{2} =O_{p}\left(1\right).
	\end{equation*}
	%----------------------------------------------------------------------------%
}
\end{lem}

\subsection{Proofs of Lemmas}

\begin{proof}[\textbf{Proof of Lemma \protect\ref{Lemma:a_i_hat}}]
Consider $\mathbf{\widehat{\Sigma}}_{\mathbf{x}}=\left( NT\right)
^{-1}\sum\nolimits_{t=1}^{T}\mathbf{x}_{t}\mathbf{x}_{t}^{\prime }$, and $\mathbf{\widehat{H}}=\left(\mathbf{GG}^{\prime}/T\right)\left(\mathbf{A}%
^{\prime }\mathbf{\widehat{A}}/N\right)\mathbf{\widehat{V}} ^{-1}$ as defined in \eqref{eq:hat_H}. By the
definition of eigenvectors and eigenvalues, $\mathbf{\widehat{\Sigma}}_{\mathbf{x%
}}\mathbf{\widehat{A}}=\mathbf{\widehat{A}}\mathbf{\widehat{V}}$, where $%
\mathbf{\widehat{V}}$ is the $\bar{r}\times \bar{r}$ diagonal matrix of the first $\bar{r}=\left(r_1+r_2\right)$
largest eigenvalues of $\mathbf{\widehat{\Sigma}}_{\mathbf{x}}$ in decreasing
order, and $\mathbf{\widehat{A}}$ is $\sqrt{N}$ times the $N\times \bar{r}$ matrix of
eigenvectors of $\mathbf{\widehat{\Sigma}}_{\mathbf{x}}$ corresponding to its $\bar{r}$
largest eigenvalues. Note that $\left\Vert \mathbf{\widehat{V}} \right\Vert=O_{p}\left(1\right)$ and $\left\Vert \mathbf{\widehat{H}}\right\Vert \leq \left\Vert \mathbf{GG}^{\prime
}\left/ T\right. \right\Vert \left\Vert \mathbf{AA}^{\prime }\left/ N\right.
\right\Vert ^{1\left/ 2\right. }\left\Vert \mathbf{\widehat{A}\widehat{A}}^{\prime
}\left/ N\right. \right\Vert ^{1\left/ 2\right. }\left\Vert \mathbf{\widehat{V}}%
^{-1}\right\Vert =O_{p}\left( 1\right)$ by Assumptions \ref{assum:F} and \ref{assum:FL}. We then have
%----------------------------------------------------------------------------%
\begin{equation*}
\left( \mathbf{\widehat{A}}-\mathbf{A\widehat{H}}\right) \mathbf{\widehat{V}}
= \mathbf{\widehat{A}}\mathbf{\widehat{V}}-\mathbf{A\widehat{H}}\mathbf{\widehat{V}}
= \mathbf{\widehat{A}}\mathbf{\widehat{V}}-\mathbf{A}\dfrac{\mathbf{GG}%
	^{\prime }}{T}\dfrac{\mathbf{A}^{\prime }\mathbf{\widehat{A}}}{N},%
\end{equation*}
%----------------------------------------------------------------------------%
which implies
%----------------------------------------------------------------------------%
\begin{equation*}
\mathbf{\widehat{V}}\mathbf{\widehat{A}}^{\prime }-\dfrac{\mathbf{\widehat{A}} ^{\prime }\mathbf{A}}{N}\dfrac{\mathbf{GG}^{\prime }}{T}\mathbf{%
	A}^{\prime } = \mathbf{\widehat{A}} ^{\prime }\mathbf{\widehat{%
		\Sigma}}_{\mathbf{x}}-\dfrac{\mathbf{\widehat{A}}^{\prime }%
	\mathbf{A}}{N}\dfrac{\mathbf{GG}^{\prime }}{T}\mathbf{A}^{\prime } = 
 \mathbf{\widehat{A}} ^{\prime }\dfrac{1}{NT}\left[ \left(
\sum\limits_{t=1}^{T}\mathbf{x}_{t}\mathbf{x}_{t}^{\prime }\right) -\mathbf{%
	AGG}^{\prime }\mathbf{A}^{\prime }\right] .%
\end{equation*}
%----------------------------------------------------------------------------%
Taking into account (\ref{eq:aux_quant}), after some algebra we have
%----------------------------------------------------------------------------%
\begin{equation}
\label{eq:a__hat-a_i}
\begin{array}{rcl}
\mathbf{\widehat{V}}\left( \mathbf{\widehat{a}}_{i}-\mathbf{\widehat{H}} ^{\prime }\mathbf{a}_{i}\right) & = & \mathbf{\widehat{A}} ^{\prime }\dfrac{1}{NT}\left[ \left( \sum\limits_{t=1}^{T}%
\mathbf{x}_{t}x_{it}\right) -\mathbf{AGG}^{\prime }\mathbf{a}_{i}\right] \\ 
& = & \left[ \sum\limits_{j=1}^{2}%
\left( \dfrac{1}{N}\sum\limits_{l=1}^{N}\mathbf{\widehat{a}}_{l}\sigma
_{jil}+\dfrac{1}{N}\sum\limits_{l=1}^{N}\mathbf{\widehat{a}}_{l}\chi _{jil}+%
\dfrac{1}{N}\sum\limits_{l=1}^{N}\mathbf{\widehat{a}}_{l}\varphi _{jil}+%
\dfrac{1}{N}\sum\limits_{l=1}^{N}\mathbf{\widehat{a}}_{l}\varphi
_{jli}\right) \right]. \\ 
\end{array}%
\end{equation}%
%----------------------------------------------------------------------------%
It follows that%
%----------------------------------------------------------------------------%
\begin{equation}
\label{eq:Lemma_1_1}
\dfrac{1}{N}\sum\limits_{i=1}^{N}\left\Vert \mathbf{\widehat{a}}_{i}-\mathbf{\widehat{H}} ^{\prime }\mathbf{a}_{i}\right\Vert ^{2}\leq 8\left\Vert \mathbf{\widehat{V}} ^{-1}\right\Vert
^{2}\sum\limits_{j=1}^{2}\left( \dfrac{1}{N}\sum\limits_{i=1}^{N}\widehat{%
	\sigma}_{ji\cdot }+\dfrac{1}{N}\sum\limits_{i=1}^{N}\widehat{\chi}_{ji\cdot }+%
\dfrac{1}{N}\sum\limits_{i=1}^{N}\widehat{\varphi}_{ji\cdot }+\dfrac{1}{N}%
\sum\limits_{i=1}^{N}\widehat{\varphi}_{j\cdot i}\right) ,
\end{equation}
%----------------------------------------------------------------------------%
where
%----------------------------------------------------------------------------%
\begin{equation*}
\begin{array}{cccc}
\widehat{\sigma}_{ji\cdot }=\dfrac{1}{N^{2}}\left\Vert \sum\limits_{l=1}^{N}%
\mathbf{\widehat{a}}_{l}\sigma _{jil}\right\Vert ^{2}, & \widehat{\chi}_{ji\cdot
}=\dfrac{1}{N^{2}}\left\Vert \sum\limits_{l=1}^{N}\mathbf{\widehat{a}}%
_{l}\chi _{jil}\right\Vert ^{2}, & \widehat{\varphi}_{ji\cdot }=\dfrac{1}{%
	N^{2}}\left\Vert \sum\limits_{l=1}^{N}\mathbf{\widehat{a}}_{l}\varphi
_{jil}\right\Vert ^{2}, & \widehat{\varphi}_{j\cdot i}=\dfrac{1}{N^{2}}%
\left\Vert \sum\limits_{l=1}^{N}\mathbf{\widehat{a}}_{l}\varphi
_{jli}\right\Vert ^{2}.%
\end{array}%
\end{equation*}
%----------------------------------------------------------------------------%
Consider $\widehat{\sigma}_{ji\cdot }$ and note that
%----------------------------------------------------------------------------%
\begin{equation*}
\left\Vert \sum\limits_{l=1}^{N}\mathbf{\widehat{a}}_{l}\sigma
_{jil}\right\Vert ^{2}\leq \left( \sum\limits_{l=1}^{N}\left\Vert \mathbf{%
	\widehat{a}}_{l}\right\Vert ^{2}\right) \left( \sum\limits_{l=1}^{N}\sigma
_{jil}^{2}\right)
\end{equation*}
%----------------------------------------------------------------------------%
so that
%----------------------------------------------------------------------------%
\begin{equation*}
\dfrac{1}{N}\sum\limits_{i=1}^{N}\widehat{\sigma}_{ji\cdot }=\dfrac{1}{N}%
\sum\limits_{i=1}^{N}\left( \dfrac{1}{N^{2}}\left\Vert
\sum\limits_{l=1}^{N}\mathbf{\widehat{a}}_{l}\sigma _{jil}\right\Vert
^{2}\right) \leq \dfrac{1}{N}\left( \dfrac{1}{N}\sum\limits_{l=1}^{N}\left%
\Vert \mathbf{\widehat{a}}_{l}\right\Vert ^{2}\right) \dfrac{1}{N}\left(
\sum\limits_{i=1}^{N}\sum\limits_{l=1}^{N}\sigma _{jil}^{2}\right) :
\end{equation*}
%----------------------------------------------------------------------------%
given Assumption \ref{assum:TCSDH}(b), $N^{-1}\left(
\sum\nolimits_{i=1}^{N}\sum\nolimits_{l=1}^{N}\sigma _{jil}^{2}\right)
\leq M$ by Lemma A.1(a) in \cite{Massacci_2017_JoE}, which implies that
%----------------------------------------------------------------------------%
\begin{equation}
\label{eq:Lemma_1_2}
\frac{1}{N}\sum\limits_{i=1}^{N}\widehat{\sigma}_{ji\cdot }=O_{p}\left( \dfrac{1%
}{N}\right) .
\end{equation}
%----------------------------------------------------------------------------%
Consider now,
%----------------------------------------------------------------------------% 
\begin{equation*}
\begin{array}{rcl}
\sum\limits_{i=1}^{N}\widehat{\chi}_{ji\cdot } & = & \dfrac{1}{N^{2}}%
\sum\limits_{i=1}^{N}\left\Vert \sum\limits_{l=1}^{N}\mathbf{\widehat{a}}%
_{l}\chi _{jil}\right\Vert ^{2} \\ 
& = & \dfrac{1}{N^{2}}\sum\limits_{i=1}^{N}\sum\limits_{l=1}^{N}\sum%
\limits_{q=1}^{N}\mathbf{\widehat{a}}_{l} ^{\prime }\mathbf{%
	\widehat{a}}_{q}\chi _{jil}\chi _{jiq} \\ 
& \leq & \left[\dfrac{1}{N^{2}}\sum\limits_{l=1}^{N}\sum\limits_{q=1}^{N}%
\left(\mathbf{\widehat{a}}_{l} ^{\prime }\mathbf{\widehat{a}}%
_{q}\right)^{2}\right]^{1\left/ 2\right. }\left[ \dfrac{1}{N^{2}}%
\sum\limits_{l=1}^{N}\sum\limits_{q=1}^{N}\left(
\sum\limits_{i=1}^{N}\chi _{jil}\chi _{jiq}\right) ^{2}\right] ^{1\left/
	2\right. } \\ 
& \leq & \left( \dfrac{1}{N}\sum\limits_{l=1}^{N}\left\Vert \mathbf{\widehat{a}}%
_{l}\right\Vert ^{2}\right) \left[ \dfrac{1}{N^{2}}\sum%
\limits_{l=1}^{N}\sum\limits_{q=1}^{N}\left( \sum\limits_{i=1}^{N}\chi
_{jil}\chi _{jiq}\right) ^{2}\right] ^{1\left/ 2\right. };%
\end{array}%
\end{equation*}
%----------------------------------------------------------------------------%
since
%----------------------------------------------------------------------------%
\begin{equation*}
\E\left[ \left( \sum\limits_{i=1}^{N}\chi _{jil}\chi _{jiq}\right)
^{2}\right] =\E\left(
\sum\limits_{i=1}^{N}\sum\limits_{u=1}^{N}\chi _{jil}\chi _{jiq}\chi
_{jul}\chi _{juq}\right) \leq N^{2}\max_{i,l}\E\left( \left\vert
\chi _{jil}\right\vert ^{4}\right)
\end{equation*}
%----------------------------------------------------------------------------%
and%
\begin{equation*}
\begin{array}{rcl}
\E\left( \left\vert \chi _{jil}\right\vert ^{4}\right) & = & \E\left[ \left\vert \dfrac{1}{T}\sum\limits_{t=1}^{T}\mathbb{I}%
_{jt}e_{it}e_{lt}-\E\left( \dfrac{1}{T}\sum\limits_{t=1}^{T}\mathbb{%
	I}_{jt}e_{it}e_{lt}\right) \right\vert ^{4}\right] \\ 
& = & \dfrac{1}{T^{2}}\E\left\{ \left\vert \dfrac{1}{\sqrt{T}}\left[
\sum\limits_{t=1}^{T}\mathbb{I}_{jt}e_{it}e_{lt}-\E\left(
\sum\limits_{t=1}^{T}\mathbb{I}_{jt}e_{it}e_{lt}\right) \right] \right\vert
^{4}\right\} \\ 
& \leq & \dfrac{1}{T^{2}}M%
\end{array}%
\end{equation*}
%----------------------------------------------------------------------------%
by Assumption \ref{assum:TCSDH}(c), then%
%----------------------------------------------------------------------------%
\begin{equation*}
\sum\limits_{i=1}^{N}\widehat{\chi}_{ji\cdot }\leq O_{p}\left( 1\right) \sqrt{%
	\dfrac{N^{2}}{T^{2}}}=O_{p}\left( \dfrac{N}{T}\right)
\end{equation*}%
%----------------------------------------------------------------------------%
and%
%----------------------------------------------------------------------------%
\begin{equation}
\label{eq:Lemma_1_3}
\dfrac{1}{N}\sum\limits_{i=1}^{N}\widehat{\chi}_{ji\cdot }=O_{p}\left( \dfrac{1%
}{T}\right) .
\end{equation}
%----------------------------------------------------------------------------%
Also
%----------------------------------------------------------------------------%
\begin{equation*}
\begin{array}{rcl}
\widehat{\varphi}_{ji\cdot } & = & \dfrac{1}{N^{2}}\left\Vert
\sum\limits_{l=1}^{N}\mathbf{\widehat{a}}_{l}\varphi _{jil}\right\Vert ^{2}
\\ 
& = & \dfrac{1}{N^{2}}\left\Vert \sum\limits_{l=1}^{N}\mathbf{\widehat{a}}%
_{l}\left( \dfrac{1}{T}\sum\limits_{t=1}^{T}\mathbb{I}_{jt}\bm{%
	\lambda }_{ji}^{\prime }\mathbf{f}_{jt}e_{lt}\right) \right\Vert ^{2} \\ 
& = & \dfrac{1}{N^{2}}\left\Vert \sum\limits_{l=1}^{N}\mathbf{\widehat{a}}%
_{l}\bm{\lambda }_{ji}^{\prime }\left( \dfrac{1}{T}%
\sum\limits_{t=1}^{T}\mathbb{I}_{jt}\mathbf{f}_{jt}e_{lt}\right)
\right\Vert ^{2} \\ 
& \leq & \left[ \dfrac{1}{N}\sum\limits_{l=1}^{N}\left( \dfrac{1}{T^{2}}%
\left\Vert \sum\limits_{t=1}^{T}\mathbb{I}_{jt}\mathbf{f}%
_{jt}e_{lt}\right\Vert ^{2}\right) \right] \left\Vert \bm{\lambda }%
_{ji}\right\Vert ^{2}\left( \dfrac{1}{N}\sum\limits_{l=1}^{N}\left\Vert 
\mathbf{\widehat{a}}_{l}\right\Vert ^{2}\right)%
\end{array}%
\end{equation*}
%----------------------------------------------------------------------------%
and
%----------------------------------------------------------------------------%
\begin{equation}
\label{eq:Lemma_1_4}
\begin{array}{rcl}
\dfrac{1}{N}\sum\limits_{i=1}^{N}\widehat{\varphi}_{ji\cdot } & = & \left[ 
\dfrac{1}{N}\sum\limits_{l=1}^{N}\left( \dfrac{1}{T^{2}}\left\Vert
\sum\limits_{t=1}^{T}\mathbb{I}_{jt}\mathbf{f}_{jt}e_{lt}\right\Vert
^{2}\right) \right] \left( \dfrac{1}{N}\sum\limits_{i=1}^{N}\left\Vert 
\bm{\lambda }_{ji}\right\Vert ^{2}\right) \left( \dfrac{1}{N}%
\sum\limits_{l=1}^{N}\left\Vert \mathbf{\widehat{a}}_{l}\right\Vert
^{2}\right) \\ 
& = & \dfrac{1}{T}\left( \dfrac{1}{N}\sum\limits_{l=1}^{N}\left\Vert \dfrac{%
	1}{\sqrt{T}}\sum\limits_{t=1}^{T}\mathbb{I}_{jt}\mathbf{f}%
_{jt}e_{lt}\right\Vert ^{2}\right) \left( \dfrac{1}{N}\sum\limits_{i=1}^{N}%
\left\Vert \bm{\lambda }_{ji}\right\Vert ^{2}\right) \left( \dfrac{1}{N}%
\sum\limits_{l=1}^{N}\left\Vert \mathbf{\widehat{a}}_{l}\right\Vert
^{2}\right) \\ 
& = & O_{p}\left( \dfrac{1}{T}\right)%
\end{array}%
\end{equation}
%----------------------------------------------------------------------------%
by Assumptions \ref{assum:FL} and \ref{assum:WD}. Finally,
%----------------------------------------------------------------------------%
\begin{equation*}
\begin{array}{rcl}
\widehat{\varphi}_{j\cdot i} & = & \dfrac{1}{N^{2}}\left\Vert
\sum\limits_{l=1}^{N}\mathbf{\widehat{a}}_{l}\varphi _{jli}\right\Vert ^{2}
\\ 
& = & \dfrac{1}{N^{2}}\left\Vert \sum\limits_{l=1}^{N}\mathbf{\widehat{a}}%
_{l}\left( \dfrac{1}{T}\sum\limits_{t=1}^{T}\mathbb{I}_{jt}\bm{%
	\lambda }_{jl}^{\prime }\mathbf{f}_{jt}e_{it}\right) \right\Vert ^{2} \\ 
& = & \dfrac{1}{N^{2}}\left\Vert \sum\limits_{l=1}^{N}\mathbf{\widehat{a}}%
_{l}\bm{\lambda }_{jl}^{\prime }\left( \dfrac{1}{T}%
\sum\limits_{t=1}^{T}\mathbb{I}_{jt}\mathbf{f}_{jt}e_{it}\right)
\right\Vert ^{2} \\ 
& \leq & \dfrac{1}{N^{2}}\left\Vert \sum\limits_{l=1}^{N}\mathbf{\widehat{a}}%
_{l}\bm{\lambda }_{jl}^{\prime }\right\Vert ^{2}\left\Vert \dfrac{1}{%
	T}\sum\limits_{t=1}^{T}\mathbb{I}_{jt}\mathbf{f}_{jt}e_{it}\right\Vert ^{2}
\\ 
& \leq & \dfrac{1}{T}\left\Vert \dfrac{1}{\sqrt{T}}\sum\limits_{t=1}^{T}%
\mathbb{I}_{jt}\mathbf{f}_{jt}e_{it}\right\Vert ^{2}\left( \dfrac{1}{N}%
\sum\limits_{l=1}^{N}\left\Vert \bm{\lambda }_{jl}\right\Vert
^{2}\right) \left( \dfrac{1}{N}\sum\limits_{l=1}^{N}\left\Vert \mathbf{\widehat{%
		a}}_{l}\right\Vert ^{2}\right)%
\end{array}%
\end{equation*}
%----------------------------------------------------------------------------%
and
%----------------------------------------------------------------------------%
\begin{equation}
\label{eq:Lemma_1_5}
\dfrac{1}{N}\sum\limits_{i=1}^{N}\widehat{\varphi}_{j\cdot i}\leq \dfrac{1}{T}%
\left( \dfrac{1}{N}\sum\limits_{i=1}^{N}\left\Vert \dfrac{1}{\sqrt{T}}%
\sum\limits_{t=1}^{T}\mathbb{I}_{jt}\mathbf{f}_{jt}e_{it}\right\Vert
^{2}\right) \left( \dfrac{1}{N}\sum\limits_{l=1}^{N}\left\Vert \bm{%
	\lambda }_{jl}\right\Vert ^{2}\right) \left( \dfrac{1}{N}\sum%
\limits_{l=1}^{N}\left\Vert \mathbf{\widehat{a}}_{l}\right\Vert ^{2}\right)
=O_{p}\left( \dfrac{1}{T}\right)
\end{equation}
%----------------------------------------------------------------------------%
by Assumptions \ref{assum:FL} and \ref{assum:WD}. By combining (\ref{eq:Lemma_1_1}) - (\ref{eq:Lemma_1_5}), and since $\left\Vert \mathbf{\widehat{V}} ^{-1}\right\Vert=O_{p}\left(1\right)$, then
%----------------------------------------------------------------------------%
\begin{equation*}
\dfrac{1}{N}\sum\limits_{i=1}^{N}\left\Vert \mathbf{\widehat{a}}_{i}-\mathbf{\widehat{H}} ^{\prime }\mathbf{a}_{i}\right\Vert
^{2}=O_{p}\left( \dfrac{1}{N}\right) +O_{p}\left( \dfrac{1}{T}\right)
\end{equation*}
%----------------------------------------------------------------------------%
and the result stated in the lemma follows.
\end{proof}

\begin{proof}[\textbf{Proof of Lemma \protect\ref{Lemma:a_i_hat_components}}]
	Starting from $\left( a\right) $, consider%
	\begin{equation*}
	\dfrac{1}{N}\sum\limits_{l=1}^{N}\mathbf{\widehat{a}}_{l}\sigma _{jil}=\dfrac{1%
	}{N}\sum\limits_{l=1}^{N}\left( \mathbf{\widehat{a}}_{l}-\mathbf{\widehat{H}}%
	^{\prime }\mathbf{a}_{l}+\mathbf{\widehat{H}}^{\prime }\mathbf{a}_{l}\right)
	\sigma _{jil}=\dfrac{1}{N}\sum\limits_{l=1}^{N}\left( \mathbf{\widehat{a}}_{l}-%
	\mathbf{\widehat{H}}^{\prime }\mathbf{a}_{l}\right) \sigma _{jil}+\mathbf{\widehat{H}%
	}^{\prime }\dfrac{1}{N}\sum\limits_{l=1}^{N}\mathbf{a}_{l}\sigma _{jil}.
	\end{equation*}%
	Note that%
	\begin{equation*}
	\left\Vert \sum\limits_{l=1}^{N}\mathbf{a}_{l}\sigma _{jil}\right\Vert \leq
	\left( \max_{l}\left\Vert \mathbf{a}_{l}\right\Vert \right) \left(
	\sum\limits_{l=1}^{N}\left\vert \sigma _{jil}\right\vert \right) \leq \left[
	\max_{l}\left( \left\Vert \bm{\lambda }_{1l}\right\Vert +\left\Vert 
	\bm{\lambda }_{2l}\right\Vert \right) \right] \left(
	\sum\limits_{l=1}^{N}\left\vert \sigma _{jil}\right\vert \right) \leq 2\bar{%
		\lambda}M
	\end{equation*}%
	by Assumption \ref{assum:FL} and Assumption \ref{assum:TCSDH}(b), so that%
	\begin{equation*}
	\dfrac{1}{N}\sum\limits_{l=1}^{N}\mathbf{a}_{l}\sigma _{jil}=O\left( \dfrac{%
		1}{N}\right) .
	\end{equation*}%
	Further%
	\begin{equation*}
	\begin{array}{rcl}
	\left\Vert \dfrac{1}{N}\sum\limits_{l=1}^{N}\left( \mathbf{\widehat{a}}_{l}-%
	\mathbf{\widehat{H}}^{\prime }\mathbf{a}_{l}\right) \sigma _{jil}\right\Vert & 
	\leq & \left( \dfrac{1}{N}\sum\limits_{l=1}^{N}\left\Vert \mathbf{\widehat{a}}%
	_{l}-\mathbf{\widehat{H}}^{\prime }\mathbf{a}_{l}\right\Vert ^{2}\right)
	^{1\left/ 2\right. }\dfrac{1}{\sqrt{N}}\left(
	\sum\limits_{l=1}^{N}\left\vert \sigma _{jil}\right\vert ^{2}\right)
	^{1\left/ 2\right. } \\ 
	& = & \left[ O_{p}\left( \dfrac{1}{C_{NT}^{2}}\right) \right] ^{1\left/
		2\right. }O_{p}\left( \dfrac{1}{\sqrt{N}}\right) \\ 
	& = & O_{p}\left( \dfrac{1}{\sqrt{N}C_{NT}}\right)%
	\end{array}%
	\end{equation*}%
	by Lemma \ref{Lemma:a_i_hat} and Assumption \ref{assum:TCSDH}(b). It thus follows that%
	\begin{equation*}
	\dfrac{1}{N}\sum\limits_{l=1}^{N}\mathbf{\widehat{a}}_{l}\sigma
	_{jil}=O_{p}\left( \dfrac{1}{\sqrt{N}C_{NT}}\right) +O_{p}\left( \dfrac{1}{N}%
	\right) =O_{p}\left( \dfrac{1}{\sqrt{N}C_{NT}}\right) .
	\end{equation*}%
	Moving on to $\left( b\right) $, we have%
	\begin{equation*}
	\dfrac{1}{N}\sum\limits_{l=1}^{N}\mathbf{\widehat{a}}_{l}\chi _{jil}=\dfrac{1}{N%
	}\sum\limits_{l=1}^{N}\left( \mathbf{\widehat{a}}_{l}-\mathbf{\widehat{H}}^{\prime }%
	\mathbf{a}_{l}\right) \chi _{jil}+\mathbf{\widehat{H}}^{\prime }\dfrac{1}{N}%
	\sum\limits_{l=1}^{N}\mathbf{a}_{l}\chi _{jil}.
	\end{equation*}%
	Note that%
	\begin{equation*}
	\left\Vert \dfrac{1}{N}\sum\limits_{l=1}^{N}\left( \mathbf{\widehat{a}}_{l}-%
	\mathbf{\widehat{H}}^{\prime }\mathbf{a}_{l}\right) \chi _{jil}\right\Vert \leq
	\left( \dfrac{1}{N}\sum\limits_{l=1}^{N}\left\Vert \mathbf{\widehat{a}}_{l}-%
	\mathbf{\widehat{H}}^{\prime }\mathbf{a}_{l}\right\Vert ^{2}\right) ^{1\left/
		2\right. }\left( \dfrac{1}{N}\sum\limits_{l=1}^{N}\chi _{jil}^{2}\right)
	^{1\left/ 2\right. },
	\end{equation*}%
	with%
	\begin{equation*}
	\begin{array}{rcl}
	\dfrac{1}{N}\sum\limits_{l=1}^{N}\chi _{jil}^{2} & = & \dfrac{1}{N}%
	\sum\limits_{l=1}^{N}\left[ \dfrac{1}{T}\sum\limits_{t=1}^{T}\mathbb{I}%
	_{jt}e_{it}e_{lt}-\E\left( \dfrac{1}{T}\sum\limits_{t=1}^{T}\mathbb{%
		I}_{jt}e_{it}e_{lt}\right) \right] ^{2} \\ 
	& = & \dfrac{1}{NT}\sum\limits_{l=1}^{N}\left\{ \dfrac{1}{\sqrt{T}}%
	\sum\limits_{t=1}^{T}\left[ \mathbb{I}_{jt}e_{it}e_{lt}-\E\left( 
	\mathbb{I}_{jt}e_{it}e_{lt}\right) \right] \right\} ^{2} \\ 
	& = & O_{p}\left( \dfrac{1}{T}\right)%
	\end{array}%
	\end{equation*}%
	so that%
	\begin{equation*}
	\left\Vert \dfrac{1}{N}\sum\limits_{l=1}^{N}\left( \mathbf{\widehat{a}}_{l}-%
	\mathbf{\widehat{H}}^{\prime }\mathbf{a}_{l}\right) \chi _{jil}\right\Vert
	=O_{p}\left( \dfrac{1}{C_{NT}}\right) O_{p}\left( \dfrac{1}{\sqrt{T}}\right)
	=O_{p}\left( \dfrac{1}{\sqrt{T}C_{NT}}\right) .
	\end{equation*}%
	Further%
	\begin{equation*}
	\begin{array}{rcl}
	\dfrac{1}{N}\sum\limits_{l=1}^{N}\mathbf{a}_{l}\chi _{jil} & = & \dfrac{1}{N}\sum\limits_{l=1}^{N}\mathbf{a}_{l}\left[ \dfrac{1}{T}\sum%
	\limits_{t=1}^{T}\mathbb{I}_{jt}e_{it}e_{lt}-\E\left( \dfrac{1}{T}%
	\sum\limits_{t=1}^{T}\mathbb{I}_{jt}e_{it}e_{lt}\right) \right] \\ 
	& = & \dfrac{1}{NT}\sum\limits_{l=1}^{N}\sum\limits_{t=1}^{T}%
	\mathbf{a}_{l}\left[ \mathbb{I}_{jt}e_{it}e_{lt}-\E\left( \mathbb{I}%
	_{jt}e_{it}e_{lt}\right) \right] \\ 
	& = & O_{p}\left( \dfrac{1}{\sqrt{NT}}\right)%
	\end{array}%
	\end{equation*}%
	by Assumption \ref{assum:MCLT}(a). It follows that%
	\begin{equation*}
	\dfrac{1}{N}\sum\limits_{l=1}^{N}\mathbf{\widehat{a}}_{l}\chi
	_{jil}=O_{p}\left( \dfrac{1}{\sqrt{T}C_{NT}}\right) +O_{p}\left( \dfrac{1}{%
		\sqrt{NT}}\right) =O_{p}\left( \dfrac{1}{\sqrt{T}C_{NT}}\right) .
	\end{equation*}%
	As for $\left( c\right) $, consider%
	\begin{equation*}
	\begin{array}{rcl}
	\dfrac{1}{N}\sum\limits_{l=1}^{N}\mathbf{\widehat{a}}_{l}\varphi _{jil} & = & 
	\dfrac{1}{N}\sum\limits_{l=1}^{N}\mathbf{\widehat{a}}_{l}\left( \dfrac{1}{T}%
	\sum\limits_{t=1}^{T}\mathbb{I}_{jt}\bm{\lambda }_{ji}^{\prime }\mathbf{%
		f}_{jt}e_{lt}\right) \\ 
	& = & \dfrac{1}{NT}\sum\limits_{l=1}^{N}\sum\limits_{t=1}^{T}\mathbb{I}%
	_{jt}\mathbf{\widehat{a}}_{l}e_{lt}\mathbf{f}_{jt}^{\prime }\bm{\lambda }%
	_{ji} \\ 
	& = & \dfrac{1}{NT}\sum\limits_{l=1}^{N}\sum\limits_{t=1}^{T}\mathbb{I}%
	_{jt}\left( \mathbf{\widehat{a}}_{l}-\mathbf{\widehat{H}}^{\prime }\mathbf{a}_{l}+%
	\mathbf{\widehat{H}}^{\prime }\mathbf{a}_{l}\right) e_{lt}\mathbf{f}%
	_{jt}^{\prime }\bm{\lambda }_{ji} \\ 
	& = & \dfrac{1}{NT}\sum\limits_{l=1}^{N}\sum\limits_{t=1}^{T}\mathbb{I}%
	_{jt}\left( \mathbf{\widehat{a}}_{l}-\mathbf{\widehat{H}}^{\prime }\mathbf{a}%
	_{l}\right) e_{lt}\mathbf{f}_{jt}^{\prime }\bm{\lambda }_{ji}+\mathbf{%
		\widehat{H}}^{\prime }\dfrac{1}{NT}\sum\limits_{l=1}^{N}\sum\limits_{t=1}^{T}%
	\mathbb{I}_{jt}\mathbf{a}_{l}e_{lt}\mathbf{f}_{jt}^{\prime }\bm{\lambda }%
	_{ji}.%
	\end{array}%
	\end{equation*}%
	We have%
	\begin{equation*}
	\begin{array}{rcl}
	\left\Vert \dfrac{1}{NT}\sum\limits_{l=1}^{N}\sum\limits_{t=1}^{T}\mathbb{I%
	}_{jt}\left( \mathbf{\widehat{a}}_{l}-\mathbf{\widehat{H}}^{\prime }\mathbf{a}%
	_{l}\right) e_{lt}\mathbf{f}_{jt}^{\prime }\bm{\lambda }_{ji}\right\Vert
	& \leq & \dfrac{1}{\sqrt{T}}\left( \dfrac{1}{N}\sum\limits_{l=1}^{N}\left%
	\Vert \mathbf{\widehat{a}}_{l}-\mathbf{\widehat{H}}^{\prime }\mathbf{a}%
	_{l}\right\Vert ^{2}\right) ^{1\left/ 2\right. }\\
	&   & \times \left( \dfrac{1}{N}%
	\sum\limits_{l=1}^{N}\left\Vert \dfrac{1}{\sqrt{T}}\sum\limits_{t=1}^{T}%
	\mathbb{I}_{jt}e_{lt}\mathbf{f}_{jt}\right\Vert ^{2}\right) ^{1\left/
		2\right. }\left\Vert \mathbf{\lambda }_{ji}\right\Vert \\
	& = & O\left( \dfrac{1}{\sqrt{T}}\right) O_{p}\left( \dfrac{1}{C_{NT}}%
	\right) O_{p}\left( 1\right) O\left( 1\right) \\ 
	& = & O_{p}\left( \dfrac{1}{\sqrt{T}C_{NT}}\right)%
	\end{array}%
	\end{equation*}%
	by Lemma \ref{Lemma:a_i_hat}, Assumption \ref{assum:MCLT}(c) and Assumption \ref{assum:FL}. Also,%
	\begin{equation*}
	\dfrac{1}{NT}\sum\limits_{l=1}^{N}\sum\limits_{t=1}^{T}\mathbb{I}_{jt}%
	\mathbf{a}_{l}e_{lt}\mathbf{f}_{jt}^{\prime }\bm{\lambda }_{ji}=\dfrac{1%
	}{\sqrt{NT}}\left[ \dfrac{1}{\sqrt{NT}}\sum\limits_{l=1}^{N}\sum%
	\limits_{t=1}^{T}\mathbb{I}_{jt}\left( 
	\begin{array}{c}
	\bm{\lambda }_{1l} \\ 
	\bm{\lambda }_{2l}%
	\end{array}%
	\right) e_{lt}\mathbf{f}_{jt}^{\prime }\right] \bm{\lambda }%
	_{ji}=O_{p}\left( \dfrac{1}{\sqrt{NT}}\right)
	\end{equation*}%
	by Assumption \ref{assum:MCLT}(b) and Assumption \ref{assum:FL}. It follows that%
	\begin{equation*}
	\dfrac{1}{N}\sum\limits_{l=1}^{N}\mathbf{\widehat{a}}_{l}\varphi
	_{jil}=O_{p}\left( \dfrac{1}{\sqrt{T}C_{NT}}\right) +O_{p}\left( \dfrac{1}{%
		\sqrt{NT}}\right) =O_{p}\left( \dfrac{1}{\sqrt{T}C_{NT}}\right) .
	\end{equation*}%
	Finally, for $\left( d\right) $ we have%
	\begin{equation*}
	\dfrac{1}{N}\sum\limits_{l=1}^{N}\mathbf{\widehat{a}}_{l}\varphi _{jli}=\dfrac{1%
	}{N}\sum\limits_{l=1}^{N}\left( \mathbf{\widehat{a}}_{l}-\mathbf{\widehat{H}}%
	^{\prime }\mathbf{a}_{l}\right) \varphi _{jli}+\mathbf{\widehat{H}}^{\prime }%
	\dfrac{1}{N}\sum\limits_{l=1}^{N}\mathbf{a}_{l}\varphi _{jli}.
	\end{equation*}%
	Note that%
	\begin{equation*}
	\begin{array}{rcl}
	\dfrac{1}{N}\sum\limits_{l=1}^{N}\mathbf{a}_{l}\varphi _{jli} & = & \dfrac{1%
	}{N}\sum\limits_{l=1}^{N}\mathbf{a}_{l}\left( \dfrac{1}{T}%
	\sum\limits_{t=1}^{T}\mathbb{I}_{jt}\bm{\lambda }_{jl}^{\prime }\mathbf{%
		f}_{jt}e_{it}\right) \\ 
	& = & \left( \dfrac{1}{N}\sum\limits_{l=1}^{N}\mathbf{a}_{l}\bm{\lambda 
	}_{jl}^{\prime }\right) \left( \dfrac{1}{T}\sum\limits_{t=1}^{T}\mathbb{I}%
	_{jt}\mathbf{f}_{jt}e_{it}\right) \\ 
	& = & \left[ \dfrac{1}{N}\sum\limits_{l=1}^{N}\left( 
	\begin{array}{c}
	\bm{\lambda }_{1l} \\ 
	\bm{\lambda }_{2l}%
	\end{array}%
	\right) \bm{\lambda }_{jl}^{\prime }\right] \dfrac{1}{\sqrt{T}}\left( 
	\dfrac{1}{\sqrt{T}}\sum\limits_{t=1}^{T}\mathbb{I}_{jt}\mathbf{f}%
	_{jt}e_{it}\right) \\ 
	& = & O_{p}\left( \dfrac{1}{\sqrt{T}}\right) ,%
	\end{array}%
	\end{equation*}%
	by Assumption \ref{assum:FL} and Assumption \ref{assum:MCLT}(c). Further,%
	\begin{equation*}
	\left\Vert \dfrac{1}{N}\sum\limits_{l=1}^{N}\left( \mathbf{\widehat{a}}_{l}-%
	\mathbf{\widehat{H}}^{\prime }\mathbf{a}_{l}\right) \varphi _{jli}\right\Vert
	\leq \left( \dfrac{1}{N}\sum\limits_{l=1}^{N}\left\Vert \mathbf{\widehat{a}}%
	_{l}-\mathbf{\widehat{H}}^{\prime }\mathbf{a}_{l}\right\Vert ^{2}\right)
	^{1\left/ 2\right. }\left( \dfrac{1}{N}\sum\limits_{l=1}^{N}\varphi
	_{jli}^{2}\right) ^{1\left/ 2\right. }
	\end{equation*}%
	with%
	\begin{equation*}
	\dfrac{1}{N}\sum\limits_{l=1}^{N}\varphi _{jli}^{2}=\dfrac{1}{N}%
	\sum\limits_{l=1}^{N}\left( \dfrac{1}{T}\sum\limits_{t=1}^{T}\mathbb{I}%
	_{jt}\bm{\lambda }_{jl}^{\prime }\mathbf{f}_{jt}e_{it}\right) ^{2}\leq 
	\dfrac{1}{T}\left( \dfrac{1}{N}\sum\limits_{l=1}^{N}\left\Vert \bm{%
		\lambda }_{jl}\right\Vert ^{2}\right) \left( \dfrac{1}{\sqrt{T}}%
	\sum\limits_{t=1}^{T}\mathbb{I}_{jt}\mathbf{f}_{jt}e_{it}\right) ^{2}\leq
	O_{p}\left( \dfrac{1}{T}\right) ,
	\end{equation*}%
	by Assumption \ref{assum:FL} and Assumption \ref{assum:MCLT}(c), so that taking into account Lemma \ref{Lemma:a_i_hat} we have%
	\begin{equation*}
	\dfrac{1}{N}\sum\limits_{l=1}^{N}\left( \mathbf{\widehat{a}}_{l}-\mathbf{\widehat{H}%
	}^{\prime }\mathbf{a}_{l}\right) \varphi _{jli}=O_{p}\left( \dfrac{1}{C_{NT}}%
	\right) O_{p}\left( \dfrac{1}{\sqrt{T}}\right) =O_{p}\left( \dfrac{1}{\sqrt{T%
		}C_{NT}}\right) .
	\end{equation*}%
	It follows that%
	\begin{equation*}
	\dfrac{1}{N}\sum\limits_{l=1}^{N}\mathbf{\widehat{a}}_{l}\varphi
	_{jli}=O_{p}\left( \dfrac{1}{\sqrt{T}C_{NT}}\right) +O_{p}\left( \dfrac{1}{%
		\sqrt{T}}\right) =O_{p}\left( \dfrac{1}{\sqrt{T}}\right) ,
	\end{equation*}%
	which completes the proof of the lemma.
\end{proof}

\begin{proof}[\textbf{Proof of Lemma \protect\ref{Lemma:3}}]
	Consider%
%----------------------------------------------------------------------------%
	\begin{equation}
	\label{eq:Lemma_3}
	\begin{array}{rcl}
	N^{-1}\left( \mathbf{\widehat{A}}-\mathbf{A\widehat{H}}\right) ^{\prime }\mathbf{%
		\widehat{A}} & = & N^{-1}\left( \mathbf{\widehat{A}}-\mathbf{A\widehat{H}}\right)
	^{\prime }\mathbf{\widehat{A}}-N^{-1}\left( \mathbf{\widehat{A}}-\mathbf{A\widehat{H}}%
	\right) ^{\prime }\mathbf{A\widehat{H}}+N^{-1}\left( \mathbf{\widehat{A}}-\mathbf{A%
		\widehat{H}}\right) ^{\prime }\mathbf{A\widehat{H}} \\ 
	& = & N^{-1}\left( \mathbf{\widehat{A}}-\mathbf{A\widehat{H}}\right) ^{\prime }%
	\mathbf{A\widehat{H}}+N^{-1}\left( \mathbf{\widehat{A}}-\mathbf{A\widehat{H}}\right)
	^{\prime }\left( \mathbf{\widehat{A}}-\mathbf{A\widehat{H}}\right) .%
	\end{array}%
	\end{equation}
%----------------------------------------------------------------------------%
	Using the identity in (\ref{eq:a__hat-a_i}), we have%
%----------------------------------------------------------------------------%
	\begin{equation}
	\label{eq:Lemma_3.1}
	\begin{array}{rcl}
	N^{-1}\left( \mathbf{\widehat{A}}-\mathbf{A\widehat{H}}\right) ^{\prime }\mathbf{A}
	& = & \mathbf{\widehat{V}}^{-1}\dfrac{1}{N}\sum\limits_{i=1}^{N}\left( \mathbf{%
		\widehat{a}}_{i}-\mathbf{\widehat{H}}^{\prime }\mathbf{a}_{i}\right) \mathbf{a}_{i}
	\\ 
	& = & \mathbf{\widehat{V}}^{-1}\left\{ 
	\begin{array}{c}
	\sum\limits_{j=1}^{2}\left[ \dfrac{1}{N}\sum\limits_{i=1}^{N}\left( \dfrac{1%
	}{N}\sum\limits_{l=1}^{N}\mathbf{\widehat{a}}_{l}\sigma _{jil}\right) 
	\mathbf{a}_{i}^{\prime }\right] +\sum\limits_{j=1}^{2}\left[ \dfrac{1}{N}%
	\sum\limits_{i=1}^{N}\left( \dfrac{1}{N}\sum\limits_{l=1}^{N}\mathbf{\widehat{a}}%
	_{l}\chi _{jil}\right) \mathbf{a}_{i}^{\prime }\right] \\ 
	+\sum\limits_{j=1}^{2}\left[ \dfrac{1}{N}\sum\limits_{i=1}^{N}\left( \dfrac{1%
	}{N}\sum\limits_{l=1}^{N}\mathbf{\widehat{a}}_{l}\varphi _{jil}\right) 
	\mathbf{a}_{i}^{\prime }\right] +\sum\limits_{j=1}^{2}\left[ \dfrac{1}{N}%
	\sum\limits_{i=1}^{N}\left( \dfrac{1}{N}\sum\limits_{l=1}^{N}\mathbf{\widehat{a}}%
	_{l}\varphi _{jli}\right) \mathbf{a}_{i}^{\prime }\right]%
	\end{array}%
	\right\}.%
	\end{array}%
	\end{equation}
%----------------------------------------------------------------------------%
	Consider
%----------------------------------------------------------------------------%
	\begin{equation*}
	\dfrac{1}{N}\sum\limits_{i=1}^{N}\left( \dfrac{1}{N}\sum\limits_{l=1}^{N}%
	\mathbf{\widehat{a}}_{l}\sigma _{jil}\right) \mathbf{a}_{i}^{\prime }=\dfrac{1}{N%
	}\sum\limits_{i=1}^{N}\left[ \dfrac{1}{N}\sum\limits_{l=1}^{N}\left( \mathbf{%
		\widehat{a}}_{l}-\mathbf{\widehat{H}}^{\prime }\mathbf{a}_{l}\right) \sigma _{jil}%
	\right] \mathbf{a}_{i}^{\prime }+\mathbf{\widehat{H}}^{\prime }\dfrac{1}{N}%
	\sum\limits_{i=1}^{N}\dfrac{1}{N}\sum\limits_{l=1}^{N}\mathbf{a}_{l}\mathbf{a%
	}_{i}^{\prime }\sigma _{jil}.
	\end{equation*}
%----------------------------------------------------------------------------%
	We have
%----------------------------------------------------------------------------%
	\begin{equation*}
	\begin{array}{rcl}
	\left\Vert \dfrac{1}{N}\sum\limits_{i=1}^{N}\left( \dfrac{1}{N}%
	\sum\limits_{l=1}^{N}\mathbf{\widehat{a}}_{l}\sigma _{jil}\right) \mathbf{a}%
	_{i}^{\prime }\right\Vert & \leq & \dfrac{1}{\sqrt{N}}\left( \dfrac{1}{N}%
	\sum\limits_{l=1}^{N}\left\Vert \mathbf{\widehat{a}}_{l}-\mathbf{\widehat{H}}%
	^{\prime }\mathbf{a}_{l}\right\Vert ^{2}\right) ^{1\left/ 2\right. }\left( 
	\dfrac{1}{N}\sum\limits_{i=1}^{N}\sum\limits_{l=1}^{N}\left\vert \sigma
	_{jil}\right\vert ^{2}\right) ^{1\left/ 2\right. }\left( \dfrac{1}{N}%
	\sum\limits_{i=1}^{N}\left\Vert \mathbf{a}_{i}\right\Vert ^{2}\right)
	^{1\left/ 2\right. } \\ 
	& = & \dfrac{1}{\sqrt{N}}O_{p}\left( \dfrac{1}{C_{NT}}\right) O_{p}\left(
	1\right) O_{p}\left( 1\right) \\ 
	& = & O_{p}\left( \dfrac{1}{\sqrt{N}C_{NT}}\right) ,%
	\end{array}%
	\end{equation*}%
%----------------------------------------------------------------------------%
	by Lemma \ref{Lemma:a_i_hat}, Assumption \ref{assum:FL}, and the fact that, given $\rho _{jil}=\sigma
	_{jil}\left/ \left( \sigma _{jii}\sigma _{jll}\right) ^{1\left/ 2\right.
	}\right. $, by Assumption \ref{assum:TCSDH}(b) we have%
%----------------------------------------------------------------------------%
	\begin{equation}
	\label{eq:bound_M^sq}
	\dfrac{1}{N}\sum\limits_{i=1}^{N}\sum\limits_{l=1}^{N}\left\vert \sigma
	_{jil}\right\vert ^{2}=\dfrac{1}{N}\sum\limits_{i=1}^{N}\sum%
	\limits_{l=1}^{N}\sigma _{jii}\sigma _{jll}\rho _{jil}^{2}\leq M\dfrac{1}{N}%
	\sum\limits_{i=1}^{N}\sum\limits_{l=1}^{N}\left\vert \sigma _{jii}\sigma
	_{jll}\right\vert ^{1\left/ 2\right. }\left\vert \rho _{jil}\right\vert =M%
	\dfrac{1}{N}\sum\limits_{i=1}^{N}\sum\limits_{l=1}^{N}\left\vert \sigma
	_{jil}\right\vert \leq M^{2}.  
	\end{equation}
%----------------------------------------------------------------------------%
	Further
%----------------------------------------------------------------------------%
	\begin{equation*}
	\left\Vert \dfrac{1}{N}\sum\limits_{i=1}^{N}\dfrac{1}{N}\sum\limits_{l=1}^{N}%
	\mathbf{a}_{l}\mathbf{a}_{i}^{\prime }\sigma _{jil}\right\Vert \leq \dfrac{1%
	}{N}\left( \dfrac{1}{N}\sum\limits_{i=1}^{N}\sum\limits_{l=1}^{N}\left\Vert 
	\mathbf{a}_{l}\right\Vert \left\Vert \mathbf{a}_{i}\right\Vert \left\vert
	\sigma _{jil}\right\vert \right) =O\left( \dfrac{1}{N}\right)
	\end{equation*}%
%----------------------------------------------------------------------------%
	by Assumptions \ref{assum:FL} and \ref{assum:TCSDH}(b). Therefore,%
%----------------------------------------------------------------------------%
	\begin{equation}
	\label{eq:Lemma_3.1.1}
	\dfrac{1}{N}\sum\limits_{i=1}^{N}\left( \dfrac{1}{N}\sum\limits_{l=1}^{N}%
	\mathbf{\widehat{a}}_{l}\sigma _{jil}\right) \mathbf{a}_{i}^{\prime
	}=O_{p}\left( \dfrac{1}{\sqrt{N}C_{NT}}\right) +O\left( \dfrac{1}{N}\right)
	=O_{p}\left( \dfrac{1}{\sqrt{N}C_{NT}}\right) .
	\end{equation}%
%----------------------------------------------------------------------------%
	Consider now%
%----------------------------------------------------------------------------%
	\begin{equation*}
	\dfrac{1}{N}\sum\limits_{i=1}^{N}\left( \dfrac{1}{N}\sum\limits_{l=1}^{N}%
	\mathbf{\widehat{a}}_{l}\chi _{jil}\right) \mathbf{a}_{i}^{\prime }=\dfrac{1}{N}%
	\sum\limits_{i=1}^{N}\left[ \dfrac{1}{N}\sum\limits_{l=1}^{N}\left( \mathbf{%
		\widehat{a}}_{l}-\mathbf{\widehat{H}}^{\prime }\mathbf{a}_{l}\right) \chi _{jil}%
	\right] \mathbf{a}_{i}^{\prime }+\mathbf{\widehat{H}}^{\prime }\dfrac{1}{N}%
	\sum\limits_{i=1}^{N}\dfrac{1}{N}\sum\limits_{l=1}^{N}\mathbf{a}_{l}\mathbf{a%
	}_{i}^{\prime }\chi _{jil}.
	\end{equation*}%
%----------------------------------------------------------------------------%
	We have%
%----------------------------------------------------------------------------%
	\begin{equation*}
	\left\Vert \dfrac{1}{N}\sum\limits_{i=1}^{N}\left[ \dfrac{1}{N}%
	\sum\limits_{l=1}^{N}\left( \mathbf{\widehat{a}}_{l}-\mathbf{\widehat{H}}^{\prime }%
	\mathbf{a}_{l}\right) \chi _{jil}\right] \mathbf{a}_{i}^{\prime }\right\Vert
	\leq \dfrac{1}{N}\sum\limits_{i=1}^{N}\left\Vert \dfrac{1}{N}%
	\sum\limits_{l=1}^{N}\left( \mathbf{\widehat{a}}_{l}-\mathbf{\widehat{H}}^{\prime }%
	\mathbf{a}_{l}\right) \chi _{jil}\right\Vert \left\Vert \mathbf{a}%
	_{i}\right\Vert
	\end{equation*}
%----------------------------------------------------------------------------%
	and consider
%----------------------------------------------------------------------------%
	\begin{equation*}
	\left\Vert \dfrac{1}{N}\sum\limits_{l=1}^{N}\left( \mathbf{\widehat{a}}_{l}-%
	\mathbf{\widehat{H}}^{\prime }\mathbf{a}_{l}\right) \chi _{jil}\right\Vert \leq
	\left( \dfrac{1}{N}\sum\limits_{l=1}^{N}\left\Vert \mathbf{\widehat{a}}_{l}-%
	\mathbf{\widehat{H}}^{\prime }\mathbf{a}_{l}\right\Vert ^{2}\right) ^{1\left/
		2\right. }\left( \dfrac{1}{N}\sum\limits_{l=1}^{N}\left\vert \chi
	_{jil}\right\vert ^{2}\right) ^{1\left/ 2\right. }
	\end{equation*}
%----------------------------------------------------------------------------%
	with
%----------------------------------------------------------------------------%
	\begin{equation*}
	\begin{array}{rcl}
	\left( \dfrac{1}{N}\sum\limits_{l=1}^{N}\left\vert \chi _{jil}\right\vert
	^{2}\right) ^{1\left/ 2\right. } & = & \left[ \dfrac{1}{N}%
	\sum\limits_{l=1}^{N}\left\vert \dfrac{1}{T}\sum\limits_{t=1}^{T}\mathbb{I}%
	_{jt}e_{it}e_{lt}-\E\left( \dfrac{1}{T}\sum\limits_{t=1}^{T}\mathbb{I%
	}_{jt}e_{it}e_{lt}\right) \right\vert ^{2}\right] ^{1\left/ 2\right. } \\ 
	& = & \dfrac{1}{\sqrt{T}}\left[ \dfrac{1}{N}\sum\limits_{l=1}^{N}\left\vert 
	\dfrac{1}{\sqrt{T}}\sum\limits_{t=1}^{T}\mathbb{I}_{jt}e_{it}e_{lt}-\E\left( \dfrac{1}{T}\sum\limits_{t=1}^{T}\mathbb{I}_{jt}e_{it}e_{lt}\right)
	\right\vert ^{2}\right] ^{1\left/ 2\right. } \\ 
	& = & O_{p}\left( \dfrac{1}{\sqrt{T}}\right)%
	\end{array}%
	\end{equation*}
%----------------------------------------------------------------------------%
	by Assumption \ref{assum:TCSDH}(c). Therefore, taking into account Lemma \ref{Lemma:a_i_hat},
%----------------------------------------------------------------------------%
	\begin{equation*}
	\left\Vert \dfrac{1}{N}\sum\limits_{l=1}^{N}\left( \mathbf{\widehat{a}}_{l}-%
	\mathbf{\widehat{H}}^{\prime }\mathbf{a}_{l}\right) \chi _{jil}\right\Vert
	=O_{p}\left( \dfrac{1}{C_{NT}}\right) O_{p}\left( \dfrac{1}{\sqrt{T}}\right)
	=O_{p}\left( \dfrac{1}{\sqrt{T}C_{NT}}\right) .
	\end{equation*}%
%----------------------------------------------------------------------------%
	Further,%
%----------------------------------------------------------------------------%
	\begin{equation*}
	\begin{array}{rcl}
	\left\Vert \dfrac{1}{N}\sum\limits_{i=1}^{N}\dfrac{1}{N}\sum\limits_{l=1}^{N}%
	\mathbf{a}_{l}\mathbf{a}_{i}^{\prime }\chi _{jil}\right\Vert & = & 
	\left\Vert \dfrac{1}{N}\sum\limits_{i=1}^{N}\dfrac{1}{N}\sum\limits_{l=1}^{N}%
	\mathbf{a}_{l}\mathbf{a}_{i}^{\prime }\left[ \dfrac{1}{T}\sum%
	\limits_{t=1}^{T}\mathbb{I}_{jt}e_{it}e_{lt}-\E\left( \dfrac{1}{T}%
	\sum\limits_{t=1}^{T}\mathbb{I}_{jt}e_{it}e_{lt}\right) \right] \right\Vert
	\\ 
	& \leq & \dfrac{1}{\sqrt{NT}}\left\{ \dfrac{1}{N}\sum\limits_{l=1}^{N}\left%
	\Vert \mathbf{a}_{l}\right\Vert \left\Vert \dfrac{1}{\sqrt{NT}}%
	\sum\limits_{i=1}^{N}\sum\limits_{t=1}^{T}\mathbf{a}_{i}\left[ \mathbb{I}%
	_{jt}e_{it}e_{lt}-\E\left( \mathbb{I}_{jt}e_{it}e_{lt}\right) \right]
	\right\Vert \right\} \\ 
	& \leq & \dfrac{1}{\sqrt{NT}}\left( \dfrac{1}{N}\sum\limits_{l=1}^{N}\left%
	\Vert \mathbf{a}_{l}\right\Vert ^{2}\right) ^{1\left/ 2\right. }\left\{ 
	\dfrac{1}{N}\sum\limits_{l=1}^{N}\left\Vert \dfrac{1}{\sqrt{NT}}%
	\sum\limits_{i=1}^{N}\sum\limits_{t=1}^{T}\mathbf{a}_{i}\left[ \mathbb{I}%
	_{jt}e_{it}e_{lt}-\E\left( \mathbb{I}_{jt}e_{it}e_{lt}\right) \right]
	\right\Vert ^{2}\right\} ^{1\left/ 2\right. } \\ 
	& = & O_{p}\left( \dfrac{1}{\sqrt{NT}}\right)%
	\end{array}%
	\end{equation*}%
%----------------------------------------------------------------------------%
	by Assumptions \ref{assum:FL} and \ref{assum:MCLT}(a). Therefore,%
%----------------------------------------------------------------------------%
	\begin{equation}
	\label{eq:Lemma_3.1.2}
	\dfrac{1}{N}\sum\limits_{i=1}^{N}\left( \dfrac{1}{N}\sum\limits_{l=1}^{N}%
	\mathbf{\widehat{a}}_{l}\chi _{jil}\right) \mathbf{a}_{i}^{\prime }=O_{p}\left( 
	\dfrac{1}{\sqrt{T}C_{NT}}\right) +O_{p}\left( \dfrac{1}{\sqrt{NT}}\right)
	=O_{p}\left( \dfrac{1}{\sqrt{T}C_{NT}}\right) .
	\end{equation}%
%----------------------------------------------------------------------------%
	Consider now%
%----------------------------------------------------------------------------%
	\begin{equation*}
	\dfrac{1}{N}\sum\limits_{i=1}^{N}\left( \dfrac{1}{N}\sum\limits_{l=1}^{N}%
	\mathbf{\widehat{a}}_{l}\varphi _{jil}\right) \mathbf{a}_{i}^{\prime }=\dfrac{1}{%
		N}\sum\limits_{i=1}^{N}\left[ \dfrac{1}{N}\sum\limits_{l=1}^{N}\left( 
	\mathbf{\widehat{a}}_{l}-\mathbf{\widehat{H}}^{\prime }\mathbf{a}_{l}\right) \varphi
	_{jil}\right] \mathbf{a}_{i}^{\prime }+\mathbf{\widehat{H}}^{\prime }\dfrac{1}{N}%
	\sum\limits_{i=1}^{N}\left( \dfrac{1}{N}\sum\limits_{l=1}^{N}\mathbf{a}_{l}%
	\mathbf{a}_{i}^{\prime }\varphi _{jil}\right) .
	\end{equation*}%
%----------------------------------------------------------------------------%
	We have%
%----------------------------------------------------------------------------%
	\begin{equation*}
	\left\Vert \dfrac{1}{N}\sum\limits_{i=1}^{N}\left[ \dfrac{1}{N}%
	\sum\limits_{l=1}^{N}\left( \mathbf{\widehat{a}}_{l}-\mathbf{\widehat{H}}^{\prime }%
	\mathbf{a}_{l}\right) \varphi _{jil}\right] \mathbf{a}_{i}^{\prime
	}\right\Vert \leq \left( \dfrac{1}{N}\sum\limits_{l=1}^{N}\left\Vert \mathbf{%
		\widehat{a}}_{l}-\mathbf{\widehat{H}}^{\prime }\mathbf{a}_{l}\right\Vert ^{2}\right)
	^{1\left/ 2\right. }\left( \dfrac{1}{N}\sum\limits_{i=1}^{N}\left\Vert 
	\dfrac{1}{N}\sum\limits_{l=1}^{N}\varphi _{jil}\mathbf{a}_{i}\right\Vert
	^{2}\right) ^{1\left/ 2\right. }
	\end{equation*}%
%----------------------------------------------------------------------------%
	and%
%----------------------------------------------------------------------------%
	\begin{equation*}
	\begin{array}{rcl}
	\left( \dfrac{1}{N}\sum\limits_{i=1}^{N}\left\Vert \dfrac{1}{N}%
	\sum\limits_{l=1}^{N}\varphi _{jil}\mathbf{a}_{i}\right\Vert ^{2}\right)
	^{1\left/ 2\right. } & = & \left[ \dfrac{1}{N}\sum\limits_{i=1}^{N}\left%
	\Vert \dfrac{1}{N}\sum\limits_{l=1}^{N}\left( \dfrac{1}{T}%
	\sum\limits_{t=1}^{T}\mathbb{I}_{jt}\bm{\lambda }_{ji}^{\prime }\mathbf{f%
	}_{jt}e_{lt}\right) \mathbf{a}_{i}\right\Vert ^{2}\right] ^{1\left/ 2\right.
	} \\ 
	& = & \dfrac{1}{\sqrt{NT}}\left[ \dfrac{1}{N}\sum\limits_{i=1}^{N}\left\Vert
	\left( \dfrac{1}{\sqrt{NT}}\sum\limits_{l=1}^{N}\sum\limits_{t=1}^{T}\mathbb{%
		I}_{jt}\bm{\lambda }_{ji}^{\prime }\mathbf{f}_{jt}e_{lt}\right) \mathbf{a%
	}_{i}\right\Vert ^{2}\right] ^{1\left/ 2\right. } \\ 
	& \leq & \dfrac{1}{\sqrt{NT}}\left( \dfrac{1}{N}\sum\limits_{i=1}^{N}\left%
	\Vert \dfrac{1}{\sqrt{NT}}\sum\limits_{l=1}^{N}\sum\limits_{t=1}^{T}\mathbb{I%
	}_{jt}\bm{\lambda }_{ji}^{\prime }\mathbf{f}_{jt}e_{lt}\right\Vert
	^{2}\left\Vert \mathbf{a}_{i}\right\Vert ^{2}\right) ^{1\left/ 2\right. } \\ 
	& = & O_{p}\left( \dfrac{1}{\sqrt{NT}}\right)%
	\end{array}%
	\end{equation*}%
%----------------------------------------------------------------------------%
	by Assumptions \ref{assum:FL} and \ref{assum:MCLT}(b). Therefore,%
%----------------------------------------------------------------------------%
	\begin{equation*}
	\left\Vert \dfrac{1}{N}\sum\limits_{i=1}^{N}\left[ \dfrac{1}{N}%
	\sum\limits_{l=1}^{N}\left( \mathbf{\widehat{a}}_{l}-\mathbf{\widehat{H}}^{\prime }%
	\mathbf{a}_{l}\right) \varphi _{jil}\right] \mathbf{a}_{i}^{\prime
	}\right\Vert =O_{p}\left( \dfrac{1}{C_{NT}}\right) O_{p}\left( \dfrac{1}{%
		\sqrt{NT}}\right) =O_{p}\left( \dfrac{1}{\sqrt{NT}C_{NT}}\right)
	\end{equation*}%
%----------------------------------------------------------------------------%
	by Lemma \ref{Lemma:a_i_hat}. Further%
%----------------------------------------------------------------------------%
	\begin{equation*}
	\begin{array}{rcl}
	\left\Vert \dfrac{1}{N}\sum\limits_{i=1}^{N}\left( \dfrac{1}{N}%
	\sum\limits_{l=1}^{N}\mathbf{a}_{l}\mathbf{a}_{i}^{\prime }\varphi
	_{jil}\right) \right\Vert  & = & \left\Vert \dfrac{1}{N^{2}}%
	\sum\limits_{i=1}^{N}\sum\limits_{l=1}^{N}\mathbf{a}_{l}\mathbf{a}%
	_{i}^{\prime }\left( \dfrac{1}{T}\sum\limits_{t=1}^{T}\mathbb{I}_{jt}\bm{%
		\lambda }_{ji}^{\prime }\mathbf{f}_{jt}e_{lt}\right) \right\Vert  \\ 
	& = & \left\Vert \dfrac{1}{N^{2}}\sum\limits_{i=1}^{N}\sum%
	\limits_{l=1}^{N}\left( 
	\begin{array}{c}
	\bm{\lambda }_{1l} \\ 
	\bm{\lambda }_{2l}%
	\end{array}%
	\right) \left( \dfrac{1}{T}\sum\limits_{t=1}^{T}\mathbb{I}_{jt}\bm{%
		\lambda }_{ji}^{\prime }\mathbf{f}_{jt}e_{lt}\right) \left( 
	\begin{array}{c}
	\bm{\lambda }_{1i} \\ 
	\bm{\lambda }_{2i}%
	\end{array}%
	\right)^{\prime } \right\Vert  \\  
	& \leq  & \dfrac{1}{\sqrt{NT}}\dfrac{1}{N}\sum\limits_{i=1}^{N}\left\Vert 
	\dfrac{1}{\sqrt{NT}}\sum\limits_{l=1}^{N}\left( 
	\begin{array}{c}
	\bm{\lambda }_{1l} \\ 
	\bm{\lambda }_{2l}%
	\end{array}%
	\right) \left( \dfrac{1}{T}\sum\limits_{t=1}^{T}\mathbb{I}_{jt}\mathbf{f}%
	_{jt}^{\prime }e_{lt}\right) \right\Vert \left\Vert \bm{\lambda }%
	_{ji}\right\Vert \left\Vert \left( 
	\begin{array}{c}
	\bm{\lambda }_{1i} \\ 
	\bm{\lambda }_{2i}%
	\end{array}%
	\right)^{\prime } \right\Vert  \\ 
	& = & O_{p}\left( \dfrac{1}{\sqrt{NT}}\right) 
	\end{array}%
	\end{equation*}
%----------------------------------------------------------------------------%
	by Assumptions \ref{assum:FL} and \ref{assum:MCLT}(b). Therefore%
%----------------------------------------------------------------------------%
	\begin{equation}
	\label{eq:Lemma_3.1.3}
	\dfrac{1}{N}\sum\limits_{i=1}^{N}\left( \dfrac{1}{N}\sum\limits_{l=1}^{N}%
	\mathbf{\widehat{a}}_{l}\varphi _{jil}\right) \mathbf{a}_{i}^{\prime
	}=O_{p}\left( \dfrac{1}{\sqrt{NT}C_{NT}}\right) +O_{p}\left( \dfrac{1}{\sqrt{%
			NT}}\right) =O_{p}\left( \dfrac{1}{\sqrt{NT}}\right) .
	\end{equation}%
%----------------------------------------------------------------------------%
	Finally,%
%----------------------------------------------------------------------------%
	\begin{equation*}
	\dfrac{1}{N}\sum\limits_{i=1}^{N}\left( \dfrac{1}{N}\sum\limits_{l=1}^{N}%
	\mathbf{\widehat{a}}_{l}\varphi _{jli}\right) \mathbf{a}_{i}^{\prime }=\dfrac{1}{%
		N}\sum\limits_{i=1}^{N}\left[ \dfrac{1}{N}\sum\limits_{l=1}^{N}\left( 
	\mathbf{\widehat{a}}_{l}-\mathbf{\widehat{H}}^{\prime }\mathbf{a}_{l}\right) \varphi
	_{jli}\right] \mathbf{a}_{i}^{\prime }+\mathbf{\widehat{H}}^{\prime }\dfrac{1}{N}%
	\sum\limits_{i=1}^{N}\left( \dfrac{1}{N}\sum\limits_{l=1}^{N}\mathbf{a}_{l}%
	\mathbf{a}_{i}^{\prime }\varphi _{jli}\right) .
	\end{equation*}%
%----------------------------------------------------------------------------%
	We have%
%----------------------------------------------------------------------------%
	\begin{equation*}
	\left\Vert \dfrac{1}{N}\sum\limits_{i=1}^{N}\left[ \dfrac{1}{N}%
	\sum\limits_{l=1}^{N}\left( \mathbf{\widehat{a}}_{l}-\mathbf{\widehat{H}}^{\prime }%
	\mathbf{a}_{l}\right) \varphi _{jli}\right] \mathbf{a}_{i}^{\prime
	}\right\Vert \leq \left( \dfrac{1}{N}\sum\limits_{l=1}^{N}\left\Vert \mathbf{%
		\widehat{a}}_{l}-\mathbf{\widehat{H}}^{\prime }\mathbf{a}_{l}\right\Vert ^{2}\right)
	^{1\left/ 2\right. }\left( \dfrac{1}{N}\sum\limits_{i=1}^{N}\left\Vert 
	\dfrac{1}{N}\sum\limits_{l=1}^{N}\varphi _{jli}\mathbf{a}_{i}\right\Vert
	^{2}\right) ^{1\left/ 2\right. }
	\end{equation*}%
%----------------------------------------------------------------------------%
	with%
%----------------------------------------------------------------------------%
	\begin{equation*}
	\begin{array}{rcl}
	\left( \dfrac{1}{N}\sum\limits_{i=1}^{N}\left\Vert \dfrac{1}{N}%
	\sum\limits_{l=1}^{N}\varphi _{jli}\mathbf{a}_{i}\right\Vert ^{2}\right)
	^{1\left/ 2\right. } & = & \left[ \dfrac{1}{N}\sum\limits_{i=1}^{N}\left%
	\Vert \dfrac{1}{N}\sum\limits_{l=1}^{N}\left( \dfrac{1}{T}%
	\sum\limits_{t=1}^{T}\mathbb{I}_{jt}\bm{\lambda }_{jl}^{\prime }\mathbf{f%
	}_{jt}e_{it}\right) \mathbf{a}_{i}\right\Vert ^{2}\right] ^{1\left/ 2\right.
	} \\ 
	& = & \dfrac{1}{\sqrt{NT}}\left[ \dfrac{1}{N}\sum\limits_{i=1}^{N}\left\Vert
	\left( \dfrac{1}{\sqrt{NT}}\sum\limits_{l=1}^{N}\sum\limits_{t=1}^{T}\mathbb{%
		I}_{jt}\bm{\lambda }_{jl}^{\prime }\mathbf{f}_{jt}e_{it}\right) \mathbf{a%
	}_{i}\right\Vert ^{2}\right] ^{1\left/ 2\right. } \\ 
	& \leq & \dfrac{1}{\sqrt{NT}}\left( \dfrac{1}{N}\sum\limits_{i=1}^{N}\left%
	\Vert \dfrac{1}{\sqrt{NT}}\sum\limits_{l=1}^{N}\sum\limits_{t=1}^{T}\mathbb{I%
	}_{jt}\bm{\lambda }_{jl}^{\prime }\mathbf{f}_{jt}e_{it}\right\Vert
	^{2}\left\Vert \mathbf{a}_{i}\right\Vert ^{2}\right) ^{1\left/ 2\right. } \\ 
	& = & O_{p}\left( \dfrac{1}{\sqrt{NT}}\right)%
	\end{array}%
	\end{equation*}%
%----------------------------------------------------------------------------%
	by Assumptions \ref{assum:FL} and \ref{assum:MCLT}(b). Further%
%----------------------------------------------------------------------------%
	\begin{equation*}
	\begin{array}{rcl}
	\left\Vert \dfrac{1}{N}\sum\limits_{i=1}^{N}\left( \dfrac{1}{N}%
	\sum\limits_{l=1}^{N}\mathbf{a}_{l}\mathbf{a}_{i}^{\prime }\varphi
	_{jli}\right) \right\Vert  & = & \left\Vert \dfrac{1}{N^{2}}%
	\sum\limits_{i=1}^{N}\sum\limits_{l=1}^{N}\mathbf{a}_{l}\mathbf{a}%
	_{i}^{\prime }\left( \dfrac{1}{T}\sum\limits_{t=1}^{T}\mathbb{I}_{jt}\bm{%
		\lambda }_{jl}^{\prime }\mathbf{f}_{jt}e_{it}\right) \right\Vert  \\ 
	& = & \left\Vert \dfrac{1}{N^{2}}\sum\limits_{i=1}^{N}\sum\limits_{l=1}^{N}%
	\left( 
	\begin{array}{c}
	\bm{\lambda }_{1l} \\ 
	\bm{\lambda }_{2l}%
	\end{array}%
	\right) \left( \dfrac{1}{T}\sum\limits_{t=1}^{T}\mathbb{I}_{jt}\bm{%
		\lambda }_{jl}^{\prime }\mathbf{f}_{jt}e_{it}\right) \left( 
	\begin{array}{c}
	\bm{\lambda }_{1i}^{\prime } \\ 
	\bm{\lambda }_{2i}^{\prime }%
	\end{array}%
	\right) \right\Vert  \\ 
	& \leq  & \dfrac{1}{\sqrt{NT}}\dfrac{1}{N}\sum\limits_{i=1}^{N}\left\Vert 
	\dfrac{1}{\sqrt{NT}}\sum\limits_{l=1}^{N}\left( 
	\begin{array}{c}
	\bm{\lambda }_{1l} \\ 
	\bm{\lambda }_{2l}%
	\end{array}%
	\right) \left( \dfrac{1}{T}\sum\limits_{t=1}^{T}\mathbb{I}_{jt}\mathbf{f}%
	_{jt}^{\prime }e_{it}\right) \right\Vert \left\Vert \bm{\lambda }%
	_{jl}\right\Vert \left\Vert \left( 
	\begin{array}{c}
	\bm{\lambda }_{1i}^{\prime } \\ 
	\bm{\lambda }_{2i}^{\prime }%
	\end{array}%
	\right) \right\Vert  \\ 
	& = & O_{p}\left( \dfrac{1}{\sqrt{NT}}\right) 
	\end{array}%
	\end{equation*}
%----------------------------------------------------------------------------%
by Assumptions \ref{assum:FL} and \ref{assum:MCLT}(b). Therefore,%
%----------------------------------------------------------------------------%
	\begin{equation}
	\label{eq:Lemma_3.1.4}
	\dfrac{1}{N}\sum\limits_{i=1}^{N}\left( \dfrac{1}{N}\sum\limits_{l=1}^{N}%
	\mathbf{\widehat{a}}_{l}\varphi _{jli}\right) \mathbf{a}_{i}^{\prime
	}=O_{p}\left( \dfrac{1}{\sqrt{NT}C_{NT}}\right) +O_{p}\left( \dfrac{1}{\sqrt{%
			NT}}\right) =O_{p}\left( \dfrac{1}{\sqrt{NT}}\right) .
	\end{equation}%
%----------------------------------------------------------------------------%
	Combining equations (\ref{eq:Lemma_3.1}) through (\ref{eq:Lemma_3.1.4}), we obtain%
%----------------------------------------------------------------------------%
	\begin{equation}
	\label{eq:Lemma_3.5}
	N^{-1}\left( \mathbf{\widehat{A}}-\mathbf{A\widehat{H}}\right) ^{\prime }\mathbf{A}%
	=O_{p}\left( \dfrac{1}{\sqrt{N}C_{NT}}\right) +O_{p}\left( \dfrac{1}{\sqrt{T}%
		C_{NT}}\right) +O_{p}\left( \dfrac{1}{\sqrt{NT}}\right) +O_{p}\left( \dfrac{1%
	}{\sqrt{NT}}\right) =O_{p}\left( \dfrac{1}{C_{NT}^{2}}\right) .
	\end{equation}%
%----------------------------------------------------------------------------%
	From (\ref{eq:Lemma_3}), (\ref{eq:Lemma_3.5}) and Lemma \ref{Lemma:a_i_hat}, we obtain
%----------------------------------------------------------------------------%
	\begin{equation*}
	N^{-1}\left( \mathbf{\widehat{A}}-\mathbf{A\widehat{H}}\right) ^{\prime }\mathbf{%
		\widehat{A}}=O_{p}\left( \dfrac{1}{C_{NT}^{2}}\right) +O_{p}\left( \dfrac{1}{%
		C_{NT}^{2}}\right) =O_{p}\left( \dfrac{1}{C_{NT}^{2}}\right) .
	\end{equation*}
%----------------------------------------------------------------------------%
	 which completes the proof of the lemma.
	
\end{proof}

\begin{proof}[\textbf{Proof of Lemma \protect\ref{Lemma:4}}]
	Given the identity in (\ref{eq:a__hat-a_i}), we can write
%----------------------------------------------------------------------------%
	\begin{equation}
	\label{eq:Lemma_4}
	\begin{array}{rcl}
	N^{-1}\left( \mathbf{\widehat{A}}-\mathbf{A\widehat{H}}\right) ^{\prime }\mathbf{e}%
	_{t} & = & \mathbf{\widehat{V}}^{-1}\dfrac{1}{N}\sum\limits_{i=1}^{N}\left( 
	\mathbf{\widehat{a}}_{i}-\mathbf{\widehat{H}}^{\prime }\mathbf{a}_{i}\right) e_{it}
	\\ 
	& = & \mathbf{\widehat{V}}^{-1}\left\{ 
	\begin{array}{c}
	\sum\limits_{j=1}^{2}\left[ \dfrac{1}{N}\sum\limits_{i=1}^{N}\left( \dfrac{1%
	}{N}\sum\limits_{l=1}^{N}\mathbf{\widehat{a}}_{l}\sigma _{jil}\right) e_{it}%
	\right] +\sum\limits_{j=1}^{2}\left[ \dfrac{1}{N}\sum\limits_{i=1}^{N}\left( 
	\dfrac{1}{N}\sum\limits_{l=1}^{N}\mathbf{\widehat{a}}_{l}\chi _{jil}\right)
	e_{it}\right]  \\ 
	+\sum\limits_{j=1}^{2}\left[ \dfrac{1}{N}\sum\limits_{i=1}^{N}\left( \dfrac{1%
	}{N}\sum\limits_{l=1}^{N}\mathbf{\widehat{a}}_{l}\varphi _{jil}\right) e_{it}%
	\right] +\sum\limits_{j=1}^{2}\left[ \dfrac{1}{N}\sum\limits_{i=1}^{N}\left( 
	\dfrac{1}{N}\sum\limits_{l=1}^{N}\mathbf{\widehat{a}}_{l}\varphi
	_{jli}\right) e_{it}\right] 
	\end{array}%
	\right\} .%
	\end{array}%
	\end{equation}
%----------------------------------------------------------------------------%
	Consider%
%----------------------------------------------------------------------------%
\begin{equation*}
	\dfrac{1}{N}\sum\limits_{i=1}^{N}\left( \dfrac{1}{N}\sum\limits_{l=1}^{N}%
	\widehat{\mathbf{a}}_{l}\sigma _{jil}\right) e_{it}=\dfrac{1}{N}%
	\sum\limits_{i=1}^{N}\left[ \dfrac{1}{N}\sum\limits_{l=1}^{N}\left( 
	\widehat{\mathbf{a}}_{l}-\widehat{\mathbf{H}}^{\prime }\mathbf{a}_{l}\right)
	\sigma _{jil}\right] e_{it}+\widehat{\mathbf{H}}^{\prime }\dfrac{1}{N}%
	\sum\limits_{i=1}^{N}\left( \dfrac{1}{N}\sum\limits_{l=1}^{N}\mathbf{a}%
	_{l}\sigma _{jil}e_{it}\right),
\end{equation*}
%----------------------------------------------------------------------------%
where
%----------------------------------------------------------------------------%
	\begin{equation*}
	\begin{array}{rcl}
	\left\Vert \dfrac{1}{N}%
	\sum\limits_{i=1}^{N}\left[ \dfrac{1}{N}\sum\limits_{l=1}^{N}\left( 
	\widehat{\mathbf{a}}_{l}-\widehat{\mathbf{H}}^{\prime }\mathbf{a}_{l}\right)
	\sigma _{jil}\right] e_{it}\right\Vert  & \leq  & \dfrac{1}{\sqrt{N}}\left( \dfrac{1}{N}%
	\sum\limits_{l=1}^{N}\left\Vert \mathbf{\widehat{a}}_{l}-\mathbf{\widehat{H}}%
	^{\prime }\mathbf{a}_{l}\right\Vert ^{2}\right) ^{1\left/ 2\right. }\left( 
	\dfrac{1}{N}\sum\limits_{i=1}^{N}\sum\limits_{l=1}^{N}\left\vert \sigma
	_{jil}\right\vert ^{2}\right) ^{1\left/ 2\right. } \\
	&   & \times \left( \dfrac{1}{N}%
	\sum\limits_{i=1}^{N}\left\vert e_{it}\right\vert ^{2}\right) ^{1\left/2\right. }\\
	& = & \dfrac{1}{\sqrt{N}}O_{p}\left( \dfrac{1}{C_{NT}}\right) O_{p}\left(
	1\right) O_{p}\left( 1\right)  \\ 
	& = & O_{p}\left( \dfrac{1}{\sqrt{N}C_{NT}}\right) 
	\end{array}%
	\end{equation*}%
%----------------------------------------------------------------------------%
	by Lemma \ref{Lemma:a_i_hat}, equation (\ref{eq:bound_M^sq}), and Assumption \ref{assum:TCSDH}(a), and
%----------------------------------------------------------------------------%
	\begin{equation*}
	\begin{array}{rcl}
		\left\Vert \dfrac{1}{N}\sum\limits_{i=1}^{N}\left( \dfrac{1}{N}%
		\sum\limits_{l=1}^{N}\mathbf{a}_{l}\sigma _{jil}e_{it}\right) \right\Vert 
		& = & \left\Vert \dfrac{1}{N}\sum\limits_{i=1}^{N}\dfrac{1}{N}%
		\sum\limits_{l=1}^{N}\mathbf{a}_{l}\E\left( \dfrac{1}{T}%
		\sum\limits_{t=1}^{T}\mathbb{I}_{jt}e_{it}e_{lt}\right) e_{it}\right\Vert 
		\\ 
		& = & \left\Vert \dfrac{1}{T}\sum\limits_{t=1}^{T}\dfrac{1}{N}%
		\sum\limits_{i=1}^{N}\dfrac{1}{N}\sum\limits_{l=1}^{N}\mathbf{a}_{l}%
		\E\left( \mathbb{I}_{jt}e_{it}e_{lt}\right) e_{it}\right\Vert  \\ 
		& \leq  & \dfrac{1}{N}\dfrac{1}{T}\sum\limits_{t=1}^{T}\left[ \dfrac{1}{N}%
		\sum\limits_{i=1}^{N}\sum\limits_{l=1}^{N}\left\vert \E\left( 
		\mathbb{I}_{jt}e_{it}e_{lt}\right) \right\vert \right] \left\Vert \mathbf{a}%
		_{l}\right\Vert \left\vert e_{it}\right\vert  \\ 
		& = & O_{p}\left( \dfrac{1}{N}\right)
	\end{array}%
	\end{equation*}
%----------------------------------------------------------------------------%
by Assumptions \ref{assum:FL}(a), Assumption \ref{assum:TCSDH}(a), and Assumption \ref{assum:TCSDH}(b), so that
%----------------------------------------------------------------------------%
	\begin{equation}
		\label{eq:Lemma_4.1}
		\begin{array}{rcl}
			\left\Vert \dfrac{1}{N}\sum\limits_{i=1}^{N}\left( \dfrac{1}{N}%
			\sum\limits_{l=1}^{N}\widehat{\mathbf{a}}_{l}\sigma _{jil}\right)
			e_{it}\right\Vert  & \leq  & \left\Vert \dfrac{1}{N}\sum\limits_{i=1}^{N}%
			\left[ \dfrac{1}{N}\sum\limits_{l=1}^{N}\left( \widehat{\mathbf{a}}_{l}-%
			\widehat{\mathbf{H}}^{\prime }\mathbf{a}_{l}\right) \sigma _{jil}\right]
			e_{it}\right\Vert   \\ 
			&   & +\left\Vert \widehat{\mathbf{H}}^{\prime }\dfrac{1}{N}%
			\sum\limits_{i=1}^{N}\left( \dfrac{1}{N}\sum\limits_{l=1}^{N}\mathbf{a}%
			_{l}\sigma _{jil}e_{it}\right) \right\Vert \\
			& = & O_{p}\left( \dfrac{1}{\sqrt{N}C_{NT}}\right) .%
		\end{array}%
	\end{equation}
%----------------------------------------------------------------------------%
	Consider now%
%----------------------------------------------------------------------------%
	\begin{equation*}
	\dfrac{1}{N}\sum\limits_{i=1}^{N}\left( \dfrac{1}{N}\sum\limits_{l=1}^{N}%
	\mathbf{\widehat{a}}_{l}\chi _{jil}\right) e_{it}=\dfrac{1}{N}%
	\sum\limits_{i=1}^{N}\left[ \dfrac{1}{N}\sum\limits_{l=1}^{N}\left( \mathbf{%
		\widehat{a}}_{l}-\mathbf{\widehat{H}}^{\prime }\mathbf{a}_{l}\right) \chi _{jil}%
	\right] e_{it}+\mathbf{\widehat{H}}^{\prime }\dfrac{1}{N}\sum\limits_{i=1}^{N}%
	\dfrac{1}{N}\sum\limits_{l=1}^{N}\mathbf{a}_{l}e_{it}\chi _{jil}.
	\end{equation*}%
%----------------------------------------------------------------------------%
	We have%
%----------------------------------------------------------------------------%
	\begin{equation*}
	\begin{array}{rcl}
	\left\Vert \dfrac{1}{N}\sum\limits_{i=1}^{N}\left[ \dfrac{1}{N}%
	\sum\limits_{l=1}^{N}\left( \mathbf{\widehat{a}}_{l}-\mathbf{\widehat{H}}^{\prime }%
	\mathbf{a}_{l}\right) \chi _{jil}\right] e_{it}\right\Vert  & \leq  & \dfrac{%
		1}{N}\sum\limits_{l=1}^{N}\left\Vert \left( \mathbf{\widehat{a}}_{l}-\mathbf{%
		\widehat{H}}^{\prime }\mathbf{a}_{l}\right) \right\Vert \left( \dfrac{1}{N}%
	\sum\limits_{i=1}^{N}\left\vert \chi _{jil}e_{it}\right\vert \right)  \\ 
	& \leq  & \left[ \dfrac{1}{N}\sum\limits_{l=1}^{N}\left\Vert \left( \mathbf{%
		\widehat{a}}_{l}-\mathbf{\widehat{H}}^{\prime }\mathbf{a}_{l}\right) \right\Vert ^{2}%
	\right] ^{1\left/ 2\right. }\left[ \dfrac{1}{N}\sum\limits_{l=1}^{N}\left( 
	\dfrac{1}{N}\sum\limits_{i=1}^{N}\left\vert \chi _{jil}e_{it}\right\vert
	\right) ^{2}\right] ^{1\left/ 2\right. }%
	\end{array}%
	\end{equation*}%
%----------------------------------------------------------------------------%
	with%
%----------------------------------------------------------------------------%
	\begin{equation*}
	\begin{array}{rcl}
	\dfrac{1}{N}\sum\limits_{i=1}^{N}\left\vert \chi _{jil}e_{it}\right\vert  & =
	& \dfrac{1}{N}\sum\limits_{i=1}^{N}\left\vert \left[ \dfrac{1}{T}%
	\sum\limits_{t=1}^{T}\mathbb{I}_{jt}e_{it}e_{lt}-\E\left( \dfrac{1}{T%
	}\sum\limits_{t=1}^{T}\mathbb{I}_{jt}e_{it}e_{lt}\right) \right]
	e_{it}\right\vert  \\ 
	& = & \dfrac{1}{\sqrt{T}}\dfrac{1}{N}\sum\limits_{i=1}^{N}\left\vert \left[ 
	\dfrac{1}{\sqrt{T}}\sum\limits_{t=1}^{T}\mathbb{I}_{jt}e_{it}e_{lt}-\E\left( \dfrac{1}{\sqrt{T}}\sum\limits_{t=1}^{T}\mathbb{I}%
	_{jt}e_{it}e_{lt}\right) \right] e_{it}\right\vert  \\ 
	& = & O_{p}\left( \dfrac{1}{\sqrt{T}}\right) 
	\end{array}%
	\end{equation*}%
%----------------------------------------------------------------------------%
	by Assumptions \ref{assum:TCSDH}(a) and \ref{assum:TCSDH}(c). Therefore, taking into account Lemma \ref{Lemma:a_i_hat},
%----------------------------------------------------------------------------%
	\begin{equation*}
	\left\Vert \dfrac{1}{N}\sum\limits_{i=1}^{N}\left[ \dfrac{1}{N}%
	\sum\limits_{l=1}^{N}\left( \mathbf{\widehat{a}}_{l}-\mathbf{\widehat{H}}^{\prime }%
	\mathbf{a}_{l}\right) \chi _{jil}\right] e_{it}\right\Vert =O_{p}\left( 
	\dfrac{1}{C_{NT}}\right) O_{p}\left( \dfrac{1}{\sqrt{T}}\right) =O_{p}\left( 
	\dfrac{1}{\sqrt{T}C_{NT}}\right) .
	\end{equation*}%
%----------------------------------------------------------------------------%
	Further,%
%----------------------------------------------------------------------------%
	\begin{equation*}
	\begin{array}{rcl}
	\left\Vert \dfrac{1}{N}\sum\limits_{i=1}^{N}\dfrac{1}{N}\sum\limits_{l=1}^{N}%
	\mathbf{a}_{l}\chi _{jil}e_{it}\right\Vert  & = & \left\Vert \dfrac{1}{N}%
	\sum\limits_{i=1}^{N}\dfrac{1}{N}\sum\limits_{l=1}^{N}\mathbf{a}_{l}\left[ 
	\dfrac{1}{T}\sum\limits_{t=1}^{T}\mathbb{I}_{jt}e_{it}e_{lt}-\E%
	\left( \dfrac{1}{T}\sum\limits_{t=1}^{T}\mathbb{I}_{jt}e_{it}e_{lt}\right) %
	\right] e_{it}\right\Vert  \\ 
	& = & \dfrac{1}{\sqrt{NT}}\dfrac{1}{N}\sum\limits_{i=1}^{N}\left\Vert \dfrac{%
		1}{\sqrt{NT}}\sum\limits_{l=1}^{N}\sum\limits_{t=1}^{T}\mathbf{a}_{l}\left[ 
	\mathbb{I}_{jt}e_{it}e_{lt}-\E\left( \mathbb{I}_{jt}e_{it}e_{lt}%
	\right) \right] \right\Vert \left\vert e_{it}\right\vert  \\ 
	& \leq  & \dfrac{1}{\sqrt{NT}}\left\{ \dfrac{1}{N}\sum\limits_{i=1}^{N}\left%
	\Vert \dfrac{1}{\sqrt{NT}}\sum\limits_{l=1}^{N}\sum\limits_{t=1}^{T}\mathbf{a%
	}_{l}\left[ \mathbb{I}_{jt}e_{it}e_{lt}-\E\left( \mathbb{I}%
	_{jt}e_{it}e_{lt}\right) \right] \right\Vert ^{2}\right\} ^{1\left/ 2\right.
	}\left( \dfrac{1}{N}\sum\limits_{i=1}^{N}\left\vert e_{it}\right\vert
	^{2}\right) ^{1\left/ 2\right. } \\ 
	& = & O_{p}\left( \dfrac{1}{\sqrt{NT}}\right) 
	\end{array}%
	\end{equation*}%
%----------------------------------------------------------------------------%
	by Assumptions \ref{assum:TCSDH}(a) and \ref{assum:MCLT}(a). Therefore,%
%----------------------------------------------------------------------------%
	\begin{equation}
	\label{eq:Lemma_4.2}
	\dfrac{1}{N}\sum\limits_{i=1}^{N}\left( \dfrac{1}{N}\sum\limits_{l=1}^{N}%
	\mathbf{\widehat{a}}_{l}\chi _{jil}\right) e_{it}=O_{p}\left( \dfrac{1}{\sqrt{T}%
		C_{NT}}\right) +O_{p}\left( \dfrac{1}{\sqrt{NT}}\right) =O_{p}\left( \dfrac{1%
	}{\sqrt{T}C_{NT}}\right) .
	\end{equation}%
%----------------------------------------------------------------------------%
	Consider now%
%----------------------------------------------------------------------------%
	\begin{equation*}
	\dfrac{1}{N}\sum\limits_{i=1}^{N}\left( \dfrac{1}{N}\sum\limits_{l=1}^{N}%
	\mathbf{\widehat{a}}_{l}\varphi _{jil}\right) e_{it}=\dfrac{1}{N}%
	\sum\limits_{i=1}^{N}\left[ \dfrac{1}{N}\sum\limits_{l=1}^{N}\left( \mathbf{%
		\widehat{a}}_{l}-\mathbf{\widehat{H}}^{\prime }\mathbf{a}_{l}\right) \varphi _{jil}%
	\right] e_{it}+\dfrac{1}{N}\sum\limits_{i=1}^{N}\dfrac{1}{N}%
	\sum\limits_{l=1}^{N}\mathbf{\widehat{H}}^{\prime }\mathbf{a}_{l}\varphi
	_{jil}e_{it}.
	\end{equation*}%
%----------------------------------------------------------------------------%
	We have%
%----------------------------------------------------------------------------%
	\begin{equation*}
	\begin{array}{rcl}
	\left\Vert \dfrac{1}{N}\sum\limits_{i=1}^{N}\left[ \dfrac{1}{N}%
	\sum\limits_{l=1}^{N}\left( \mathbf{\widehat{a}}_{l}-\mathbf{\widehat{H}}^{\prime }%
	\mathbf{a}_{l}\right) \varphi _{jil}\right] e_{it}\right\Vert  & \leq  & 
	\dfrac{1}{N}\sum\limits_{l=1}^{N}\left\Vert \mathbf{\widehat{a}}_{l}-\mathbf{%
		\widehat{H}}^{\prime }\mathbf{a}_{l}\right\Vert \left( \dfrac{1}{N}%
	\sum\limits_{i=1}^{N}\left\vert \varphi _{jil}e_{it}\right\vert \right)  \\ 
	& \leq  & \left( \dfrac{1}{N}\sum\limits_{l=1}^{N}\left\Vert \mathbf{\widehat{a}}%
	_{l}-\mathbf{\widehat{H}}^{\prime }\mathbf{a}_{l}\right\Vert ^{2}\right)
	^{1\left/ 2\right. }\left[ \dfrac{1}{N}\sum\limits_{l=1}^{N}\left( \dfrac{1}{%
		N}\sum\limits_{i=1}^{N}\left\vert \varphi _{jil}e_{it}\right\vert \right)
	^{2}\right] ^{1\left/ 2\right. },%
	\end{array}%
	\end{equation*}%
%----------------------------------------------------------------------------%
	with
%----------------------------------------------------------------------------%
	\begin{equation*}
	\begin{array}{rcl}
	\dfrac{1}{N}\sum\limits_{i=1}^{N}\left\vert \varphi _{jil}e_{it}\right\vert 
	& = & \dfrac{1}{N}\sum\limits_{i=1}^{N}\left\vert \left( \dfrac{1}{T}%
	\sum\limits_{t=1}^{T}\mathbb{I}_{jt}\bm{\lambda }_{ji}^{\prime }\mathbf{f%
	}_{jt}e_{lt}\right) e_{it}\right\vert  \\ 
	& \leq  & \dfrac{1}{\sqrt{T}}\dfrac{1}{N}\sum\limits_{i=1}^{N}\left\Vert 
	\bm{\lambda }_{ji}\right\Vert \left\Vert \dfrac{1}{\sqrt{T}}%
	\sum\limits_{t=1}^{T}\mathbb{I}_{jt}\mathbf{f}_{jt}e_{it}\right\Vert
	\left\vert e_{lt}\right\vert  \\ 
	& \leq  & \bar{\lambda}\dfrac{1}{\sqrt{T}}\left\vert e_{lt}\right\vert
	\left( \dfrac{1}{N}\sum\limits_{i=1}^{N}\left\Vert \dfrac{1}{\sqrt{T}}%
	\sum\limits_{t=1}^{T}\mathbb{I}_{jt}\mathbf{f}_{jt}e_{it}\right\Vert
	^{2}\right) ^{1\left/ 2\right. } \\ 
	& = & O_{p}\left( \dfrac{1}{\sqrt{T}}\right) 
	\end{array}%
	\end{equation*}%
%----------------------------------------------------------------------------%
	by Assumptions \ref{assum:FL}, \ref{assum:TCSDH}(a) and \ref{assum:WD}. Taking into account Lemma \ref{Lemma:a_i_hat},
%----------------------------------------------------------------------------%
	\begin{equation*}
	\dfrac{1}{N}\sum\limits_{i=1}^{N}\left[ \dfrac{1}{N}\sum\limits_{l=1}^{N}%
	\left( \mathbf{\widehat{a}}_{l}-\mathbf{\widehat{H}}^{\prime }\mathbf{a}_{l}\right)
	\varphi _{jil}\right] e_{it}=O_{p}\left( \dfrac{1}{C_{NT}}\right)
	O_{p}\left( \dfrac{1}{\sqrt{T}}\right) =O_{p}\left( \dfrac{1}{\sqrt{T}C_{NT}}%
	\right) .
	\end{equation*}%
%----------------------------------------------------------------------------%
	Further,%
%----------------------------------------------------------------------------%
	\begin{equation*}
	\begin{array}{rcl}
	\left\Vert \dfrac{1}{N}\sum\limits_{i=1}^{N}\dfrac{1}{N}\sum\limits_{l=1}^{N}%
	\mathbf{a}_{l}\varphi _{jil}e_{it}\right\Vert  & = & \dfrac{1}{\sqrt{NT}}%
	\left\Vert \dfrac{1}{N}\sum\limits_{l=1}^{N}\mathbf{a}_{l}\left( \dfrac{1}{%
		\sqrt{NT}}\sum\limits_{i=1}^{N}\sum\limits_{t=1}^{T}\mathbb{I}_{jt}\bm{%
		\lambda }_{ji}^{\prime }\mathbf{f}_{jt}e_{it}\right) e_{lt}\right\Vert  \\ 
	& \leq  & \dfrac{1}{\sqrt{NT}}\left( \dfrac{1}{N}\sum\limits_{l=1}^{N}\left%
	\Vert \mathbf{a}_{l}\right\Vert \left\vert e_{lt}\right\vert \right)
	\left\Vert \dfrac{1}{\sqrt{NT}}\sum\limits_{i=1}^{N}\sum\limits_{t=1}^{T}%
	\mathbb{I}_{jt}\bm{\lambda }_{ji}^{\prime }\mathbf{f}_{jt}e_{it}\right%
	\Vert  \\ 
	& = & \dfrac{1}{\sqrt{NT}}\left( \dfrac{1}{N}\sum\limits_{l=1}^{N}\left\Vert 
	\mathbf{a}_{l}\right\Vert ^{2}\right) ^{1\left/ 2\right. }\left( \dfrac{1}{N}%
	\sum\limits_{l=1}^{N}\left\vert e_{lt}\right\vert ^{2}\right) ^{1\left/
		2\right. }\left\Vert \dfrac{1}{\sqrt{NT}}\sum\limits_{i=1}^{N}\sum%
	\limits_{t=1}^{T}\mathbb{I}_{jt}\bm{\lambda }_{ji}^{\prime }\mathbf{f}%
	_{jt}e_{it}\right\Vert  \\ 
	& = & O_{p}\left( \dfrac{1}{\sqrt{NT}}\right) ,%
	\end{array}%
	\end{equation*}%
%----------------------------------------------------------------------------%
	by Assumptions \ref{assum:FL}, \ref{assum:TCSDH}(a) and \ref{assum:MCLT}(a). Therefore,%
%----------------------------------------------------------------------------%
	\begin{equation}
	\label{eq:Lemma_4.3}
	\dfrac{1}{N}\sum\limits_{i=1}^{N}\left( \dfrac{1}{N}\sum\limits_{l=1}^{N}%
	\mathbf{\widehat{a}}_{l}\varphi _{jil}\right) e_{it}=O_{p}\left( \dfrac{1}{\sqrt{%
			T}C_{NT}}\right) +O_{p}\left( \dfrac{1}{\sqrt{NT}}\right) =O_{p}\left( 
	\dfrac{1}{\sqrt{T}C_{NT}}\right) .
	\end{equation}%
%----------------------------------------------------------------------------%
	Finally,
%----------------------------------------------------------------------------%
\begin{equation*}
	\dfrac{1}{N}\sum\limits_{i=1}^{N}\left( \dfrac{1}{N}\sum\limits_{l=1}^{N}%
	\widehat{\mathbf{a}}_{l}\varphi _{jli}\right) e_{it}=\dfrac{1}{N}%
	\sum\limits_{i=1}^{N}\left[ \dfrac{1}{N}\sum\limits_{l=1}^{N}\left( 
	\widehat{\mathbf{a}}_{l}-\widehat{\mathbf{H}}^{\prime }\mathbf{a}_{l}\right)
	\varphi _{jli}\right] e_{it}+\widehat{\mathbf{H}}^{\prime }\dfrac{1}{N}%
	\sum\limits_{i=1}^{N}\dfrac{1}{N}\sum\limits_{l=1}^{N}\mathbf{a}%
	_{l}\varphi _{jli}e_{it}.
\end{equation*}
%----------------------------------------------------------------------------%
Consider first
%----------------------------------------------------------------------------%
\begin{equation*}
	\begin{array}{rcl}
		\left\Vert \dfrac{1}{N}\sum\limits_{i=1}^{N}\left[ \dfrac{1}{N}%
		\sum\limits_{l=1}^{N}\left( \widehat{\mathbf{a}}_{l}-\widehat{\mathbf{H}}%
		^{\prime }\mathbf{a}_{l}\right) \varphi _{jli}\right] e_{it}\right\Vert  & 
		\leq  & \dfrac{1}{N}\sum\limits_{l=1}^{N}\left[ \left\Vert \widehat{\mathbf{%
				a}}_{l}-\widehat{\mathbf{H}}^{\prime }\mathbf{a}_{l}\right\Vert \left( 
		\dfrac{1}{N}\sum\limits_{i=1}^{N}\left\vert \varphi _{jli}e_{it}\right\vert
		\right) \right]  \\ 
		& \leq  & \left( \dfrac{1}{N}\sum\limits_{l=1}^{N}\left\Vert \widehat{%
			\mathbf{a}}_{l}-\widehat{\mathbf{H}}^{\prime }\mathbf{a}_{l}\right\Vert
		^{2}\right) ^{1\left/ 2\right. }\left[ \dfrac{1}{N}\sum\limits_{l=1}^{N}%
		\left( \dfrac{1}{N}\sum\limits_{i=1}^{N}\left\vert \varphi
		_{jli}e_{it}\right\vert \right) ^{2}\right] ^{1\left/ 2\right. },
	\end{array}%
\end{equation*}
%----------------------------------------------------------------------------%
with
%----------------------------------------------------------------------------%
\begin{equation*}
	\begin{array}{rcl}
		\dfrac{1}{N}\sum\limits_{i=1}^{N}\left\vert \varphi _{jli}e_{it}\right\vert 
		& = & \dfrac{1}{N}\sum\limits_{i=1}^{N}\left\vert \left( \dfrac{1}{T}%
		\sum\limits_{t=1}^{T}\mathbb{I}_{jt}\bm{\lambda }_{jl}^{\prime }\mathbf{%
			f}_{jt}e_{it}\right) e_{it}\right\vert  \\ 
		& \leq  & \left\Vert \bm{\lambda }_{jl}\right\Vert \dfrac{1}{\sqrt{T}}%
		\dfrac{1}{N}\sum\limits_{i=1}^{N}\left( \left\Vert \dfrac{1}{\sqrt{T}}%
		\sum\limits_{t=1}^{T}\mathbb{I}_{jt}\mathbf{f}_{jt}e_{it}\right\Vert
		\left\vert e_{it}\right\vert \right)  \\ 
		& \leq  & \bar{\lambda}\dfrac{1}{\sqrt{T}}\left( \dfrac{1}{N}%
		\sum\limits_{i=1}^{N}\left\Vert \dfrac{1}{\sqrt{T}}\sum\limits_{t=1}^{T}%
		\mathbb{I}_{jt}\mathbf{f}_{jt}e_{it}\right\Vert ^{2}\right) ^{1\left/
			2\right. }\left( \dfrac{1}{N}\sum\limits_{i=1}^{N}\left\vert
		e_{it}\right\vert ^{2}\right) ^{1\left/ 2\right. } \\ 
		& = & O_{p}\left( \dfrac{1}{\sqrt{T}}\right) 
	\end{array}
\end{equation*}
%----------------------------------------------------------------------------%
by Assumptions \ref{assum:FL}(a), \ref{assum:TCSDH}(a), and \ref{assum:WD}, so that
%----------------------------------------------------------------------------%
\begin{equation*}
	\left\Vert \dfrac{1}{N}\sum\limits_{i=1}^{N}\left[ \dfrac{1}{N}%
	\sum\limits_{l=1}^{N}\left( \widehat{\mathbf{a}}_{l}-\widehat{\mathbf{H}}%
	^{\prime }\mathbf{a}_{l}\right) \varphi _{jli}\right] e_{it}\right\Vert
	=O_{p}\left( \dfrac{1}{\sqrt{T}C_{NT}}\right) .
\end{equation*}
%----------------------------------------------------------------------------%
Also,
%----------------------------------------------------------------------------%
\begin{equation*}
	\begin{array}{rcl}
		\left\Vert \dfrac{1}{N}\sum\limits_{i=1}^{N}\dfrac{1}{N}\sum%
		\limits_{l=1}^{N}\mathbf{a}_{l}\varphi _{jli}e_{it}\right\Vert  & = & 
		\left\Vert \dfrac{1}{N}\sum\limits_{i=1}^{N}\dfrac{1}{N}\sum%
		\limits_{l=1}^{N}\mathbf{a}_{l}\left( \dfrac{1}{T}\sum\limits_{v=1}^{T}%
		\mathbb{I}_{jv}\bm{\lambda }_{jl}^{\prime }\mathbf{f}_{jv}e_{iv}\right)
		e_{it}\right\Vert  \\ 
		& = & \left\Vert \left( \dfrac{1}{N}\sum\limits_{l=1}^{N}\mathbf{a}_{l}%
		\bm{\lambda }_{jl}^{\prime }\right) \left( \dfrac{1}{NT}%
		\sum\limits_{i=1}^{N}\sum\limits_{v=1}^{T}\mathbb{I}_{jv}\mathbf{f}%
		_{jv}e_{iv}e_{it}\right) \right\Vert  \\ 
		& \leq  & \left\Vert \dfrac{1}{N}\sum\limits_{l=1}^{N}\mathbf{a}_{l}\bm{%
			\lambda }_{jl}^{\prime }\right\Vert \left\Vert \dfrac{1}{NT}%
		\sum\limits_{i=1}^{N}\sum\limits_{v=1}^{T}\mathbb{I}_{jv}\mathbf{f}%
		_{jv}e_{iv}e_{it}\right\Vert  \\ 
		& = & \left\Vert \dfrac{1}{NT}\sum\limits_{i=1}^{N}\sum\limits_{v=1}^{T}%
		\mathbf{f}_{jv}\left[ \mathbb{I}_{jv}e_{iv}e_{it}-\mathrm{E}\left( \mathbb{I}%
		_{jv}e_{iv}e_{it}\right) +\mathrm{E}\left( \mathbb{I}_{jv}e_{iv}e_{it}%
		\right) \right] \right\Vert O\left( 1\right)  \\ 
		& \leq  & \dfrac{1}{NT}\sum\limits_{i=1}^{N}\sum\limits_{v=1}^{T}\left\{
		\left\Vert \mathbf{f}_{jv}\right\Vert \left[ \left\vert \mathbb{I}%
		_{jv}e_{iv}e_{it}-\mathrm{E}\left( \mathbb{I}_{jv}e_{iv}e_{it}\right)
		\right\vert +\left\vert \mathrm{E}\left( \mathbb{I}_{jv}e_{iv}e_{it}\right)
		\right\vert \right] \right\} O\left( 1\right)  \\ 
		& \leq  & \dfrac{1}{NT}\sum\limits_{i=1}^{N}\sum\limits_{v=1}^{T}\left[
		\left\vert \mathbb{I}_{jv}e_{iv}e_{it}-\mathrm{E}\left( \mathbb{I}%
		_{jv}e_{iv}e_{it}\right) \right\vert +\left\vert \mathrm{E}\left( \mathbb{I}%
		_{jv}e_{iv}e_{it}\right) \right\vert \right] O_{p}\left( 1\right)  \\ 
		& \leq  & \left[ \dfrac{1}{NT}\sum\limits_{i=1}^{N}\sum\limits_{v=1}^{T}%
		\left\vert \mathbb{I}_{jv}e_{iv}e_{it}-\mathrm{E}\left( \mathbb{I}%
		_{jv}e_{iv}e_{it}\right) \right\vert \right] O_{p}\left( 1\right)  \\ 
		&  & +\left[ \dfrac{1}{NT}\sum\limits_{i=1}^{N}\sum\limits_{v=1}^{T}\left%
		\vert \mathrm{E}\left( \mathbb{I}_{jv}e_{iv}e_{it}\right) \right\vert \right]
		O_{p}\left( 1\right)  \\ 
		& = & O_{p}\left( \dfrac{1}{T}\right)
	\end{array}%
\end{equation*}
%----------------------------------------------------------------------------%
by Assumption \ref{assum:TCSDH}(c). Therefore,
%----------------------------------------------------------------------------%
\begin{equation}
	\label{eq:Lemma_4.4}
	\dfrac{1}{N}\sum\limits_{i=1}^{N}\left( \dfrac{1}{N}\sum\limits_{l=1}^{N}%
	\widehat{\mathbf{a}}_{l}\varphi _{jli}\right) e_{it}=O_{p}\left( \dfrac{1}{%
		\sqrt{T}C_{NT}}\right) +O_{p}\left( \dfrac{1}{T}\right) =O_{p}\left( \dfrac{1%
	}{\sqrt{T}C_{NT}}\right) .
\end{equation}
%----------------------------------------------------------------------------%
By combining (\ref{eq:Lemma_4}) through (\ref{eq:Lemma_4.4}), we have
%----------------------------------------------------------------------------%
	\begin{equation*}
	N^{-1}\left( \mathbf{\widehat{A}}-\mathbf{A\widehat{H}}\right) ^{\prime }\mathbf{e}%
	_{t}=O_{p}\left( \dfrac{1}{\sqrt{N}C_{NT}}\right)+O_{p}\left( \dfrac{1%
	}{\sqrt{T}C_{NT}}\right)=O_{p}\left( \dfrac{1}{C^{2}_{NT}}\right),
	\end{equation*}	
%----------------------------------------------------------------------------%
which completes the proof of the lemma.
\end{proof}

\begin{proof}[\textbf{Proof of Lemma \protect\ref{Lemma:avg}}]
Starting from (a), and taking into account (\ref{eq:linear_model}), consider
%----------------------------------------------------------------------------%
\begin{equation*}
\mathbf{\widehat{g}}_{t}=N^{-1}\mathbf{\widehat{A}}^{\prime }\mathbf{x}_{t}=N^{-1}%
\mathbf{\widehat{A}}^{\prime }\left( \mathbf{Ag}_{t}+\mathbf{e}_{t}\right)
=N^{-1}\mathbf{\widehat{A}}^{\prime }\mathbf{Ag}_{t}+N^{-1}\mathbf{\widehat{A}}%
^{\prime }\mathbf{e}_{t}
\end{equation*}%
%----------------------------------------------------------------------------%
and note that%
%----------------------------------------------------------------------------%
\begin{equation*}
\mathbf{A}=\mathbf{A}-\mathbf{\widehat{A}\widehat{H}}^{-1}+\mathbf{\widehat{A}\widehat{H}}^{-1},
\end{equation*}%
%----------------------------------------------------------------------------%
so that we have%
%----------------------------------------------------------------------------%
\begin{equation*}
\begin{array}{rcl}
\mathbf{\widehat{g}}_{t} & = & N^{-1}\mathbf{\widehat{A}}^{\prime }\left( \mathbf{A}-%
\mathbf{\widehat{A}\widehat{H}}^{-1}+\mathbf{\widehat{A}\widehat{H}}^{-1}\right) \mathbf{g}%
_{t}+N^{-1}\mathbf{\widehat{A}}^{\prime }\mathbf{e}_{t} \\ 
& = & N^{-1}\mathbf{\widehat{A}}^{\prime }\left( \mathbf{A}-\mathbf{\widehat{A}\widehat{H%
}}^{-1}+\mathbf{\widehat{A}\widehat{H}}^{-1}\right) \mathbf{g}_{t}+N^{-1}\mathbf{%
	\widehat{A}}^{\prime }\mathbf{e}_{t}+N^{-1}\left( \mathbf{A\widehat{H}}\right)
^{\prime }\mathbf{e}_{t}-N^{-1}\left( \mathbf{A\widehat{H}}\right) ^{\prime }%
\mathbf{e}_{t} \\ 
& = & N^{-1}\mathbf{\widehat{A}}^{\prime }\left( \mathbf{A}-\mathbf{\widehat{A}\widehat{H%
}}^{-1}\right) \mathbf{g}_{t}+N^{-1}\mathbf{\widehat{A}}^{\prime }\mathbf{\widehat{A}%
	\widehat{H}}^{-1}\mathbf{g}_{t}+N^{-1}\left( \mathbf{\widehat{A}}-\mathbf{A\widehat{H}}%
\right) ^{\prime }\mathbf{e}_{t}+N^{-1}\left( \mathbf{A\widehat{H}}\right)
^{\prime }\mathbf{e}_{t},
\end{array}%
\end{equation*}%
%----------------------------------------------------------------------------%
which leads to
%----------------------------------------------------------------------------%
\begin{equation}
\label{eq:Lemma_5.0}
\mathbf{\widehat{g}}_{t}-\mathbf{\widehat{H}}^{-1}\mathbf{g}_{t} = N^{-1}\left( 
\mathbf{A\widehat{H}}\right) ^{\prime }\mathbf{e}_{t}+N^{-1}\mathbf{\widehat{A}}%
^{\prime }\left( \mathbf{A}-\mathbf{\widehat{A}\widehat{H}}^{-1}\right) \mathbf{g}%
_{t}+N^{-1}\left( \mathbf{\widehat{A}}-\mathbf{A\widehat{H}}\right) ^{\prime }%
\mathbf{e}_{t}.%
\end{equation}
%----------------------------------------------------------------------------%
The result in (a) follows by taking into account Assumption \ref{assum:MCLT}(d), Lemma \ref{Lemma:3} and Lemma \ref{Lemma:4}. As for (b), adding and subtracting terms we have
%----------------------------------------------------------------------------%
\begin{equation}
\label{eq:Lemma_5.1}
\begin{array}{rcl}
\dfrac{1}{T}\sum\limits_{t=1}^{T}\left( \widehat{\mathbf{g}}_{t}-\widehat{%
	\mathbf{H}}^{-1}\mathbf{g}_{t}\right) \mathbf{\widehat{g}}_{t}^{\prime } & = & 
\dfrac{1}{T}\sum\limits_{t=1}^{T}\left( \widehat{\mathbf{g}}_{t}-\widehat{%
	\mathbf{H}}^{-1}\mathbf{g}_{t}\right) \left( \widehat{\mathbf{g}}_{t}-%
\widehat{\mathbf{H}}^{-1}\mathbf{g}_{t}\right) ^{\prime } \\ 
&  & +\dfrac{1}{T}\sum\limits_{t=1}^{T}\left( \widehat{\mathbf{g}}_{t}-%
\widehat{\mathbf{H}}^{-1}\mathbf{g}_{t}\right) \mathbf{g}_{t}^{\prime
}\left( \widehat{\mathbf{H}}^{-1}\right) ^{\prime }.%
\end{array}%
\end{equation}
%----------------------------------------------------------------------------%
Taking into account the results in (a), it follows that
%----------------------------------------------------------------------------%
\begin{equation}
\label{eq:Lemma_5.2}
\dfrac{1}{T}\sum\limits_{t=1}^{T}\left( \widehat{\mathbf{g}}_{t}-\widehat{%
	\mathbf{H}}^{-1}\mathbf{g}_{t}\right) \left( \widehat{\mathbf{g}}_{t}-%
\widehat{\mathbf{H}}^{-1}\mathbf{g}_{t}\right) ^{\prime }=O_{p}\left( \dfrac{%
	1}{N}\right) +O_{p}\left( \dfrac{1}{\sqrt{N}C_{NT}^{2}}\right) +O_{p}\left( 
\dfrac{1}{C_{NT}^{4}}\right).
\end{equation}%
%----------------------------------------------------------------------------%
From (\ref{eq:Lemma_5.0}),  we also have that
%----------------------------------------------------------------------------%
\begin{equation*}
\begin{array}{rcl}
\dfrac{1}{T}\sum\limits_{t=1}^{T}\left( \widehat{\mathbf{g}}_{t}-\widehat{%
	\mathbf{H}}^{-1}\mathbf{g}_{t}\right) \mathbf{g}_{t}^{\prime } & = & \dfrac{1%
}{T}\sum\limits_{t=1}^{T}\left[ \dfrac{1}{N}\left( \mathbf{A}\widehat{%
	\mathbf{H}}\right) ^{\prime }\mathbf{e}_{t}+\dfrac{1}{N}\mathbf{\widehat{A}}%
^{\prime }\left( \mathbf{A-}\widehat{\mathbf{A}}\widehat{\mathbf{H}}%
^{-1}\right) \mathbf{g}_{t}+\dfrac{1}{N}\left( \widehat{\mathbf{A}}-\mathbf{A%
}\widehat{\mathbf{H}}\right) ^{\prime }\mathbf{e}_{t}\right] \mathbf{g}%
_{t}^{\prime } \\ 
& = & \dfrac{1}{\sqrt{NT}}\widehat{\mathbf{H}}^{\prime }\dfrac{\mathbf{A}%
	^{\prime }}{\sqrt{N}}\left( \dfrac{1}{\sqrt{T}}\sum\limits_{t=1}^{T}\mathbf{%
	e}_{t}\mathbf{g}_{t}^{\prime }\right) +\dfrac{\mathbf{\widehat{A}}^{\prime
	}\left( \mathbf{A-}\widehat{\mathbf{A}}\widehat{\mathbf{H}}^{-1}\right) }{N}%
\dfrac{1}{T}\sum\limits_{t=1}^{T}\mathbf{g}_{t}\mathbf{g}_{t}^{\prime } \\ 
&  & +\dfrac{1}{\sqrt{NT}}\left( \dfrac{\widehat{\mathbf{A}}-\mathbf{A}%
	\widehat{\mathbf{H}}}{\sqrt{N}}\right) ^{\prime }\left( \dfrac{1}{\sqrt{T}}%
\sum\limits_{t=1}^{T}\mathbf{e}_{t}\mathbf{g}_{t}^{\prime }\right),
\end{array}%
\end{equation*}%
%----------------------------------------------------------------------------%
and taking into account Assumptions \ref{assum:FL} and \ref{assum:MCLT}(c), and Lemma \ref{Lemma:3},
%----------------------------------------------------------------------------%
\begin{equation}
\label{eq:Lemma_5.3}
\begin{array}{rcl}
\left\Vert \dfrac{1}{T}\sum\limits_{t=1}^{T}\left( \widehat{\mathbf{g}}_{t}-%
\widehat{\mathbf{H}}^{-1}\mathbf{g}_{t}\right) \mathbf{g}_{t}^{\prime
}\right\Vert  & = & \dfrac{1}{\sqrt{NT}}\left\Vert \widehat{\mathbf{H}}%
\right\Vert \left\Vert \dfrac{\mathbf{A}}{\sqrt{N}}\right\Vert \left\Vert 
\dfrac{1}{\sqrt{T}}\sum\limits_{t=1}^{T}\mathbf{e}_{t}\mathbf{g}%
_{t}^{\prime }\right\Vert  \\ 
&  & +\left\Vert \dfrac{\mathbf{\widehat{A}}^{\prime }\left( \mathbf{A-}\widehat{%
		\mathbf{A}}\widehat{\mathbf{H}}^{-1}\right) }{N}\right\Vert \left\Vert 
\dfrac{1}{T}\sum\limits_{t=1}^{T}\mathbf{g}_{t}\mathbf{g}_{t}^{\prime
}\right\Vert  \\ 
&  & +\dfrac{1}{\sqrt{NT}}\left\Vert \dfrac{\widehat{\mathbf{A}}-\mathbf{A}%
	\widehat{\mathbf{H}}}{\sqrt{N}}\right\Vert \left\Vert \dfrac{1}{\sqrt{T}}%
\sum\limits_{t=1}^{T}\mathbf{e}_{t}\mathbf{g}_{t}^{\prime }\right\Vert  \\ 
& = & O_{p}\left( \dfrac{1}{\sqrt{NT}}\right) +O_{p}\left( \dfrac{1}{%
	C_{NT}^{2}}\right) +O_{p}\left( \dfrac{1}{\sqrt{NT}}\right)  \\ 
& = & O_{p}\left( \dfrac{1}{C_{NT}^{2}}\right).
\end{array}%
\end{equation}%
%----------------------------------------------------------------------------%
Combining (\ref{eq:Lemma_5.1}) through (\ref{eq:Lemma_5.3}), it follows that
%----------------------------------------------------------------------------%
\begin{equation*}
\begin{array}{rcl}
\dfrac{1}{T}\sum\limits_{t=1}^{T}\left( \widehat{\mathbf{g}}_{t}-\widehat{%
\mathbf{H}}^{-1}\mathbf{g}_{t}\right) \mathbf{\widehat{g}}_{t}^{\prime }&=&O_{p}\left( \dfrac{1}{N}\right) +O_{p}\left( \dfrac{1}{\sqrt{N}C_{NT}^{2}}\right) +O_{p}\left( 
\dfrac{1}{C_{NT}^{4}}\right)+O_{p}\left( \dfrac{1}{C_{NT}^{2}}\right)\\
&=&O_{p}\left( \dfrac{1}{C_{NT}^{2}}\right),
\end{array}
\end{equation*}
%----------------------------------------------------------------------------%
which shows (b) and completes the proof of the lemma.
\end{proof}

\begin{proof}[\textbf{Proof of Lemma \protect\ref{Lemma:var_hat}}] We proceed by following steps analogous to those in the proof of Proposition 1 in \cite{bai03}, and we develop the proof of the lemma for the sake of completeness. Given $\mathbf{\widehat{H}}=\left(\mathbf{GG}^{\prime}/T\right)\left(\mathbf{A}^{\prime }\mathbf{\widehat{A}}/N\right)\mathbf{\widehat{V}}^{-1}$, pre-multiply both sides of the identity $\left(1/NT\right)\mathbf{X}^{\prime}\mathbf{X\widehat{A}}=\mathbf{\widehat{A}\widehat{V}}$
by $\left( \mathbf{GG}^{\prime }\left/
T\right. \right) ^{1\left/ 2\right. }N^{-1}\mathbf{A}^{\prime }$ to obtain
%----------------------------------------------------------------------------%
\begin{equation*}
\frac{1}{N}\left( \dfrac{\mathbf{GG}^{\prime }}{T}\right) ^{1\left/ 2\right.
}\mathbf{A}^{\prime }\left( \frac{\mathbf{X}^{\prime }\mathbf{X}}{NT}\right) 
\mathbf{\widehat{A}}=\left( \dfrac{\mathbf{GG}^{\prime }}{T}\right) ^{1\left/
	2\right. }\left( \dfrac{\mathbf{A}^{\prime }\mathbf{\widehat{A}}}{N}\right) 
\mathbf{\widehat{V}.}
\end{equation*}%
%----------------------------------------------------------------------------%
Given $\left(\ref{eq:linear_model}\right)$, write
$\mathbf{X}=\mathbf{G}^{\prime }\mathbf{A}^{\prime }+\mathbf{E}$
with $\mathbf{X}=\left(\mathbf{x}_{1},\ldots,\mathbf{x}_{T}\right)^{\prime}$ and $\mathbf{E}=\left(\mathbf{e}_{1},\ldots,\mathbf{e}_{T}\right)^{\prime}$. We thus have
%----------------------------------------------------------------------------%
\begin{equation}
\label{eq:lemma_6_1}
\dfrac{1}{N}\left( \dfrac{\mathbf{GG}^{\prime }}{T}%
\right) ^{1\left/ 2\right. }\mathbf{A}^{\prime }\left( \dfrac{\mathbf{AGG}%
	^{\prime }\mathbf{A}^{\prime }}{NT}\right) \mathbf{\widehat{A}}+\mathbf{\widehat{D}}
 = \left( \dfrac{\mathbf{GG}^{\prime }}{T}\right) ^{1\left/ 2\right.
}\left( \dfrac{\mathbf{A}^{\prime }\mathbf{\widehat{A}}}{N}\right) \mathbf{\widehat{V}},
\end{equation}%
%----------------------------------------------------------------------------%
where%
%----------------------------------------------------------------------------%
\begin{equation*}
\mathbf{\widehat{D}}=\dfrac{1}{N}\left( \dfrac{\mathbf{GG}^{\prime }}{T}\right)
^{1\left/ 2\right. }\mathbf{A}^{\prime }\left( \dfrac{\mathbf{AGE+E}^{\prime
	}\mathbf{G}^{\prime }\mathbf{A}^{\prime }+\mathbf{E}^{\prime }\mathbf{E}}{NT}%
\right) \mathbf{\widehat{A}}=o_{p}\left( 1\right)
\end{equation*}%
%----------------------------------------------------------------------------%
by Lemma \ref{Lemma:a_i_hat_components}.
%----------------------------------------------------------------------------%
Let%
%----------------------------------------------------------------------------%
\begin{equation*}
\begin{array}{ll}
\mathbf{W}=\left( \dfrac{\mathbf{GG}^{\prime }}{T}\right) ^{1\left/ 2\right.
}\left( \dfrac{\mathbf{A}^{\prime }\mathbf{A}}{N}\right) \left( \dfrac{%
	\mathbf{GG}^{\prime }}{T}\right) ^{1\left/ 2\right. }, & \mathbf{\widehat{Z}}%
=\left( \dfrac{\mathbf{GG}^{\prime }}{T}\right) ^{1\left/ 2\right. }\left( 
\dfrac{\mathbf{A}^{\prime }\mathbf{\widehat{A}}}{N}\right),
\end{array}%
\end{equation*}%
%----------------------------------------------------------------------------%
so that we can write (\ref{eq:lemma_6_1}) as
%----------------------------------------------------------------------------%
\begin{equation*}
\left( \mathbf{W}+\mathbf{\widehat{D}\widehat{Z}}^{-1}\right) \mathbf{\widehat{Z}}=%
\mathbf{\widehat{Z}\widehat{V}}.
\end{equation*}%
%----------------------------------------------------------------------------%
Therefore, each column of $\mathbf{\widehat{Z}}$ is an eigenvector of $\left( \mathbf{W}+\mathbf{\widehat{D}\widehat{Z}}^{-1}\right) $, with length different from unity. Let $\mathbf{\widehat{V}}^{\ast }$ be the diagonal matrix of the diagonal elements of $\mathbf{\widehat{Z}}^{\prime }\mathbf{\widehat{Z}}$. Define $\mathbf{\widehat{\Psi}}=\mathbf{\widehat{Z}}\left( \mathbf{\widehat{V}%
}^{\ast }\right) ^{-1\left/ 2\right. }$\ so that each column of $\mathbf{\widehat{\Psi}}$ has unit length. We thus get%
%----------------------------------------------------------------------------%
\begin{equation*}
\left( \mathbf{W}+\mathbf{\widehat{D}\widehat{Z}}^{-1}\right) \mathbf{\widehat{\Psi}}=%
\mathbf{\widehat{\Psi}\widehat{V}},
\end{equation*}%
%----------------------------------------------------------------------------%
where $\mathbf{\widehat{\Psi}}$ is the eigenvector matrix of $\left( \mathbf{W}+%
\mathbf{\widehat{D}\widehat{Z}}^{-1}\right) $. Consider
%----------------------------------------------------------------------------%
\begin{equation*}
\mathbf{W}+\mathbf{\widehat{D}\widehat{Z}}^{-1}=\left( \dfrac{\mathbf{GG}^{\prime }}{%
	T}\right) ^{1\left/ 2\right. }\left( \dfrac{\mathbf{A}^{\prime }\mathbf{A}}{N%
}\right) \left( \dfrac{\mathbf{GG}^{\prime }}{T}\right) ^{1\left/ 2\right. }+%
\mathbf{\widehat{D}\widehat{Z}}^{-1},
\end{equation*}
%----------------------------------------------------------------------------%
and note that%
%----------------------------------------------------------------------------%
\begin{equation*}
\dfrac{\mathbf{GG}^{\prime }}{T}=\dfrac{1}{T}\sum\limits_{t=1}^{T}\left( 
\begin{array}{c}
\mathbb{I}_{1t}\mathbf{f}_{1t} \\ 
\mathbb{I}_{2t}\mathbf{f}_{2t}%
\end{array}%
\right) \left( 
\begin{array}{c}
\mathbb{I}_{1t}\mathbf{f}_{1t} \\ 
\mathbb{I}_{2t}\mathbf{f}_{2t}%
\end{array}%
\right) ^{\prime }=\dfrac{1}{T}\sum\limits_{t=1}^{T}\left( 
\begin{array}{cc}
\mathbb{I}_{1t}\mathbf{f}_{1t}\mathbf{f}_{1t}^{\prime } & \mathbf{0} \\ 
\mathbf{0} & \mathbb{I}_{2t}\mathbf{f}_{2t}\mathbf{f}_{2t}^{\prime }%
\end{array}%
\right) \overset{p}{\rightarrow }\left( 
\begin{array}{cc}
\mathbf{\Sigma }_{\mathbf{f}_{1}} & \mathbf{0} \\ 
\mathbf{0} & \mathbf{\Sigma }_{\mathbf{f}_{2}}%
\end{array}%
\right) =\mathbf{\Sigma }_{\mathbf{g}}
\end{equation*}%
%----------------------------------------------------------------------------%
by Assumption 1. Further, $\left( \mathbf{A}^{\prime }\mathbf{A}\left/
N\right. \right) \rightarrow \mathbf{\Sigma }_{\mathbf{A}}$ by Assumption 2.
Therefore, by Assumptions 1 and 2, $\mathbf{W}+\mathbf{\widehat{D}\widehat{Z}}^{-1}%
\overset{p}{\rightarrow }\mathbf{\Sigma }_{\mathbf{g}}^{1\left/ 2\right. }%
\mathbf{\Sigma }_{\mathbf{A}}\mathbf{\Sigma }_{\mathbf{g}%
}^{1\left/ 2\right. }$. Because the eigenvalues of $\mathbf{\Sigma }_{\mathbf{g}}^{1\left/ 2\right. }\mathbf{\Sigma }_{\mathbf{A}}\mathbf{\Sigma }_{\mathbf{g}}^{1\left/ 2\right. }$ are distinct by
Assumption \ref{assum:eigenvalues}, the eigenvalues of $\mathbf{W}+\mathbf{\widehat{D}\widehat{Z}}^{-1}$ are also distinct for large $N$ and $T$, by the continuity of eigenvalues. This implies that the eigenvector matrix of $\mathbf{W}+\mathbf{\widehat{D}\widehat{Z}}^{-1}$ is unique except for the fact that each column can be replaced by its negative value. Further, the $p-th$ column of $\mathbf{\widehat{Z}}$ depends on $\mathbf{\widehat{A}}$ only through the $p-th$ column of $\mathbf{\widehat{A}}$, for $p=1,\ldots ,r$. This implies that the sign of each column in 
$\mathbf{\widehat{Z}}$, and thus in $\mathbf{\widehat{\Psi}}=\mathbf{\widehat{Z}}\left( 
\mathbf{\widehat{V}}^{\ast }\right) ^{-1\left/ 2\right. }$, is determined by the
sign of the corresponding column of $\mathbf{\widehat{A}}$. Therefore, the
column sign of $\mathbf{\widehat{A}}$ and $\mathbf{\widehat{\Psi}}$ are uniquely
determined. By the eigenvector perturbation theory, which requires the
eigenvalues to be distinct, there exists a unique eigenvector matrix $%
\mathbf{\Psi }$ of $\mathbf{\Sigma }_{\mathbf{A}}^{1\left/ 2\right. }\mathbf{%
	\Sigma }_{\mathbf{g}}^{1\left/ 2\right. }\mathbf{\Sigma }_{\mathbf{A}%
}^{1\left/ 2\right. }$ such that $\left\Vert \mathbf{\widehat{\Psi}}-\mathbf{%
	\Psi }\right\Vert =o_{p}\left( 1\right) $. Since $\mathbf{\widehat{\Psi}}=%
\mathbf{\widehat{Z}}\left( \mathbf{\widehat{V}}^{\ast }\right) ^{-1\left/ 2\right. }$
and $\mathbf{\widehat{Z}}=\left( \mathbf{GG}^{\prime }\left/ T\right. \right)
^{1\left/ 2\right. }\left( \mathbf{A}^{\prime }\mathbf{\widehat{A}}\left/
N\right. \right) $ then $\mathbf{\widehat{\Psi}}=\left( \mathbf{GG}^{\prime
}\left/ T\right. \right) ^{1\left/ 2\right. }\left( \mathbf{A}^{\prime }%
\mathbf{\widehat{A}}\left/ N\right. \right) \left( \mathbf{\widehat{V}}^{\ast
}\right) ^{-1\left/ 2\right. }$, which implies that%
%----------------------------------------------------------------------------%
\begin{equation*}
\dfrac{\mathbf{A}^{\prime }\mathbf{\widehat{A}}}{N}=\left( \dfrac{\mathbf{GG}%
	^{\prime }}{T}\right) ^{-1\left/ 2\right. }\mathbf{\widehat{\Psi}}\left( \mathbf{%
	\widehat{V}}^{\ast }\right) ^{1\left/ 2\right. }\overset{p}{\rightarrow }\mathbf{%
	\Sigma }_{\mathbf{g}}^{-1\left/ 2\right. }\mathbf{\Psi V}^{1\left/ 2\right. }
\end{equation*}%
%----------------------------------------------------------------------------%
by Assumption 1 and since $\mathbf{\widehat{V}}^{\ast }\overset{p}{\rightarrow }%
\mathbf{V}$, the latter following from arguments analogous to those in the proof of Proposition 1 in \cite{bai03}. This completes the proof of the lemma.
\end{proof}

\begin{proof}[\textbf{Proof of Lemma \protect\ref{Lemma:Q_j}}]
From Lemma \ref{Lemma:var_hat}, and taking into account (\ref{eq:SG}), we have
%----------------------------------------------------------------------------%
\begin{equation*}
\begin{array}{rcl}
\mathbf{Q} & = & \left( 
\begin{array}{c}
\mathbf{Q}_{1} \\ 
\mathbf{Q}_{2}%
\end{array}%
\right)  \\ 
& = & \mathbf{\Sigma }_{\mathbf{g}}^{-1\left/ 2\right. }\mathbf{\Psi V}%
^{1\left/ 2\right. } \\ 
& = & \left( 
\begin{array}{cc}
\mathbf{\Sigma }_{\mathbf{f}1} & \mathbf{0} \\ 
\mathbf{0} & \mathbf{\Sigma }_{\mathbf{f}2}%
\end{array}%
\right) ^{-1\left/ 2\right. }\mathbf{\Psi V}^{1\left/ 2\right. } \\ 
& = & \left( 
\begin{array}{cc}
\mathbf{\Sigma }_{\mathbf{f}1}^{-1\left/ 2\right. } & \mathbf{0} \\ 
\mathbf{0} & \mathbf{\Sigma }_{\mathbf{f}2}^{-1\left/ 2\right. }%
\end{array}%
\right) \left( 
\begin{array}{c}
\mathbf{\Psi }_{1} \\ 
\mathbf{\Psi }_{2}%
\end{array}%
\right) \mathbf{V}^{1\left/ 2\right. } \\ 
& = & \left( 
\begin{array}{c}
\mathbf{\Sigma }_{\mathbf{f}1}^{-1\left/ 2\right. }\mathbf{\Psi }_{1}\mathbf{%
	V}^{1\left/ 2\right. } \\ 
\mathbf{\Sigma }_{\mathbf{f}2}^{-1\left/ 2\right. }\mathbf{\Psi }_{2}\mathbf{%
	V}^{1\left/ 2\right. }%
\end{array}%
\right) 
\end{array},
\end{equation*}
%----------------------------------------------------------------------------%
which completes the proof of the lemma.
\end{proof}

\begin{proof}[\textbf{Proof of Lemma \protect\ref{Lemma:Hat_V_NT}}]
Given the equivalent linear representation in (\ref{eq:linear_model}), we can write
%----------------------------------------------------------------------------%
\begin{equation}
\label{eq:Lemma_8_Proof_1}
\begin{array}{rcl}
\dfrac{1}{NT}\sum\limits_{t=1}^{T}\mathbf{x}_{t}\mathbf{x}_{t} & = & \dfrac{%
	1}{NT}\sum\limits_{t=1}^{T}\left( \mathbf{Ag}_{t}+\mathbf{e}_{t}\right)
\left( \mathbf{Ag}_{t}+\mathbf{e}_{t}\right) ^{\prime } \\ 
& = & \dfrac{\mathbf{A}}{\sqrt{N}}\left( \dfrac{1}{T}\sum\limits_{t=1}^{T}%
\mathbf{g}_{t}\mathbf{g}_{t}^{\prime }\right) \dfrac{\mathbf{A}}{\sqrt{N}}%
^{\prime }+\dfrac{\mathbf{A}}{N}\left( \dfrac{1}{T}\sum\limits_{t=1}^{T}%
\mathbf{g}_{t}\mathbf{e}_{t}^{\prime }\right)  \\ 
&  & +\left( \dfrac{1}{T}\sum\limits_{t=1}^{T}\mathbf{e}_{t}\mathbf{g}%
_{t}^{\prime }\right) \dfrac{\mathbf{A}^{\prime }}{N}+\dfrac{1}{NT}%
\sum\limits_{t=1}^{T}\mathbf{e}_{t}\mathbf{e}_{t}^{\prime }.%
\end{array}%
\end{equation}%
%----------------------------------------------------------------------------%
Taking into account Assumption \ref{assum:FL}(b) and Assumption \ref{assum:WD}, it follows that
%----------------------------------------------------------------------------%
\begin{equation}
\label{eq:Lemma_8_Proof_2}
\begin{array}{rcl}
\left\Vert \dfrac{\mathbf{A}}{N}\left( \dfrac{1}{T}\sum\limits_{t=1}^{T}%
\mathbf{g}_{t}\mathbf{e}_{t}^{\prime }\right) \right\Vert  & \leq  & \dfrac{1%
}{\sqrt{NT}}\left\Vert \dfrac{\mathbf{A}}{\sqrt{N}}\right\Vert \left\Vert 
\dfrac{1}{\sqrt{T}}\sum\limits_{t=1}^{T}\left( 
\begin{array}{c}
\mathbb{I}_{1t}\mathbf{f}_{1t}\mathbf{e}_{t}^{\prime } \\ 
\mathbb{I}_{2t}\mathbf{f}_{2t}\mathbf{e}_{t}^{\prime }%
\end{array}%
\right) \right\Vert  \\ 
& = & \dfrac{1}{\sqrt{NT}}O_{p}\left( 1\right) O_{p}\left( \sqrt{N}\right) 
\\ 
& = & O_{p}\left( \dfrac{1}{\sqrt{T}}\right).
\end{array}%
\end{equation}
%----------------------------------------------------------------------------%
Similarly, we can prove that
%----------------------------------------------------------------------------%
\begin{equation}
\label{eq:Lemma_8_Proof_3}
\dfrac{1}{N}\mathbf{A}\left( \dfrac{1}{T}\sum\limits_{t=1}^{T}\mathbf{e}_{t}%
\mathbf{g}_{t}^{\prime }\right) =O_{p}\left( \dfrac{1}{\sqrt{T}}\right).
\end{equation}
%----------------------------------------------------------------------------%
Finally, by the weak dependence condition in Assumption (\ref{assum:TCSDH}),
%----------------------------------------------------------------------------%
\begin{equation}
\label{eq:Lemma_8_Proof_4}
\left\Vert \dfrac{1}{NT}\sum\limits_{t=1}^{T}\mathbf{e}_{t}\mathbf{e}%
_{t}^{\prime }\right\Vert =o_{p}\left( 1\right).
\end{equation}
%----------------------------------------------------------------------------%
By combining (\ref{eq:Lemma_8_Proof_1}) through (\ref{eq:Lemma_8_Proof_4}), we then have
%----------------------------------------------------------------------------%
\begin{equation*}
\dfrac{1}{NT}\sum\limits_{t=1}^{T}\mathbf{x}_{t}\mathbf{x}_{t} = \dfrac{%
\mathbf{A}}{\sqrt{N}}\left( \dfrac{1}{T}\sum\limits_{t=1}^{T}\mathbf{g}_{t}%
\mathbf{g}_{t}^{\prime }\right) \dfrac{\mathbf{A}^{\prime }}{\sqrt{N}}%
+o_{p}\left( 1\right) = \dfrac{\mathbf{A}}{\sqrt{N}}\dfrac{\mathbf{GG}^{\prime }}{T}\dfrac{\mathbf{A}^{\prime }}{\sqrt{N}}+o_{p}\left( 1\right).
\end{equation*}%
%----------------------------------------------------------------------------%
The result in the lemma follows from Assumptions (\ref{assum:F}) and (\ref{assum:FL}) by noting that the eigenvalues of  
$\left( \mathbf{A}\left/ \sqrt{N}\right. \right) \left( \mathbf{GG}^{\prime
}\left/ T\right. \right) \left( \mathbf{A}^{\prime }\left/ \sqrt{N}\right.
\right)$ are the same as those of
$\left( \mathbf{G}%
^{\prime }\left/ \sqrt{T}\right. \right) \left( \mathbf{A}^{\prime }\mathbf{A%
}\left/ N\right. \right) \left( \mathbf{G}\left/ \sqrt{T}\right. \right) $.
\end{proof}

\begin{proof}[\textbf{Proof of Lemma \protect\ref{Lemma:Hat_I_xi}}]
From the definition of $\mathbf{\widehat{I}}_{\mathbf{\widehat{\xi}}k_{1}}$ in (\ref{eq:hat_I_xi}), and taking into account Lemma (\ref{Lemma:avg})(a), we have
%----------------------------------------------------------------------------%
\begin{equation*}
	\begin{array}{rcl}
		\mathbf{\widehat{I}}_{\mathbf{\widehat{\xi}}k_{1}} & = & \left( \sum\limits_{t=1}^{T}\widehat{%
			\mathbf{\xi }}_{j,t\left\vert T\right. }\mathbb{I}_{jt}\widehat{\mathbf{g}}%
		_{t}\widehat{\mathbf{g}}_{t}^{\prime }\right) \left( \sum\limits_{t=1}^{T}%
		\widehat{\mathbf{\xi }}_{j,t\left\vert T\right. }\widehat{\mathbf{g}}_{t}%
		\widehat{\mathbf{g}}_{t}^{\prime }\right) ^{-1} \\ 
		& = & \left\{ \sum\limits_{t=1}^{T}\widehat{\mathbf{\xi }}_{j,t\left\vert
			T\right. }\mathbb{I}_{jt}\left\{ 
		\begin{array}{c}
			\left[ \widehat{\mathbf{H}}^{-1}\mathbf{g}_{t}+O_{p}\left( \dfrac{1}{\sqrt{N}%
			}\right) +O_{p}\left( \dfrac{1}{C_{NT}^{2}}\right) \right]  \\ 
			\times \left[ \widehat{\mathbf{H}}^{-1}\mathbf{g}_{t}+O_{p}\left( \dfrac{1}{%
				\sqrt{N}}\right) +O_{p}\left( \dfrac{1}{C_{NT}^{2}}\right) \right] ^{\prime }%
		\end{array}%
		\right\} \right\}  \\ 
		&  & \times \left\{ \sum\limits_{t=1}^{T}\widehat{\mathbf{\xi }}%
		_{j,t\left\vert T\right. }\left\{ 
		\begin{array}{c}
			\left[ \widehat{\mathbf{H}}^{-1}\mathbf{g}_{t}+O_{p}\left( \dfrac{1}{\sqrt{N}%
			}\right) +O_{p}\left( \dfrac{1}{C_{NT}^{2}}\right) \right]  \\ 
			\times \left[ \widehat{\mathbf{H}}^{-1}\mathbf{g}_{t}+O_{p}\left( \dfrac{1}{%
				\sqrt{N}}\right) +O_{p}\left( \dfrac{1}{C_{NT}^{2}}\right) \right] ^{\prime }%
		\end{array}%
		\right\} \right\} ^{-1} \\ 
		& = & \left\{ \dfrac{1}{T}\sum\limits_{t=1}^{T}\widehat{\mathbf{\xi }}%
		_{j,t\left\vert T\right. }\mathbb{I}_{jt}\left[ 
		\begin{array}{l}
			\widehat{\mathbf{H}}^{-1}\mathbf{g}_{t}\mathbf{g}_{t}^{\prime }\left( 
			\widehat{\mathbf{H}}^{-1}\right) ^{\prime }+O_{p}\left( \dfrac{1}{\sqrt{N}}%
			\right) +O_{p}\left( \dfrac{1}{C_{NT}^{2}}\right)  \\ 
			+O_{p}\left( \dfrac{1}{\sqrt{N}C_{NT}^{2}}\right) +O_{p}\left( \dfrac{1}{%
				C_{NT}^{4}}\right) 
		\end{array}%
		\right] \right\}  \\ 
		&  & \times \left\{ \dfrac{1}{T}\sum\limits_{t=1}^{T}\widehat{\mathbf{\xi }}%
		_{j,t\left\vert T\right. }\left[ 
		\begin{array}{l}
			\widehat{\mathbf{H}}^{-1}\mathbf{g}_{t}\mathbf{g}_{t}^{\prime }\left( 
			\widehat{\mathbf{H}}^{-1}\right) ^{\prime }+O_{p}\left( \dfrac{1}{\sqrt{N}}%
			\right) +O_{p}\left( \dfrac{1}{C_{NT}^{2}}\right)  \\ 
			+O_{p}\left( \dfrac{1}{\sqrt{N}C_{NT}^{2}}\right) +O_{p}\left( \dfrac{1}{%
				C_{NT}^{4}}\right) 
		\end{array}%
		\right] \right\} ^{-1} \\ 
		& = & \left[ \widehat{\mathbf{H}}^{-1}\left( \dfrac{1}{T}\sum%
		\limits_{t=1}^{T}\widehat{\mathbf{\xi }}_{j,t\left\vert T\right. }\mathbb{I}%
		_{jt}\mathbf{g}_{t}\mathbf{g}_{t}^{\prime }\right) \left( \widehat{\mathbf{H}%
		}^{-1}\right) ^{\prime }+O_{p}\left( \dfrac{1}{\sqrt{N}}\right) +O_{p}\left( 
		\dfrac{1}{C_{NT}^{2}}\right) \right]  \\ 
		&  & \times \left[ \widehat{\mathbf{H}}^{-1}\left( \dfrac{1}{T}%
		\sum\limits_{t=1}^{T}\widehat{\mathbf{\xi }}_{j,t\left\vert T\right. }%
		\mathbf{g}_{t}\mathbf{g}_{t}^{\prime }\right) \left( \widehat{\mathbf{H}}%
		^{-1}\right) ^{\prime }+O_{p}\left( \dfrac{1}{\sqrt{N}}\right) +O_{p}\left( 
		\dfrac{1}{C_{NT}^{2}}\right) \right] ^{-1} \\ 
		& = & \widehat{\mathbf{H}}^{-1}\left( \dfrac{1}{T}\sum\limits_{t=1}^{T}%
		\widehat{\mathbf{\xi }}_{j,t\left\vert T\right. }\mathbb{I}_{jt}\mathbf{g}%
		_{t}\mathbf{g}_{t}^{\prime }\right) \left( \dfrac{1}{T}\sum\limits_{t=1}^{T}%
		\widehat{\mathbf{\xi }}_{j,t\left\vert T\right. }\mathbf{g}_{t}\mathbf{g}%
		_{t}^{\prime }\right) ^{-1}\widehat{\mathbf{H}}+o_{p}\left( 1\right).
	\end{array}
\end{equation*}
%----------------------------------------------------------------------------%
Taking further into account the definition of $\mathbf{g}_{t}$ in (\ref{eq:g_t}), it follows that
%----------------------------------------------------------------------------%
\begin{equation*}
	\begin{array}{rcl}
		\mathbf{\widehat{I}}_{\mathbf{\widehat{\xi}}k_{1}} & = & \widehat{\mathbf{H}}^{-1}\left( 
		\dfrac{1}{T}\sum\limits_{t=1}^{T}\widehat{\mathbf{\xi }}_{j,t\left\vert
			T\right. }\mathbb{I}_{jt}\mathbf{g}_{t}\mathbf{g}_{t}^{\prime }\right)
		\left( \dfrac{1}{T}\sum\limits_{t=1}^{T}\widehat{\mathbf{\xi }}%
		_{j,t\left\vert T\right. }\mathbf{g}_{t}\mathbf{g}_{t}^{\prime }\right) ^{-1}%
		\widehat{\mathbf{H}}+o_{p}\left( 1\right)  \\ 
		& = & \widehat{\mathbf{H}}^{-1}\left[ \dfrac{1}{T}\sum\limits_{t=1}^{T}%
		\widehat{\mathbf{\xi }}_{j,t\left\vert T\right. }\mathbb{I}_{jt}\left( 
		\begin{array}{c}
			\mathbb{I}_{1t}\mathbf{f}_{1t} \\ 
			\mathbb{I}_{2t}\mathbf{f}_{2t}%
		\end{array}%
		\right) \left( 
		\begin{array}{c}
			\mathbb{I}_{1t}\mathbf{f}_{1t} \\ 
			\mathbb{I}_{2t}\mathbf{f}_{2t}%
		\end{array}%
		\right) ^{\prime }\right]  \\ 
		&  & \times \left[ \dfrac{1}{T}\sum\limits_{t=1}^{T}\widehat{\mathbf{\xi }}%
		_{j,t\left\vert T\right. }\left( 
		\begin{array}{c}
			\mathbb{I}_{1t}\mathbf{f}_{1t} \\ 
			\mathbb{I}_{2t}\mathbf{f}_{2t}%
		\end{array}%
		\right) \left( 
		\begin{array}{c}
			\mathbb{I}_{1t}\mathbf{f}_{1t} \\ 
			\mathbb{I}_{2t}\mathbf{f}_{2t}%
		\end{array}%
		\right) ^{\prime }\right] ^{-1}\widehat{\mathbf{H}}+o_{p}\left( 1\right)  \\ 
		& = & \widehat{\mathbf{H}}^{-1}\left[ \dfrac{1}{T}\sum\limits_{t=1}^{T}%
		\left( 
		\begin{array}{cc}
			\widehat{\mathbf{\xi }}_{j,t\left\vert T\right. }\mathbb{I}_{jt}\mathbb{I}%
			_{1t}\mathbf{f}_{1t}\mathbf{f}_{1t}^{\prime } & \mathbf{0} \\ 
			\mathbf{0} & \widehat{\mathbf{\xi }}_{j,t\left\vert T\right. }\mathbb{I}_{jt}%
			\mathbb{I}_{2t}\mathbf{f}_{2t}\mathbf{f}_{2t}^{\prime }%
		\end{array}%
		\right) \right]  \\ 
		&  & \times \left[ \dfrac{1}{T}\sum\limits_{t=1}^{T}\left( 
		\begin{array}{cc}
			\widehat{\mathbf{\xi }}_{j,t\left\vert T\right. }\mathbb{I}_{1t}\mathbf{f}%
			_{1t}\mathbf{f}_{1t}^{\prime } & \mathbf{0} \\ 
			\mathbf{0} & \widehat{\mathbf{\xi }}_{j,t\left\vert T\right. }\mathbb{I}_{2t}%
			\mathbf{f}_{2t}\mathbf{f}_{2t}^{\prime }%
		\end{array}%
		\right) \right] ^{-1}\widehat{\mathbf{H}}+o_{p}\left( 1\right)  \\ 
		& = & \mathbf{H}^{-1}\left[ 
		\begin{array}{cc}
			\mathbb{I}\left( j=1\right) \mathbf{I}_{r_{1}} & \mathbf{0} \\ 
			\mathbf{0} & \mathbb{I}\left( j=2\right) \mathbf{I}_{r_{2}}%
		\end{array}%
		\right] \mathbf{H}+o_{p}\left( 1\right),
	\end{array}
\end{equation*}
%----------------------------------------------------------------------------%
where the last equality follows from \eqref{eq:Hlim}. Therefore,
%----------------------------------------------------------------------------%
\begin{equation*}
	\mathbf{\widehat{I}}_{\mathbf{\widehat{\xi}}k_{1}}\overset{p}{\rightarrow }\mathbf{H}^{-1}\left[ 
	\begin{array}{cc}
		\mathbb{I}\left( j=1\right) \mathbf{I}_{r_{1}} & \mathbf{0} \\ 
		\mathbf{0} & \mathbb{I}\left( j=2\right) \mathbf{I}_{r_{2}}%
	\end{array}%
	\right] \mathbf{H},
\end{equation*}
%----------------------------------------------------------------------------%
which completes the proof of the lemma.
\end{proof}

\begin{proof}[\textbf{Proof of Lemma \protect\ref{Lemma:V_hat_j}}]

From the definitions of eigenvectors and eigenvalues, for $j=1,2$ it follows that	
%----------------------------------------------------------------------------%
\begin{equation*}
	\widehat{\mathbf{\Sigma }}_{\widehat{\xi},\mathbf{x}j}\widehat{\mathbf{\Lambda }}%
	_{\widehat{\xi},j}^{\left( p\right) }=\widehat{\mathbf{\Lambda }}_{\widehat{\xi},j}^{\left( p\right) }%
	\widehat{\mathbf{V}}_{\widehat{\xi},j}^{\left( p\right) },
\end{equation*}
%----------------------------------------------------------------------------%
and, given the definition of $\widehat{\mathbf{\Sigma }}_{\widehat{\xi},\mathbf{x}j}$ in (\ref{eq:sigma_hat_x_j}), we can write
%----------------------------------------------------------------------------%
\begin{equation}
\label{eq:lemma_V_000}
	\dfrac{\sum\nolimits_{t=1}^{T}\widehat{\xi }_{jt\left\vert T\right. }%
		\mathbf{x}_{t}\mathbf{x}_{t}^{\prime }}{N\sum\nolimits_{t=1}^{T}\widehat{%
			\xi }_{jt\left\vert T\right. }}\widehat{\mathbf{\Lambda }}_{\widehat{\xi},j}^{\left(
		p\right) }=\widehat{\mathbf{\Lambda }}_{\widehat{\xi},j}^{\left( p\right) }\widehat{%
		\mathbf{V}}_{\widehat{\xi},j}^{\left( p\right) }.
\end{equation}
%----------------------------------------------------------------------------%
The normalisation constraint
%----------------------------------------------------------------------------%
\begin{equation}
	\label{eq:lemma_V_0000}
	\dfrac{\widehat{\mathbf{\Lambda }}_{\widehat{\xi},j}^{\left( p\right) \prime }\widehat{%
			\mathbf{\Lambda }}_{\widehat{\xi},j}^{\left( p\right) }}{N}=\mathbf{I}_{p}
\end{equation}
%----------------------------------------------------------------------------%
allows us to obtain
%----------------------------------------------------------------------------%
\begin{equation*}
	\dfrac{\widehat{\mathbf{\Lambda }}_{\widehat{\xi},j}^{\left( p\right) \prime }}{N}\dfrac{%
		\sum\nolimits_{t=1}^{T}\widehat{\xi }_{jt\left\vert T\right. }\mathbf{x}_{t}%
		\mathbf{x}_{t}^{\prime }}{N\sum\nolimits_{t=1}^{T}\widehat{\xi }%
		_{jt\left\vert T\right. }}\widehat{\mathbf{\Lambda }}_{\widehat{\xi},j}^{\left( p\right) }=%
	\widehat{\mathbf{V}}_{\widehat{\xi},j}^{\left( p\right) }.
\end{equation*}
%----------------------------------------------------------------------------%
Taking into account Assumption \ref{assum:FL}(b), we then have
%----------------------------------------------------------------------------%
\begin{equation}
\label{eq:lemma_V_00}
	\begin{array}{rcl}
		\left\Vert \dfrac{\widehat{\mathbf{\Lambda }}_{\widehat{\xi},j}^{\left( p\right) \prime }}{%
			N}\dfrac{\sum\nolimits_{t=1}^{T}\widehat{\xi }_{jt\left\vert T\right. }%
			\mathbf{x}_{t}\mathbf{x}_{t}^{\prime }}{N\sum\nolimits_{t=1}^{T}\widehat{%
				\xi }_{jt\left\vert T\right. }}\widehat{\mathbf{\Lambda }}_{\widehat{\xi},j}^{\left(
			p\right) }\right\Vert  & \leq  & \left\Vert \dfrac{\widehat{\mathbf{\Lambda }%
			}_{\widehat{\xi},j}^{\left( p\right) }}{\sqrt{N}}\right\Vert \left\Vert \dfrac{%
			\sum\nolimits_{t=1}^{T}\widehat{\xi }_{jt\left\vert T\right. }\mathbf{x}_{t}%
			\mathbf{x}_{t}^{\prime }}{N\sum\nolimits_{t=1}^{T}\widehat{\xi }%
			_{jt\left\vert T\right. }}\right\Vert \left\Vert \dfrac{\widehat{\mathbf{%
					\Lambda }}_{\widehat{\xi},j}^{\left( p\right) }}{\sqrt{N}}\right\Vert  \\ 
		& = & \left\Vert \dfrac{\sum\nolimits_{t=1}^{T}\widehat{\xi }_{jt\left\vert
				T\right. }\mathbf{x}_{t}\mathbf{x}_{t}^{\prime }}{N\sum\nolimits_{t=1}^{T}%
			\widehat{\xi }_{jt\left\vert T\right. }}\right\Vert O_{p}\left( 1\right).
	\end{array}%
\end{equation}
%----------------------------------------------------------------------------%
Consider now
%----------------------------------------------------------------------------%
\begin{equation}
\label{eq:lemma_V_0}
	\begin{array}{rcl}
		\dfrac{\sum\nolimits_{t=1}^{T}\widehat{\xi }_{jt\left\vert T\right. }%
			\mathbf{x}_{t}\mathbf{x}_{t}^{\prime }}{N\sum\nolimits_{t=1}^{T}\widehat{%
				\xi }_{jt\left\vert T\right. }} & = & \dfrac{\sum\nolimits_{t=1}^{T}%
			\widehat{\xi }_{jt\left\vert T\right. }\mathbf{x}_{t}\mathbf{x}_{t}^{\prime }%
		}{N\sum\nolimits_{t=1}^{T}\widehat{\xi }_{jt\left\vert T\right. }} \\ 
		& = & \dfrac{\sum\nolimits_{t=1}^{T}\widehat{\xi }_{jt\left\vert T\right.
			}\left( \mathbb{I}_{1t}\mathbf{\Lambda }_{1}\mathbf{f}_{1t}+\mathbb{I}_{2t}%
			\mathbf{\Lambda }_{2}\mathbf{f}_{2t}+\mathbf{e}_{t}\right) \left( \mathbb{I}%
			_{1t}\mathbf{\Lambda }_{1}\mathbf{f}_{1t}+\mathbb{I}_{2t}\mathbf{\Lambda }%
			_{2}\mathbf{f}_{2t}+\mathbf{e}_{t}\right) ^{\prime }}{N\sum%
			\nolimits_{t=1}^{T}\widehat{\xi }_{jt\left\vert T\right. }} \\ 
		& = & \dfrac{\mathbf{\Lambda }_{1}\left( \sum\nolimits_{t=1}^{T}\mathbb{I}%
			_{1t}\widehat{\xi }_{jt\left\vert T\right. }\mathbf{f}_{1t}\mathbf{f}%
			_{1t}^{\prime }\right) \mathbf{\Lambda }_{1}^{\prime }}{N\sum%
			\nolimits_{t=1}^{T}\widehat{\xi }_{jt\left\vert T\right. }}+\dfrac{\mathbf{%
				\Lambda }_{1}\left( \sum\nolimits_{t=1}^{T}\mathbb{I}_{1t}\widehat{\xi }%
			_{jt\left\vert T\right. }\mathbf{f}_{1t}\mathbf{e}_{t}^{\prime }\right) }{%
			N\sum\nolimits_{t=1}^{T}\widehat{\xi }_{jt\left\vert T\right. }} \\ 
		&  & +\dfrac{\mathbf{\Lambda }_{2}\left( \sum\nolimits_{t=1}^{T}\mathbb{I}%
			_{2t}\widehat{\xi }_{jt\left\vert T\right. }\mathbf{f}_{2t}\mathbf{f}%
			_{2t}^{\prime }\right) \mathbf{\Lambda }_{2}^{\prime }}{N\sum%
			\nolimits_{t=1}^{T}\widehat{\xi }_{jt\left\vert T\right. }}+\dfrac{\mathbf{%
				\Lambda }_{2}\left( \sum\nolimits_{t=1}^{T}\mathbb{I}_{2t}\widehat{\xi }%
			_{jt\left\vert T\right. }\mathbf{f}_{2t}\mathbf{e}_{t}^{\prime }\right) }{%
			N\sum\nolimits_{t=1}^{T}\widehat{\xi }_{jt\left\vert T\right. }} \\ 
		&  & +\dfrac{\left( \sum\nolimits_{t=1}^{T}\mathbb{I}_{1t}\widehat{\xi }%
			_{jt\left\vert T\right. }\mathbf{e}_{t}\mathbf{f}_{1t}^{\prime }\right) 
			\mathbf{\Lambda }_{1}^{\prime }}{N\sum\nolimits_{t=1}^{T}\widehat{\xi }%
			_{jt\left\vert T\right. }}+\dfrac{\left( \sum\nolimits_{t=1}^{T}\mathbb{I}%
			_{2t}\widehat{\xi }_{jt\left\vert T\right. }\mathbf{e}_{t}\mathbf{f}%
			_{2t}^{\prime }\right) \mathbf{\Lambda }_{2}^{\prime }}{N\sum%
			\nolimits_{t=1}^{T}\widehat{\xi }_{jt\left\vert T\right. }}+\dfrac{%
			\sum\nolimits_{t=1}^{T}\widehat{\xi }_{jt\left\vert T\right. }\mathbf{e}_{t}%
			\mathbf{e}_{t}^{\prime }}{N\sum\nolimits_{t=1}^{T}\widehat{\xi }%
			_{jt\left\vert T\right. }}.
	\end{array}%
\end{equation}
%----------------------------------------------------------------------------%
By Assumptions \ref{assum:F}(b) and \ref{assum:FL}(b), it follows that
%----------------------------------------------------------------------------%
\begin{equation}
\label{eq:lemma_V_01}
	\begin{array}{rcl}
		\left\Vert \dfrac{\mathbf{\Lambda }_{1}\left( \sum\nolimits_{t=1}^{T}%
			\mathbb{I}_{1t}\widehat{\xi }_{jt\left\vert T\right. }\mathbf{f}_{1t}\mathbf{%
				f}_{1t}^{\prime }\right) \mathbf{\Lambda }_{1}^{\prime }}{%
			N\sum\nolimits_{t=1}^{T}\widehat{\xi }_{jt\left\vert T\right. }}\right\Vert 
		& = & \left\Vert \dfrac{\mathbf{\Lambda }_{1}^{\prime }\mathbf{\Lambda }_{1}%
		}{N}\dfrac{T}{\sum\nolimits_{t=1}^{T}\widehat{\xi }_{jt\left\vert T\right. }%
		}\dfrac{\left( \sum\nolimits_{t=1}^{T}\mathbb{I}_{1t}\widehat{\xi }%
			_{jt\left\vert T\right. }\mathbf{f}_{1t}\mathbf{f}_{1t}^{\prime }\right) }{T}%
		\right\Vert  \\ 
		& \leq  & \dfrac{T}{\sum\nolimits_{t=1}^{T}\widehat{\xi }_{jt\left\vert
				T\right. }}\left\Vert \dfrac{\mathbf{\Lambda }_{1}^{\prime }\mathbf{\Lambda }%
			_{1}}{N}\right\Vert \left\Vert \dfrac{\sum\nolimits_{t=1}^{T}\mathbb{I}_{1t}%
			\widehat{\xi }_{jt\left\vert T\right. }\mathbf{f}_{1t}\mathbf{f}%
			_{1t}^{\prime }}{T}\right\Vert  \\ 
		& = & O_{p}\left( 1\right) .%
	\end{array}%
\end{equation}
%----------------------------------------------------------------------------%
In a similar way, it can be proved that
%----------------------------------------------------------------------------%
\begin{equation}
\label{eq:lemma_V_02}
	\left\Vert \dfrac{\mathbf{\Lambda }_{1}\left( \sum\nolimits_{t=1}^{T}%
		\mathbb{I}_{2t}\widehat{\xi }_{jt\left\vert T\right. }\mathbf{f}_{2t}\mathbf{%
			f}_{2t}^{\prime }\right) \mathbf{\Lambda }_{2}^{\prime }}{%
		N\sum\nolimits_{t=1}^{T}\widehat{\xi }_{jt\left\vert T\right. }}\right\Vert = O_{p}\left(1\right).
\end{equation}
%----------------------------------------------------------------------------%
Assumptions \ref{assum:FL}(b) implies that 
%----------------------------------------------------------------------------%
\begin{equation}
\label{eq:lemma_V_1}
	\begin{array}{rcl}
		\left\Vert \dfrac{\mathbf{\Lambda }_{1}\left( \sum\nolimits_{t=1}^{T}%
			\mathbb{I}_{1t}\widehat{\xi }_{jt\left\vert T\right. }\mathbf{f}_{1t}\mathbf{%
				e}_{t}^{\prime }\right) }{N\sum\nolimits_{t=1}^{T}\widehat{\xi }%
			_{jt\left\vert T\right. }}\right\Vert  & \leq  & \dfrac{1}{\sqrt{T}}\dfrac{T%
		}{\sum\nolimits_{t=1}^{T}\widehat{\xi }_{jt\left\vert T\right. }}\left\Vert 
		\dfrac{\mathbf{\Lambda }_{1}}{\sqrt{N}}\right\Vert \left\Vert \dfrac{%
			\sum\nolimits_{t=1}^{T}\mathbb{I}_{1t}\widehat{\xi }_{jt\left\vert T\right.
			}\mathbf{f}_{1t}\mathbf{e}_{t}^{\prime }}{\sqrt{NT}}\right\Vert  \\ 
		& = & \dfrac{1}{\sqrt{T}}\left\Vert \dfrac{\sum\nolimits_{t=1}^{T}\mathbb{I}%
			_{1t}\widehat{\xi }_{jt\left\vert T\right. }\mathbf{f}_{1t}\mathbf{e}%
			_{t}^{\prime }}{\sqrt{NT}}\right\Vert O_{p}\left( 1\right), 
	\end{array}%
\end{equation}
%----------------------------------------------------------------------------%
and, taking into account Assumption \ref{assum:WD},
%----------------------------------------------------------------------------%
\begin{equation}
	\label{eq:lemma_V_2}
	\begin{array}{rcl}
		\left\Vert \dfrac{\sum\nolimits_{t=1}^{T}\mathbb{I}_{1t}\widehat{\xi }%
			_{jt\left\vert T\right. }\mathbf{f}_{1t}\mathbf{e}_{t}^{\prime }}{\sqrt{NT}}%
		\right\Vert  & = & \left\{ \mathrm{tr}\left[ \left( \dfrac{%
			\sum\nolimits_{t=1}^{T}\mathbb{I}_{1t}\widehat{\xi }_{jt\left\vert T\right.
			}\mathbf{f}_{1t}\mathbf{e}_{t}^{\prime }}{\sqrt{NT}}\right) \left( \dfrac{%
			\sum\nolimits_{t=1}^{T}\mathbb{I}_{1t}\widehat{\xi }_{jt\left\vert T\right.
			}\mathbf{f}_{1t}\mathbf{e}_{t}^{\prime }}{\sqrt{NT}}\right) ^{\prime }\right]
		\right\} ^{1\left/ 2\right. } \\ 
		& = & \left\{ \mathrm{tr}\left[ \left( \dfrac{\sum\nolimits_{t=1}^{T}%
			\mathbb{I}_{1t}\widehat{\xi }_{jt\left\vert T\right. }\mathbf{f}_{1t}\mathbf{%
				e}_{t}^{\prime }}{\sqrt{NT}}\right) ^{\prime }\left( \dfrac{%
			\sum\nolimits_{t=1}^{T}\mathbb{I}_{1t}\widehat{\xi }_{jt\left\vert T\right.
			}\mathbf{f}_{1t}\mathbf{e}_{t}^{\prime }}{\sqrt{NT}}\right) \right] \right\}
		^{1\left/ 2\right. } \\ 
		& = & \left\{ \mathrm{tr}\left[ \left( \dfrac{\sum\nolimits_{t=1}^{T}%
			\mathbb{I}_{1t}\widehat{\xi }_{jt\left\vert T\right. }\mathbf{e}_{t}\mathbf{f}
			_{1t}^{\prime }}{\sqrt{NT}}\right) \left( \dfrac{%
			\sum\nolimits_{t=1}^{T}\mathbb{I}_{1t}\widehat{\xi }_{jt\left\vert T\right.
			}\mathbf{f}_{1t}\mathbf{e}_{t}^{\prime }}{\sqrt{NT}}\right) \right] \right\}
		^{1\left/ 2\right. } \\ 
		& = & \left\{ \mathrm{tr}\left[ 
		\begin{array}{c}
			\left( 
			\begin{array}{c}
				\dfrac{\sum\nolimits_{t=1}^{T}\mathbb{I}_{1t}\widehat{\xi }_{jt\left\vert
						T\right. }\mathbf{f}_{1t}^{\prime }e_{1t}}{\sqrt{NT}} \\ 
				\vdots  \\ 
				\dfrac{\sum\nolimits_{t=1}^{T}\mathbb{I}_{1t}\widehat{\xi }_{jt\left\vert
						T\right. }\mathbf{f}_{1t}^{\prime }e_{Nt}}{\sqrt{NT}}%
			\end{array}%
			\right)  \\ 
			\times \left( 
			\begin{array}{ccc}
				\dfrac{\sum\nolimits_{t=1}^{T}\mathbb{I}_{1t}\widehat{\xi }_{jt\left\vert
						T\right. }\mathbf{f}_{1t}e_{1t}}{\sqrt{NT}} & \cdots  & \dfrac{%
					\sum\nolimits_{t=1}^{T}\mathbb{I}_{1t}\widehat{\xi }_{jt\left\vert T\right.
					}\mathbf{f}_{1t}e_{Nt}}{\sqrt{NT}}%
			\end{array}%
			\right) 
		\end{array}%
		\right] \right\} ^{1\left/ 2\right. } \\ 
		& = & \left[ \sum\limits_{i=1}^{N}\left( \dfrac{\sum\nolimits_{t=1}^{T}%
			\mathbb{I}_{1t}\widehat{\xi }_{jt\left\vert T\right. }\mathbf{f}%
			_{1t}^{\prime }e_{it}}{\sqrt{NT}}\right) \left( \dfrac{\sum%
			\nolimits_{t=1}^{T}\mathbb{I}_{1t}\widehat{\xi }_{jt\left\vert T\right. }%
			\mathbf{f}_{it}e_{it}}{\sqrt{NT}}\right) \right] ^{1\left/ 2\right. } \\ 
		& = & \left[ \dfrac{1}{N}\sum\limits_{i=1}^{N}\left( \dfrac{%
			\sum\nolimits_{t=1}^{T}\mathbb{I}_{1t}\widehat{\xi }_{jt\left\vert T\right.
			}\mathbf{f}_{1t}^{\prime }e_{it}}{\sqrt{T}}\right) \left( \dfrac{%
			\sum\nolimits_{t=1}^{T}\mathbb{I}_{1t}\widehat{\xi }_{jt\left\vert T\right.
			}\mathbf{f}_{it}e_{it}}{\sqrt{T}}\right) \right] ^{1\left/ 2\right. } \\ 
		& = & \left[ \dfrac{1}{N}\sum\limits_{i=1}^{N}\left\Vert \dfrac{1}{\sqrt{T}}%
		\left( \sum\limits_{t=1}^{T}\mathbb{I}_{1t}\widehat{\xi }_{jt\left\vert
			T\right. }\mathbf{f}_{1t}e_{it}\right) \right\Vert ^{2}\right] ^{1\left/
			2\right. } \\ 
		& = & O_{p}\left( 1\right),
	\end{array}%
\end{equation}
%----------------------------------------------------------------------------%
and taking into account (\ref{eq:lemma_V_1}) and (\ref{eq:lemma_V_2}), 
%----------------------------------------------------------------------------% 
\begin{equation}
\label{eq:lemma_V_3}
	\left\Vert \dfrac{\mathbf{\Lambda }_{1}\left( \sum\nolimits_{t=1}^{T}\mathbb{I}_{1t}%
		\widehat{\xi }_{jt\left\vert T\right. }\mathbf{f}_{1t}\mathbf{e}_{t}^{\prime
		}\right) }{N\sum\nolimits_{t=1}^{T}\widehat{\xi }_{jt\left\vert T\right. }} \right\Vert
	=O_{p}\left( \dfrac{1}{\sqrt{T}}\right) .
\end{equation}
%----------------------------------------------------------------------------%
In a similar way, it can be proved that
%----------------------------------------------------------------------------%
\begin{equation}
\label{eq:lemma_V_4}
\left\Vert \dfrac{\mathbf{%
		\Lambda }_{2}\left( \sum\nolimits_{t=1}^{T}\mathbb{I}_{2t}\widehat{\xi }%
	_{jt\left\vert T\right. }\mathbf{f}_{2t}\mathbf{e}_{t}^{\prime }\right) }{%
	N\sum\nolimits_{t=1}^{T}\widehat{\xi }_{jt\left\vert T\right. }} \right\Vert = O_{p}\left( \dfrac{1}{\sqrt{T}}\right),
\end{equation}
%----------------------------------------------------------------------------%
\begin{equation}
\label{eq:lemma_V_5}
\left\Vert \dfrac{\left( \sum\nolimits_{t=1}^{T}\mathbb{I}_{1t}\widehat{\xi }%
	_{jt\left\vert T\right. }\mathbf{e}_{t}\mathbf{f}_{1t}^{\prime }\right) 
	\mathbf{\Lambda }_{1}^{\prime }}{N\sum\nolimits_{t=1}^{T}\widehat{\xi }%
	_{jt\left\vert T\right. }} \right\Vert = O_{p}\left( \dfrac{1}{\sqrt{T}}\right),
\end{equation}
%----------------------------------------------------------------------------%
and
%----------------------------------------------------------------------------%
\begin{equation}
\label{eq:lemma_V_6}
\left\Vert\dfrac{\left( \sum\nolimits_{t=1}^{T}\mathbb{I}%
	_{2t}\widehat{\xi }_{jt\left\vert T\right. }\mathbf{e}_{t}\mathbf{f}%
	_{2t}^{\prime }\right) \mathbf{\Lambda }_{2}^{\prime }}{N\sum%
	\nolimits_{t=1}^{T}\widehat{\xi }_{jt\left\vert T\right. }} \right\Vert = O_{p}\left( \dfrac{1}{\sqrt{T}}\right).
\end{equation}
%----------------------------------------------------------------------------%
Finally, by Assumption \ref{assum:TCSDH}(b),
%----------------------------------------------------------------------------%
\begin{equation}
\label{eq:lemma_V_7}
	\begin{array}{rcl}
		\left\Vert \dfrac{\sum\nolimits_{t=1}^{T}\widehat{\xi }_{jt\left\vert
				T\right. }\mathbf{e}_{t}\mathbf{e}_{t}^{\prime }}{N\sum\nolimits_{t=1}^{T}%
			\widehat{\xi }_{jt\left\vert T\right. }}\right\Vert  & \leq  & \dfrac{%
			\sum\nolimits_{t=1}^{T}\widehat{\xi }_{jt\left\vert T\right. }\left\Vert 
			\mathbf{e}_{t}\right\Vert \left\Vert \mathbf{e}_{t}\right\Vert }{%
			N\sum\nolimits_{t=1}^{T}\widehat{\xi }_{jt\left\vert T\right. }} \\ 
		& \leq  & \dfrac{\sum\nolimits_{t=1}^{T}\widehat{\xi }_{jt\left\vert
				T\right. }\left( N^{-1\left/ 2\right. }\left\Vert \mathbb{I}_{1t}\mathbf{e}%
			_{t}\right\Vert +N^{-1\left/ 2\right. }\left\Vert \mathbb{I}_{2t}\mathbf{e}%
			_{t}\right\Vert \right) ^{2}}{\sum\nolimits_{t=1}^{T}\widehat{\xi }%
			_{jt\left\vert T\right. }} \\ 
		& = & O_{p}\left( 1\right).
	\end{array}%
\end{equation}
%----------------------------------------------------------------------------%
By combining equations (\ref{eq:lemma_V_0}), (\ref{eq:lemma_V_01}), (\ref{eq:lemma_V_02}), (\ref{eq:lemma_V_3}), (\ref{eq:lemma_V_4}), (\ref{eq:lemma_V_5}), (\ref{eq:lemma_V_6}) and (\ref{eq:lemma_V_7}), it follows that
%----------------------------------------------------------------------------%
\begin{equation*}
\dfrac{\sum\nolimits_{t=1}^{T}\widehat{\xi }_{jt\left\vert T\right. }%
	\mathbf{x}_{t}\mathbf{x}_{t}^{\prime }}{N\sum\nolimits_{t=1}^{T}\widehat{%
		\xi }_{jt\left\vert T\right. }} = O_{p}\left(1\right),
\end{equation*}
%----------------------------------------------------------------------------%
which, taking into account (\ref{eq:lemma_V_00}), implies that
%----------------------------------------------------------------------------%
\begin{equation*}
\dfrac{\widehat{\mathbf{\Lambda }}_{\widehat{\xi},j}^{\left( p\right) \prime }}{%
	N}\dfrac{\sum\nolimits_{t=1}^{T}\widehat{\xi }_{jt\left\vert T\right. }%
	\mathbf{x}_{t}\mathbf{x}_{t}^{\prime }}{N\sum\nolimits_{t=1}^{T}\widehat{%
		\xi }_{jt\left\vert T\right. }}\widehat{\mathbf{\Lambda }}_{\widehat{\xi},j}^{\left(
	p\right) } = O_{p}\left(1\right).
\end{equation*}
%----------------------------------------------------------------------------%
The result stated in the lemma then follows directly from (\ref{eq:lemma_V_000}) and \eqref{eq:lemma_V_0000}.
\end{proof}

\begin{proof}[\textbf{Proof of Lemma \protect\ref{Lemma:O_p_1}}]
	
Let $\rho _{\widehat{\xi },jkil}=\sigma _{\widehat{\xi },jkil}\left/ \left(\sigma _{\widehat{\xi },jkii}\sigma _{\widehat{\xi },jkll}\right) ^{1\left/2\right. }\right. $ such that $\left\vert \rho _{\widehat{\xi },jkil}\right\vert \leq 1$. Since $\left\vert \sigma _{\widehat{\xi },jkii}\right\vert \leq M<\infty $ by Assumption \ref{assum:TCSDH}(c), then
%----------------------------------------------------------------------------% 
\begin{equation*}
	\begin{array}{rcl}
		\dfrac{1}{N}\sum\limits_{i=1}^{N}\sum\limits_{l=1}^{N}\sigma _{\widehat{%
				\xi },jkil}^{2} & = & \dfrac{1}{N}\sum\limits_{i=1}^{N}\sum%
		\limits_{l=1}^{N}\rho _{\widehat{\xi },jkil}^{2}\sigma _{\widehat{\xi },
			jkii}\sigma _{\widehat{\xi },jkll} \\ 
		& \leq  & MN^{-1}\sum\limits_{i=1}^{N}\sum\limits_{l=1}^{N}\left\vert
		\sigma _{\widehat{\xi },jkii}\sigma _{\widehat{\xi },jkll}\right\vert
		^{1\left/ 2\right. }\left\vert \rho _{\widehat{\xi },jkil}\right\vert  \\ 
		& = & MN^{-1}\sum\limits_{i=1}^{N}\sum\limits_{l=1}^{N}\left\vert \sigma _{\widehat{\xi },jkii}\right\vert  \\ 
		& \leq  & MT^{-1}\sum\limits_{t=1}^{T}\left[N^{-1}\sum\limits_{i=1}^{N}\sum\limits_{l=1}^{N} \left|\E\left( 
		\mathbb{I}_{jt}\widehat{\xi}_{kt\left\vert T\right. }e_{it}e_{lt}\right)\right| \right] \\
		& \leq  & M^{2},
	\end{array}%
\end{equation*}
%----------------------------------------------------------------------------%
by Assumption \ref{assum:TCSDH}(b), which completes the proof of the lemma.

\end{proof}

\subsection{Proof of Theorem \ref{th:asympt_dist}}

Given the specification in $\left(\ref{eq:model}\right)$, from Section \ref{section:ESSR} recall $\mathbf{B}_{1}=[\bm\Lambda _{1}\ \mathbf{0}]$ and $\mathbf{B}_{2}=[\mathbf{0}\ \bm\Lambda _{2}]$. Adding and subtracting terms, we have
%----------------------------------------------------------------------------%
\begin{equation}
	\label{eq:Th_1_0}
\begin{array}{rcl}
\mathbf{x}_{t} & = & \mathbb{I}_{1t}\mathbf{B}_{1}\mathbf{g}_{t}+\mathbb{I}%
_{2t}\mathbf{B}_{2}\mathbf{g}_{t}+\mathbf{e}_{t} \\ 
& = & \mathbb{I}_{1t}\mathbf{B}_{1}\mathbf{\widehat{H}\widehat{g}}_{t}+\mathbb{I}%
_{2t}\mathbf{B}_{2}\mathbf{\widehat{H}\widehat{g}}_{t}+\mathbb{I}_{1t}\mathbf{B}_{1}%
\mathbf{\widehat{H}}\left( \mathbf{\widehat{H}}^{-1}\mathbf{g}_{t}-\mathbf{\widehat{g}}%
_{t}\right) +\mathbb{I}_{2t}\mathbf{B}_{2}\mathbf{\widehat{H}}\left( \mathbf{%
	\widehat{H}}^{-1}\mathbf{g}_{t}-\mathbf{\widehat{g}}_{t}\right) +\mathbf{e}_{t},
\end{array}
\end{equation}
%----------------------------------------------------------------------------%
where $\mathbf{\widehat{H}}$ is defined in \eqref{eq:hat_H}, and $\mathbf{\widehat{g}}_{t}$ is the estimator for $\mathbf{g}_{t}$ given in \eqref{eq:ghatPC}. We focus upon $\mathbf{\wh{B}}_{1}=\left[\mathbf{\widehat{b}}_{11},\ldots,\mathbf{\widehat{b}}_{1N}\right]^{\prime}$ as an estimator for $\mathbf{B}_{1}=\left[\mathbf{b}_{11},\ldots,\mathbf{b}_{1N}\right]^{\prime}$: analogous arguments hold for $\mathbf{\widehat{B}}_{2}$. From \eqref{eq:B_j_hatmain}, and taking into account \eqref{eq:Th_1_0}, we have
%----------------------------------------------------------------------------%
\begin{equation*}
\begin{array}{rcl}
\mathbf{\widehat{B}}_{1} & = & \left( \sum\limits_{t=1}^{T}\widehat{\xi}%
_{1,t\left\vert T\right. }\mathbf{x}_{t}\mathbf{\widehat{g}}_{t}^{\prime }\right)
\left( \sum\limits_{t=1}^{T}\widehat{\xi}_{1,t\left\vert T\right. }\mathbf{\widehat{g}%
}_{t}\mathbf{\widehat{g}}_{t}^{\prime }\right) ^{-1} \\ 
& = & \left\{ \sum\limits_{t=1}^{T}\widehat{\xi}_{1,t\left\vert T\right. }\left[ 
\mathbb{I}_{1t}\mathbf{B}_{1}\mathbf{\widehat{H}\widehat{g}}_{t}+\mathbb{I}_{2t}%
\mathbf{B}_{2}\mathbf{\widehat{H}\widehat{g}}_{t}+\mathbb{I}_{1t}\mathbf{B}_{1}%
\mathbf{\widehat{H}}\left( \mathbf{\widehat{H}}^{-1}\mathbf{g}_{t}-\mathbf{\widehat{g}}%
_{t}\right) +\mathbb{I}_{2t}\mathbf{B}_{2}\mathbf{\widehat{H}}\left( \mathbf{%
	\widehat{H}}^{-1}\mathbf{g}_{t}-\mathbf{\widehat{g}}_{t}\right) +\mathbf{e}_{t}%
\right] \mathbf{\widehat{g}}_{t}^{\prime }\right\}  \\ 
&  & \times \left( \sum\limits_{t=1}^{T}\widehat{\xi}_{1,t\left\vert T\right. }%
\mathbf{\widehat{g}}_{t}\mathbf{\widehat{g}}_{t}^{\prime }\right) ^{-1} \\ 
& = & \mathbf{B}_{1}\mathbf{\widehat{H}}\left( \sum\limits_{t=1}^{T}\widehat{\xi}%
_{1,t\left\vert T\right. }\mathbb{I}_{1t}\mathbf{\widehat{g}}_{t}\mathbf{\widehat{g}}%
_{t}^{\prime }\right) \left( \sum\limits_{t=1}^{T}\widehat{\xi}_{1,t\left\vert
	T\right. }\mathbf{\widehat{g}}_{t}\mathbf{\widehat{g}}_{t}^{\prime }\right) ^{-1}+%
\mathbf{B}_{2}\mathbf{\widehat{H}}\left( \sum\limits_{t=1}^{T}\widehat{\xi}%
_{1,t\left\vert T\right. }\mathbb{I}_{2t}\mathbf{\widehat{g}}_{t}\mathbf{\widehat{g}}%
_{t}^{\prime }\right) \left( \sum\limits_{t=1}^{T}\widehat{\xi}_{1,t\left\vert
	T\right. }\mathbf{\widehat{g}}_{t}\mathbf{\widehat{g}}_{t}^{\prime }\right) ^{-1}
\\ 
&  & +\mathbf{B}_{1}\mathbf{\widehat{H}}\left[ \sum\limits_{t=1}^{T}\widehat{\xi}%
_{1,t\left\vert T\right. }\mathbb{I}_{1t}\left( \mathbf{\widehat{H}}^{-1}\mathbf{g%
}_{t}-\mathbf{\widehat{g}}_{t}\right) \mathbf{\widehat{g}}_{t}^{\prime }\right]
\left( \sum\limits_{t=1}^{T}\widehat{\xi}_{1,t\left\vert T\right. }\mathbf{\widehat{g}%
}_{t}\mathbf{\widehat{g}}_{t}^{\prime }\right) ^{-1} \\ 
&  & +\mathbf{B}_{2}\mathbf{\widehat{H}}\left[ \sum\limits_{t=1}^{T}\widehat{\xi}%
_{1,t\left\vert T\right. }\mathbb{I}_{2t}\left( \mathbf{\widehat{H}}^{-1}\mathbf{g%
}_{t}-\mathbf{\widehat{g}}_{t}\right) \mathbf{\widehat{g}}_{t}^{\prime }\right]
\left( \sum\limits_{t=1}^{T}\widehat{\xi}_{1,t\left\vert T\right. }\mathbf{\widehat{g}%
}_{t}\mathbf{\widehat{g}}_{t}^{\prime }\right) ^{-1} \\ 
&  & +\left( \sum\limits_{t=1}^{T}\widehat{\xi}_{1,t\left\vert T\right. }\mathbf{e%
}_{t}\mathbf{\widehat{g}}_{t}^{\prime }\right) \left( \sum\limits_{t=1}^{T}\widehat{%
	\xi}_{1,t\left\vert T\right. }\mathbf{\widehat{g}}_{t}\mathbf{\widehat{g}}%
_{t}^{\prime }\right) ^{-1}.%
\end{array}%
\end{equation*}%
%----------------------------------------------------------------------------%
Since $\mathbb{I}_{2t}=1-\mathbb{I}_{1t}$, and recalling the definition of $\mathbf{\widehat{I}}_{\mathbf{\widehat{\xi}}1}$ in \eqref{eq:hat_I_xi}, after some algebra we get
%----------------------------------------------------------------------------%
\begin{equation}
\label{eq:Th_1_1}
\begin{array}{rrl}
\sqrt{T}\left[ \mathbf{\widehat{B}}_{1}-\mathbf{B}_{1}\mathbf{\widehat{H}\widehat{I}}_{%
	\mathbf{\widehat{\xi}}1}-\mathbf{B}_{2}\mathbf{\widehat{H}}\left( \mathbf{I}-\mathbf{%
	\widehat{I}}_{\mathbf{\widehat{\xi}}1}\right) \right]  & = & \left( \dfrac{1}{\sqrt{T%
}}\sum\limits_{t=1}^{T}\widehat{\xi}_{1,t\left\vert T\right. }\mathbf{e}_{t}%
\mathbf{\widehat{g}}_{t}^{\prime }\right) \left( \dfrac{1}{T}\sum%
\limits_{t=1}^{T}\widehat{\xi}_{1,t\left\vert T\right. }\mathbf{\widehat{g}}_{t}%
\mathbf{\widehat{g}}_{t}^{\prime }\right) ^{-1} \\ 
&  & +\mathbf{B}_{1}\mathbf{\widehat{H}}\left[ \dfrac{1}{\sqrt{T}}%
\sum\limits_{t=1}^{T}\widehat{\xi}_{1,t\left\vert T\right. }\mathbb{I}_{1t}\left( 
\mathbf{\widehat{H}}^{-1}\mathbf{g}_{t}-\mathbf{\widehat{g}}_{t}\right) \mathbf{\widehat{%
		g}}_{t}^{\prime }\right]  \\ 
&  & \times \left( \dfrac{1}{T}\sum\limits_{t=1}^{T}\widehat{\xi}%
_{1,t\left\vert T\right. }\mathbf{\widehat{g}}_{t}\mathbf{\widehat{g}}_{t}^{\prime
}\right) ^{-1}\\
&  & +\mathbf{B}_{2}\mathbf{\widehat{H}}\left[ \dfrac{1}{\sqrt{T}}%
\sum\limits_{t=1}^{T}\widehat{\xi}_{1,t\left\vert T\right. }\mathbb{I}_{2t}\left( 
\mathbf{\widehat{H}}^{-1}\mathbf{g}_{t}-\mathbf{\widehat{g}}_{t}\right) \mathbf{\widehat{%
		g}}_{t}^{\prime }\right] \\
&  & \times \left( \dfrac{1}{T}\sum\limits_{t=1}^{T}\widehat{\xi}%
_{1,t\left\vert T\right. }\mathbf{\widehat{g}}_{t}\mathbf{\widehat{g}}_{t}^{\prime
}\right) ^{-1}	.%
\end{array}%
\end{equation}%
%----------------------------------------------------------------------------%
For $0<M<\infty$, and taking into account Lemma \ref{Lemma:avg}(b), for $j=1,2$ we have that,
%----------------------------------------------------------------------------%
\begin{equation}
\dfrac{1}{T}\sum\limits_{t=1}^{T}\widehat{\xi }%
_{1,t\left\vert T\right. }\mathbb{I}_{jt}\left( \mathbf{\widehat{H}}^{-1}%
\mathbf{g}_{t}-\mathbf{\widehat{g}}_{t}\right) \mathbf{\widehat{g}}%
_{t}^{\prime } 
\leq M \left[\dfrac{1}{T}\sum\limits_{t=1}^{T}\left( \mathbf{\widehat{H}}^{-1}%
\mathbf{g}_{t}-\mathbf{\widehat{g}}_{t}\right) \mathbf{\widehat{g}}%
_{t}^{\prime }\right]=O_{p}\left( \dfrac{1}{C_{NT}^{2}}\right) .
\label{eq:Th_1_2}
\end{equation}
%----------------------------------------------------------------------------%
From (\ref{eq:Th_1_1}) and (\ref{eq:Th_1_2}), and taking into account Assumption \ref{assum:rates}, it follows that
%----------------------------------------------------------------------------%
\begin{equation*}
\sqrt{T}\left[ \mathbf{\widehat{B}}_{1}-\mathbf{B}_{1}\mathbf{\widehat{H}\widehat{I}}_{%
	\mathbf{\widehat{\xi}}1}-\mathbf{B}_{2}\mathbf{\widehat{H}}\left( \mathbf{I}-\mathbf{%
	\widehat{I}}_{\mathbf{\widehat{\xi}}1}\right) \right] =\left( \dfrac{1}{\sqrt{T}}%
\sum\limits_{t=1}^{T}\widehat{\xi}_{1,t\left\vert T\right. }\mathbf{e}_{t}\mathbf{%
	\widehat{g}}_{t}^{\prime }\right) \left( \dfrac{1}{T}\sum\limits_{t=1}^{T}\widehat{%
	\xi}_{1,t\left\vert T\right. }\mathbf{\widehat{g}}_{t}\mathbf{\widehat{g}}%
_{t}^{\prime }\right) ^{-1}+o_{p}\left( 1\right) .
\end{equation*}
%----------------------------------------------------------------------------%
Since $\mathbf{\widehat{g}}_{t}=N^{-1}\mathbf{\widehat{A}}^{\prime }\mathbf{x}_{t}$ and $\mathbf{x}_{t}=\mathbb{I}_{1t}\mathbf{\Lambda }_{1}\mathbf{f}_{1t}+\mathbb{I}%
_{2t}\mathbf{\Lambda }_{2}\mathbf{f}_{2t}+\mathbf{e}_{t}$ then $\mathbf{\widehat{g}}_{t}=N^{-1}\left(\mathbb{I}_{1t}\mathbf{%
	\widehat{A}}^{\prime }\mathbf{\Lambda }_{1}\mathbf{f}_{1t}+\mathbb{I}_{2t}%
\mathbf{\widehat{A}}^{\prime }\mathbf{\Lambda }_{2}\mathbf{f}_{2t}+\mathbf{\widehat{A%
}}^{\prime }\mathbf{e}_{t}\right)$. After some algebra, we have
%----------------------------------------------------------------------------%
\begin{equation}
\label{eq:Th_1_4}
\begin{array}{cl}
& \sqrt{T}\left[ \mathbf{\widehat{B}}_{1}-\mathbf{B}_{1}\mathbf{\widehat{H}\widehat{I}}_{%
	\mathbf{\widehat{\xi}}1}-\mathbf{B}_{2}\mathbf{\widehat{H}}\left( \mathbf{I}-\mathbf{%
	\widehat{I}}_{\mathbf{\widehat{\xi}}1}\right) \right]  \\  
= & \left\{ \left( \dfrac{1}{\sqrt{T}}\sum\limits_{t=1}^{T}\mathbb{I}_{1t}%
\widehat{\xi}_{1,t\left\vert T\right. }\mathbf{e}_{t}\mathbf{f}_{1t}^{\prime
}\right) \dfrac{\mathbf{\Lambda }_{1}^{\prime }\mathbf{\widehat{A}}}{N}+\left( 
\dfrac{1}{\sqrt{T}}\sum\limits_{t=1}^{T}\mathbb{I}_{2t}\widehat{\xi}%
_{1,t\left\vert T\right. }\mathbf{e}_{t}\mathbf{f}_{2t}^{\prime }\right) 
\dfrac{\mathbf{\Lambda }_{2}^{\prime }\mathbf{\widehat{A}}}{N}+\left[ \dfrac{1}{%
	\sqrt{T}}\sum\limits_{t=1}^{T}\widehat{\xi}_{1,t\left\vert T\right. }\mathbf{e}%
_{t}\left( \mathbf{e}_{t}^{\prime }\dfrac{\mathbf{\widehat{A}}}{N}\right) \right]
\right\}  \\ 
& \times \left[ 
\begin{array}{l}
\dfrac{\mathbf{\widehat{A}}^{\prime }\mathbf{\Lambda }_{1}}{N}\left( \dfrac{1}{T}%
\sum\limits_{t=1}^{T}\mathbb{I}_{1t}\widehat{\xi}_{1,t\left\vert T\right. }%
\mathbf{f}_{1t}\mathbf{f}_{1t}^{\prime }\right) \dfrac{\mathbf{\Lambda }%
	_{1}^{\prime }\mathbf{\widehat{A}}}{N}+\dfrac{\mathbf{\widehat{A}}^{\prime }\mathbf{%
		\Lambda }_{2}}{N}\left( \dfrac{1}{T}\sum\limits_{t=1}^{T}\mathbb{I}_{2t}\widehat{%
	\xi}_{1,t\left\vert T\right. }\mathbf{f}_{2t}\mathbf{f}_{2t}^{\prime }\right) 
\dfrac{\mathbf{\Lambda }_{2}^{\prime }\mathbf{\widehat{A}}}{N} \\ 
+\dfrac{\mathbf{\widehat{A}}^{\prime }\mathbf{\Lambda }_{1}}{N}\left( \dfrac{1}{T%
}\sum\limits_{t=1}^{T}\mathbb{I}_{1t}\widehat{\xi}_{1,t\left\vert T\right. }%
\mathbf{f}_{1t}\mathbf{e}_{t}^{\prime }\right) \dfrac{\mathbf{\widehat{A}}}{N}+%
\dfrac{\mathbf{\widehat{A}}^{\prime }}{N}\left( \dfrac{1}{T}\sum\limits_{t=1}^{T}%
\mathbb{I}_{1t}\widehat{\xi}_{1,t\left\vert T\right. }\mathbf{e}_{t}\mathbf{f}%
_{1t}^{\prime }\right) \dfrac{\mathbf{\Lambda }_{1}^{\prime }\mathbf{\widehat{A}}%
}{N} \\ 
+\dfrac{\mathbf{\widehat{A}}^{\prime }\mathbf{\Lambda }_{2}}{N}\left( \dfrac{1}{T%
}\sum\limits_{t=1}^{T}\mathbb{I}_{2t}\widehat{\xi}_{1,t\left\vert T\right. }%
\mathbf{f}_{2t}\mathbf{e}_{t}^{\prime }\right) \dfrac{\mathbf{\widehat{A}}}{N}+%
\dfrac{\mathbf{\widehat{A}}^{\prime }}{N}\left( \dfrac{1}{T}\sum\limits_{t=1}^{T}%
\mathbb{I}_{2t}\widehat{\xi}_{1,t\left\vert T\right. }\mathbf{e}_{t}\mathbf{f}%
_{2t}^{\prime }\right) \dfrac{\mathbf{\Lambda }_{2}^{\prime }\mathbf{\widehat{A}}%
}{N} \\ 
+\dfrac{\mathbf{\widehat{A}}^{\prime }}{N}\left( \dfrac{1}{T}\sum%
\limits_{t=1}^{T}\widehat{\xi}_{1,t\left\vert T\right. }\mathbf{e}_{t}\mathbf{e}%
_{t}^{\prime }\right) \dfrac{\mathbf{\widehat{A}}}{N}%
\end{array}%
\right] ^{-1}+o_{p}\left(1\right).%
\end{array}%
\end{equation}%
%----------------------------------------------------------------------------%
By Lemma \ref{Lemma:a_i_hat_components}, and taking into account the identity in \eqref{eq:a__hat-a_i}, it follows that%
%----------------------------------------------------------------------------%
\begin{equation}
\label{eq:Th_1_5}
\mathbf{\widehat{A}}^{\prime }-\mathbf{\widehat{H}}^{\prime }\mathbf{A}^{\prime
}=O_{p}\left( \dfrac{1}{\sqrt{N}C_{NT}}\right) +O_{p}\left( \dfrac{1}{\sqrt{T%
	}C_{NT}}\right) +O_{p}\left( \dfrac{1}{\sqrt{T}}\right) ,
\end{equation}%
%----------------------------------------------------------------------------%
which implies that%
%----------------------------------------------------------------------------%
\begin{equation}
\label{eq:Th_1_6}
\mathbf{\widehat{A}}-\mathbf{A\widehat{H}}=O_{p}\left( \dfrac{1}{\sqrt{N}C_{NT}}%
\right) +O_{p}\left( \dfrac{1}{\sqrt{T}C_{NT}}\right) +O_{p}\left( \dfrac{1}{%
	\sqrt{T}}\right) .
\end{equation}%
%----------------------------------------------------------------------------%
From (\ref{eq:Th_1_4}) through (\ref{eq:Th_1_6}), it follows that%
%----------------------------------------------------------------------------%
\begin{equation*}
\begin{array}{rl}
& \sqrt{T}\left[ \mathbf{\widehat{b}}_{1i}-\mathbf{\widehat{I}}^{\prime}_{\mathbf{\widehat{\xi}}1}%
\mathbf{\widehat{H}}^{\prime }\mathbf{b}_{1i}-\left( \mathbf{I}-\mathbf{\widehat{I}}
_{\mathbf{\widehat{\xi}}1}\right)^{\prime} \mathbf{\widehat{H}}^{\prime }\mathbf{b}_{2i}%
\right]  \\ 
= & \left[ \dfrac{\mathbf{\widehat{A}}^{\prime }\mathbf{\Lambda }_{1}}{N}\left( 
\dfrac{1}{T}\sum\limits_{t=1}^{T}\mathbb{I}_{1t}\widehat{\xi}_{1,t\left\vert
	T\right. }\mathbf{f}_{1t}\mathbf{f}_{1t}^{\prime }\right) \dfrac{\mathbf{%
		\Lambda }_{1}^{\prime }\mathbf{\widehat{A}}}{N}+\dfrac{\mathbf{\widehat{A}}^{\prime }%
	\mathbf{\Lambda }_{2}}{N}\left( \dfrac{1}{T}\sum\limits_{t=1}^{T}\mathbb{I}%
_{2t}\widehat{\xi}_{1,t\left\vert T\right. }\mathbf{f}_{2t}\mathbf{f}%
_{2t}^{\prime }\right) \dfrac{\mathbf{\Lambda }_{2}^{\prime }\mathbf{\widehat{A}}%
}{N}\right] ^{-1} \\ 
& \times \left[ \dfrac{\mathbf{\widehat{A}}^{\prime }\mathbf{\Lambda }_{1}}{N}%
\left( \dfrac{1}{\sqrt{T}}\sum\limits_{t=1}^{T}\mathbb{I}_{1t}\widehat{\xi}%
_{1,t\left\vert T\right. }\mathbf{f}_{1t}e_{it}\right) +\dfrac{\mathbf{\widehat{A}%
	}^{\prime }\mathbf{\Lambda }_{2}}{N}\left( \dfrac{1}{\sqrt{T}}%
\sum\limits_{t=1}^{T}\mathbb{I}_{2t}\widehat{\xi}_{1,t\left\vert T\right. }%
\mathbf{f}_{2t}e_{it}\right) \right] +o_{p}\left( 1\right),
\end{array}%
\end{equation*}%
%----------------------------------------------------------------------------%
and the result stated in the theorem follows by Assumption \ref{assum:F} and Lemma \ref{Lemma:var_hat}, and by noting that, by Assumption \ref{assum:MCLT}(c), $\left( T^{-1/2}\sum_{t=1}^{T}\mathbb{I}_{1t}\widehat{\xi}%
_{1,t\left\vert T\right. }\mathbf{f}_{1t}e_{it}\right)$ and $\left( T^{-1/2}
\sum_{t=1}^{T}\mathbb{I}_{2t}\widehat{\xi}_{1,t\left\vert T\right. }%
\mathbf{f}_{2t}e_{it}\right)$ converge in distribution to two independent Normal random variables.

\subsection{Proof of Theorem \ref{th:asympt_dist_factors}}

Given the representation in (\ref{eq:state_space_mes}), we can write
%----------------------------------------------------------------------------%
\begin{equation*}
\mathbf{x}_{t} = \left( \mathbf{B}_{1}~\mathbf{B}_{2}\right) \left( 
\mathbf{\xi }_{t}\otimes \mathbf{g}_{t}\right) +\mathbf{e}_{t} = \left( \mathbf{B}_{1}~\mathbf{B}_{2}\right) \left(\xi _{1t}\mathbf{g}_{t}~\xi _{2t}\mathbf{g}_{t}\right)^{\prime}+\mathbf{e}_{t}.
\end{equation*}%
%----------------------------------------------------------------------------%
Recall also the estimators $\mathbf{\widehat{B}}_{1}$ and $\mathbf{\widehat{B}}_{2}$ defined according to (\ref{eq:B_j_hat}), with $\mathbf{\widehat{B}}_{j}\equiv\mathbf{\widehat{B}}_{j}^{\left(k^{*}+1\right)}$, where $k^{*}$ is the last iteration of the EM algorithm detailed in Section \ref{Appendix:Est}. The estimators $\widehat{\xi}_{1,t\left\vert T\right. }\mathbf{\wh{g}}_{t}$ and $\widehat{\xi}_{2,t\left\vert T\right. }\mathbf{\wh{g}}_{t}$ for $\xi_{1t}\mathbf{g}_{t}$ and $\xi_{2t}\mathbf{g}_{t}$, respectively, are obtained as
%----------------------------------------------------------------------------%
\begin{equation*}
\begin{array}{rcl}
\left( 
\begin{array}{c}
\widehat{\xi}_{1,t\left\vert T\right. }\mathbf{\wh{g}}_{t} \\ 
\widehat{\xi}_{2,t\left\vert T\right. }\mathbf{\wh{g}}_{t}
\end{array}%
\right)  & = & \left[ \left( \mathbf{\widehat{B}}_{1}~\mathbf{\widehat{B}}%
_{2}\right) ^{\prime }\left( \mathbf{\widehat{B}}_{1}~\mathbf{\widehat{B}}%
_{2}\right) \right] ^{-1}\left( \mathbf{\widehat{B}}_{1}~\mathbf{\widehat{B}}%
_{2}\right) ^{\prime }\mathbf{x}_{t} \\    
& = & \left( 
\begin{array}{cc}
\mathbf{\widehat{B}}_{1}^{\prime }\mathbf{\widehat{B}}_{1} & \mathbf{\widehat{B}}%
_{1}^{\prime }\mathbf{\widehat{B}}_{2} \\ 
\mathbf{\widehat{B}}_{2}^{\prime }\mathbf{\widehat{B}}_{1} & \mathbf{\widehat{B}}%
_{2}^{\prime }\mathbf{\widehat{B}}_{2}%
\end{array}%
\right) ^{-1}\left( 
\begin{array}{c}
\mathbf{\widehat{B}}_{1}^{\prime } \\ 
\mathbf{\widehat{B}}_{2}^{\prime }%
\end{array}%
\right) \left( \mathbf{B}_{1}~\mathbf{B}_{2}\right) \left( 
\begin{array}{c}
\xi_{1t }\mathbf{g}_{t} \\ 
\xi_{2t }\mathbf{g}_{t}
\end{array}%
\right)  \\ 
&  & +\left( 
\begin{array}{cc}
\mathbf{\widehat{B}}_{1}^{\prime }\mathbf{\widehat{B}}_{1} & \mathbf{\widehat{B}}%
_{1}^{\prime }\mathbf{\widehat{B}}_{2} \\ 
\mathbf{\widehat{B}}_{2}^{\prime }\mathbf{\widehat{B}}_{1} & \mathbf{\widehat{B}}%
_{2}^{\prime }\mathbf{\widehat{B}}_{2}%
\end{array}%
\right) ^{-1}\left( 
\begin{array}{c}
\mathbf{\widehat{B}}_{1}^{\prime } \\ 
\mathbf{\widehat{B}}_{2}^{\prime }%
\end{array}%
\right) \mathbf{e}_{t}.%
\end{array}%
\end{equation*}%
%----------------------------------------------------------------------------%
Adding and subtracting terms, it follows that
%----------------------------------------------------------------------------%
\begin{equation*}
\begin{array}{rcl}
\left( 
\begin{array}{c}
\widehat{\xi}_{1,t\left\vert T\right. }\mathbf{\wh{g}}_{t} \\ 
\widehat{\xi}_{2,t\left\vert T\right. }\mathbf{\wh{g}}_{t}
\end{array}%
\right)  & = & \left( 
\begin{array}{cc}
\mathbf{\widehat{B}}_{1}^{\prime }\mathbf{\widehat{B}}_{1} & \mathbf{%
	\widehat{B}}_{1}^{\prime }\mathbf{\widehat{B}}_{2} \\ 
\mathbf{\widehat{B}}_{2}^{\prime }\mathbf{\widehat{B}}_{1} & \mathbf{%
	\widehat{B}}_{2}^{\prime }\mathbf{\widehat{B}}_{2}%
\end{array}%
\right) ^{-1}\left( 
\begin{array}{c}
\mathbf{\widehat{B}}_{1}^{\prime } \\ 
\mathbf{\widehat{B}}_{2}^{\prime }%
\end{array}%
\right) \left( \mathbf{B}_{1}~\mathbf{B}_{2}\right) \left( 
\begin{array}{c}
\xi_{1t }\mathbf{g}_{t} \\ 
\xi_{2t }\mathbf{g}_{t}
\end{array}%
\right)  \\ 
&  & +\left( 
\begin{array}{cc}
\mathbf{\widehat{B}}_{1}^{\prime }\mathbf{\widehat{B}}_{1} & \mathbf{%
	\widehat{B}}_{1}^{\prime }\mathbf{\widehat{B}}_{2} \\ 
\mathbf{\widehat{B}}_{2}^{\prime }\mathbf{\widehat{B}}_{1} & \mathbf{%
	\widehat{B}}_{2}^{\prime }\mathbf{\widehat{B}}_{2}%
\end{array}%
\right) ^{-1}\left( 
\begin{array}{c}
\mathbf{\widehat{B}}_{1}^{\prime } \\ 
\mathbf{\widehat{B}}_{2}^{\prime }%
\end{array}%
\right) \left( \mathbf{\widehat{B}}_{1}~\mathbf{\widehat{B}}_{2}\right) 
\mathbf{\widehat{H}}_{\mathbf{\xi }}^{-1}\left( 
\begin{array}{c}
\xi_{1t }\mathbf{g}_{t} \\ 
\xi_{2t }\mathbf{g}_{t}
\end{array}%
\right)  \\ 
&  & -\left( 
\begin{array}{cc}
\mathbf{\widehat{B}}_{1}^{\prime }\mathbf{\widehat{B}}_{1} & \mathbf{%
	\widehat{B}}_{1}^{\prime }\mathbf{\widehat{B}}_{2} \\ 
\mathbf{\widehat{B}}_{2}^{\prime }\mathbf{\widehat{B}}_{1} & \mathbf{%
	\widehat{B}}_{2}^{\prime }\mathbf{\widehat{B}}_{2}%
\end{array}%
\right) ^{-1}\left( 
\begin{array}{c}
\mathbf{\widehat{B}}_{1}^{\prime } \\ 
\mathbf{\widehat{B}}_{2}^{\prime }%
\end{array}%
\right) \left( \mathbf{\widehat{B}}_{1}~\mathbf{\widehat{B}}_{2}\right) 
\mathbf{\widehat{H}}_{\mathbf{\xi }}^{-1}\left( 
\begin{array}{c}
\xi_{1t }\mathbf{g}_{t} \\ 
\xi_{2t }\mathbf{g}_{t}
\end{array}%
\right)  \\ 
&  & +\left( 
\begin{array}{cc}
\mathbf{\widehat{B}}_{1}^{\prime }\mathbf{\widehat{B}}_{1} & \mathbf{%
	\widehat{B}}_{1}^{\prime }\mathbf{\widehat{B}}_{2} \\ 
\mathbf{\widehat{B}}_{2}^{\prime }\mathbf{\widehat{B}}_{1} & \mathbf{%
	\widehat{B}}_{2}^{\prime }\mathbf{\widehat{B}}_{2}%
\end{array}%
\right) ^{-1}\left( 
\begin{array}{c}
\mathbf{\widehat{B}}_{1}^{\prime } \\ 
\mathbf{\widehat{B}}_{2}^{\prime }%
\end{array}%
\right) \mathbf{e}_{t} \\ 
&  & +\left( 
\begin{array}{cc}
\mathbf{\widehat{B}}_{1}^{\prime }\mathbf{\widehat{B}}_{1} & \mathbf{%
	\widehat{B}}_{1}^{\prime }\mathbf{\widehat{B}}_{2} \\ 
\mathbf{\widehat{B}}_{2}^{\prime }\mathbf{\widehat{B}}_{1} & \mathbf{%
	\widehat{B}}_{2}^{\prime }\mathbf{\widehat{B}}_{2}%
\end{array}%
\right) ^{-1}\mathbf{\widehat{H}}_{\mathbf{\xi }}^{^{\prime }}\left( 
\begin{array}{c}
\mathbf{B}_{1}^{\prime } \\ 
\mathbf{B}_{2}^{\prime }%
\end{array}%
\right) \mathbf{e}_{t} \\ 
&  & -\left( 
\begin{array}{cc}
\mathbf{\widehat{B}}_{1}^{\prime }\mathbf{\widehat{B}}_{1} & \mathbf{%
	\widehat{B}}_{1}^{\prime }\mathbf{\widehat{B}}_{2} \\ 
\mathbf{\widehat{B}}_{2}^{\prime }\mathbf{\widehat{B}}_{1} & \mathbf{%
	\widehat{B}}_{2}^{\prime }\mathbf{\widehat{B}}_{2}%
\end{array}%
\right) ^{-1}\mathbf{\widehat{H}}_{\mathbf{\xi }}^{^{\prime }}\left( 
\begin{array}{c}
\mathbf{B}_{1}^{\prime } \\ 
\mathbf{B}_{2}^{\prime }%
\end{array}%
\right) \mathbf{e}_{t},
\end{array}%
\end{equation*}%
%----------------------------------------------------------------------------%
or equivalently
%----------------------------------------------------------------------------%
\begin{equation}
\label{eq:factor_1}
\begin{array}{cl}
& \left[ \left( 
\begin{array}{c}
\widehat{\xi}_{1,t\left\vert T\right. }\mathbf{\wh{g}}_{t} \\ 
\widehat{\xi}_{2,t\left\vert T\right. }\mathbf{\wh{g}}_{t}
\end{array}%
\right) -\mathbf{\widehat{H}}_{\mathbf{\xi }}^{-1}\left( 
\begin{array}{c}
\xi_{1t }\mathbf{g}_{t} \\ 
\xi_{2t }\mathbf{g}_{t}
\end{array}%
\right) \right]  \\ 
= & \left[ N^{-1}\left( 
\begin{array}{cc}
\mathbf{\widehat{B}}_{1}^{\prime }\mathbf{\widehat{B}}_{1} & \mathbf{%
	\widehat{B}}_{1}^{\prime }\mathbf{\widehat{B}}_{2} \\ 
\mathbf{\widehat{B}}_{2}^{\prime }\mathbf{\widehat{B}}_{1} & \mathbf{%
	\widehat{B}}_{2}^{\prime }\mathbf{\widehat{B}}_{2}%
\end{array}%
\right) \right] ^{-1}\mathbf{\widehat{H}}_{\mathbf{\xi }}^{^{\prime }}\left[
N^{-1}\left( 
\begin{array}{c}
\mathbf{B}_{1}^{\prime } \\ 
\mathbf{B}_{2}^{\prime }%
\end{array}%
\right) \mathbf{e}_{t}\right]  \\ 
& +\left[ N^{-1}\left( 
\begin{array}{cc}
\mathbf{\widehat{B}}_{1}^{\prime }\mathbf{\widehat{B}}_{1} & \mathbf{%
	\widehat{B}}_{1}^{\prime }\mathbf{\widehat{B}}_{2} \\ 
\mathbf{\widehat{B}}_{2}^{\prime }\mathbf{\widehat{B}}_{1} & \mathbf{%
	\widehat{B}}_{2}^{\prime }\mathbf{\widehat{B}}_{2}%
\end{array}%
\right) \right] \left\{ N^{-1}\left( 
\begin{array}{c}
\mathbf{\widehat{B}}_{1}^{\prime } \\ 
\mathbf{\widehat{B}}_{2}^{\prime }%
\end{array}%
\right) \left[ \left( \mathbf{B}_{1}~\mathbf{B}_{2}\right) -\left( \mathbf{\widehat{B}}_{1}~\mathbf{\widehat{B}}%
_{2}\right) \mathbf{\widehat{H}}_{\mathbf{\xi }}^{-1}\right] \right\} \left( 
\begin{array}{c}
\xi_{1t }\mathbf{g}_{t} \\ 
\xi_{2t }\mathbf{g}_{t}
\end{array}%
\right)  \\ 
& +\left[ N^{-1}\left( 
\begin{array}{cc}
\mathbf{\widehat{B}}_{1}^{\prime }\mathbf{\widehat{B}}_{1} & \mathbf{%
	\widehat{B}}_{1}^{\prime }\mathbf{\widehat{B}}_{2} \\ 
\mathbf{\widehat{B}}_{2}^{\prime }\mathbf{\widehat{B}}_{1} & \mathbf{%
	\widehat{B}}_{2}^{\prime }\mathbf{\widehat{B}}_{2}%
\end{array}%
\right) \right] ^{-1}\left\{ N^{-1}\left[ \left( 
\begin{array}{c}
\mathbf{\widehat{B}}_{1}^{\prime } \\ 
\mathbf{\widehat{B}}_{2}^{\prime }%
\end{array}%
\right) -\mathbf{\widehat{H}}_{\mathbf{\xi }}^{^{\prime }}\left( 
\begin{array}{c}
\mathbf{B}_{1}^{\prime } \\ 
\mathbf{B}_{2}^{\prime }%
\end{array}%
\right) \right] \mathbf{e}_{t}\right\}.
\end{array}%
\end{equation}
%----------------------------------------------------------------------------%
Consider first
%----------------------------------------------------------------------------%
\begin{equation*}
\begin{array}{rl}
& \dfrac{1}{N}\left( 
\begin{array}{c}
\mathbf{\widehat{B}}_{1}^{\prime } \\ 
\mathbf{\widehat{B}}_{2}^{\prime }%
\end{array}%
\right) \left[ \left( \mathbf{B}_{1}~\mathbf{B}_{2}\right) -\left( \mathbf{%
	\widehat{B}}_{1}~\mathbf{\widehat{B}}_{2}\right) \mathbf{\widehat{H}}_{\mathbf{\xi }%
}^{-1}\right] \left( 
\begin{array}{c}
\xi_{1t }\mathbf{g}_{t} \\ 
\xi_{2t }\mathbf{g}_{t}
\end{array}%
\right)  \\ 
= & \dfrac{1}{N}\left( 
\begin{array}{c}
\mathbf{\widehat{B}}_{1}^{\prime } \\ 
\mathbf{\widehat{B}}_{2}^{\prime }%
\end{array}%
\right) \left[ \left( \mathbf{B}_{1}~\mathbf{B}_{2}\right) \mathbf{%
\widehat{H}}_{\mathbf{\xi }}-\left( \mathbf{\widehat{B}}_{1}~\mathbf{\widehat{B}}%
_{2}\right) \right] \mathbf{\widehat{H}}_{\mathbf{\xi }}^{-1}\left( 
\begin{array}{c}
\xi_{1t }\mathbf{g}_{t} \\ 
\xi_{2t }\mathbf{g}_{t}
\end{array}%
\right),
\end{array}%
\end{equation*}
%----------------------------------------------------------------------------%
so that from (\ref{eq:Th_1_1}) and (\ref{eq:Th_1_2}), and taking into account Assumption \ref{assum:FL}, it follows that
%----------------------------------------------------------------------------%
\begin{equation}
\label{eq:factor_2}
\begin{array}{rl}
& \left\Vert \dfrac{1}{N}\left( 
\begin{array}{c}
\mathbf{\widehat{B}}_{1}^{\prime } \\ 
\mathbf{\widehat{B}}_{2}^{\prime }%
\end{array}%
\right) \left[ \left( \mathbf{B}_{1}~\mathbf{B}_{2}\right) -\left( \mathbf{%
	\widehat{B}}_{1}~\mathbf{\widehat{B}}_{2}\right) \mathbf{\widehat{H}}_{\mathbf{\xi }%
}^{-1}\right] \left( 
\begin{array}{c}
\xi_{1t }\mathbf{g}_{t} \\ 
\xi_{2t }\mathbf{g}_{t}
\end{array}%
\right) \right\Vert  \\ 
\leq  & \left\Vert \dfrac{1}{\sqrt{N}}\left( 
\begin{array}{c}
\mathbf{\widehat{B}}_{1}^{\prime } \\ 
\mathbf{\widehat{B}}_{2}^{\prime }%
\end{array}%
\right) \right\Vert \left\Vert \dfrac{1}{\sqrt{N}} \left[ \left( \mathbf{B}_{1}~\mathbf{B}%
_{2}\right) \mathbf{\widehat{H}}_{\mathbf{\xi }}-\left( \mathbf{\widehat{B}}_{1}~%
\mathbf{\widehat{B}}_{2}\right) \right] \right\Vert \left\Vert \mathbf{\widehat{H}}_{%
	\mathbf{\xi }}\right\Vert \left\Vert \left( 
\begin{array}{c}
\xi_{1t }\mathbf{g}_{t} \\ 
\xi_{2t }\mathbf{g}_{t}
\end{array}%
\right) \right\Vert  \\ 
= & O_{p}\left( \dfrac{1}{\sqrt{NT}}\right) + O_{p}\left( \dfrac{1}{\sqrt{N}C^{2}_{NT}}\right).
\end{array}%
\end{equation}%
%----------------------------------------------------------------------------%
By (\ref{eq:Th_1_1}) and (\ref{eq:Th_1_2}), and taking into account Assumption \ref{assum:TCSDH}(b), we also have that,
%----------------------------------------------------------------------------%
\begin{equation}
\label{eq:factor_3}
\begin{array}{rcl}
\left\Vert \dfrac{1}{N}\left[ \left( 
\begin{array}{c}
\mathbf{\widehat{B}}_{1}^{\prime } \\ 
\mathbf{\widehat{B}}_{2}^{\prime }%
\end{array}%
\right) -\mathbf{\widehat{H}}_{\mathbf{\xi }}^{^{\prime }}\left( 
\begin{array}{c}
\mathbf{B}_{1}^{\prime } \\ 
\mathbf{B}_{2}^{\prime }%
\end{array}%
\right) \right] \mathbf{e}_{t}\right\Vert  & \leq  & \dfrac{\left\Vert 
	\mathbf{e}_{t}\right\Vert }{\sqrt{N}}\left\Vert \dfrac{1}{\sqrt{N}}\left[
\left( 
\begin{array}{c}
\mathbf{\widehat{B}}_{1}^{\prime } \\ 
\mathbf{\widehat{B}}_{2}^{\prime }%
\end{array}%
\right) -\mathbf{\widehat{H}}_{\mathbf{\xi }}^{^{\prime }}\left( 
\begin{array}{c}
\mathbf{B}_{1}^{\prime } \\ 
\mathbf{B}_{2}^{\prime }%
\end{array}%
\right) \right] \right\Vert  \\ 
& = & O_{p}\left( \dfrac{1}{\sqrt{NT}}\right) +O_{p}\left( \dfrac{1}{\sqrt{N}%
	C_{NT}^{2}}\right).
\end{array}%
\end{equation}%
%----------------------------------------------------------------------------%
Therefore, taking into account (\ref{eq:factor_1}), (\ref{eq:factor_2}) and (\ref{eq:factor_3}), and by Assumption \ref{assum:rates}, we have
%----------------------------------------------------------------------------%
\begin{equation*}
\sqrt{N}\left[ \left( 
\begin{array}{c}
\widehat{\xi}_{1,t\left\vert T\right. }\mathbf{\wh{g}}_{t} \\ 
\widehat{\xi}_{2,t\left\vert T\right. }\mathbf{\wh{g}}_{t}
\end{array}%
\right) -\mathbf{\widehat{H}}_{\mathbf{\xi }}^{-1}\left( 
\begin{array}{c}
\xi_{1t }\mathbf{g}_{t} \\ 
\xi_{2t }\mathbf{g}_{t}
\end{array}%
\right) \right] =\left( 
\begin{array}{cc}
\dfrac{\mathbf{\widehat{B}}_{1}^{\prime }\mathbf{\widehat{B}}_{1}}{N} & \dfrac{%
	\mathbf{\widehat{B}}_{1}^{\prime }\mathbf{\widehat{B}}_{2}}{N} \\ 
\dfrac{\mathbf{\widehat{B}}_{2}^{\prime }\mathbf{\widehat{B}}_{1}}{N} & \dfrac{%
	\mathbf{\widehat{B}}_{2}^{\prime }\mathbf{\widehat{B}}_{2}}{N}%
\end{array}%
\right) ^{-1}\dfrac{1}{\sqrt{N}}\mathbf{\widehat{H}}_{\mathbf{\xi }}\left( 
\begin{array}{c}
\mathbf{B}_{1}^{\prime } \\ 
\mathbf{B}_{2}^{\prime }%
\end{array}%
\right) \mathbf{e}_{t}+o_{p}\left(1\right) .
\end{equation*}%
%----------------------------------------------------------------------------%
Given $\mathbf{\widehat{H}}_{\mathbf{\xi }}$, recall $\mathbf{I}_{\mathbf{\xi }j}=\textrm{p}\lim\nolimits_{N,T\rightarrow \infty }\mathbf{\widehat{I}}_{\mathbf{\xi }j}$ for $j=1,2$, where $\mathbf{I}_{\mathbf{\xi }j}$ and $\mathbf{\widehat{I}}_{\mathbf{\xi }j}$ are defined in Lemma \ref{Lemma:Hat_I_xi} and in (\ref{eq:hat_I_xi}), respectively. Also, given $\mathbf{\widehat{H}}$ defined in (\ref{eq:hat_H}), we have $\mathbf{\widehat{H}}\overset{p}{\rightarrow }\mathbf{\Sigma }_{\mathbf{g}}%
\mathbf{QV}^{-1}=\mathbf{H}$, where $\mathbf{\Sigma }_{\mathbf{g}}=\textrm{p}\lim_{N,T\rightarrow \infty }\left(\mathbf{GG}/T\right)$
by Assumption (\ref{assum:F}), and $\mathbf{Q}=\textrm{p}\lim_{N,T\rightarrow \infty }\left(\mathbf{A}^{\prime }\mathbf{\widehat{A}}/N\right)$ by Lemma \ref{Lemma:var_hat}. By Theorem \ref{th:asympt_dist}, we then have $\left( 
\mathbf{\widehat{B}}_{1}~\mathbf{\widehat{B}}_{2}\right)^{\prime} \overset{p}{\rightarrow }\mathbf{H}_{\mathbf{\xi }}\left( \mathbf{B}_{1}~
\mathbf{B}_{2}\right)^{\prime} $. Therefore%
%----------------------------------------------------------------------------%
\begin{equation*}
\textrm{p}\lim_{N,T\rightarrow \infty }\left( 
\begin{array}{cc}
\dfrac{\mathbf{\widehat{B}}_{1}^{\prime }\mathbf{\widehat{B}}_{1}}{N} & \dfrac{%
	\mathbf{\widehat{B}}_{1}^{\prime }\mathbf{\widehat{B}}_{2}}{N} \\ 
\dfrac{\mathbf{\widehat{B}}_{2}^{\prime }\mathbf{\widehat{B}}_{1}}{N} & \dfrac{%
	\mathbf{\widehat{B}}_{1}^{\prime }\mathbf{\widehat{B}}_{2}}{N}%
\end{array}%
\right) =\mathbf{H}_{\mathbf{\xi }}\left( 
\begin{array}{cc}
\mathbf{\Sigma }_{\mathbf{B}1} & \mathbf{\Sigma }_{\mathbf{B}12} \\ 
\mathbf{\Sigma }_{\mathbf{B2}1} & \mathbf{\Sigma }_{\mathbf{B}2}%
\end{array}%
\right) \mathbf{H}_{\mathbf{\xi }}^{\prime },
\end{equation*}
%----------------------------------------------------------------------------%
where, by Assumption \ref{assum:FL}, $\left\Vert \left(\mathbf{B}_{j}^{\prime }\mathbf{B}_{j}/N\right) -\mathbf{\Sigma }_{\mathbf{B}j}\right\Vert \rightarrow 0$ and $\left\Vert \left(\mathbf{B}%
_{j}^{\prime }\mathbf{B}_{k}/N\right)-\mathbf{\Sigma }_{\mathbf{B}jk}\right\Vert
\rightarrow 0$, for $j,k=1,2$ with $j\neq k$ as $N\rightarrow\infty$ . The result stated in the theorem follows by noting that
%----------------------------------------------------------------------------%
\begin{equation*}
\dfrac{1}{\sqrt{N}}\left( 
\begin{array}{c}
\mathbf{B}_{1}^{\prime } \\ 
\mathbf{B}_{2}^{\prime }%
\end{array}%
\right) \mathbf{e}_{t}\overset{d}{\rightarrow }\mathcal{N}\left( \mathbf{0},%
\mathbf{\Sigma }_{\mathbf{Be}t}\right).
\end{equation*}
%----------------------------------------------------------------------------%
by Assumption \ref{assum:MCLT}(d), which concludes the proof.

\subsection{Proof of Theorem \ref{th:asympt_dist_fjt}}

Given $r_{1}=r_{2}$, consider $j=1$: analogous arguments hold for $j=2$. We can then partition the vector $\widehat{\mathbf{b}}_{1i}$ in \eqref{eq:b_1i_hat} as
%----------------------------------------------------------------------------%
\begin{equation*}
	\widehat{\mathbf{b}}_{1i}=\left( 
	\begin{array}{c}
		\widehat{\mathbf{b}}_{1i}^{\left( 1\right) } \\ 
		\widehat{\mathbf{b}}_{1i}^{\left( 2\right) }%
	\end{array}%
	\right).
\end{equation*}
%----------------------------------------------------------------------------%
In this way, \eqref{eq:b_1i_hat} itself may be written as
%----------------------------------------------------------------------------%
\begin{equation*}
	\begin{array}{rl}
		& \sqrt{T}\left\{ \widehat{\mathbf{b}}_{1i}^{\prime }-\bm{\lambda }%
		_{1i}^{\prime }\left[ \widehat{\mathbf{R}}_{1,11},\widehat{\mathbf{R}}_{1,12}%
		\right] -\bm{\lambda }_{2i}^{\prime }\left[ \left( \widehat{\mathbf{H}}%
		_{21}-\widehat{\mathbf{R}}_{1,21}\right) ,\left( \widehat{\mathbf{H}}_{22}-%
		\widehat{\mathbf{R}}_{1,22}\right) \right] \right\} \\ 
		= & \sqrt{T}\left\{ \left( \widehat{\mathbf{b}}_{1i}^{\left( 1\right) \prime
		},\widehat{\mathbf{b}}_{1i}^{\left( 2\right) \prime }\right) ^{\prime }-%
		\bm{\lambda }_{1i}^{\prime }\left[ \widehat{\mathbf{R}}_{1,11},\widehat{%
			\mathbf{R}}_{1,12}\right] -\bm{\lambda }_{2i}^{\prime }\left[ \left( 
		\widehat{\mathbf{H}}_{21}-\widehat{\mathbf{R}}_{1,21}\right) ,\left( 
		\widehat{\mathbf{H}}_{22}-\widehat{\mathbf{R}}_{1,22}\right) \right] \right\}
		\\ 
		= & \sqrt{T}\left\{ \widehat{\mathbf{b}}_{1i}^{\left( 1\right) \prime }-%
		\bm{\lambda }_{1i}^{\prime }\widehat{\mathbf{R}}_{1,11}-\bm{\lambda }%
		_{2i}^{\prime }\left( \widehat{\mathbf{H}}_{21}-\widehat{\mathbf{R}}%
		_{1,21}\right) ,\widehat{\mathbf{b}}_{1i}^{\left( 2\right) \prime }-\bm{%
			\lambda }_{1i}^{\prime }\widehat{\mathbf{R}}_{1,12}-\bm{\lambda }%
		_{2i}^{\prime }\left( \widehat{\mathbf{H}}_{22}-\widehat{\mathbf{R}}%
		_{1,22}\right) \right\} .%
	\end{array}%
\end{equation*}%
%----------------------------------------------------------------------------%
Since it is known that $r_{1}=r_{2}$, the estimator $\widehat{\bm{\lambda }}_{1i}$ for $\bm{\lambda }_{1i}$ is equal to $\widehat{\mathbf{b}}_{1i}^{\left( 1\right) }$.
Formally, for $i=1,\ldots ,N$, it follows that%
%----------------------------------------------------------------------------%
\begin{equation*}
	\sqrt{T}\left[ \widehat{\mathbf{b}}_{1i}^{\left( 1\right) \prime }-\bm{%
		\lambda }_{1i}^{\prime }\widehat{\mathbf{R}}_{1,11}-\bm{\lambda }%
	_{2i}^{\prime }\left( \widehat{\mathbf{H}}_{21}-\widehat{\mathbf{R}}%
	_{1,21}\right) \right] =\sqrt{T}\left[ \widehat{\bm{\lambda }}_{1i}^{\prime }-\bm{\lambda }_{1i}^{\prime }\widehat{\mathbf{R}}_{1,11}-%
	\bm{\lambda }_{2i}^{\prime }\left( \widehat{\mathbf{H}}_{21}-\widehat{%
		\mathbf{R}}_{1,21}\right) \right] .
\end{equation*}
%----------------------------------------------------------------------------%
Given $\widehat{\mathbf{\Lambda }}_{1}=\left(\widehat{\bm{\lambda }}_{11},\ldots ,\widehat{\bm{\lambda }}_{1N}\right) ^{\prime }$,
from (\ref{eq:fjhat}) interest lies in
%----------------------------------------------------------------------------%
\begin{equation}
	\begin{array}{rcl}
		\widehat{\mathbf{f}}_{1t} & = & \widehat{\xi }_{1,t\left\vert T\right.
		}\left( \widehat{\mathbf{\Lambda }}_{1}^{\prime }\widehat{\mathbf{\Lambda }}%
		_{1}\right) ^{-1}\left( \widehat{\mathbf{\Lambda }}_{1}^{\prime }\mathbf{x}%
		_{t}\right)  \\ 
		& = & \left( \widehat{\mathbf{\Lambda }}_{1}^{\prime }\widehat{\mathbf{%
				\Lambda }}_{1}\right) ^{-1}\left( \widehat{\mathbf{\Lambda }}_{1}^{\prime }%
		\widehat{\xi }_{1,t\left\vert T\right. }\mathbf{x}_{t}\right)  \\ 
		& = & \left( \widehat{\mathbf{\Lambda }}_{1}^{\prime }\widehat{\mathbf{%
				\Lambda }}_{1}\right) ^{-1}\left[ \widehat{\mathbf{\Lambda }}_{1}^{\prime }%
		\widehat{\xi }_{1,t\left\vert T\right. }\left( \mathbf{\Lambda }_{1}\mathbf{f%
		}_{1t}\mathbb{I}_{1t}+\mathbf{\Lambda }_{2}\mathbf{f}_{2t}\mathbb{I}_{2t}+%
		\mathbf{e}_{t}\right) \right]  \\ 
		& = & \left( \widehat{\mathbf{\Lambda }}_{1}^{\prime }\widehat{\mathbf{%
				\Lambda }}_{1}\right) ^{-1}\left( \widehat{\mathbf{\Lambda }}_{1}^{\prime }%
		\mathbf{\Lambda }_{1}\right) \left( \widehat{\xi }_{1,t\left\vert T\right. }%
		\mathbb{I}_{1t}\mathbf{f}_{1t}\right) +\left( \widehat{\mathbf{\Lambda }}%
		_{1}^{\prime }\widehat{\mathbf{\Lambda }}_{1}\right) ^{-1}\left( \widehat{%
			\mathbf{\Lambda }}_{1}^{\prime }\mathbf{\Lambda }_{2}\right) \left( \widehat{%
			\xi }_{1,t\left\vert T\right. }\mathbb{I}_{2t}\mathbf{f}_{2t}\right)  \\ 
		&  & +\left( \widehat{\mathbf{\Lambda }}_{1}^{\prime }\widehat{\mathbf{%
				\Lambda }}_{1}\right) ^{-1}\left( \widehat{\mathbf{\Lambda }}_{1}^{\prime }%
		\widehat{\xi }_{1,t\left\vert T\right. }\mathbf{e}_{t}\right)  \\ 
		& = & \left( \dfrac{\widehat{\mathbf{\Lambda }}_{1}^{\prime }\widehat{%
				\mathbf{\Lambda }}_{1}}{N}\right) ^{-1}\left( \dfrac{\widehat{\mathbf{%
					\Lambda }}_{1}^{\prime }\mathbf{\Lambda }_{1}}{N}\right) \left( \widehat{\xi 
		}_{1,t\left\vert T\right. }\mathbb{I}_{1t}\mathbf{f}_{1t}\right) +\left( 
		\dfrac{\widehat{\mathbf{\Lambda }}_{1}^{\prime }\widehat{\mathbf{\Lambda }}%
			_{1}}{N}\right) ^{-1}\left( \dfrac{\widehat{\mathbf{\Lambda }}_{1}^{\prime }%
			\mathbf{\Lambda }_{2}}{N}\right) \left( \widehat{\xi }_{1,t\left\vert
			T\right. }\mathbb{I}_{2t}\mathbf{f}_{2t}\right)  \\ 
		&  & +\left( \dfrac{\widehat{\mathbf{\Lambda }}_{1}^{\prime }\widehat{%
				\mathbf{\Lambda }}_{1}}{N}\right) ^{-1}\left( \dfrac{\widehat{\mathbf{%
					\Lambda }}_{1}^{\prime }\widehat{\xi }_{1,t\left\vert T\right. }\mathbf{e}%
			_{t}}{N}\right) .%
	\end{array}
	\label{eq:hat_f_1t}
\end{equation}
%----------------------------------------------------------------------------%
Adding and subtracting terms, we have
%----------------------------------------------------------------------------%
\begin{equation*}
	\widehat{\mathbf{\Lambda }}_{1}=\widehat{\mathbf{\Lambda }}_{1}-\mathbf{%
		\Lambda }_{1}\widehat{\mathbf{R}}_{1,11}-\mathbf{\Lambda }_{2}\left( 
	\widehat{\mathbf{H}}_{21}-\widehat{\mathbf{R}}_{1,21}\right) +\mathbf{%
		\Lambda }_{1}\widehat{\mathbf{R}}_{1,11}+\mathbf{\Lambda }_{2}\left( 
	\widehat{\mathbf{H}}_{21}-\widehat{\mathbf{R}}_{1,21}\right) ,
\end{equation*}
%----------------------------------------------------------------------------%
which implies that
%----------------------------------------------------------------------------%
\begin{equation*}
	\begin{array}{rcl}
		\dfrac{\widehat{\mathbf{\Lambda }}_{1}^{\prime }\mathbf{\Lambda }_{1}}{N} & =
		& \dfrac{\left[ \widehat{\mathbf{\Lambda }}_{1}-\mathbf{\Lambda }_{1}%
			\widehat{\mathbf{R}}_{1,11}-\mathbf{\Lambda }_{2}\left( \widehat{\mathbf{H}}%
			_{21}-\widehat{\mathbf{R}}_{1,21}\right) \right] ^{\prime }\mathbf{\Lambda }%
			_{1}}{N} \\ 
		& = & +\dfrac{\left[ \mathbf{\Lambda }_{1}\widehat{\mathbf{R}}_{1,11}+%
			\mathbf{\Lambda }_{2}\left( \widehat{\mathbf{H}}_{21}-\widehat{\mathbf{R}}%
			_{1,21}\right) \right] ^{\prime }\mathbf{\Lambda }_{1}}{N}.%
	\end{array}%
\end{equation*}
%----------------------------------------------------------------------------%
Note that$\left. \left[ \widehat{\mathbf{\Lambda }}_{1}-\mathbf{\Lambda }_{1}%
\widehat{\mathbf{R}}_{1,11}-\mathbf{\Lambda }_{2}\left( \widehat{\mathbf{H}}%
_{21}-\widehat{\mathbf{R}}_{1,21}\right) \right] \right/ N$ is of the same
order as $\left. \left( \widehat{\mathbf{A}}-\mathbf{A}\widehat{\mathbf{H}}%
\right) \right/ N$. Therefore, by (\ref{eq:Lemma_3.5})  it follows that
%----------------------------------------------------------------------------%
\begin{equation*}
	\dfrac{\left[ \widehat{\mathbf{\Lambda }}_{1}-\mathbf{\Lambda }_{1}\widehat{%
			\mathbf{R}}_{1,11}-\mathbf{\Lambda }_{2}\left( \widehat{\mathbf{H}}_{21}-%
		\widehat{\mathbf{R}}_{1,21}\right) \right] ^{\prime }\mathbf{\Lambda }_{1}}{N%
	}=O_{p}\left( \dfrac{1}{C_{NT}^{2}}\right) ,
\end{equation*}
%----------------------------------------------------------------------------%
so that
%----------------------------------------------------------------------------%
\begin{equation}
	\label{eq:hat_f_1t_1}
	\dfrac{\widehat{\mathbf{\Lambda }}_{1}^{\prime }\mathbf{\Lambda }_{1}}{N}=%
	\dfrac{\left[ \mathbf{\Lambda }_{1}\widehat{\mathbf{R}}_{1,11}+\mathbf{%
			\Lambda }_{2}\left( \widehat{\mathbf{H}}_{21}-\widehat{\mathbf{R}}%
		_{1,21}\right) \right] ^{\prime }\mathbf{\Lambda }_{1}}{N}+O_{p}\left( 
	\dfrac{1}{C_{NT}^{2}}\right) .
\end{equation}
%----------------------------------------------------------------------------%
Similarly,
%----------------------------------------------------------------------------%
\begin{equation}
	\label{eq:hat_f_1t_2}
	\dfrac{\widehat{\mathbf{\Lambda }}_{1}^{\prime }\mathbf{\Lambda }_{2}}{N}=%
	\dfrac{\left[ \mathbf{\Lambda }_{1}\widehat{\mathbf{R}}_{1,11}+\mathbf{%
			\Lambda }_{2}\left( \widehat{\mathbf{H}}_{21}-\widehat{\mathbf{R}}%
		_{1,21}\right) \right] ^{\prime }\mathbf{\Lambda }_{2}}{N}+O_{p}\left( 
	\dfrac{1}{C_{NT}^{2}}\right) .
\end{equation}
%----------------------------------------------------------------------------%
Also,
%----------------------------------------------------------------------------%
\begin{equation}
	\label{eq:hat_f_1t_3}
	\begin{array}{rcl}
		\dfrac{\widehat{\mathbf{\Lambda }}_{1}^{\prime }\widehat{\mathbf{\Lambda }}%
			_{1}}{N} & = & \left[ \dfrac{\widehat{\mathbf{\Lambda }}_{1}-\mathbf{\Lambda 
			}_{1}\widehat{\mathbf{R}}_{1,11}-\mathbf{\Lambda }_{2}\left( \widehat{%
				\mathbf{H}}_{21}-\widehat{\mathbf{R}}_{1,21}\right) +\mathbf{\Lambda }_{1}%
			\widehat{\mathbf{R}}_{1,11}+\mathbf{\Lambda }_{2}\left( \widehat{\mathbf{H}}%
			_{21}-\widehat{\mathbf{R}}_{1,21}\right) }{\sqrt{N}}\right] ^{\prime } \\ 
		&  & \times \left[ \dfrac{\widehat{\mathbf{\Lambda }}_{1}-\mathbf{\Lambda }%
			_{1}\widehat{\mathbf{R}}_{1,11}-\mathbf{\Lambda }_{2}\left( \widehat{\mathbf{%
					H}}_{21}-\widehat{\mathbf{R}}_{1,21}\right) +\mathbf{\Lambda }_{1}\widehat{%
				\mathbf{R}}_{1,11}+\mathbf{\Lambda }_{2}\left( \widehat{\mathbf{H}}_{21}-%
			\widehat{\mathbf{R}}_{1,21}\right) }{\sqrt{N}}\right]  \\ 
		& = & \left[ \dfrac{\mathbf{\Lambda }_{1}\widehat{\mathbf{R}}_{1,11}+\mathbf{%
				\Lambda }_{2}\left( \widehat{\mathbf{H}}_{21}-\widehat{\mathbf{R}}%
			_{1,21}\right) }{\sqrt{N}}+O_{p}\left( \dfrac{\sqrt{N}}{C_{NT}^{2}}\right) %
		\right] ^{\prime } \\ 
		&  & \times \left[ \dfrac{\mathbf{\Lambda }_{1}\widehat{\mathbf{R}}_{1,11}+%
			\mathbf{\Lambda }_{2}\left( \widehat{\mathbf{H}}_{21}-\widehat{\mathbf{R}}%
			_{1,21}\right) }{\sqrt{N}}+O_{p}\left( \dfrac{\sqrt{N}}{C_{NT}^{2}}\right) %
		\right]  \\ 
		& = & \dfrac{\left[ \mathbf{\Lambda }_{1}\widehat{\mathbf{R}}_{1,11}+\mathbf{%
				\Lambda }_{2}\left( \widehat{\mathbf{H}}_{21}-\widehat{\mathbf{R}}%
			_{1,21}\right) \right] ^{\prime }\left[ \mathbf{\Lambda }_{1}\widehat{%
				\mathbf{R}}_{1,11}+\mathbf{\Lambda }_{2}\left( \widehat{\mathbf{H}}_{21}-%
			\widehat{\mathbf{R}}_{1,21}\right) \right] }{N}+O_{p}\left( \dfrac{\sqrt{N}}{%
			C_{NT}^{2}}\right) .%
	\end{array}%
\end{equation}
%----------------------------------------------------------------------------%
Therefore, taking into account (\ref{eq:hat_f_1t}) through (\ref{eq:hat_f_1t_3}) we have
%----------------------------------------------------------------------------%
\begin{equation*}
	\begin{array}{rcl}
		\widehat{\mathbf{f}}_{1t} & = & \left\{ \dfrac{\left[ \mathbf{\Lambda }_{1}%
			\widehat{\mathbf{R}}_{1,11}+\mathbf{\Lambda }_{2}\left( \widehat{\mathbf{H}}%
			_{21}-\widehat{\mathbf{R}}_{1,21}\right) \right] ^{\prime }\left[ \mathbf{%
				\Lambda }_{1}\widehat{\mathbf{R}}_{1,11}+\mathbf{\Lambda }_{2}\left( 
			\widehat{\mathbf{H}}_{21}-\widehat{\mathbf{R}}_{1,21}\right) \right] }{N}%
		+O_{p}\left( \dfrac{\sqrt{N}}{C_{NT}^{2}}\right) \right\} ^{-1} \\ 
		&  & \times \left\{ \dfrac{\left[ \mathbf{\Lambda }_{1}\widehat{\mathbf{R}}%
			_{1,11}+\mathbf{\Lambda }_{2}\left( \widehat{\mathbf{H}}_{21}-\widehat{%
				\mathbf{R}}_{1,21}\right) \right] ^{\prime }\mathbf{\Lambda }_{1}}{N}%
		\right\} \left( \widehat{\xi }_{1,t\left\vert T\right. }\mathbb{I}_{1t}%
		\mathbf{f}_{1t}\right)  \\ 
		&  & +\left\{ \dfrac{\left[ \mathbf{\Lambda }_{1}\widehat{\mathbf{R}}_{1,11}+%
			\mathbf{\Lambda }_{2}\left( \widehat{\mathbf{H}}_{21}-\widehat{\mathbf{R}}%
			_{1,21}\right) \right] ^{\prime }\left[ \mathbf{\Lambda }_{1}\widehat{%
				\mathbf{R}}_{1,11}+\mathbf{\Lambda }_{2}\left( \widehat{\mathbf{H}}_{21}-%
			\widehat{\mathbf{R}}_{1,21}\right) \right] }{N}+O_{p}\left( \dfrac{\sqrt{N}}{%
			C_{NT}^{2}}\right) \right\} ^{-1} \\ 
		&  & \times \left\{ \dfrac{\left[ \mathbf{\Lambda }_{1}\widehat{\mathbf{R}}%
			_{1,11}+\mathbf{\Lambda }_{2}\left( \widehat{\mathbf{H}}_{21}-\widehat{%
				\mathbf{R}}_{1,21}\right) \right] ^{\prime }\mathbf{\Lambda }_{2}}{N}%
		\right\} \left( \widehat{\xi }_{1,t\left\vert T\right. }\mathbb{I}_{2t}%
		\mathbf{f}_{2t}\right)  \\ 
		&  & +\left\{ \dfrac{\left[ \mathbf{\Lambda }_{1}\widehat{\mathbf{R}}_{1,11}+%
			\mathbf{\Lambda }_{2}\left( \widehat{\mathbf{H}}_{21}-\widehat{\mathbf{R}}%
			_{1,21}\right) \right] ^{\prime }\left[ \mathbf{\Lambda }_{1}\widehat{%
				\mathbf{R}}_{1,11}+\mathbf{\Lambda }_{2}\left( \widehat{\mathbf{H}}_{21}-%
			\widehat{\mathbf{R}}_{1,21}\right) \right] }{N}+O_{p}\left( \dfrac{\sqrt{N}}{%
			C_{NT}^{2}}\right) \right\} ^{-1} \\ 
		&  & \times \left\{ \dfrac{\left[ \mathbf{\Lambda }_{1}\widehat{\mathbf{R}}%
			_{1,11}+\mathbf{\Lambda }_{2}\left( \widehat{\mathbf{H}}_{21}-\widehat{%
				\mathbf{R}}_{1,21}\right) \right] ^{\prime }\widehat{\xi }_{1,t\left\vert
				T\right. }\mathbf{e}_{t}}{N}\right\}  \\ 
		&  & +O_{p}\left( \dfrac{1}{C_{NT}^{2}}\right) .%
	\end{array}%
\end{equation*}
%----------------------------------------------------------------------------%
It follows that,
%----------------------------------------------------------------------------%
\begin{equation}
	\label{eq:hat_f_asym_1}
	\begin{array}{cl}
		& \sqrt{N}\left\{ \widehat{\mathbf{f}}_{1t}-\left\{ 
		\begin{array}{c}
			\left\{ \dfrac{\left[ \mathbf{\Lambda }_{1}\widehat{\mathbf{R}}_{1,11}+%
				\mathbf{\Lambda }_{2}\left( \widehat{\mathbf{H}}_{21}-\widehat{\mathbf{R}}%
				_{1,21}\right) \right] ^{\prime }\left[ \mathbf{\Lambda }_{1}\widehat{%
					\mathbf{R}}_{1,11}+\mathbf{\Lambda }_{2}\left( \widehat{\mathbf{H}}_{21}-%
				\widehat{\mathbf{R}}_{1,21}\right) \right] }{N}\right\} ^{-1} \\ 
			\times \dfrac{\left[ \mathbf{\Lambda }_{1}\widehat{\mathbf{R}}_{1,11}+%
				\mathbf{\Lambda }_{2}\left( \widehat{\mathbf{H}}_{21}-\widehat{\mathbf{R}}%
				_{1,21}\right) \right] ^{\prime }\widehat{\xi }_{1,t\left\vert T\right.
				}\left( \mathbb{I}_{1t}\mathbf{\Lambda }_{1}\mathbf{f}_{1t}+\mathbb{I}_{2t}%
				\mathbf{\Lambda }_{2}\mathbf{f}_{2t}\right) }{N}%
		\end{array}%
		\right\} \right\}  \\ 
		= & \widehat{\xi }_{1,t\left\vert T\right. }\dfrac{\left[ \mathbf{\Lambda }%
			_{1}\widehat{\mathbf{R}}_{1,11}+\mathbf{\Lambda }_{2}\left( \widehat{\mathbf{%
					H}}_{21}-\widehat{\mathbf{R}}_{1,21}\right) \right] ^{\prime }\mathbf{e}_{t}%
		}{\sqrt{N}}+O_{p}\left( \dfrac{\sqrt{N}}{C_{NT}^{2}}\right) .%
	\end{array}%
\end{equation}
%----------------------------------------------------------------------------%
Consider%
%----------------------------------------------------------------------------%
\begin{equation}
	\label{eq:hat_f_asym_2}
	\begin{array}{cl}
		& \widehat{\xi }_{1,t\left\vert T\right. }\dfrac{\left[ \mathbf{\Lambda }_{1}%
			\widehat{\mathbf{R}}_{1,11}+\mathbf{\Lambda }_{2}\left( \widehat{\mathbf{H}}%
			_{21}-\widehat{\mathbf{R}}_{1,21}\right) \right] ^{\prime }\mathbf{e}_{t}}{%
			\sqrt{N}} \\ 
		= & \widehat{\xi }_{1,t\left\vert T\right. }\left[ \widehat{\mathbf{R}}%
		_{1,11}^{\prime }\dfrac{1}{\sqrt{N}}\mathbf{\Lambda }_{1}^{\prime }\mathbf{e}%
		_{t}+\left( \widehat{\mathbf{H}}_{21}-\widehat{\mathbf{R}}_{1,21}\right)
		^{\prime }\dfrac{1}{\sqrt{N}}\mathbf{\Lambda }_{2}^{\prime }\mathbf{e}_{t}%
		\right]  \\ 
		= & \widehat{\xi }_{1,t\left\vert T\right. }\left[ \widehat{\mathbf{R}}%
		_{1,11}^{\prime }\dfrac{1}{\sqrt{N}}\sum\limits_{i=1}^{N}\bm{\lambda }%
		_{1i}e_{it}+\left( \widehat{\mathbf{H}}_{21}-\widehat{\mathbf{R}}%
		_{1,21}\right) ^{\prime }\dfrac{1}{\sqrt{N}}\sum\limits_{i=1}^{N}\bm{%
			\lambda }_{2i}e_{it}\right] .%
	\end{array}%
\end{equation}%
%----------------------------------------------------------------------------%
and let
%----------------------------------------------------------------------------%
\begin{equation}
	\label{eq:plim_xi}
	\xi _{1,t}^{\ast }=p\lim_{N,T\rightarrow \infty }\widehat{\xi }%
	_{1,t\left\vert T\right. }.
\end{equation}
%----------------------------------------------------------------------------%
Further, from \eqref{eq_R_k_H} recall that for $j=1,2$,
%----------------------------------------------------------------------------%
\begin{equation*}
	\widehat{\mathbf{R}}_{j}=\widehat{\mathbf{H}}\widehat{\mathbf{I}}_{\widehat{%
			\mathbf{\xi }}j}=\left( 
	\begin{array}{cc}
		\widehat{\mathbf{R}}_{j,11} & \widehat{\mathbf{R}}_{j,12} \\ 
		\widehat{\mathbf{R}}_{j,21} & \widehat{\mathbf{R}}_{j,22}%
	\end{array}%
	\right) ,
\end{equation*}
%----------------------------------------------------------------------------%
where $\widehat{\mathbf{H}}$ and $\widehat{\mathbf{I}}_{\widehat{\mathbf{\xi}}j}$ are defined in (\ref{eq:hat_H}) and (\ref{eq:hat_I_xi}), respectively. Taking into account (\ref{eq:Hlim}) and Lemma (\ref{Lemma:Hat_I_xi}), it follows that
%----------------------------------------------------------------------------%
\begin{equation*}
	p\lim_{N,T\rightarrow \infty }\widehat{\mathbf{R}}_{j}=\mathbf{H\cdot I}_{%
		\mathbf{\xi }j}=\mathbf{HH}^{-1}\left[ 
	\begin{array}{cc}
		\mathbb{I}\left( j=1\right) \mathbb{I}_{r_{1}} & \mathbf{0} \\ 
		\mathbf{0} & \mathbb{I}\left( j=2\right) \mathbb{I}_{r_{2}}%
	\end{array}%
	\right] \mathbf{H=}\left[ 
	\begin{array}{cc}
		\mathbb{I}\left( j=1\right) \mathbb{I}_{r_{1}} & \mathbf{0} \\ 
		\mathbf{0} & \mathbb{I}\left( j=2\right) \mathbb{I}_{r_{2}}%
	\end{array}%
	\right] \mathbf{H.}
\end{equation*}
%----------------------------------------------------------------------------%
Given (\ref{eq:Hlim}), from (\ref{eq:SG}) and (\ref{eq:QQQ}),
recall the definitions of $\mathbf{\Sigma }_{\mathbf{g}}$ and $\mathbf{Q}$, respectively. We then have%
%----------------------------------------------------------------------------%
\begin{equation*}
	\mathbf{H}=\mathbf{\Sigma }_{\mathbf{g}}\mathbf{QV}^{-1}=\mathbf{\Sigma }_{%
		\mathbf{g}}\left( \mathbf{\Sigma }_{\mathbf{g}}^{-1\left/ 2\right. }\mathbf{%
		\Psi V}^{1\left/ 2\right. }\right) \mathbf{V}^{-1}=\mathbf{\Sigma }_{\mathbf{%
			g}}^{1\left/ 2\right. }\mathbf{\Psi V}^{-1\left/ 2\right. },
\end{equation*}
%----------------------------------------------------------------------------%
which implies that
%----------------------------------------------------------------------------%
\begin{equation*}
	\begin{array}{rcl}
		\mathbf{H} & = & \mathbf{\Sigma }_{\mathbf{g}}^{1\left/ 2\right. }\mathbf{%
			\Psi V}^{-1\left/ 2\right. } \\ 
		& = & \left( 
		\begin{array}{cc}
			\mathbf{\Sigma }_{\mathbf{f}_{1}}^{1\left/ 2\right. } & \mathbf{0} \\ 
			\mathbf{0} & \mathbf{\Sigma }_{\mathbf{f}_{2}}^{1\left/ 2\right. }%
		\end{array}%
		\right) \left( 
		\begin{array}{cc}
			\mathbf{\Psi }_{11} & \mathbf{\Psi }_{12} \\ 
			\mathbf{\Psi }_{21} & \mathbf{\Psi }_{22}%
		\end{array}%
		\right) \left( 
		\begin{array}{cc}
			\mathbf{V}_{1}^{-1\left/ 2\right. } & \mathbf{0} \\ 
			\mathbf{0} & \mathbf{V}_{2}^{-1\left/ 2\right. }%
		\end{array}%
		\right)  \\ 
		& = & \left( 
		\begin{array}{cc}
			\mathbf{\Sigma }_{\mathbf{f}_{1}}^{1\left/ 2\right. }\mathbf{\Psi }_{11} & 
			\mathbf{\Sigma }_{\mathbf{f}_{1}}^{1\left/ 2\right. }\mathbf{\Psi }_{12} \\ 
			\mathbf{\Sigma }_{\mathbf{f}_{2}}^{1\left/ 2\right. }\mathbf{\Psi }_{21} & 
			\mathbf{\Sigma }_{\mathbf{f}_{2}}^{1\left/ 2\right. }\mathbf{\Psi }_{22}%
		\end{array}%
		\right) \left( 
		\begin{array}{cc}
			\mathbf{V}_{1}^{-1\left/ 2\right. } & \mathbf{0} \\ 
			\mathbf{0} & \mathbf{V}_{2}^{-1\left/ 2\right. }%
		\end{array}%
		\right)  \\ 
		& = & \left( 
		\begin{array}{cc}
			\mathbf{\Sigma }_{\mathbf{f}_{1}}^{1\left/ 2\right. }\mathbf{\Psi }_{11}%
			\mathbf{V}_{1}^{-1\left/ 2\right. } & \mathbf{\Sigma }_{\mathbf{f}%
				_{1}}^{1\left/ 2\right. }\mathbf{\Psi }_{12}\mathbf{V}_{2}^{-1\left/
				2\right. } \\ 
			\mathbf{\Sigma }_{\mathbf{f}_{2}}^{1\left/ 2\right. }\mathbf{\Psi }_{21}%
			\mathbf{V}_{1}^{-1\left/ 2\right. } & \mathbf{\Sigma }_{\mathbf{f}%
				_{2}}^{1\left/ 2\right. }\mathbf{\Psi }_{22}\mathbf{V}_{2}^{-1\left/
				2\right. }%
		\end{array}%
		\right)  \\ 
		& = & \left( 
		\begin{array}{cc}
			\mathbf{H}_{11} & \mathbf{H}_{12} \\ 
			\mathbf{H}_{21} & \mathbf{H}_{22}%
		\end{array}%
		\right) ,%
	\end{array}%
\end{equation*}
%----------------------------------------------------------------------------%
where $\mathbf{H}_{jk}=p\lim_{N,T\rightarrow \infty }\mathbf{\widehat{H}}_{jk}$. Therefore,
%----------------------------------------------------------------------------%
\begin{equation*}
	\begin{array}{rcl}
		p\lim\limits_{N,T\rightarrow \infty }\widehat{\mathbf{R}}_{j} & = & \left[ 
		\begin{array}{cc}
			\mathbb{I}\left( j=1\right) \mathbb{I}_{r_{1}} & \mathbf{0} \\ 
			\mathbf{0} & \mathbb{I}\left( j=2\right) \mathbb{I}_{r_{2}}%
		\end{array}%
		\right] \left( 
		\begin{array}{cc}
			\mathbf{\Sigma }_{\mathbf{f}_{1}}^{1\left/ 2\right. }\mathbf{\Psi }_{11}%
			\mathbf{V}_{1}^{-1\left/ 2\right. } & \mathbf{\Sigma }_{\mathbf{f}%
				_{1}}^{1\left/ 2\right. }\mathbf{\Psi }_{12}\mathbf{V}_{2}^{-1\left/
				2\right. } \\ 
			\mathbf{\Sigma }_{\mathbf{f}_{2}}^{1\left/ 2\right. }\mathbf{\Psi }_{21}%
			\mathbf{V}_{1}^{-1\left/ 2\right. } & \mathbf{\Sigma }_{\mathbf{f}%
				_{2}}^{1\left/ 2\right. }\mathbf{\Psi }_{22}\mathbf{V}_{2}^{-1\left/
				2\right. }%
		\end{array}%
		\right)  \\ 
		& = & \left[ 
		\begin{array}{cc}
			\mathbb{I}\left( j=1\right) \mathbf{\Sigma }_{\mathbf{f}_{1}}^{1\left/
				2\right. }\mathbf{\Psi }_{11}\mathbf{V}_{1}^{-1\left/ 2\right. } & \mathbb{I}%
			\left( j=1\right) \mathbf{\Sigma }_{\mathbf{f}_{1}}^{1\left/ 2\right. }%
			\mathbf{\Psi }_{12}\mathbf{V}_{2}^{-1\left/ 2\right. } \\ 
			\mathbb{I}\left( j=2\right) \mathbf{\Sigma }_{\mathbf{f}_{2}}^{1\left/
				2\right. }\mathbf{\Psi }_{21}\mathbf{V}_{1}^{-1\left/ 2\right. } & \mathbb{I}%
			\left( j=2\right) \mathbf{\Sigma }_{\mathbf{f}_{2}}^{1\left/ 2\right. }%
			\mathbf{\Psi }_{22}\mathbf{V}_{2}^{-1\left/ 2\right. }%
		\end{array}%
		\right].
	\end{array}%
\end{equation*}
%----------------------------------------------------------------------------%
Therefore, we have $\widehat{\mathbf{R}}_{1,11}=\mathbf{H}_{11}+o_{p}\left(1\right)$ and $\widehat{\mathbf{R}}_{1,21}=o_{p}\left( 1\right)$. Taking this into account in (\ref{eq:hat_f_asym_1}) and (\ref{eq:hat_f_asym_2}), and recalling (\ref{eq:plim_xi}), it follows that
%----------------------------------------------------------------------------%
\begin{equation*}
	\begin{array}{cl}
		& \sqrt{N}\left\{ \widehat{\mathbf{f}}_{1t}^{\left( 1\right) }-\left\{ 
		\begin{array}{c}
			\left[ \dfrac{\left( \mathbf{\Lambda }_{1}\widehat{\mathbf{H}}_{11}+\mathbf{%
					\Lambda }_{2}\widehat{\mathbf{H}}_{21}\right) ^{\prime }\left( \mathbf{%
					\Lambda }_{1}\widehat{\mathbf{H}}_{11}+\mathbf{\Lambda }_{2}\widehat{\mathbf{%
						H}}_{21}\right) }{N}\right] ^{-1} \\ 
			\times \dfrac{\left( \mathbf{\Lambda }_{1}\widehat{\mathbf{H}}_{11}+\mathbf{%
					\Lambda }_{2}\widehat{\mathbf{H}}_{21}\right) ^{\prime }\widehat{\xi }%
				_{1,t\left\vert T\right. }\left( \mathbb{I}_{1t}\mathbf{\Lambda }_{1}\mathbf{%
					f}_{1t}+\mathbb{I}_{2t}\mathbf{\Lambda }_{2}\mathbf{f}_{2t}\right) }{N}%
		\end{array}%
		\right\} \right\}  \\ 
		= & \xi _{1,t}^{\ast }\left( \mathbf{H}_{11}^{\prime }\dfrac{1}{\sqrt{N}}%
		\sum\limits_{i=1}^{N}\bm{\lambda }_{1i}e_{it}+\mathbf{H}_{21}^{\prime }%
		\dfrac{1}{\sqrt{N}}\sum\limits_{i=1}^{N}\bm{\lambda }_{2i}e_{it}\right)
		+o_{p}\left( 1\right) .%
	\end{array}%
\end{equation*}
%----------------------------------------------------------------------------%
By Assumption (\ref{assum:MCLT})(d), it follows that
%----------------------------------------------------------------------------%
\begin{equation*}
	\begin{array}{cl}
		& \sqrt{N}\left\{ \widehat{\mathbf{f}}_{1t}-\left\{ 
		\begin{array}{c}
			\left[ \dfrac{\left( \mathbf{\Lambda }_{1}\widehat{\mathbf{H}}_{11}+\mathbf{%
					\Lambda }_{2}\widehat{\mathbf{H}}_{21}\right) ^{\prime }\left( \mathbf{%
					\Lambda }_{1}\widehat{\mathbf{H}}_{11}+\mathbf{\Lambda }_{2}\widehat{\mathbf{%
						H}}_{21}\right) }{N}\right] ^{-1} \\ 
			\times \dfrac{\left( \mathbf{\Lambda }_{1}\widehat{\mathbf{H}}_{11}+\mathbf{%
					\Lambda }_{2}\widehat{\mathbf{H}}_{21}\right) ^{\prime }\widehat{\xi }%
				_{1,t\left\vert T\right. }\left( \mathbb{I}_{1t}\mathbf{\Lambda }_{1}\mathbf{%
					f}_{1t}+\mathbb{I}_{2t}\mathbf{\Lambda }_{2}\mathbf{f}_{2t}\right) }{N}%
		\end{array}%
		\right\} \right\}  \\ 
		\overset{d}{\rightarrow } & \mathcal{N}\left( \mathbf{0},\mathbf{\Sigma}_{\widehat{\mathbf{f}}%
			_{1t}}\right),
	\end{array}%
\end{equation*}
%----------------------------------------------------------------------------%
where%
\begin{equation*}
	\mathbf{\Sigma}_{\widehat{\mathbf{f}}_{1t}}=\left( \xi _{1,t}^{\ast }\right)
	^{2}\left( \mathbf{H}_{11}^{\prime }\mathbf{\Phi }_{1t}\mathbf{H}_{11}+%
	\mathbf{H}_{11}^{\prime }\mathbf{\Phi }_{12t}\mathbf{H}_{21}+\mathbf{H}%
	_{21}^{\prime }\mathbf{\Phi }_{12t}^{\prime }\mathbf{H}_{11}+\mathbf{H}%
	_{22}^{\prime }\mathbf{\Phi }_{2t}\mathbf{H}_{22}\right), 
\end{equation*}
%----------------------------------------------------------------------------%
with $\mathbf{\Phi }_{12t}$ defined in Assumption \ref{assum:MCLT}(d). This  which completes the proof of the theorem.

\subsection{Proof of Theorem \ref{th:n_factors_regimes}}

For $j=1,2$, consider the covariance matrix $\mathbf{\widehat{\Sigma}}_{\widehat{\xi},\mathbf{x}j}$ defined in (\ref{eq:sigma_hat_x_j}). By definition of eigenvectors and eigenvalues, it follows that $\mathbf{\widehat{\Sigma}}_{\widehat{\xi},\mathbf{x}j}\mathbf{\widehat{\Lambda}}_{\widehat{\xi},j}^{\left( p\right) }=\mathbf{\widehat{\Lambda}}_{\widehat{\xi},j}^{\left( p\right)}\mathbf{\widehat{V}}_{\widehat{\xi},j}^{\left( p\right) }$. 
Recall the matrix $\mathbf{\widehat{H}}_{\widehat{\xi},kj}^{\left(p\right) }$ defined according to (\ref{eq:H_hat_p}). We can then write
%----------------------------------------------------------------------------%
\begin{equation*}
	\mathbf{\widehat{\Lambda}}_{\widehat{\xi},j}^{\left( p\right) }\mathbf{\widehat{V}}_{\widehat{\xi},j}^{\left(
		p\right) }-\left( \mathbf{\Lambda }_{j}\mathbf{\widehat{H}}_{\widehat{\xi}%
		,jj}^{\left( p\right) }+\mathbf{\Lambda }_{k}\mathbf{\widehat{H}}_{\widehat{\xi}%
		,kj}^{\left( p\right) }\right) \mathbf{\widehat{V}}_{\widehat{\xi},j}^{\left( p\right) }=\mathbf{\widehat{\Sigma}}_{\widehat{\xi},\mathbf{x}j}\mathbf{\widehat{\Lambda}}_{\widehat{\xi},j}^{\left(p\right) }-\left( \mathbf{\Lambda}_{j}\mathbf{\widehat{H}}_{\widehat{\xi}%
		,jj}^{\left( p\right) }+\mathbf{\Lambda }_{k}\mathbf{\widehat{H}}_{\widehat{\xi},kj}^{\left( p\right) }\right) \mathbf{\widehat{V}}_{\widehat{\xi},j}^{\left( p\right) },
\end{equation*}
%----------------------------------------------------------------------------%
which implies that
%----------------------------------------------------------------------------%
\begin{equation*}
	\mathbf{\widehat{V}}_{\widehat{\xi},j}^{\left( p\right) }\mathbf{\widehat{\Lambda}}_{\widehat{\xi},j}^{\left(p\right) \prime }-\mathbf{\widehat{V}}_{\widehat{\xi},j}^{\left( p\right) }\left( \mathbf{\widehat{%
			H}}_{\widehat{\xi},jj}^{\left( p\right) \prime }\mathbf{\Lambda }_{j}^{\prime }+\mathbf{\widehat{H}}_{\widehat{\xi},kj}^{\left( p\right) \prime }\mathbf{\Lambda }_{k}^{\prime }\right) =\mathbf{\widehat{\Lambda}}_{\widehat{\xi},j}^{\left( p\right) \prime }%
	\mathbf{\widehat{\Sigma}}_{\widehat{\xi},\mathbf{x}j}-\mathbf{\widehat{V}}_{\widehat{\xi},j}^{\left( p\right)
	}\left( \mathbf{\widehat{H}}_{\widehat{\xi},jj}^{\left( p\right) \prime }\mathbf{%
		\Lambda }_{j}^{\prime }+\mathbf{\widehat{H}}_{\widehat{\xi},kj}^{\left( p\right)
		\prime }\mathbf{\Lambda }_{k}^{\prime }\right) .
\end{equation*}
%----------------------------------------------------------------------------%
Without loss of generality, set $j=1$: the case $j=2$ can be dealt with in a similar way. Since $\mathbf{x}_{t}=\mathbb{I}_{1t}\mathbf{\Lambda }_{1}\mathbf{f}_{1t}+\mathbb{I}%
_{2t}\mathbf{\Lambda }_{2}\mathbf{f}_{2t}+\mathbf{e}_{t}$, and $x_{it}=\mathbb{I}_{1t}\bm{\lambda }_{1i}^{\prime }\mathbf{f}_{1t}+%
\mathbb{I}_{2t}\bm{\lambda }_{2i}^{\prime }\mathbf{f}_{2t}+e_{it}$, we can write
%----------------------------------------------------------------------------%
\begin{equation*}
	\begin{array}{rl}
		& \mathbf{\widehat{V}}_{\widehat{\xi},1}^{\left( p\right) }\bm{\widehat{\lambda}}_{\widehat{\xi},1i}^{\left(
			p\right) }-\mathbf{\widehat{V}}_{\widehat{\xi},1}^{\left( p\right) }\left( \mathbf{\widehat{H}}_{%
			\widehat{\xi},11}^{\left( p\right) \prime }\bm{\lambda }_{1i}+\mathbf{\widehat{H}%
		}_{\widehat{\xi},21}^{\left( p\right) \prime }\bm{\lambda }_{2i}\right) \\ 
		= & \mathbf{\widehat{\Lambda}}_{\widehat{\xi},1}^{\left( p\right) \prime }\dfrac{%
			\sum\nolimits_{t=1}^{T}\mathbf{x}_{t}x_{it}}{N\sum\nolimits_{t=1}^{T}\widehat{\xi}_{1t\left\vert T\right. }}-\mathbf{\widehat{V}}_{\widehat{\xi},1}^{\left( p\right) }\left( 
		\mathbf{\widehat{H}}_{\widehat{\xi},11}^{\left( p\right) \prime }\bm{\lambda }%
		_{1i}+\mathbf{\widehat{H}}_{\widehat{\xi},21}^{\left( p\right) \prime }\bm{\lambda }_{2i}\right) \\ 
		= & \mathbf{\widehat{\Lambda}}_{\widehat{\xi},1}^{\left( p\right) \prime }\dfrac{%
			\sum\nolimits_{t=1}^{T}\widehat{\xi}_{1t\left\vert T\right. }\left( \mathbb{I}_{1t}\mathbf{\Lambda }_{1}\mathbf{f}_{1t}+\mathbb{I}_{2t}\mathbf{\Lambda }_{2}\mathbf{f}_{2t}+\mathbf{e}_{t}\right) \left( \mathbb{I}_{1t}\bm{\lambda }_{1i}^{\prime }\mathbf{f}_{1t}+\mathbb{I}_{2t}\bm{\lambda }%
			_{2i}^{\prime }\mathbf{f}_{2t}+e_{it}\right) }{N\sum\nolimits_{t=1}^{T}\widehat{\xi}_{1t\left\vert T\right. }} \\ 
		& -\dfrac{\mathbf{\widehat{\Lambda}}_{\widehat{\xi},1}^{\left( p\right) \prime }\mathbf{\Lambda }_{1}}{N}\dfrac{\mathbf{F}_{11}\mathbf{F}_{\widehat{\xi},11}^{\prime }}{%
			\sum\nolimits_{t=1}^{T}\widehat{\xi}_{1t\left\vert T\right. }}\bm{\lambda }_{1i}-\dfrac{\mathbf{\widehat{\Lambda}}_{\widehat{\xi},1}^{\left( p\right) \prime }\mathbf{%
				\Lambda }_{2}}{N}\dfrac{\mathbf{F}_{22}\mathbf{F}_{\widehat{\xi},12}^{\prime }}{\sum\nolimits_{t=1}^{T}\widehat{\xi}_{1t\left\vert T\right. }}\bm{\lambda }_{2i} \\ 
		= & \dfrac{\mathbf{\widehat{\Lambda}}_{\widehat{\xi},1}^{\left( p\right) \prime }\mathbf{\Lambda }_{1}}{N}\left( \dfrac{\sum\nolimits_{t=1}^{T}\widehat{\xi}_{1t\left\vert T\right. }\mathbb{I}_{1t}\mathbf{f}_{1t}\mathbf{f}%
			_{1t}^{\prime }}{\sum\nolimits_{t=1}^{T}\widehat{\xi}_{1t\left\vert T\right. }}\right) \bm{\lambda }_{1i}+\dfrac{\mathbf{\widehat{\Lambda}}_{\widehat{\xi},1}^{\left(
				p\right) \prime }\mathbf{\Lambda }_{2}}{N}\left( \dfrac{\sum%
			\nolimits_{t=1}^{T}\widehat{\xi}_{1t\left\vert T\right. }\mathbb{I}_{2t}\mathbf{f%
			}_{2t}\mathbf{f}_{2t}^{\prime }}{\sum\nolimits_{t=1}^{T}\widehat{\xi}%
			_{1t\left\vert T\right. }}\right) \bm{\lambda }_{2i} \\ 
		& +\dfrac{\mathbf{\widehat{\Lambda}}_{\widehat{\xi},1}^{\left( p\right) \prime }\mathbf{\Lambda}_{1}}{N}\dfrac{\sum\nolimits_{t=1}^{T}\widehat{\xi}_{1t\left\vert T\right. }%
			\mathbb{I}_{1t}\mathbf{f}_{1t}e_{it}}{\sum\nolimits_{t=1}^{T}\widehat{\xi}%
			_{1t\left\vert T\right. }}+\dfrac{\mathbf{\widehat{\Lambda}}_{\widehat{\xi},1}^{\left(
				p\right) \prime }\mathbf{\Lambda }_{2}}{N}\dfrac{\sum\nolimits_{t=1}^{T}%
			\widehat{\xi}_{1t\left\vert T\right. }\mathbb{I}_{2t}\mathbf{f}_{2t}e_{it}}{%
			\sum\nolimits_{t=1}^{T}\widehat{\xi}_{1t\left\vert T\right. }} \\ 
		& +\dfrac{\mathbf{\widehat{\Lambda}}_{\widehat{\xi},1}^{\left( p\right) \prime }}{N}\left( 
		\dfrac{\sum\nolimits_{t=1}^{T}\widehat{\xi}_{1t\left\vert T\right. }\mathbb{I}%
			_{1t}\mathbf{e}_{t}\mathbf{f}_{1t}^{\prime }}{\sum\nolimits_{t=1}^{T}\widehat{%
				\xi}_{1t\left\vert T\right. }}\right) \bm{\lambda }_{1i}+\dfrac{\mathbf{%
				\widehat{\Lambda}}_{\widehat{\xi},1}^{\left( p\right) \prime }}{N}\left( \dfrac{%
			\sum\nolimits_{t=1}^{T}\widehat{\xi}_{1t\left\vert T\right. }\mathbb{I}_{2t}%
			\mathbf{e}_{t}\mathbf{f}_{2t}^{\prime }}{\sum\nolimits_{t=1}^{T}\widehat{\xi}%
			_{1t\left\vert T\right. }}\right) \bm{\lambda }_{2i}+\dfrac{\mathbf{\widehat{%
					\Lambda}}_{\widehat{\xi},1}^{\left( p\right) \prime }}{N}\left( \dfrac{\sum%
			\nolimits_{t=1}^{T}\widehat{\xi}_{1t\left\vert T\right. }\mathbf{e}_{t}e_{it}}{%
			\sum\nolimits_{t=1}^{T}\widehat{\xi}_{1t\left\vert T\right. }}\right) \\ 
		& -\dfrac{\mathbf{\widehat{\Lambda}}_{\widehat{\xi},1}^{\left( p\right) \prime }\mathbf{%
				\Lambda }_{1}}{N}\dfrac{\mathbf{F}_{11}\mathbf{F}_{\widehat{\xi},11}^{\prime }}{%
			\sum\nolimits_{t=1}^{T}\widehat{\xi}_{1t\left\vert T\right. }}\bm{\lambda }%
		_{1i}-\dfrac{\mathbf{\widehat{\Lambda}}_{\widehat{\xi},1}^{\left( p\right) \prime }\mathbf{%
				\Lambda }_{2}}{N}\dfrac{\mathbf{F}_{22}\mathbf{F}_{\widehat{\xi},12}^{\prime }}{%
			\sum\nolimits_{t=1}^{T}\widehat{\xi}_{1t\left\vert T\right. }}\bm{\lambda }%
		_{2i} \\ 
		= & \dfrac{1}{NT}\sum\limits_{l=1}^{N}\sum\limits_{t=1}^{T}\widehat{\xi}%
		_{1t\left\vert T\right. }\bm{\widehat{\lambda}}_{\widehat{\xi},1l}^{\left( p\right) }%
		\E\left( e_{lt}e_{it}\right) +\dfrac{1}{NT}\sum\limits_{l=1}^{N}%
		\sum\limits_{t=1}^{T}\widehat{\xi}_{1t\left\vert T\right. }\bm{\widehat{\lambda}%
		}_{\widehat{\xi},1l}^{\left( p\right) }\left[ e_{lt}e_{it}-\E\left(
		e_{lt}e_{it}\right) \right] \\ 
		& +\dfrac{\mathbf{\widehat{\Lambda}}_{\widehat{\xi},1}^{\left( p\right) \prime }\mathbf{%
				\Lambda }_{1}}{N}\dfrac{\sum\nolimits_{t=1}^{T}\widehat{\xi}_{1t\left\vert
				T\right. }\mathbb{I}_{1t}\mathbf{f}_{1t}e_{it}}{T}\dfrac{T}{%
			\sum\nolimits_{t=1}^{T}\widehat{\xi}_{1t\left\vert T\right. }}+\dfrac{\mathbf{%
				\widehat{\Lambda}}_{\widehat{\xi},1}^{\left( p\right) \prime }\mathbf{\Lambda }_{2}}{N}\dfrac{%
			\sum\nolimits_{t=1}^{T}\widehat{\xi}_{1t\left\vert T\right. }\mathbb{I}_{2t}%
			\mathbf{f}_{2t}e_{it}}{T}\dfrac{T}{\sum\nolimits_{t=1}^{T}\widehat{\xi}%
			_{1t\left\vert T\right. }} \\ 
		& +\dfrac{\mathbf{\widehat{\Lambda}}_{\widehat{\xi},1}^{\left( p\right) \prime }}{N}\left( 
		\dfrac{\sum\nolimits_{t=1}^{T}\widehat{\xi}_{1t\left\vert T\right. }\mathbb{I}%
			_{1t}\mathbf{e}_{t}\mathbf{f}_{1t}^{\prime }}{T}\right) \bm{\lambda }%
		_{1i}\dfrac{T}{\sum\nolimits_{t=1}^{T}\widehat{\xi}_{1t\left\vert T\right. }}+%
		\dfrac{\mathbf{\widehat{\Lambda}}_{\widehat{\xi},1}^{\left( p\right) \prime }}{N}\left( \dfrac{%
			\sum\nolimits_{t=1}^{T}\widehat{\xi}_{1t\left\vert T\right. }\mathbb{I}_{2t}%
			\mathbf{e}_{t}\mathbf{f}_{2t}^{\prime }}{T}\right) \bm{\lambda }_{2i}%
		\dfrac{T}{\sum\nolimits_{t=1}^{T}\widehat{\xi}_{1t\left\vert T\right. }},
	\end{array}%
\end{equation*}
%----------------------------------------------------------------------------%
or equivalently
%----------------------------------------------------------------------------%
\begin{equation*}
	\begin{array}{rl}
		& \mathbf{\widehat{V}}_{\widehat{\xi},1}^{\left( p\right) }\left[ \bm{\widehat{\lambda}}%
		_{\widehat{\xi},1i}^{\left( p\right) }-\left( \mathbf{\widehat{H}}_{\widehat{\xi},11}^{\left(
			p\right) \prime }\bm{\lambda }_{1i}+\mathbf{\widehat{H}}_{\widehat{\xi}%
			,21}^{\left( p\right) \prime }\bm{\lambda }_{2i}\right) \right] \\ 
		= & \dfrac{1}{NT}\sum\limits_{l=1}^{N}\sum\limits_{t=1}^{T}\bm{\widehat{%
				\lambda}}_{\widehat{\xi},1l}^{\left( p\right) }\E\left( \mathbb{I}_{1t}\widehat{\xi}%
		_{1t\left\vert T\right. }e_{lt}e_{it}\right) +\dfrac{1}{NT}%
		\sum\limits_{l=1}^{N}\sum\limits_{t=1}^{T}\bm{\widehat{\lambda}}%
		_{\widehat{\xi},1l}^{\left( p\right) }\E\left( \mathbb{I}_{2t}\widehat{\xi}%
		_{1t\left\vert T\right. }e_{lt}e_{it}\right) \\ 
		& +\dfrac{1}{NT}\sum\limits_{l=1}^{N}\sum\limits_{t=1}^{T}\bm{\widehat{%
				\lambda}}_{\widehat{\xi},1l}^{\left( p\right) }\left[ \mathbb{I}_{1t}\widehat{\xi}%
		_{1t\left\vert T\right. }e_{lt}e_{it}-\E\left( \mathbb{I}_{1t}\widehat{%
			\xi}_{1t\left\vert T\right. }e_{lt}e_{it}\right) \right] \\ 
		& +\dfrac{1}{NT}\sum\limits_{l=1}^{N}\sum\limits_{t=1}^{T}\bm{\widehat{%
				\lambda}}_{\widehat{\xi},1l}^{\left( p\right) }\left[ \mathbb{I}_{2t}\widehat{\xi}%
		_{1t\left\vert T\right. }e_{lt}e_{it}-\E\left( \mathbb{I}_{2t}\widehat{%
			\xi}_{1t\left\vert T\right. }e_{lt}e_{it}\right) \right] \\ 
		& +\dfrac{\mathbf{\widehat{\Lambda}}_{\widehat{\xi},1}^{\left( p\right) \prime }\mathbf{\Lambda }_{1}}{N}\dfrac{\sum\nolimits_{t=1}^{T}\widehat{\xi}_{1t\left\vert
				T\right. }\mathbb{I}_{1t}\mathbf{f}_{1t}e_{it}}{T}\dfrac{T}{%
			\sum\nolimits_{t=1}^{T}\widehat{\xi}_{1t\left\vert T\right. }}+\dfrac{\mathbf{%
				\widehat{\Lambda}}_{\widehat{\xi},1}^{\left( p\right) \prime }\mathbf{\Lambda }_{2}}{N}\dfrac{%
			\sum\nolimits_{t=1}^{T}\widehat{\xi}_{1t\left\vert T\right. }\mathbb{I}_{2t}%
			\mathbf{f}_{2t}e_{it}}{T}\dfrac{T}{\sum\nolimits_{t=1}^{T}\widehat{\xi}%
			_{1t\left\vert T\right. }} \\ 
		& +\dfrac{\mathbf{\widehat{\Lambda}}_{\widehat{\xi},1}^{\left( p\right) \prime }}{N}\left( 
		\dfrac{\sum\nolimits_{t=1}^{T}\widehat{\xi}_{1t\left\vert T\right. }\mathbb{I}_{1t}\mathbf{e}_{t}\mathbf{f}_{1t}^{\prime }}{T}\right) \bm{\lambda }_{1i}\dfrac{T}{\sum\nolimits_{t=1}^{T}\widehat{\xi}_{1t\left\vert T\right. }}+%
		\dfrac{\mathbf{\widehat{\Lambda}}_{\widehat{\xi},1}^{\left( p\right) \prime }}{N}\left( \dfrac{%
			\sum\nolimits_{t=1}^{T}\widehat{\xi}_{1t\left\vert T\right. }\mathbb{I}_{2t}%
			\mathbf{e}_{t}\mathbf{f}_{2t}^{\prime }}{T}\right) \bm{\lambda }_{2i}%
		\dfrac{T}{\sum\nolimits_{t=1}^{T}\widehat{\xi}_{1t\left\vert T\right. }},
	\end{array}%
\end{equation*}
%----------------------------------------------------------------------------%
which is also equal to
%----------------------------------------------------------------------------%
\begin{equation*}
	\begin{array}{rrl}
		\mathbf{\widehat{V}}_{\widehat{\xi},1}^{\left( p\right) }\left[ \bm{\widehat{\lambda}}%
		_{\widehat{\xi},1i}^{\left( p\right) }-\left( \mathbf{\widehat{H}}_{\widehat{\xi},11}^{\left(
			p\right) \prime }\bm{\lambda }_{1i}+\mathbf{\widehat{H}}_{\widehat{\xi}%
			,21}^{\left( p\right) \prime }\bm{\lambda }_{2i}\right) \right] & = & 
		\dfrac{1}{N}\sum\limits_{l=1}^{N}\bm{\widehat{\lambda}}_{\widehat{\xi},1l}^{\left(
			p\right) }\left[ \dfrac{1}{T}\sum\limits_{t=1}^{T}\E\left( \mathbb{I%
		}_{1t}\widehat{\xi}_{1t\left\vert T\right. }e_{lt}e_{it}\right) \right] \\ 
		&  & +\dfrac{1}{N}\sum\limits_{l=1}^{N}\bm{\widehat{\lambda}}_{\widehat{\xi},1l}^{\left(
			p\right) }\left[ \dfrac{1}{T}\sum\limits_{t=1}^{T}\E\left( \mathbb{I%
		}_{2t}\widehat{\xi}_{1t\left\vert T\right. }e_{lt}e_{it}\right) \right] \\ 
		&  & +\dfrac{1}{N}\sum\limits_{l=1}^{N}\bm{\widehat{\lambda}}_{\widehat{\xi},1l}^{\left(
			p\right) }\left\{ \dfrac{1}{T}\sum\limits_{t=1}^{T}\left[ \mathbb{I}_{1t}%
		\widehat{\xi}_{1t\left\vert T\right. }e_{lt}e_{it}-\E\left( \mathbb{I}%
		_{1t}\widehat{\xi}_{1t\left\vert T\right. }e_{lt}e_{it}\right) \right] \right\}
		\\ 
		&  & +\dfrac{1}{N}\sum\limits_{l=1}^{N}\bm{\widehat{\lambda}}_{\widehat{\xi},1l}^{\left(
			p\right) }\left\{ \dfrac{1}{T}\sum\limits_{t=1}^{T}\left[ \mathbb{I}_{2t}\widehat{\xi}_{1t\left\vert T\right. }e_{lt}e_{it}-\E\left( \mathbb{I}%
		_{2t}\widehat{\xi}_{1t\left\vert T\right. }e_{lt}e_{it}\right) \right] \right\}
		\\ 
		&  & +\dfrac{1}{N}\sum\limits_{i=1}^{N}\bm{\widehat{\lambda}}_{\widehat{\xi},1l}^{\left(
			p\right) }\left( \dfrac{1}{T}\sum\limits_{t=1}^{T}\bm{\lambda }%
		_{1l}^{\prime }\widehat{\xi}_{1t\left\vert T\right. }\mathbb{I}_{1t}\mathbf{f}%
		_{1t}e_{it}\right) \dfrac{T}{\sum\nolimits_{t=1}^{T}\widehat{\xi}_{1t\left\vert
				T\right. }} \\ 
		&  & +\dfrac{1}{N}\sum\limits_{i=1}^{N}\bm{\widehat{\lambda}}_{\widehat{\xi},1l}^{\left(
			p\right) }\left( \dfrac{1}{T}\sum\limits_{t=1}^{T}\bm{\lambda }%
		_{2l}^{\prime }\widehat{\xi}_{1t\left\vert T\right. }\mathbb{I}_{2t}\mathbf{f}_{2t}e_{it}\right) \dfrac{T}{\sum\nolimits_{t=1}^{T}\widehat{\xi}_{1t\left\vert
				T\right. }} \\ 
		&  & +\dfrac{1}{N}\sum\limits_{i=1}^{N}\bm{\widehat{\lambda}}_{\widehat{\xi},1l}^{\left(
			p\right) }\left( \dfrac{1}{T}\sum\limits_{t=1}^{T}\bm{\lambda }%
		_{1i}^{\prime }\widehat{\xi}_{1t\left\vert T\right. }\mathbb{I}_{1t}\mathbf{f}_{1t}e_{lt}\right) \dfrac{T}{\sum\nolimits_{t=1}^{T}\widehat{\xi}_{1t\left\vert
				T\right. }} \\ 
		&  & +\dfrac{1}{N}\sum\limits_{i=1}^{N}\bm{\widehat{\lambda}}_{\widehat{\xi},1l}^{\left(
			p\right) }\left( \dfrac{1}{T}\sum\limits_{t=1}^{T}\bm{\lambda }%
		_{2i}^{\prime }\widehat{\xi}_{1t\left\vert T\right. }\mathbb{I}_{2t}\mathbf{f}%
		_{2t}e_{lt}\right) \dfrac{T}{\sum\nolimits_{t=1}^{T}\widehat{\xi}_{1t\left\vert
				T\right. }}.%
	\end{array}%
\end{equation*}
%----------------------------------------------------------------------------%
In general, for $j,k=1,2$ define
%----------------------------------------------------------------------------%
\begin{equation*}
	\begin{array}{rcl}
		\sigma _{\widehat{\xi},jkil}=\dfrac{1}{T}\sum\limits_{t=1}^{T}\E\left( 
		\mathbb{I}_{jt}\widehat{\xi}_{kt\left\vert T\right. }e_{it}e_{lt}\right), &  & 
		\chi _{\widehat{\xi},jkil}=\dfrac{1}{T}\sum\limits_{t=1}^{T}\left[ \mathbb{I}%
		_{jt}\widehat{\xi}_{kt\left\vert T\right. }e_{lt}e_{it}-\E\left( \mathbb{%
			I}_{jt}\widehat{\xi}_{kt\left\vert T\right. }e_{lt}e_{it}\right) \right], \\ 
		&  &  \\ 
		\varphi _{\widehat{\xi},jkil}=\dfrac{1}{T}\sum\limits_{t=1}^{T}\bm{\lambda }%
		_{ji}^{\prime }\mathbf{f}_{jt}\mathbb{I}_{jt}\widehat{\xi}_{kt\left\vert
			T\right. }e_{lt}, &  & \varphi _{\widehat{\xi},jkli}=\dfrac{1}{T}%
		\sum\limits_{t=1}^{T}\bm{\lambda }_{jl}^{\prime }\mathbf{f}_{jt}\mathbb{%
			I}_{jt}\widehat{\xi}_{kt\left\vert T\right. }e_{it}.
	\end{array}%
\end{equation*}
%----------------------------------------------------------------------------%
We can then write
%----------------------------------------------------------------------------%
\begin{equation*}
	\begin{array}{rl}
		& \mathbf{\widehat{V}}_{\widehat{\xi},1}^{\left( p\right) }\left[ \bm{\widehat{\lambda}}%
		_{\widehat{\xi},1i}^{\left( p\right) }-\left( \mathbf{\widehat{H}}_{\widehat{\xi},11}^{\left(
			p\right) \prime }\bm{\lambda }_{1i}+\mathbf{\widehat{H}}_{\widehat{\xi}%
			,21}^{\left( p\right) \prime }\bm{\lambda }_{2i}\right) \right] \\ 
		= & \dfrac{1}{N}\sum\limits_{l=1}^{N}\bm{\widehat{\lambda}}_{\widehat{\xi},1l}^{\left(
			p\right) }\sigma _{\widehat{\xi},11il}+\dfrac{1}{N}\sum\limits_{l=1}^{N}\bm{
			\widehat{\lambda}}_{\widehat{\xi},1l}^{\left( p\right) }\sigma _{\widehat{\xi},21il} \\ 
		& +\dfrac{1}{N}\sum\limits_{l=1}^{N}\bm{\widehat{\lambda}}_{\widehat{\xi},1l}^{\left(
			p\right) }\chi _{\widehat{\xi},11il}+\dfrac{1}{N}\sum\limits_{l=1}^{N}\bm{
			\widehat{\lambda}}_{\widehat{\xi},1l}^{\left( p\right) }\chi _{\widehat{\xi},21il} \\ 
		& +\left( \dfrac{1}{N}\sum\limits_{l=1}^{N}\bm{\widehat{\lambda}}%
		_{\widehat{\xi},1l}^{\left( p\right) }\varphi _{\widehat{\xi},11il}\right) \dfrac{T}{%
			\sum\nolimits_{t=1}^{T}\widehat{\xi}_{1t\left\vert T\right. }}+\left( \dfrac{1}{%
			N}\sum\limits_{l=1}^{N}\bm{\widehat{\lambda}}_{\widehat{\xi},1l}^{\left( p\right)
		}\varphi _{\widehat{\xi},21il}\right) \dfrac{T}{\sum\nolimits_{t=1}^{T}\widehat{\xi}%
			_{1t\left\vert T\right. }} \\ 
		& +\left( \dfrac{1}{N}\sum\limits_{l=1}^{N}\bm{\widehat{\lambda}}%
		_{\widehat{\xi},1l}^{\left( p\right) }\varphi _{\widehat{\xi},11li}\right) \dfrac{T}{%
			\sum\nolimits_{t=1}^{T}\widehat{\xi}_{1t\left\vert T\right. }}+\left( \dfrac{1}{N}\sum\limits_{l=1}^{N}\bm{\widehat{\lambda}}_{\widehat{\xi},1l}^{\left( p\right)
		}\varphi _{\widehat{\xi},21li}\right) \dfrac{T}{\sum\nolimits_{t=1}^{T}\widehat{\xi}%
			_{1t\left\vert T\right. }}.
	\end{array}%
\end{equation*}
%----------------------------------------------------------------------------%
For $j,k=1,2$ note that 
%----------------------------------------------------------------------------%
\begin{equation*}
	\left\Vert \mathbf{\widehat{V}}_{\widehat{\xi},j}^{\left( p\right) }\mathbf{\widehat{H}}_{\widehat{\xi},kj}^{\left( p\right) }\right\Vert \leq \left\Vert \dfrac{\mathbf{F}_{\widehat{\xi},kj}\mathbf{F}_{jj}^{\prime }}{\sum\nolimits_{t=1}^{T}\widehat{\xi}_{jt\left\vert T\right. }}\dfrac{\mathbf{\Lambda }_{j}^{\prime }\mathbf{\widehat{\Lambda}}_{\widehat{\xi},j}^{\left( p\right) }}{N}\right\Vert \leq \dfrac{T}{\sum\nolimits_{t=1}^{T}\widehat{\xi}_{jt\left\vert T\right. }}\left\Vert \dfrac{\mathbf{F}_{\widehat{\xi},kj}\mathbf{F}_{jj}^{\prime }}{T}\right\Vert \left\Vert 
	\dfrac{\mathbf{\Lambda }_{j}^{\prime }\mathbf{\Lambda }_{j}}{N}\right\Vert
	^{1\left/ 2\right. }\left\Vert \dfrac{\bm{\widehat{\Lambda}}_{\widehat{\xi},j}^{\left(
			p\right) \prime }\bm{\widehat{\Lambda}}_{\widehat{\xi},j}^{\left( p\right) }}{N}%
	\right\Vert ^{1\left/ 2\right. }=O_{p}\left( 1\right)
\end{equation*}
%----------------------------------------------------------------------------%
by Assumptions \ref{assum:F}(b) and \ref{assum:FL}(b). Since $\left\Vert \mathbf{\widehat{V}}_{j}^{\left( p\right) }\right\Vert =O_{p}\left( 1\right) $ by Lemma \ref{Lemma:V_hat_j},
then $\left\Vert \mathbf{\widehat{H}}_{\widehat{\xi},kj}^{\left( p\right) }\right\Vert =O_{p}\left( 1\right) $. It follows that
%----------------------------------------------------------------------------%
\begin{equation}
\label{eq:factor_regime_1}
	\dfrac{1}{N}\sum\limits_{i=1}^{N}\left\Vert \mathbf{\widehat{V}}_{\widehat{\xi},1}^{\left(
		p\right) }\left[ \bm{\widehat{\lambda}}_{\widehat{\xi},1i}^{\left( p\right) }-\left( 
	\mathbf{\widehat{H}}_{\widehat{\xi},11}^{\left( p\right) \prime }\bm{\lambda }%
	_{1i}+\mathbf{\widehat{H}}_{\widehat{\xi},21}^{\left( p\right) \prime }\bm{%
		\lambda }_{2i}\right) \right] \right\Vert ^{2}\leq 8\dfrac{1}{N}%
	\sum\limits_{i=1}^{N}\sum\limits_{j=1}^{2}\left( \widehat{\sigma}_{\widehat{\xi},
		j1i\cdot }+\widehat{\chi}_{\widehat{\xi},j1i\cdot }+\widehat{\varphi}_{\widehat{\xi},j1i\cdot
	}+\widehat{\varphi}_{\widehat{\xi},j1\cdot i }\right),
\end{equation}
%----------------------------------------------------------------------------%
where in general
%----------------------------------------------------------------------------%
\begin{equation*}
	\begin{array}{rcl}
		\widehat{\sigma}_{\widehat{\xi}jki\cdot }=\dfrac{1}{N^{2}}\left\Vert
		\sum\limits_{l=1}^{N}\bm{\widehat{\lambda}}_{\widehat{\xi},kl}^{\left( p\right) }\sigma _{\widehat{\xi},jkil}\right\Vert ^{2}, &  & \widehat{\chi}_{\widehat{\xi},jki\cdot }=\dfrac{1%
		}{N^{2}}\left\Vert \sum\limits_{l=1}^{N}\bm{\widehat{\lambda}}_{\widehat{\xi},kl}^{\left(
			p\right) }\chi _{\widehat{\xi},jkil}\right\Vert ^{2}, \\ 
		&  &  \\ 
		\widehat{\varphi}_{\widehat{\xi},jki\cdot }=\dfrac{1}{N^{2}}\left\Vert
		\sum\limits_{l=1}^{N}\bm{\widehat{\lambda}}_{\widehat{\xi},kl}^{\left( p\right) }\varphi
		_{\widehat{\xi},jkil}\right\Vert ^{2}, &  & \widehat{\varphi}_{\widehat{\xi}jk\cdot i}=%
		\dfrac{1}{N^{2}}\left\Vert \sum\limits_{l=1}^{N}\bm{\widehat{\lambda}}%
		_{\widehat{\xi},kl}^{\left( p\right) }\varphi _{\widehat{\xi},jkli}\right\Vert ^{2}.%
	\end{array}%
\end{equation*}
%----------------------------------------------------------------------------%
Starting from $\widehat{\sigma}_{\widehat{\xi},jki\cdot }$, consider
%----------------------------------------------------------------------------%
\begin{equation*}
	\left\Vert \sum\limits_{l=1}^{N}\bm{\widehat{\lambda}}_{\widehat{\xi},kl}^{\left(
		p\right) }\sigma _{\widehat{\xi},jkil}\right\Vert ^{2}\leq \left(
	\sum\limits_{l=1}^{N}\left\Vert \bm{\widehat{\lambda}}_{kl}^{\left(
		p\right) }\right\Vert ^{2}\right) \left( \sum\limits_{l=1}^{N}\sigma _{\widehat{\xi},jkil}^{2}\right)
\end{equation*}
%----------------------------------------------------------------------------%
and
%----------------------------------------------------------------------------%
\begin{equation}
\label{eq:factor_regime_2}
	\dfrac{1}{N}\sum\limits_{i=1}^{N}\sum\limits_{j=1}^{2}\widehat{\sigma}_{\widehat{\xi},jki\cdot }\leq \dfrac{1}{N}\sum\limits_{j=1}^{2}\left( \dfrac{1}{N}%
	\sum\limits_{l=1}^{N}\left\Vert \bm{\widehat{\lambda}}_{\widehat{\xi},kl}^{\left(
		p\right) }\right\Vert ^{2}\right) \left( \dfrac{1}{N}\sum\limits_{i=1}^{N}%
	\sum\limits_{l=1}^{N}\sigma _{\widehat{\xi},jkil}^{2}\right) =O_{p}\left( \dfrac{1}{N}\right)
\end{equation}
%----------------------------------------------------------------------------%
by Assumption \ref{assum:FL}(b) and Lemma \ref{Lemma:O_p_1}. As for $\widehat{\chi}_{\widehat{\xi}jki\cdot }$,
%----------------------------------------------------------------------------%
\begin{equation*}
	\begin{array}{rcl}
		\sum\limits_{i=1}^{N}\widehat{\chi}_{\widehat{\xi},jki\cdot } & = & \dfrac{1}{N^{2}}%
		\sum\limits_{i=1}^{N}\left\Vert \sum\limits_{l=1}^{N}\bm{\widehat{\lambda}}%
		_{\widehat{\xi},kl}^{\left( p\right) }\chi _{\widehat{\xi},jkil}\right\Vert ^{2} \\ 
		& = & \dfrac{1}{N^{2}}\sum\limits_{i=1}^{N}\sum\limits_{l=1}^{N}\sum%
		\limits_{q=1}^{N}\bm{\widehat{\lambda}}_{\widehat{\xi},kl}^{\left( p\right) \prime }%
		\bm{\widehat{\lambda}}_{\widehat{\xi},kq}^{\left( p\right) }\chi _{\widehat{\xi},jkil}\chi _{\widehat{\xi},jkiq} \\ 
		& \leq & \left[ \dfrac{1}{N^{2}}\sum\limits_{l=1}^{N}\sum\limits_{q=1}^{N}%
		\left( \bm{\widehat{\lambda}}_{\widehat{\xi},kl}^{\left( p\right) \prime }\bm{\widehat{\lambda}}_{\widehat{\xi},kq}^{\left( p\right) }\right) ^{2}\right] ^{1\left/ 2\right. }%
		\left[ \dfrac{1}{N^{2}}\sum\limits_{l=1}^{N}\sum\limits_{q=1}^{N}\left(
		\sum\limits_{i=1}^{N}\chi _{\widehat{\xi},jkil}\chi _{\widehat{\xi},jkiq}\right) ^{2}%
		\right] ^{1\left/ 2\right. } \\ 
		& \leq & \left( \dfrac{1}{N^{2}}\sum\limits_{l=1}^{N}\left\Vert \bm{		\widehat{\lambda}}_{\widehat{\xi},kl}^{\left( p\right) }\right\Vert ^{2}\right) \left[ \dfrac{1%
		}{N^{2}}\sum\limits_{l=1}^{N}\sum\limits_{q=1}^{N}\left(
		\sum\limits_{i=1}^{N}\chi _{\widehat{\xi},jkil}\chi _{\widehat{\xi},jkiq}\right) ^{2}%
		\right] ^{1\left/ 2\right. }%
	\end{array}%
\end{equation*}
%----------------------------------------------------------------------------%
where
%----------------------------------------------------------------------------%
\begin{equation*}
	\E\left[ \left( \sum\limits_{i=1}^{N}\chi _{\widehat{\xi},jkil}\chi _{\widehat{\xi},jkiq}\right) ^{2}\right] =\E\left(
	\sum\limits_{i=1}^{N}\sum\limits_{q=1}^{N}\chi _{\widehat{\xi},jkil}\chi _{\widehat{%
			\xi},jkiq}\chi _{\widehat{\xi},jkul}\chi _{\widehat{\xi},jkuq}\right) \leq N^{2}\cdot
	\max_{i,l}\E\left\vert \chi _{\widehat{\xi},jkil}\right\vert ^{4},
\end{equation*}
%----------------------------------------------------------------------------%
and since
%----------------------------------------------------------------------------%
\begin{equation*}
	\begin{array}{rcl}
		\E\left\vert \chi _{\widehat{\xi},jkil}\right\vert ^{4} & = & \E \left\vert \dfrac{1}{T}\sum\limits_{t=1}^{T}\left[ \mathbb{I}_{jt}\widehat{\xi}%
		_{kt\left\vert T\right. }e_{lt}e_{it}-\mathrm{E}\left( \mathbb{I}_{jt}\widehat{\xi}_{kt\left\vert T\right. }e_{lt}e_{it}\right) \right] \right\vert ^{4} \\ 
		& = & \dfrac{1}{T^{2}}\E\left\vert \dfrac{1}{\sqrt{T}}%
		\sum\limits_{t=1}^{T}\left[ \mathbb{I}_{jt}\widehat{\xi}_{kt\left\vert T\right.
		}e_{lt}e_{it}-\E\left( \mathbb{I}_{jt}\widehat{\xi}_{kt\left\vert
			T\right. }e_{lt}e_{it}\right) \right] \right\vert ^{4} \\ 
		& \leq & \dfrac{1}{T^{2}}M%
	\end{array}%
\end{equation*}
%----------------------------------------------------------------------------%
by Assumption \ref{assum:TCSDH}(c), and taking into account Assumption \ref{assum:FL}(b),
%----------------------------------------------------------------------------%
\begin{equation*}
	\sum\limits_{i=1}^{N}\widehat{\chi}_{\widehat{\xi},jki\cdot }\leq O_{p}\left(1\right) \cdot \sqrt{\dfrac{N^{2}}{T^{2}}}=O_{p}\left( \dfrac{N}{T}\right) ,
\end{equation*}
%----------------------------------------------------------------------------%
which implies that
%----------------------------------------------------------------------------%
\begin{equation}
\label{eq:factor_regime_3}
	\dfrac{1}{N}\sum\limits_{i=1}^{N}\sum\limits_{j=1}^{2}\widehat{\chi}_{\widehat{\xi},
		jki\cdot }=\dfrac{1}{N}O_{p}\left( \dfrac{N}{T}\right) =O_{p}\left( \dfrac{1%
	}{T}\right) .
\end{equation}
%----------------------------------------------------------------------------%
Further,
%----------------------------------------------------------------------------%
\begin{equation}
\label{eq:factor_regime_4}
\begin{array}{rcl}
	\widehat{\varphi }_{\widehat{\xi },jki\cdot } & = & \dfrac{1}{N^{2}}%
	\left\Vert \sum\limits_{l=1}^{N}\widehat{\bm{\lambda }}_{\widehat{\xi},kl}^{\left(
		p\right) }\varphi _{\widehat{\xi }jkil}\right\Vert ^{2} \\ 
	& = & \dfrac{1}{N^{2}}\left\Vert \sum\limits_{l=1}^{N}\widehat{\bm{%
			\lambda }}_{\widehat{\xi},kl}^{\left( p\right) }\left( \dfrac{1}{T}\sum\limits_{t=1}^{T}%
	\bm{\lambda }_{ji}^{\prime }\mathbf{f}_{jt}\mathbb{I}_{jt}\widehat{\xi }%
	_{kt\left\vert T\right. }e_{lt}\right) \right\Vert ^{2} \\ 
	& = & \dfrac{1}{N^{2}}\left\Vert \sum\limits_{l=1}^{N}\widehat{\bm{%
			\lambda }}_{\widehat{\xi},kl}^{\left( p\right) }\bm{\lambda }_{ji}^{\prime }\left( 
	\dfrac{1}{T}\sum\limits_{t=1}^{T}\mathbf{f}_{jt}\mathbb{I}_{jt}\widehat{\xi 
	}_{kt\left\vert T\right. }e_{lt}\right) \right\Vert ^{2} \\ 
	& \leq & \dfrac{1}{T}\left\Vert \bm{\lambda }_{ji}\right\Vert ^{2}\left( 
	\dfrac{1}{N}\sum\limits_{l=1}^{N}\left\Vert \widehat{\bm{\lambda }}%
	_{\widehat{\xi},kl}^{\left( p\right) }\right\Vert ^{2}\right) \left(\dfrac{1}{N}%
	\sum\limits_{l=1}^{N}\left\Vert \dfrac{1}{\sqrt{T}}\sum\limits_{t=1}^{T}\mathbf{f}_{jt}\mathbb{I}_{jt}\widehat{\xi }%
	_{kt\left\vert T\right. }e_{lt} \right\Vert ^{2}\right)  \\ 
	& = & O_{p}\left( \dfrac{1}{T}\right) 
\end{array}%
\end{equation}
%----------------------------------------------------------------------------%
by Assumptions \ref{assum:FL}(a), \ref{assum:FL}(b) and \ref{assum:WD}. Finally,
%----------------------------------------------------------------------------%
\begin{equation}
\label{eq:factor_regime_5}
	\begin{array}{rcl}
		\widehat{\varphi }_{\widehat{\xi },jk\cdot i} & = & \dfrac{1}{N^{2}}%
		\left\Vert \sum\limits_{l=1}^{N}\widehat{\bm{\lambda }}_{\widehat{\xi},kl}^{\left(p\right) }\varphi _{\widehat{\xi },jkli}\right\Vert ^{2} \\ 
		& = & \dfrac{1}{N^{2}}\left\Vert \sum\limits_{l=1}^{N}\widehat{\bm{\lambda }}_{\widehat{\xi},kl}^{\left( p\right) }\left( \dfrac{1}{T}\sum\limits_{t=1}^{T}%
		\bm{\lambda }_{jl}^{\prime }\mathbf{f}_{jt}\mathbb{I}_{jt}\widehat{\xi }%
		_{kt\left\vert T\right. }e_{it}\right) \right\Vert ^{2} \\ 
		& = & \dfrac{1}{N^{2}}\left\Vert \sum\limits_{l=1}^{N}\widehat{\bm{%
		\lambda }}_{\widehat{\xi},kl}^{\left( p\right) }\bm{\lambda }_{jl}^{\prime }\left( 
		\dfrac{1}{T}\sum\limits_{t=1}^{T}\mathbf{f}_{jt}\mathbb{I}_{jt}\widehat{\xi 
		}_{kt\left\vert T\right. }e_{it}\right) \right\Vert ^{2} \\ 
		& \leq  & \dfrac{1}{T}\left( \dfrac{1}{N}\sum\limits_{l=1}^{N}\left\Vert 
		\widehat{\bm{\lambda }}_{\widehat{\xi},kl}^{\left( p\right) }\bm{\lambda }%
		_{jl}^{\prime }\right\Vert \left\Vert \dfrac{1}{\sqrt{T}}\sum%
		\limits_{t=1}^{T}\mathbf{f}_{jt}\mathbb{I}_{jt}\widehat{\xi }_{kt\left\vert
			T\right. }e_{it}\right\Vert \right) ^{2} \\ 
		& \leq  & \dfrac{1}{T}\left[ \left( \dfrac{1}{N}\sum\limits_{l=1}^{N}\left%
		\Vert \widehat{\bm{\lambda }}_{\widehat{\xi},kl}^{\left( p\right) }\right\Vert
		^{2}\right) ^{1\left/ 2\right. }\left( \dfrac{1}{N}\sum\limits_{l=1}^{N}%
		\left\Vert \bm{\lambda }_{jl}\right\Vert ^{2}\right) ^{1\left/ 2\right.
		}O_{p}\left( 1\right) \right]  \\ 
		& = & O_{p}\left( \dfrac{1}{T}\right) 
	\end{array}%
\end{equation}
%----------------------------------------------------------------------------%
by Assumptions \ref{assum:FL}(b) and \ref{assum:WD}. From equations (\ref{eq:factor_regime_1}) through (\ref{eq:factor_regime_5}) it follows that
%----------------------------------------------------------------------------%
\begin{equation*}
	\dfrac{1}{N}\sum\limits_{i=1}^{N}\left\Vert \mathbf{\widehat{V}}_{\widehat{\xi},1}^{\left(
		p\right) }\left[ \bm{\widehat{\lambda}}_{\widehat{\xi},1i}^{\left( p\right) }-\left( 
	\mathbf{\widehat{H}}_{\widehat{\xi},11}^{\left( p\right) \prime }\bm{\lambda }%
	_{1i}+\mathbf{\widehat{H}}_{\widehat{\xi},21}^{\left( p\right) \prime }\bm{%
		\lambda }_{2i}\right) \right] \right\Vert ^{2} = O_{p}\left(\dfrac{1}{N}\right)+O_{p}\left(\dfrac{1}{T}\right),
\end{equation*}
%----------------------------------------------------------------------------%
and since $\left\Vert \mathbf{\widehat{V}}_{\widehat{\xi},1}^{\left( p\right) }\right\Vert =O_{p}\left( 1\right) $ by Lemma \ref{Lemma:V_hat_j} the result stated in the theorem follows.

\setcounter{equation}{0}
\numberwithin{equation}{section}

\section{Proof of result \eqref{eq:contradiction} \label{Appendix:Under_model}}

Consider
%----------------------------------------------------------------------------%
\begin{align*}
%	\begin{array}{rl}
		 \dfrac{1}{NT}\log f\left( \mathbf{X};\widehat{\mathbf{q }}
		\right) -\dfrac{1}{NT}\log f\left( \mathbf{X};\mathbf{%
			q }^{\left( 1\right) }\right)  =&\,  \dfrac{1}{NT}\log f\left( \mathbf{X};\widehat{\mathbf{q 
		}}\right) -\dfrac{1}{NT}\log f\left( \mathbf{X};
		\mathbf{q}^{\left( 1\right) }\right)  \\ 
		& -\E\left[ \dfrac{1}{NT}\log
		f\left( \mathbf{X};\widehat{\mathbf{q}}\right)\right] +\E\left[ \dfrac{1}{NT}\log f\left( \mathbf{X};\widehat{\mathbf{q}}\right) \right]  \\ 
		& +\E\left[ \dfrac{1}{NT}\log
		f\left( \mathbf{X};\mathbf{q}^{\left( 1\right) }\right)  \right] -\E\left[ 
		\dfrac{1}{NT}\log f\left( \mathbf{X};\mathbf{q}%
		^{\left( 1\right) }\right)
		\right]  \\ 
		=&\,  \left\{ \dfrac{1}{NT}\log f\left( \mathbf{X};\widehat{\mathbf{q}}\right) -\E\left[ 
		\dfrac{1}{NT}\log f\left( \mathbf{X};\widehat{\mathbf{q}}
		\right) \right] \right\}  \\ 
		& -\left\{ \dfrac{1}{NT}\log f\left( \mathbf{X};\mathbf{q}^{\left( 1\right) }
		\right) -\E\left[ \dfrac{1}{NT%
		}\log f\left( \mathbf{X};\mathbf{q}^{\left(
			1\right) }\right) \right] \right\}  \\ 
		& +\left\{ \E\left[ \dfrac{1%
		}{NT}\log f\left( \mathbf{X};\widehat{\mathbf{q }}\right)
		\right] -\E\left[ \dfrac{1}{NT}\log f\left( \mathbf{X};
		\mathbf{q}^{\left( 1\right) }\right) \right] \right\} .%
%	\end{array}%
\end{align*}
%----------------------------------------------------------------------------%
Since $\widehat{\mathbf{q}}$ is the maximum likelihood estimator, it follows that
%----------------------------------------------------------------------------%
\begin{equation*}
	\begin{array}{rl}
		& \dfrac{1}{NT}\log f\left( \mathbf{X};\widehat{\mathbf{q}}
		\right) \geq \dfrac{1}{NT}\log f\left( \mathbf{X};%
		\mathbf{q}^{\left( 1\right) }\right),
	\end{array}%
\end{equation*}
or, equivalently,		
\begin{equation*}
	\begin{array}{rl}
 & \dfrac{1}{NT}\log f\left( \mathbf{X};\widehat{\mathbf{q}} \right) -\dfrac{1}{NT}\log f\left( \mathbf{X};\mathbf{q}^{\left( 1\right) } \right) \geq 0,
	\end{array}%
\end{equation*}
%----------------------------------------------------------------------------%
which implies that
%----------------------------------------------------------------------------%
\begin{align*}
		\E\left[ \dfrac{1}{NT}\log
		f\left( \mathbf{X};\widehat{\mathbf{q}}\right)
		\right] -\E\left[ \dfrac{1}{NT}\log f\left( \mathbf{X};\mathbf{q}^{\left( 1\right) }\right) \right]  \geq&\,   \left\{ \dfrac{1}{NT}\log f\left( \mathbf{X};%
		\mathbf{q}^{\left( 1\right) }\right) -\E\left[ 
		\dfrac{1}{NT}\log f\left( \mathbf{X};\mathbf{q}^{\left( 1\right) } \right)
		\right] \right\}  \\ 
		& -\left\{ \dfrac{1}{NT}\log f\left( \mathbf{X};\widehat{\mathbf{q}} \right) -\E\left[ 
		\dfrac{1}{NT}\log f\left( \mathbf{X};\widehat{\mathbf{q}}\right) \right] \right\}\\
		=&\, o_{p}\left(1\right)-\left\{ \dfrac{1}{NT}\log f\left( \mathbf{X};\widehat{\mathbf{q}} \right) -\E\left[ 
		\dfrac{1}{NT}\log f\left( \mathbf{X};\widehat{\mathbf{q}}\right) \right] \right\},
\end{align*}
%----------------------------------------------------------------------------%
so that
%----------------------------------------------------------------------------%
\begin{equation*}
	\E\left[ \dfrac{1}{NT}\log
	f\left( \mathbf{X};\widehat{\mathbf{q}}\right)
	\right] -\E\left[ \dfrac{1}{NT}\log f\left( \mathbf{X};\mathbf{q}^{\left( 1\right) }\right) \right] \geq o_{p}\left(1\right).
\end{equation*}
%----------------------------------------------------------------------------%
If $\widehat{\mathbf{q}}$ was an estimator for $\mathbf{q}^{\left( 3\right) }$, then
%----------------------------------------------------------------------------%
\begin{equation*}
		\E\left[ \dfrac{1}{NT}\log
		f\left( \mathbf{X};\widehat{\mathbf{q}}\right)
		\right] -\E\left[ \dfrac{1}{NT}\log f\left( \mathbf{X};\mathbf{q}^{\left( 1\right) }\right) \right]  -   \E\left[ \dfrac{1}{NT}\log
		f\left( \mathbf{X};\mathbf{q}^{\left(3\right)}\right)
		\right] -\E\left[ \dfrac{1}{NT}\log f\left( \mathbf{X};\mathbf{q}^{\left( 1\right) }\right) \right] =  o_{p}\left(1\right).
\end{equation*}%
%----------------------------------------------------------------------------%
This implies that, for some $C>0$, and taking into account (\ref{eq: lik_in}),
%----------------------------------------------------------------------------%
\begin{align*}
		\E\left[ \dfrac{1}{NT}\log
		f\left( \mathbf{X};\widehat{\mathbf{q}}\right)
		\right] -\E\left[ \dfrac{1}{NT}\log f\left( \mathbf{X};\mathbf{q}^{\left( 1\right) }\right) \right]  &=  - \left\{\E\left[ \dfrac{1}{NT}\log f\left( \mathbf{X};\mathbf{q}^{\left( 1\right) }\right) \right] - \E\left[ \dfrac{1}{NT}\log
		f\left( \mathbf{X};\mathbf{q}^{\left(3\right)}\right)
		\right]\right\} + o_{p}\left(1\right) \\ 
		&=  - C + o_{p}\left(1\right),
\end{align*}
%----------------------------------------------------------------------------%
which leads to \eqref{eq:contradiction}.

%{\small {\setlength{\bibsep}{.2cm} 
%		\bibliographystyle{chicago}
%		\bibliography{BM_Biblio}
%}}

%%%%%%%%%%%%%%%%%%%%%%%%%%%%%%%%%%%%%%%%%%%%%%%%%%%%%%%%%%
%%%%%%%%%%%%%%%%%%%%%%%%%%%%%%%%%%%%%%%%%%%%%%%%%%%%%%%%%%
%%%%%%%%%%%%%%%%%%%%%%%%%%%%%%%%%%%%%%%%%%%%%%%%%%%%%%%%%%
\clearpage
\section{Additional simulation results}\label{app:sim}

\subsection{Change in the number of factors}
We simulate the latent state $\bm \xi_t$ according to \eqref{eq:trans}, with $\mbf P$ having entries $p_{11}=0.9$ and $p_{22}=0.7$, so that $p_{12}=0.1$ and $p_{21}=0.3$. This configuration corresponds to the unconditional probabilities to be equal to $\p(s_t=1)=\E[\xi_{1t}]=\frac{1-p_{22}}{2-p_{11}-p_{22}}=0.75$ and $\p(s_t=2)=\E[\xi_{2t}]=\frac{1-p_{11}}{2-p_{11}-p_{22}}=0.25$. Then, we generate the innovations $\mbf v_t$ of the VAR in \eqref{eq:trans} as follows: at each given $t$ we generate $u_t\sim \mathcal U[0,1]$ and
\begin{inparaenum}
\item [(i)] if $\xi_{1,t-1}=1$ and $u_t\le p_{11}$ then $\mbf v_t=[1\; 0]^\prime-\mbf P^\prime\bm\xi_{t-1}$;
\item [(ii)] if $\xi_{1,t-1}=1$ and $u_t> p_{11}$ then $\mbf v_t=[0\; 1]^\prime-\mbf P^\prime\bm\xi_{t-1}$;
\item [(iii)] if $\xi_{1,t-1}=0$ and $u_t\le p_{21}$ then $\mbf v_t=[1\; 0]^\prime-\mbf P^\prime\bm\xi_{t-1}$;
\item [(iv)] if $\xi_{1,t-1}=0$ and $u_t> p_{21}$ then $\mbf v_t=[0\; 1]^\prime-\mbf P^\prime\bm\xi_{t-1}$.
\end{inparaenum}

We set the number of factors as $r_1=3$ and $r_2=1$. The common component is generated according to model \eqref{eq:model}. Let $\chi_{it}=\bm\lambda_{1i}^\prime \mathbf{f}_{1t}\mathbb I(s_t=1)+\bm\lambda_{2i}^\prime \mathbf{f}_{2t}\mathbb I(s_t=2)$, $i=1,\ldots, N$, $t=1,\ldots, T$. The $r$ entries of $\bm\lambda_{1i}$ and $\bm\lambda_{2i}$ are generated from a $\mathcal N(1,1)$ distribution. The matrices $\bm\Lambda_1$ and $\bm\Lambda_2$ are then transformed in such a way that $\bm\Lambda_1^\prime \bm\Lambda_1$ and $\bm\Lambda_2^\prime \bm\Lambda_2$ are diagonal matrices. The factors are such that they satisfy $T^{-1}\sum_{t=1}^T\mbf f_{jt}\mbf f_{jt}^\prime =\mbf I_{r_j}$, $j=1,2$, where each component of $\mbf f_{jt}$ is such that $f_{j,kt}=\rho_f f_{j,k,t-1}+z_{j,kt}$, $k=1,\ldots,r_j$, $j=1,2$, with $\rho_f=\{0,0.7\}$ and $z_{j,kt}\sim \mathcal N(0,1)$.

The idiosyncratic components are generated according to  \eqref{eq:cov_mat}, where $\mbf \Sigma_{je}=\mbf \Sigma_{je,a}+\mbf \Sigma_{je,b}$, $j=1,2$, with $\mbf \Sigma_{je,a}$ diagonal and $\mbf \Sigma_{je,b}$ banded. Specifically, the entries of $\mbf \Sigma_{1e,a}$ are generated from a $\mathcal U[0.25,1.25]$ and those of $\mbf \Sigma_{2e,a}$ are generated from a $\mathcal U[0.75,1.75]$, while $\mbf \Sigma_{1e,b}$ is a Toeplitz matrix with $\tau^{k}$ on the $k$th diagonal for $k=1,2$ and zero elsewhere, and, finally $\mbf \Sigma_{2e,b}$ is a Toeplitz matrix with $\tau^{k-1}$ on the $k$th diagonal for $k=1,2,3$ and zero elsewhere. We set $\tau=\{0,0.5\}$. Moreover, each component of $\bm\nu_t$ is such that $\nu_{it}=\rho_{i}\nu_{i,t-1}+\omega_{it}$, $i=1,\ldots, N$, $t=1,\ldots,T$, with $\rho_i=\{0,\rho\}$ and $\rho\sim\mathcal U[0,0.5]$. Finally, we set the average noise-to-signal ratio across all $N$ simulated time series to be $N^{-1}\sum_{i=1}^N\frac{\sum_{t=1}^T e_{it}^2}{\sum_{t=1}^T \chi_{it}^2}=0.5$.

\begin{table}[h!]
\caption{Simulation results - change in number of factors - $r_1=3$, $r_2=1$, $\rho_f=0$, $\tau=0$, $\rho=0$.}\label{tab:numf_change1}
\vskip .2cm
\small
\centering
\begin{tabular}{ll c c c c | c c | c}
\hline
\hline
\\[-10pt]
$T$ & $N$	&	$\wh p_{11}$ &	$\wh p_{22}$ &  $\bar{\widehat{\xi}}_{1,t|T}$ &  $\bar{\widehat{\xi}}_{2,t|T}$ & $R^2_{B^*}$ &  MSE($\chi$) & avg. iter \\
\hline
250	&	100	&	0.87	&	0.53	&	0.76	&	0.24	&	0.98	&	0.04	&	17.98	\\
	&		&	\footnotesize	$(	0.04	)$	&	\footnotesize	$(	0.11	)$	&	\footnotesize	$(	0.10	)$	&	\footnotesize	$(	0.10	)$	&		&	&			\\
500	&	100	&	0.89	&	0.66	&	0.75	&	0.25	&	0.99	&	0.03	&	14.73\\
	&		&	\footnotesize	$(	0.02	)$	&	\footnotesize	$(	0.08	)$	&	\footnotesize	$(	0.04	)$	&	\footnotesize	$(	0.04	)$	&		&	&			\\
750	&	100	&	0.90	&	0.68	&	0.76	&	0.24	&	0.99	&	0.03	&	12.94	\\
	&		&	\footnotesize	$(	0.01	)$	&	\footnotesize	$(	0.03	)$	&	\footnotesize	$(	0.03	)$	&	\footnotesize	$(	0.03	)$	&		&	&		\\
1000	&	100	&	0.90	&	0.64	&	0.76	&	0.24	&	0.99	&	0.02	&	11.68	\\
	&		&	\footnotesize	$(	0.02	)$	&	\footnotesize	$(	0.17	)$	&	\footnotesize	$(	0.06	)$	&	\footnotesize	$(	0.06	)$	&		&	&				\\
250	&	200	&	0.86	&	0.54	&	0.75	&	0.25	&	0.98	&	0.03	&	15.62\\
	&		&	\footnotesize	$(	0.05	)$	&	\footnotesize	$(	0.18	)$	&	\footnotesize	$(	0.08	)$	&	\footnotesize	$(	0.08	)$	&		&	&		\\
500	&	200	&	0.89&0.65		&	0.75	&	0.25	&	0.98	&	0.02	&	10.58\\
	&		&	\footnotesize	$(	0.02	)$	&	\footnotesize	$(0.11	)$	&	\footnotesize	$(	0.05	)$	&	\footnotesize	$(	0.05	)$	&		&	&		\\
750	&	200	&	0.89	&	0.69	&	0.74	&	0.26	&	0.99	&	0.02	&	10.60	\\
	&		&	\footnotesize	$(	0.01	)$	&	\footnotesize	$(	0.03	)$	&	\footnotesize	$(	0.03	)$	&	\footnotesize	$(	0.03	)$	&		&	&	\\
1000	&	200	&	0.89	&	0.69	&	0.75	&	0.25	&	0.99	&	0.01	&	9.59	\\
	&		&	\footnotesize	$(	0.01	)$	&	\footnotesize	$(	0.03	)$	&	\footnotesize	$(	0.03	)$	&	\footnotesize	$(	0.03	)$	&		&	&	\\
\hline
\hline
\end{tabular}
\end{table}

\begin{table}[h!]
\caption{Simulation results - change in number of factors - $r_1=3$, $r_2=1$, $\rho_f=0.7$, $\tau=0.5$, $\rho=0.5$.}\label{tab:numf_change2}
\vskip .2cm
\small
\centering
\begin{tabular}{ll c c c c | c c | c}
\hline
\hline
\\[-10pt]
$T$ & $N$	&	$\wh p_{11}$ &	$\wh p_{22}$ &  $\bar{\widehat{\xi}}_{1,t|T}$ &  $\bar{\widehat{\xi}}_{2,t|T}$ & $R^2_{B^*}$ &  MSE($\chi$) & avg. iter \\
\hline
250	&	100	&	0.89	&	0.49	&	0.80	&	0.20		&	0.98	&	0.04	&	18.18	\\
	&		&	\footnotesize	$(	0.05	)$	&	\footnotesize	$(	0.24	)$	&	\footnotesize	$(	0.11	)$	&	\footnotesize	$(	0.11	)$	&		&	&			\\
500	&	100	&	0.89	&	0.65	&	0.76	&	0.24		&	0.99	&	0.03	&	19.25\\
	&		&	\footnotesize	$(	0.02	)$	&	\footnotesize	$(	0.10	)$	&	\footnotesize	$(	0.05	)$	&	\footnotesize	$(	0.05	)$	&		&	&			\\
750	&	100	&	0.90	&	0.66	&	0.76	&	0.24		&	0.99	&	0.03	&	15.88	\\
	&		&	\footnotesize	$(	0.02	)$	&	\footnotesize	$(	0.12	)$	&	\footnotesize	$(	0.05	)$	&	\footnotesize	$(	0.05	)$	&		&	&		\\
1000	&	100	&	0.91	&	0.59	&	0.78	&	0.22		&	0.99	&	0.03	&	12.97	\\
	&		&	\footnotesize	$(	0.03	)$	&	\footnotesize	$(	0.23	)$	&	\footnotesize	$(	0.08	)$	&	\footnotesize	$(	0.08	)$	&		&	&				\\
250	&	200	&	0.87	&	0.52	&	0.77	&	0.23	&	0.98	&	0.03	&	14.00		\\
	&		&	\footnotesize	$(	0.05	)$	&	\footnotesize	$(	0.21	)$	&	\footnotesize	$(	0.10	)$	&	\footnotesize	$(	0.10	)$	&		&	&		\\
500	&	200	&	0.89	&	0.66	&	0.75	&	0.25	&	0.98	&	0.02	&	12.72		\\
	&		&	\footnotesize	$(	0.02	)$	&	\footnotesize	$(	0.07	)$	&	\footnotesize	$(	0.05	)$	&	\footnotesize	$(	0.05	)$	&		&	&		\\
750	&	200	&	0.89	&	0.69	&	0.74	&	0.26	&	0.99	&	0.02	&	11.93	\\
	&		&	\footnotesize	$(	0.01	)$	&	\footnotesize	$(	0.03	)$	&	\footnotesize	$(	0.03	)$	&	\footnotesize	$(	0.03	)$	&		&	&	\\
1000	&	200	&	0.89	&	0.68	&	0.75	&	0.25	&	0.99	&	0.02	&	10.72	\\
	&		&	\footnotesize	$(	0.01	)$	&	\footnotesize	$(	0.04	)$	&	\footnotesize	$(	0.03	)$	&	\footnotesize	$(	0.03	)$	&		&	&	\\
\hline
\hline
\end{tabular}
\end{table}

\clearpage
\subsection{Change in the autocorrelation of factors}

We simulate the latent state $\bm \xi_t$ according to \eqref{eq:trans}, with $\mbf P$ having entries $p_{11}=0.9$ and $p_{22}=0.7$, so that $p_{12}=0.1$ and $p_{21}=0.3$. This configuration corresponds to the unconditional probabilities to be equal to $\p(s_t=1)=\E[\xi_{1t}]=\frac{1-p_{22}}{2-p_{11}-p_{22}}=0.75$ and $\p(s_t=2)=\E[\xi_{2t}]=\frac{1-p_{11}}{2-p_{11}-p_{22}}=0.25$. Then, we generate the innovations $\mbf v_t$ of the VAR in \eqref{eq:trans} as follows: at each given $t$ we generate $u_t\sim \mathcal U[0,1]$ and
\begin{inparaenum}
\item [(i)] if $\xi_{1,t-1}=1$ and $u_t\le p_{11}$ then $\mbf v_t=[1\; 0]^\prime-\mbf P^\prime\bm\xi_{t-1}$;
\item [(ii)] if $\xi_{1,t-1}=1$ and $u_t> p_{11}$ then $\mbf v_t=[0\; 1]^\prime-\mbf P^\prime\bm\xi_{t-1}$;
\item [(iii)] if $\xi_{1,t-1}=0$ and $u_t\le p_{21}$ then $\mbf v_t=[1\; 0]^\prime-\mbf P^\prime\bm\xi_{t-1}$;
\item [(iv)] if $\xi_{1,t-1}=0$ and $u_t> p_{21}$ then $\mbf v_t=[0\; 1]^\prime-\mbf P^\prime\bm\xi_{t-1}$.
\end{inparaenum}

We set the number of factors in each state to $r_j=r=1$,  $j=1,2$. The common component is generated according to model \eqref{eq:model}. Let $\chi_{it}=\bm\lambda_{i}^\prime \mathbf{f}_{1t}\mathbb I(s_t=1)+\bm\lambda_{i}^\prime \mathbf{f}_{2t}\mathbb I(s_t=2)$, $i=1,\ldots, N$, $t=1,\ldots, T$. The $r$ entries of $\bm\lambda_{i}$ are generated from a $\mathcal N(1,1)$ distribution. The matrix $\bm\Lambda$ is then transformed in such a way that $\bm\Lambda^\prime \bm\Lambda$ is diagonal. The factors are such that $f_{1t}=0.9 f_{1,t-1}+z_{1t}$ and $f_{2t}=z_{2t}$ with $z_{kt}\sim \mathcal N(0,1)$, $k=1,2$, then $f_{1t}$ is rescaled to have variance one.
%, and satisfy $T^{-1}\sum_{t=1}^T\mbf f_{jt}\mbf f_{jt}^\prime =\mbf I_r$, where each component of $\mbf f_t$ is such that $f_{kt}=\rho_f f_{k,t-1}+z_{kt}$, $k=1,\ldots,r$, with $\rho_f=\{0,0.7\}$ and $z_{kt}\sim \mathcal N(0,1)$.

The idiosyncratic components are generated having covariance matrix $\mbf \Sigma_{e}=\mbf \Sigma_{e,a}+\mbf \Sigma_{e,b}$, with $\mbf \Sigma_{e,a}$ diagonal and $\mbf \Sigma_{e,b}$ banded. Specifically, the entries of $\mbf \Sigma_{e,a}$ are generated from a $\mathcal U[0.25,1.25]$, while $\mbf \Sigma_{e,b}$ is a Toeplitz matrix with $\tau^{k}$ on the $k$th diagonal for $k=1,2$ and zero elsewhere. We set $\tau=\{0,0.5\}$. Moreover, each component of $\bm\nu_t$ is such that $\nu_{it}=\rho_{i}\nu_{i,t-1}+\omega_{it}$, $i=1,\ldots, N$, $t=1,\ldots,T$, with $\rho_i=\{0,\rho\}$ and $\rho\sim\mathcal U[0,0.5]$. Finally, we set the average noise-to-signal ratio across all $N$ simulated time series to be $N^{-1}\sum_{i=1}^N\frac{\sum_{t=1}^T e_{it}^2}{\sum_{t=1}^T \chi_{it}^2}=0.5$.

\begin{table}[h!]
\caption{Simulation results - change in acf of factors - $r=1$, $\tau=0$, $\rho=0$.}\label{tab:acf_change1}
\vskip .2cm
\small
\centering
\begin{tabular}{ll c c c c | c c | c}
\hline
\hline
\\[-10pt]
$T$ & $N$	&	$\wh p_{11}$ &	$\wh p_{22}$ &  $\bar{\widehat{\xi}}_{1,t|T}$ &  $\bar{\widehat{\xi}}_{2,t|T}$ & $R^2_{B^*}$ &  MSE($\chi$) & avg. iter \\
\hline
250	&	100	&	0.97	&	0.04	&	0.97	&	0.03	&	0.998	&	0.02	&	13.88	\\
	&		&	\footnotesize	$(	0.01	)$	&	\footnotesize	$(	0.09	)$	&	\footnotesize	$(	0.01	)$	&	\footnotesize	$(	0.01	)$	&		&	&			\\
500	&	100	&	0.96	&	0.04	&	0.96	&	0.04	&	0.999	&	0.02	&	10.68\\
	&		&	\footnotesize	$(	0.02	)$	&	\footnotesize	$(	0.04	)$	&	\footnotesize	$(	0.02	)$	&	\footnotesize	$(	0.02	)$	&		&	&			\\
750	&	100	&	0.97	&	0.03	&	0.97	&	0.03	&	0.999	&	0.01	&	4.48	\\
	&		&	\footnotesize	$(	0.01	)$	&	\footnotesize	$(	0.01	)$	&	\footnotesize	$(	0.01	)$	&	\footnotesize	$(	0.01	)$	&		&	&		\\
1000	&	100	&	0.97	&	0.03	&	0.97	&	0.03	&	0.999	&	0.01	&	3.00	\\
	&		&	\footnotesize	$(	1\cdot10^{-6}	)$	&	\footnotesize	$(	1\cdot10^{-6}	)$	&	\footnotesize	$(	5\cdot10^{-6}	)$	&	\footnotesize	$(	5\cdot10^{-6}	)$	&		&	&				\\
250	&	200	&	0.98	&	0.04	&	0.98	&	0.02	&	0.998	&	0.01	&	9.23		\\
	&		&	\footnotesize	$(	0.01	)$	&	\footnotesize	$(	0.10	)$	&	\footnotesize	$(	0.01	)$	&	\footnotesize	$(	0.01	)$	&		&	&		\\
500	&	200	&	0.97	&	0.16	&	0.97	&	0.03	&	0.999	&	0.01	&	10.98		\\
	&		&	\footnotesize	$(	0.01	)$	&	\footnotesize	$(	0.17	)$	&	\footnotesize	$(	0.01	)$	&	\footnotesize	$(	0.01	)$	&		&	&		\\
750	&	200	&	0.97	&	0.07	&	0.97	&	0.03	&	0.999	&	0.01	&	6.41		\\
	&		&	\footnotesize	$(	0.01	)$	&	\footnotesize	$(	0.10	)$	&	\footnotesize	$(	0.01	)$	&	\footnotesize	$(	0.01	)$	&		&	&	\\
1000	&	200	&	0.97	&	0.05	&	0.97	&	0.03	&	0.999	&	0.01	&	4.37	\\
	&		&	\footnotesize	$(	0.01	)$	&	\footnotesize	$(	0.08	)$	&	\footnotesize	$(	0.01	)$	&	\footnotesize	$(	0.01	)$	&		&	&	\\
\hline
\hline
\end{tabular}
\end{table}

\begin{table}[h!]
\caption{Simulation results - change  in acf of factors - $r=1$, $\tau=0.5$, $\rho=0.5$.}\label{tab:acf_change2}
\vskip .2cm
\small
\centering
\begin{tabular}{ll c c c c | c c | c}
\hline
\hline
\\[-10pt]
$T$ & $N$	&	$\wh p_{11}$ &	$\wh p_{22}$ &  $\bar{\widehat{\xi}}_{1,t|T}$ &  $\bar{\widehat{\xi}}_{2,t|T}$ & $R^2_{B^*}$ &  MSE($\chi$) & avg. iter \\
\hline
250	&	100	&	0.96	&	0.15	&	0.95	&	0.05	&	0.998	&	0.02	&	17.69	\\
	&		&	\footnotesize	$(	0.02	)$	&	\footnotesize	$(	0.18	)$	&	\footnotesize	$(	0.02	)$	&	\footnotesize	$(	0.02	)$	&		&	&		\\
500	&	100	&	0.96	&	0.11	&	0.95	&	0.05	&	0.999	&	0.02	&	11.24	\\
	&		&	\footnotesize	$(	0.02	)$	&	\footnotesize	$(	0.15	)$	&	\footnotesize	$(	0.02	)$	&	\footnotesize	$(	0.02	)$	&		&	&		\\
750	&	100	&	0.97	&	0.04	&	0.97	&	0.03	&	0.999	&	0.01	&	3.73		\\
	&		&	\footnotesize	$(	0.01	)$	&	\footnotesize	$(	0.05	)$	&	\footnotesize	$(	0.01	)$	&	\footnotesize	$(	0.01	)$	&		&	&	\\
1000	&	100	&	0.97	&	0.03	&	0.97	&	0.03	&	1.00	&	0.01	&	3.00		\\
	&		&	\footnotesize	$(	1\cdot10^{-6}	)$	&	\footnotesize	$(	2\cdot10^{-6}	)$	&	\footnotesize	$(	6\cdot10^{-6}	)$	&	\footnotesize	$(	6\cdot10^{-6}	)$	&		&	&	\\
250	&	200	&	0.98	&	0.03	&	0.98	&	0.02	&	0.998	&	0.01	&	8.92	\\
	&		&	\footnotesize	$(	0.01	)$	&	\footnotesize	$(	0.09	)$	&	\footnotesize	$(	0.02	)$	&	\footnotesize	$(	0.02	)$	&		&	&			\\
500	&	200	&	0.97	&	0.03	&	0.97	&	0.03	&	0.999	&	0.01	&	8.54	\\
	&		&	\footnotesize	$(	0.01	)$	&	\footnotesize	$(	0.05	)$	&	\footnotesize	$(	0.01	)$	&	\footnotesize	$(	0.01	)$	&		&	&			\\
750	&	200	&	0.97	&	0.03	&	0.97	&	0.03	&	0.999	&	0.01	&	4.15		\\
	&		&	\footnotesize	$(	0.004	)$	&	\footnotesize	$(	0.02	)$	&	\footnotesize	$(	0.004	)$	&	\footnotesize	$(	0.004	)$	&		&	&		\\
1000	&	200	&	0.97	&	0.03	&	0.97	&	0.03	&	0.999	&	0.01	&	3.15			\\
	&		&	\footnotesize	$(	0.001	)$	&	\footnotesize	$(	0.003	)$	&	\footnotesize	$(	0.001	)$	&	\footnotesize	$(	0.001)$	&		&	&				\\
\hline
\hline
\end{tabular}
\end{table}

\clearpage
\subsection{No change}

We set the number of factors to $r=2$. The common component is generated according to $\chi_{it}=\bm\lambda_{i}^\prime \mathbf{f}_{t}$, $i=1,\ldots, N$, $t=1,\ldots, T$. The $r$ entries of $\bm\lambda_{i}$ are generated from a $\mathcal N(1,1)$ distribution. The matrix $\bm\Lambda$ is then transformed in such a way that $\bm\Lambda^\prime \bm\Lambda$ is diagonal. The factors are such that $T^{-1}\sum_{t=1}^T\mbf f_t\mbf f_t^\prime =\mbf I_r$, where each component of $\mbf f_t$ is such that $f_{kt}=\rho_f f_{k,t-1}+z_{kt}$, $k=1,\ldots,r$, with $\rho_f=\{0,0.7\}$ and $z_{kt}\sim \mathcal N(0,1)$.

The idiosyncratic components are generated having covariance matrix $\mbf \Sigma_{e}=\mbf \Sigma_{e,a}+\mbf \Sigma_{e,b}$, with $\mbf \Sigma_{e,a}$ diagonal and $\mbf \Sigma_{e,b}$ banded. Specifically, the entries of $\mbf \Sigma_{e,a}$ are generated from a $\mathcal U[0.25,1.25]$, while $\mbf \Sigma_{e,b}$ is a Toeplitz matrix with $\tau^{k}$ on the $k$th diagonal for $k=1,2$ and zero elsewhere. We set $\tau=\{0,0.5\}$. Moreover, each component of $\bm\nu_t$ is such that $\nu_{it}=\rho_{i}\nu_{i,t-1}+\omega_{it}$, $i=1,\ldots, N$, $t=1,\ldots,T$, with $\rho_i=\{0,\rho\}$ and $\rho\sim\mathcal U[0,0.5]$. Finally, we set the average noise-to-signal ratio across all $N$ simulated time series to be $N^{-1}\sum_{i=1}^N\frac{\sum_{t=1}^T e_{it}^2}{\sum_{t=1}^T \chi_{it}^2}=0.5$.

In this case, we report the following multiple $R^2$ for the estimated loadings 
\[
R^2_{B} = \frac{\text{tr}\l\{
\l(\bm\Lambda^{\prime} \wh{\mbf B}_1\r)
\l(\wh{\mbf B}_1^{\prime} \wh{\mbf B}_1\r)^{-1}
\l(\wh{\mbf B}_1^\prime\bm\Lambda\r)
\r\}}
{\text{tr}\l(\bm\Lambda^\prime \bm\Lambda\r)}.
\]
No bias correction is necessary in this case, since no change is present in the true data generating process.

\begin{table}[h!]
\caption{Simulation results - no change - $r=2$, $\rho_f=0$, $\tau=0$, $\rho=0$.}\label{tab:no_change1}
\vskip .2cm
\small
\centering
\begin{tabular}{ll c c c c | c c | c}
\hline
\hline
\\[-10pt]
$T$ & $N$	&	$\wh p_{11}$ &	$\wh p_{22}$ &  $\bar{\widehat{\xi}}_{1,t|T}$ &  $\bar{\widehat{\xi}}_{2,t|T}$ & $R^2_B$ &  MSE($\chi$) & avg. iter \\
\hline
250	&	100	&			0.97	&	0.03	&	0.97	&	0.03	& 0.996 &	0.02	&	13.08	\\
	&		&	\footnotesize	$(	0.01	)$	&	\footnotesize	$(	0.08	)$	&	\footnotesize	$(	0.01	)$	&	\footnotesize	$(	0.01	)$	&		&	&			\\
500	&	100	&			0.97	&	0.04	&	0.97	&	0.03	& 0.997 &	0.01	&	6.87	\\
	&		&	\footnotesize	$(	0.01	)$	&	\footnotesize	$(	0.04	)$	&	\footnotesize	$(	0.01	)$	&	\footnotesize	$(	0.01	)$	&		&	&			\\
750	&	100	&	0.97	&	0.03	&	0.97	&	0.03	& 0.998&	0.01	 & 	3.11			\\
	&		&	\footnotesize	$(	0.0003	)$	&	\footnotesize	$(	0.002	)$	&	\footnotesize	$(	0.0003	)$	&	\footnotesize	$(	0.0003	)$	&		&	&		\\
1000	&	100	&	0.97	&	0.03	&	0.97	&	0.03	&0.999 &	0.01	& 	3.00			\\
	&		&	\footnotesize	$(	2\cdot10^{-6}	)$	&	\footnotesize	$(	2\cdot10^{-5}	)$	&	\footnotesize	$(	9\cdot10^{-6}	)$	&	\footnotesize	$(	9\cdot10^{-6}	)$	&		&	&				\\
250	&	200	&	0.98	&	0.02	&	0.98	&	0.02	& 0.996&	0.01	& 	8.75			\\
	&		&	\footnotesize	$(	0.01	)$	&	\footnotesize	$(	0.06	)$	&	\footnotesize	$(	0.01	)$	&	\footnotesize	$(	0.01	)$	&		&	&		\\
500	&	200	&	0.98	&	0.02	&	0.97	&	0.03	& 0.998 &	0.01	& 	9.15			\\
	&		&	\footnotesize	$(	0.01	)$	&	\footnotesize	$(	0.05	)$	&	\footnotesize	$(	0.01	)$	&	\footnotesize	$(	0.01	)$	&		&	&		\\
750	&	200	&	0.97	&	0.03	&	0.97	&	0.03	&0.999&	0.01	& 	4.68			\\
	&		&	\footnotesize	$(	0.003	)$	&	\footnotesize	$(	0.04	)$	&	\footnotesize	$(	0.003	)$	&	\footnotesize	$(	0.003	)$	&		&	&	\\
1000	&	200	&	0.97	&	0.03	&	0.97	&	0.03	& 0.999 &	0.01	& 	3.39			\\
	&		&	\footnotesize	$(	0.003	)$	&	\footnotesize	$(	0.004	)$	&	\footnotesize	$(	0.003	)$	&	\footnotesize	$(	0.003	)$	&		&	&	\\
\hline
\hline
\end{tabular}
\end{table}

\begin{table}[h!]
\caption{Simulation results - no change - $r=2$, $\rho_f=0.7$, $\tau=0.5$, $\rho=0.5$.}\label{tab:no_change2}
\vskip .2cm
\small
\centering
\begin{tabular}{ll c c c c | c c | c}
\hline
\hline
\\[-10pt]
$T$ & $N$	&	$\wh p_{11}$ &	$\wh p_{22}$ &  $\bar{\widehat{\xi}}_{1,t|T}$ &  $\bar{\widehat{\xi}}_{2,t|T}$ & $R^2_B$ &  MSE($\chi$) & avg. iter \\
\hline
250	&	100	&	0.97 	& 0.30		&0.95 &0.05 &0.99&0.02&15.63	\\
	&		&	\footnotesize	$(	0.01	)$	&	\footnotesize	$(	0.25	)$	&	\footnotesize	$(	0.02	)$	&	\footnotesize	$(	0.02	)$	&		&	&			\\
500	&	100	&	0.97	 &  0.15 &0.94 &0.04 &0.997&0.02 &7.70\\
	&		&	\footnotesize	$(	0.01	)$	&	\footnotesize	$(	0.23	)$	&	\footnotesize	$(	0.02	)$	&	\footnotesize	$(	0.02	)$	&		&	&	\\
750	&	100	&	0.97	 &  0.04	&	0.97	 &  0.03 &0.998& 0.01 & 3.92		\\
	&		&	\footnotesize	$(	0.01	)$	&	\footnotesize	$(	0.09	)$	&	\footnotesize	$(	0.02	)$	&	\footnotesize	$(	0.02	)$	&		&	&	\\
1000	&	100	&	0.97	 &  0.04	&	0.97	 &  0.03	&0.999& 0.01 & 3.72	\\
	&		&	\footnotesize	$(	0.01	)$	&	\footnotesize	$(	0.05	)$	&	\footnotesize	$(	0.02	)$	&	\footnotesize	$(	0.02	)$	&		&	&		\\
250	&	200	&	0.98	&	0.12	&	0.98	&	0.02	&0.996&	0.01	&	9.04			\\
	&		&	\footnotesize	$(	0.01	)$	&	\footnotesize	$(	0.16	)$	&	\footnotesize	$(	0.01	)$	&	\footnotesize	$(	0.01	)$	&		&	&		\\
500	&	200	&	0.97	&	0.18	&	0.97	&	0.03	&0.998&	0.01	&	8.30			\\
	&		&	\footnotesize	$(	0.01	)$	&	\footnotesize	$(	0.21	)$	&	\footnotesize	$(	0.01	)$	&	\footnotesize	$(	0.01	)$	&		&	&		\\
750	&	200	&	0.97	&	0.11	&	0.97	&	0.03	&0.998&	0.01	&	4.96			\\
	&		&	\footnotesize	$(	0.01	)$	&	\footnotesize	$(	0.18	)$	&	\footnotesize	$(	0.01	)$	&	\footnotesize	$(	0.01	)$	&		&	&		\\
1000	&	200	&	0.97	&	0.04	&	0.97	&	0.03	&0.999&	0.01	&	3.42			\\
	&		&	\footnotesize	$(	0.004	)$	&	\footnotesize	$(	0.06	)$	&	\footnotesize	$(	0.01	)$	&	\footnotesize	$(	0.01	)$	&		&	&		\\
\hline
\hline
\end{tabular}
\end{table}
%%%%%%%%%%%%%%%%%%%%%%%%%%%%%%%%%%%%%%%%%%%%%%%%%%%%%%%%%%

%%%%%%%%%%%%%%%%%%%%%%%%%%%%%%%%%%%%%%%%%%%%%%%%%%%%%%%%%%
%%%%%%%%%%%%%%%%%%%%%%%%%%%%%%%%%%%%%%%%%%%%%%%%%%%%%%%%%%
%%%%%%%%%%%%%%%%%%%%%%%%%%%%%%%%%%%%%%%%%%%%%%%%%%%%%%%%%%
\clearpage
\section{Estimated factors}\label{app:factors}

\begin{figure}[h!]
	\caption{Estimated factors $\wh{f}_{jk, t}$, $j=1,2$, $k=1,\ldots, r_j$ - Stock returns ($r_j=1$).}\label{fig:fhat_FF}
\centering
\begin{tabular}{cc}
\includegraphics[width=.5\textwidth]{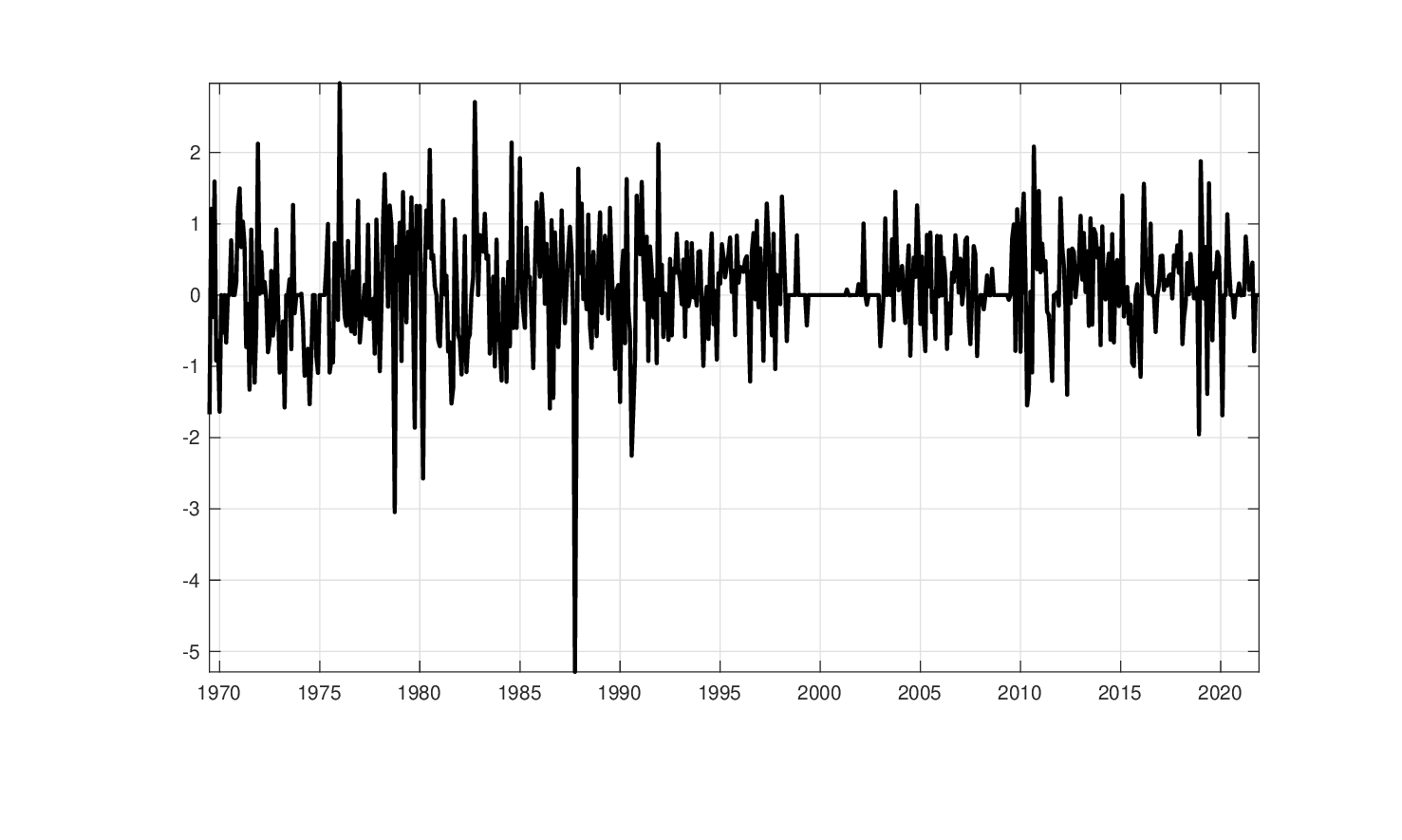}&\includegraphics[width=.5\textwidth]{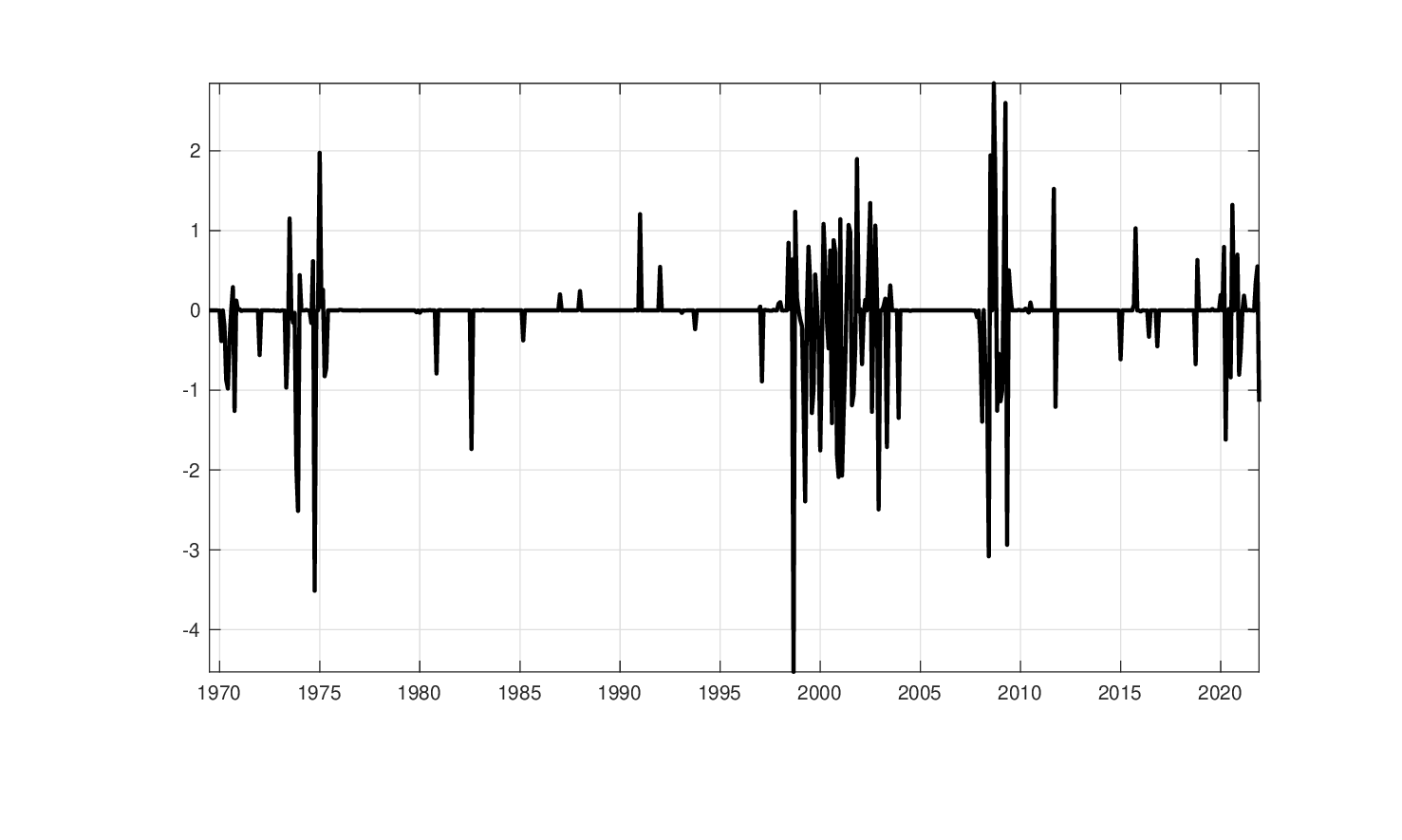}\\[-20pt]
 (a): $\wh{f}_{11,t}$ &  (b): $\wh{f}_{21,t}$
\end{tabular}
	\begin{tablenotes}
		\small
		\item This figure plots the series of estimated factors $\wh{\mbf f}_{jt}=(\wh f_{j1t}\cdots\wh  f_{jr_jt})^\prime$, obtained according to \eqref{eq:fjhat}, for regimes $j=1$ (panel (a)) and $j=2$ (panel (b)), and for $t=1,\ldots, T$, estimated from the Markov switching factor model in (\ref{eq:state_space_mes}) for the dataset of U.S. stock returns described in Section \ref{section:Empirics_SR}. The number of factors is such that $r_1=r_2=r=1$.
	\end{tablenotes}
\end{figure}

\begin{figure}[h!]
	\caption{Estimated factors $\wh{f}_{jk, t}$, $j=1,2$, $k=1,\ldots, r_j$ - Macroeconomic time series ($r_j=4$).}\label{fig:fhat_FRED}
\centering
\begin{tabular}{cc}
\includegraphics[width=.5\textwidth]{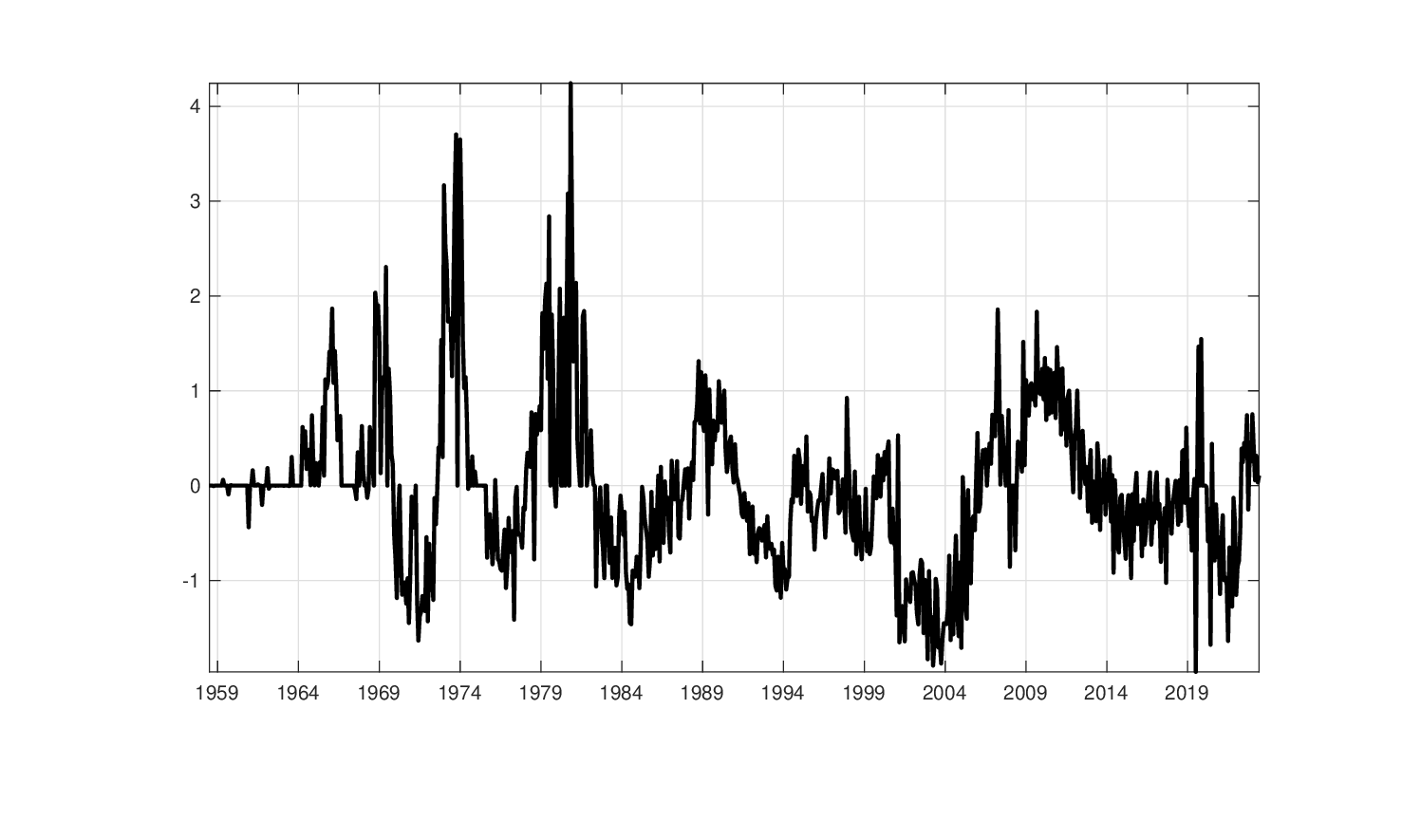}&\includegraphics[width=.5\textwidth]{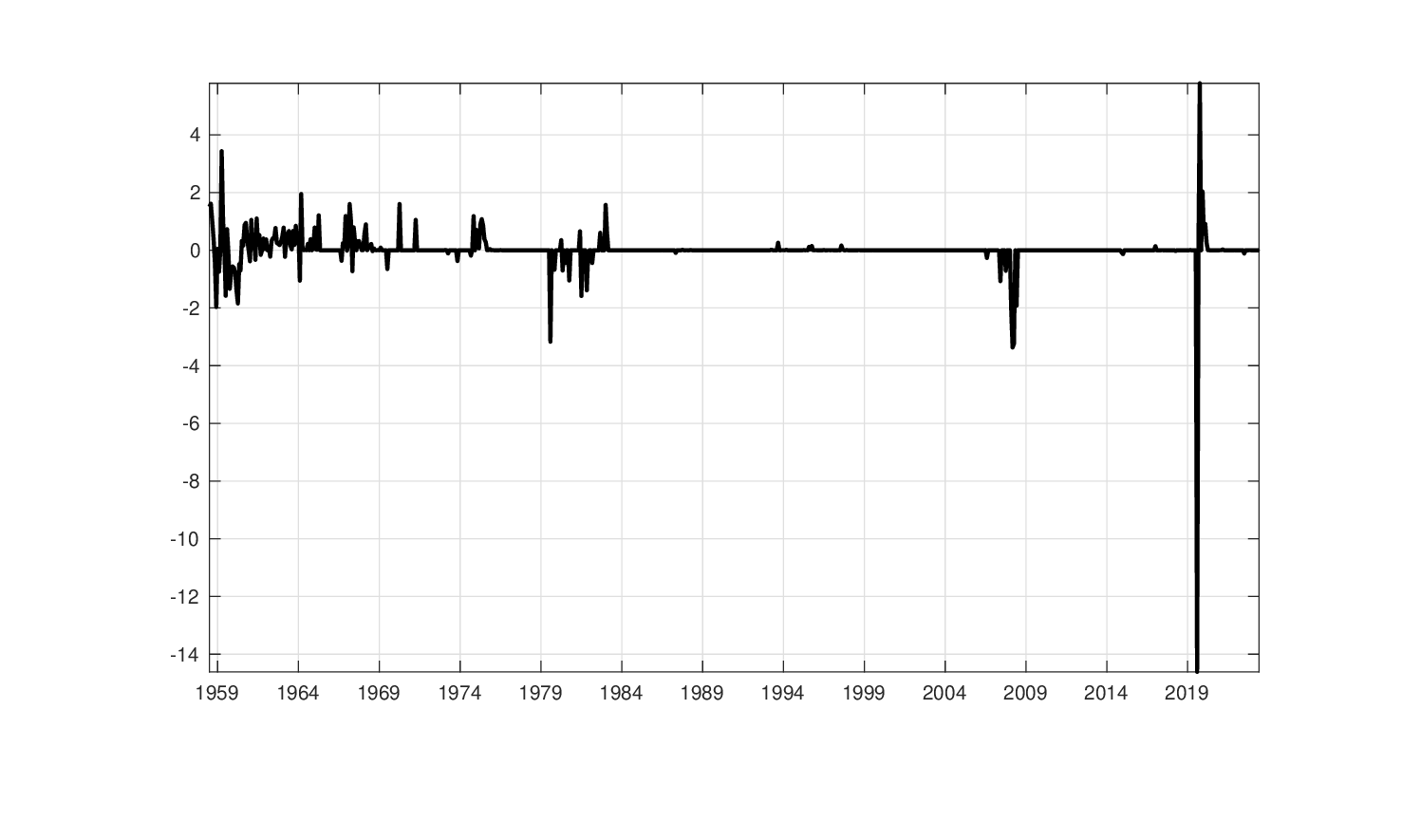}\\[-20pt]
 (a): $\wh{f}_{11,t}$ &  (b): $\wh{f}_{21,t}$\\
 \includegraphics[width=.5\textwidth]{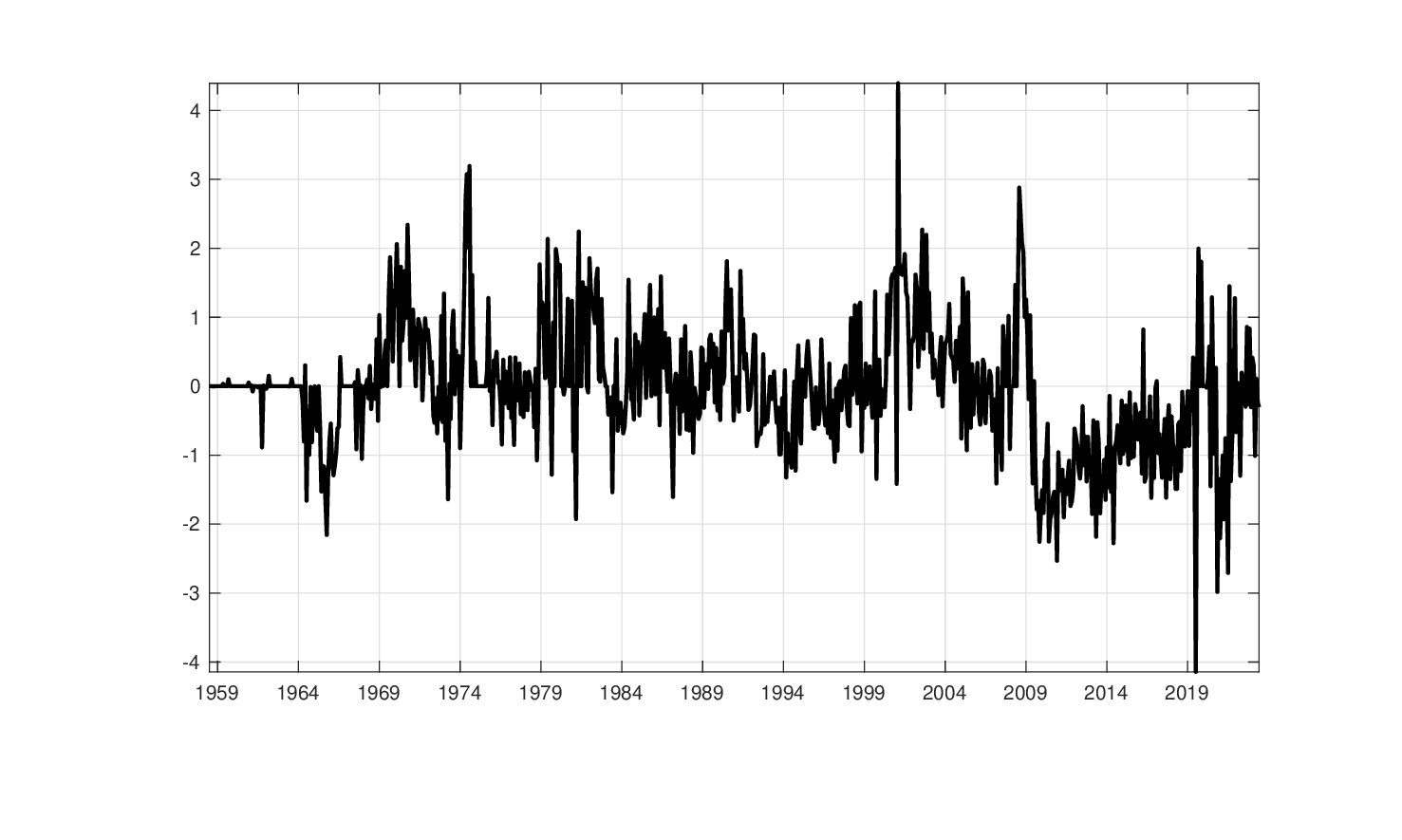}&\includegraphics[width=.5\textwidth]{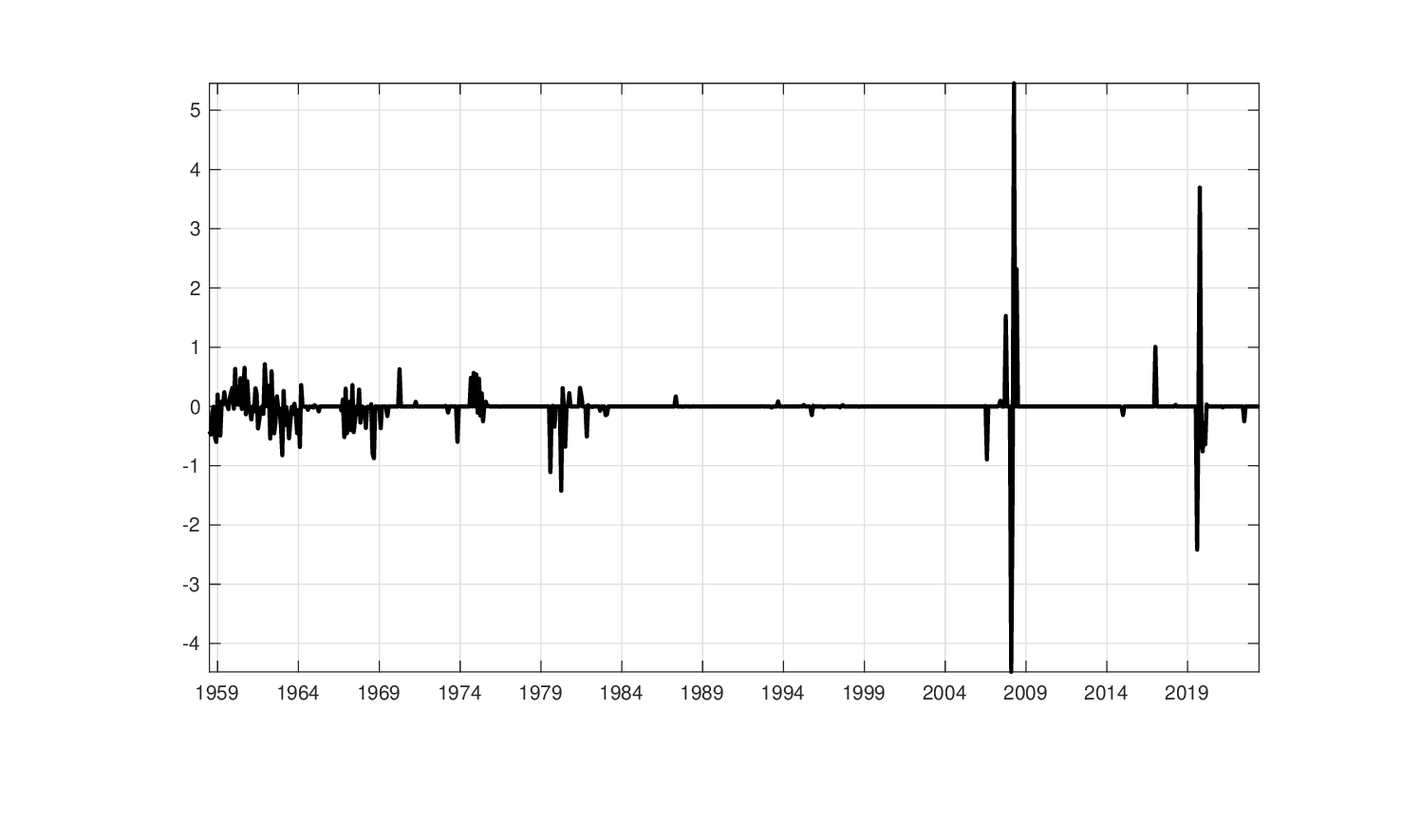}\\[-20pt]
 (a): $\wh{f}_{12,t}$ &  (b): $\wh{f}_{22,t}$\\
  \includegraphics[width=.5\textwidth]{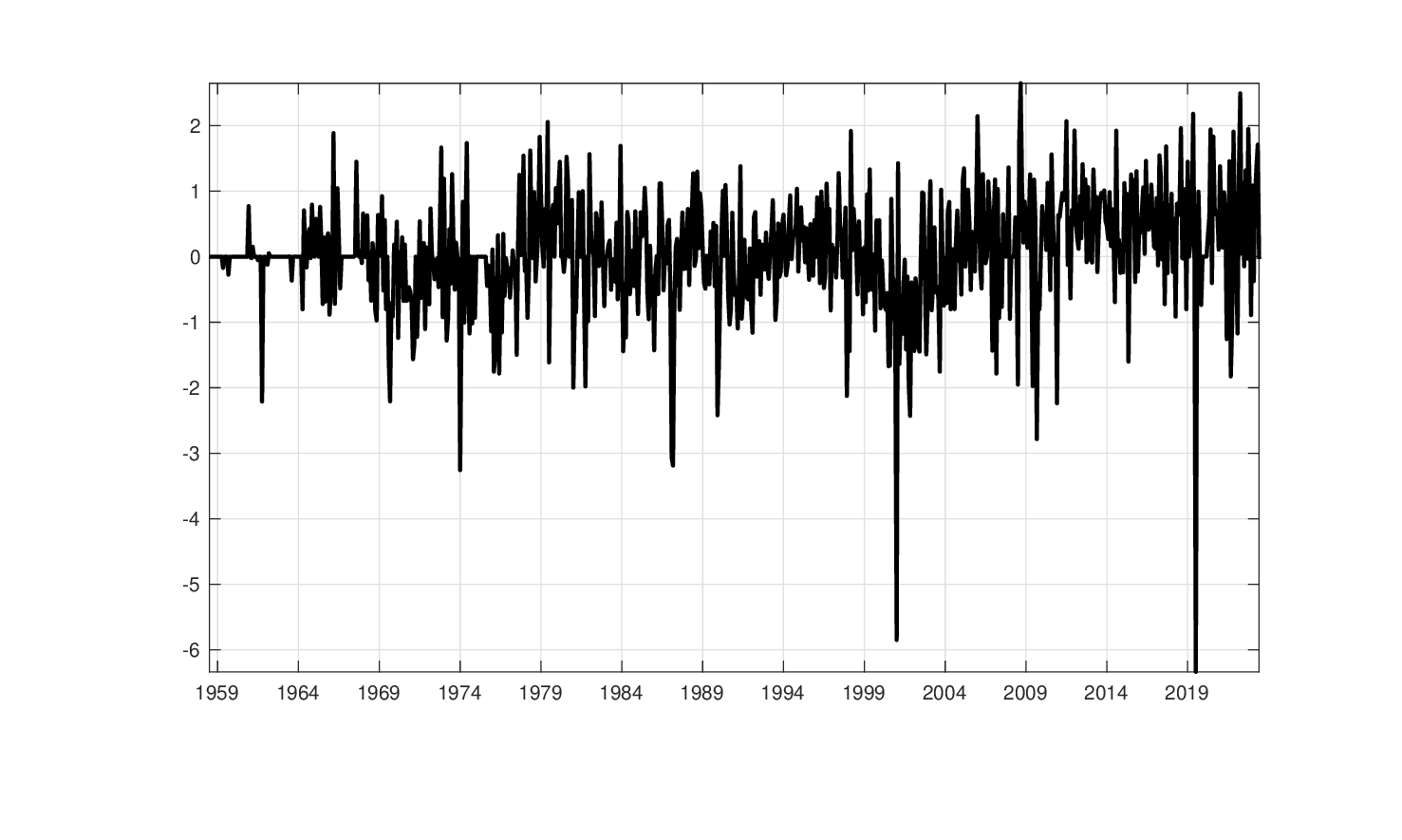}&\includegraphics[width=.5\textwidth]{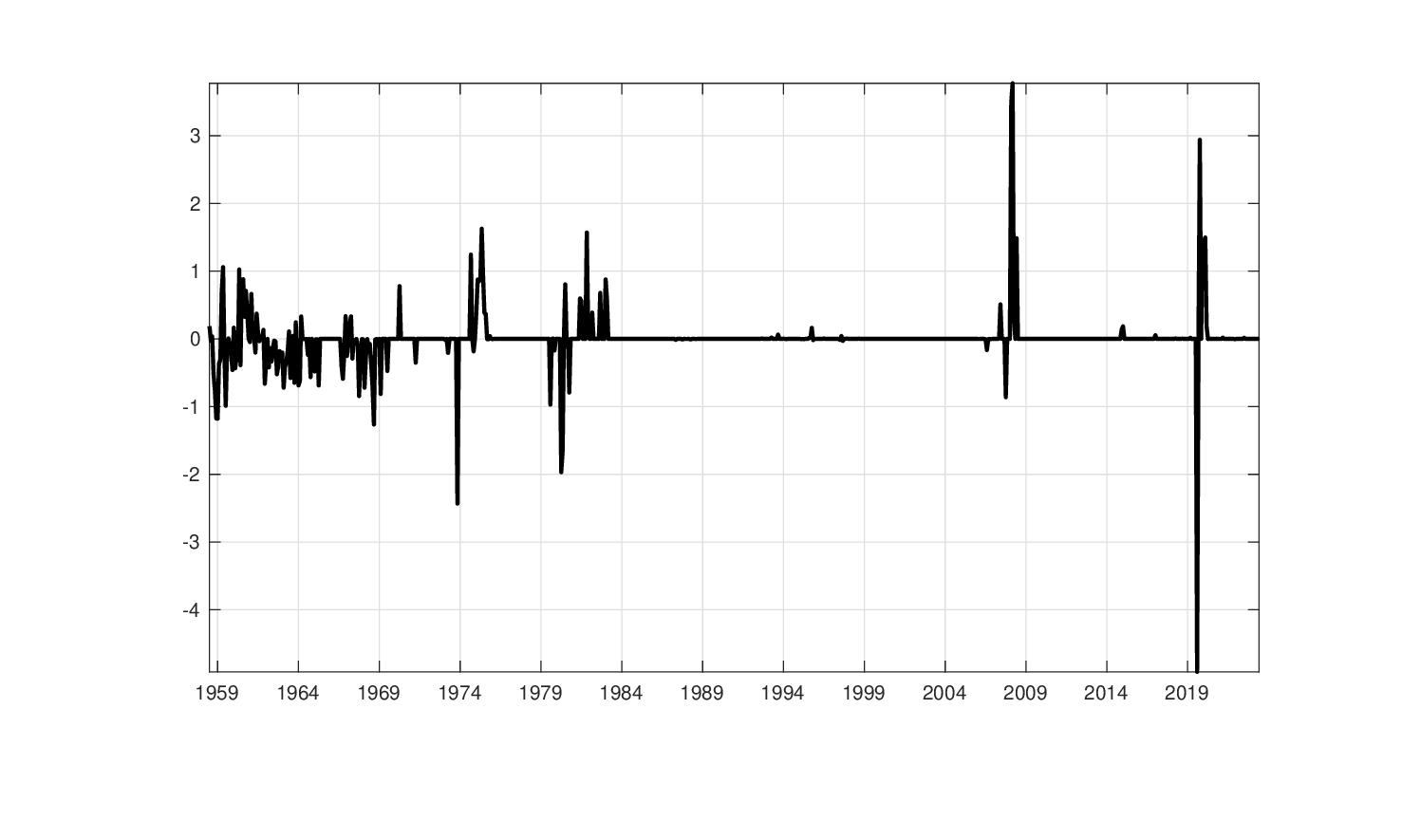}\\[-20pt]
 (a): $\wh{f}_{13,t}$ &  (b): $\wh{f}_{23,t}$\\
   \includegraphics[width=.5\textwidth]{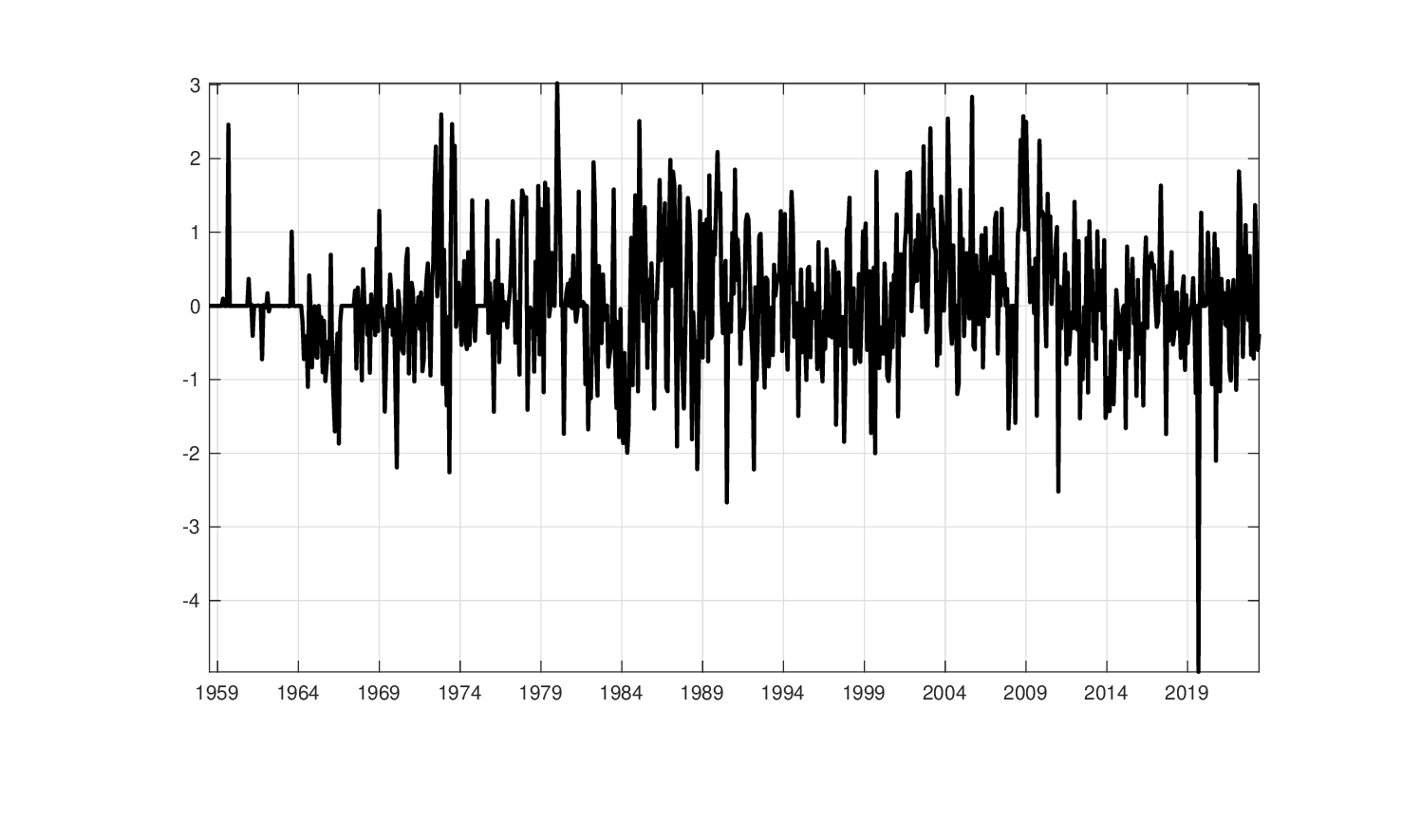}&\includegraphics[width=.5\textwidth]{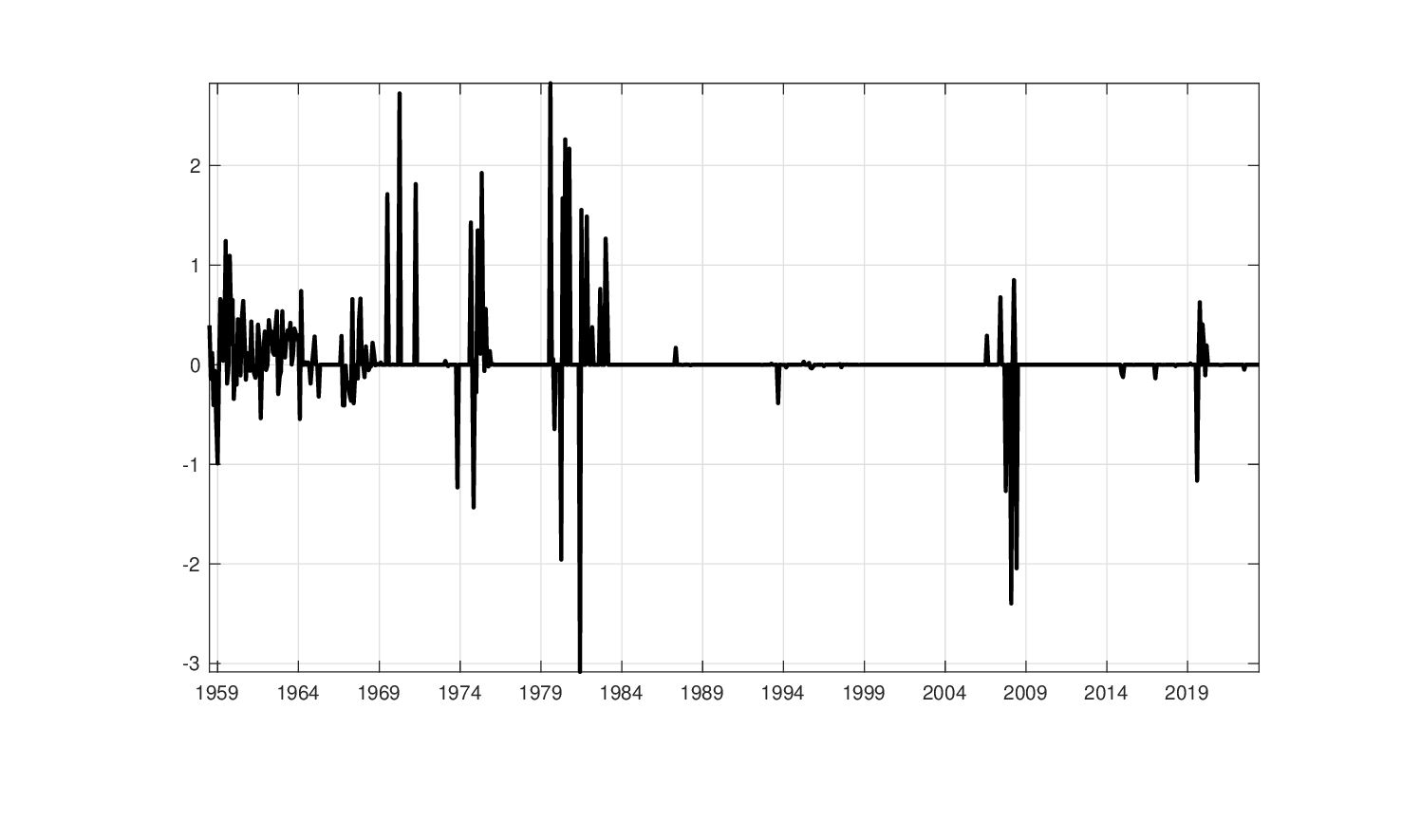}\\[-20pt]
 (a): $\wh{f}_{14,t}$ &  (b): $\wh{f}_{24,t}$
\end{tabular}
	\begin{tablenotes}
		\small
		\item This figure plots the series of estimated factors $\wh{\mbf f}_{jt}=(\wh f_{j1t}\cdots\wh  f_{jr_jt})^\prime$, obtained according to \eqref{eq:fjhat}, for regimes $j=1$ (panel (a)) and $j=2$ (panel (b)) and for $t=1,\ldots, T$, estimated from the Markov switching factor model in (\ref{eq:state_space_mes}) for the dataset of U.S. macroeconomic variables described in Section \ref{section:Empirics_macro}. The number of factors is such that $r_1=r_2=r=4$.
	\end{tablenotes}
\end{figure}

\begin{figure}[h!]
	\caption{Estimated factors $\wh{f}_{jk, t}$, $j=1,2$, $k=1,\ldots, r_j$ - Inflation indexes ($r_j=1$).}\label{fig:fhat_PCE}
	\centering
	\begin{tabular}{cc}
		\includegraphics[width=.5\textwidth]{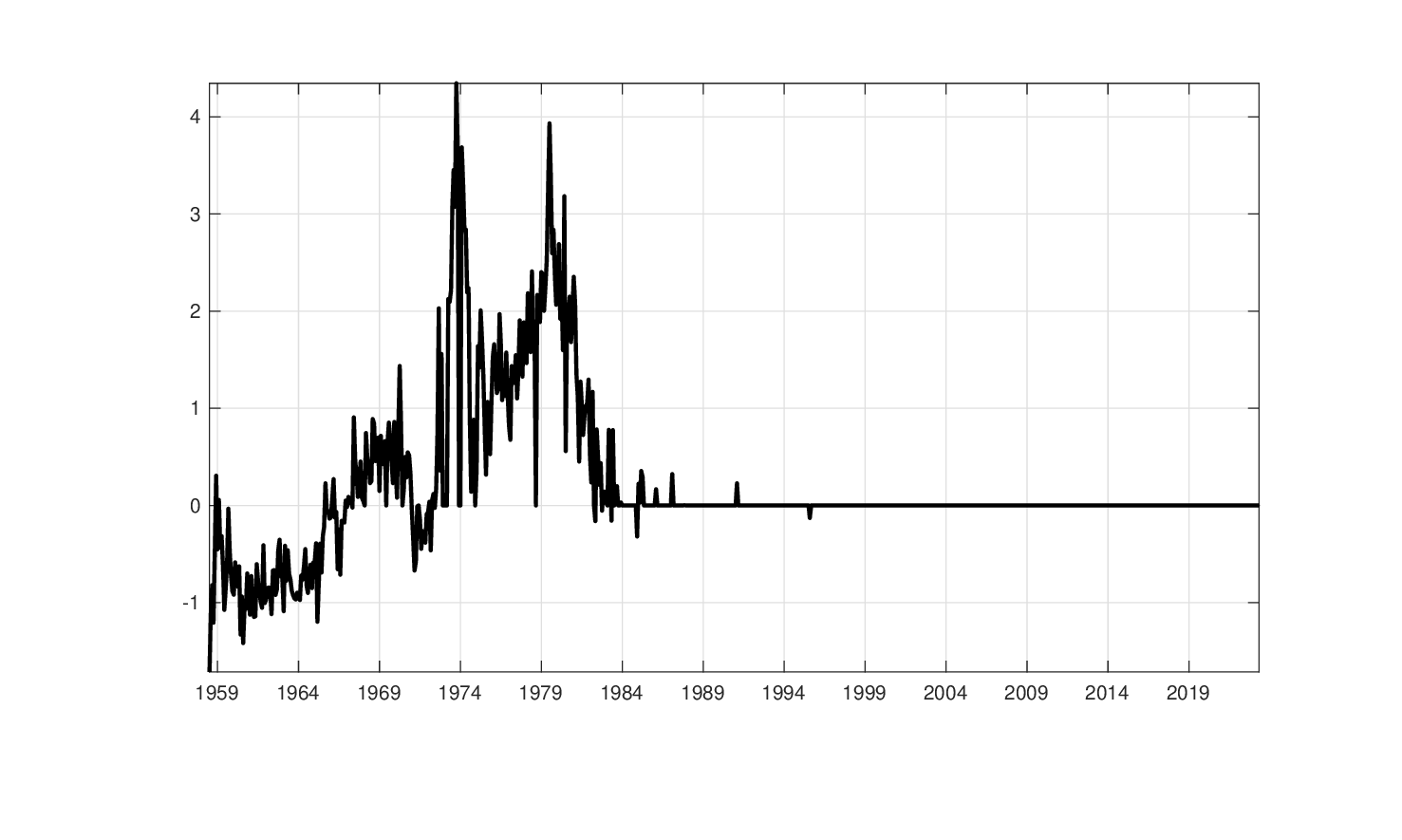}&\includegraphics[width=.5\textwidth]{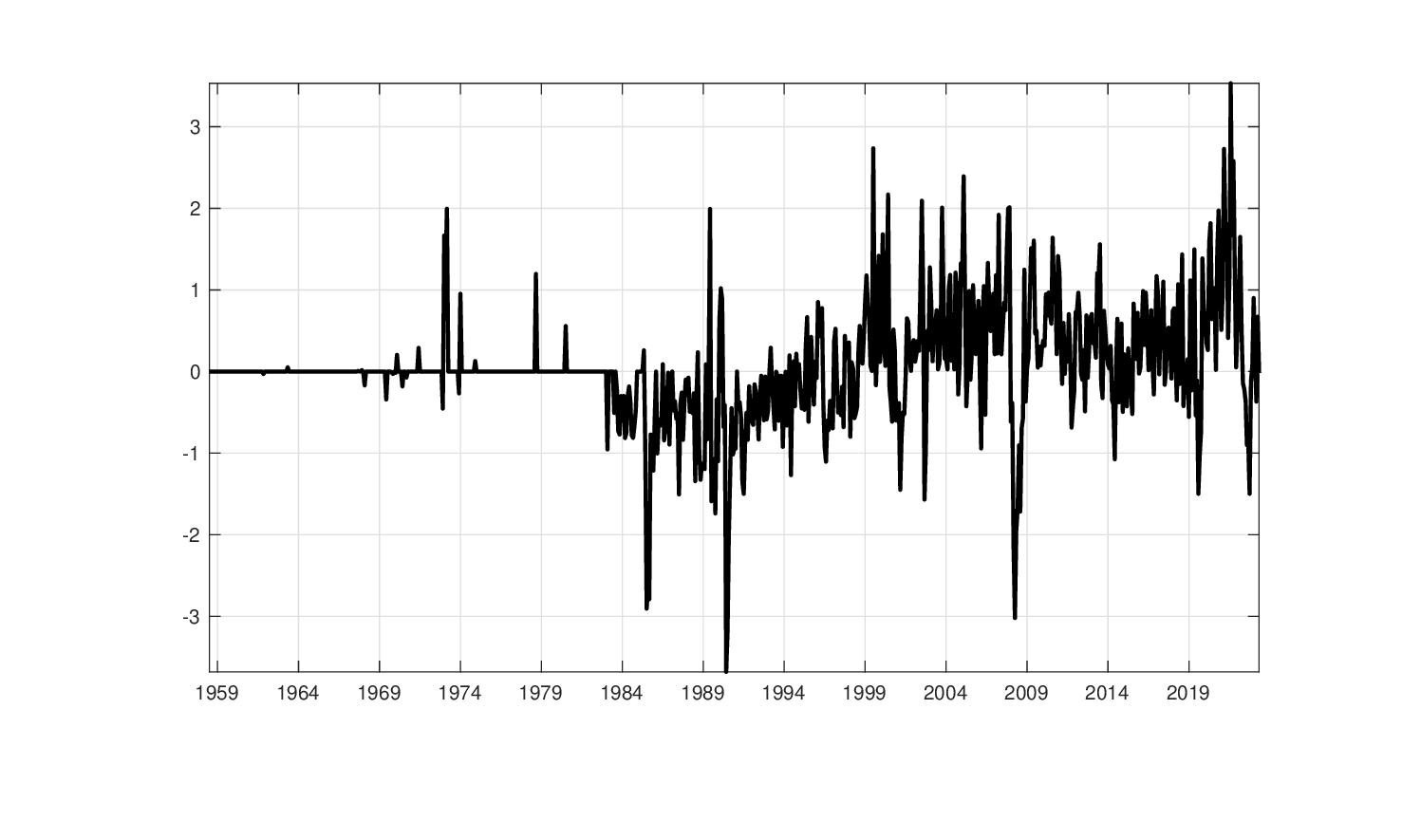}\\[-20pt]
		(a): $\wh{f}_{11,t}$ &  (b): $\wh{f}_{21,t}$
	\end{tabular}
	\begin{tablenotes}
		\small
		\item This figure plots the series of estimated factors $\wh{\mbf f}_{jt}=(\wh f_{j1t}\cdots \wh f_{jr_jt})^\prime$, obtained according to \eqref{eq:fjhat}, for regimes $j=1$ (panel (a)) and $j=2$ (panel (b)) and for $t=1,\ldots, T$, estimated from the Markov switching factor model in (\ref{eq:state_space_mes}) for the dataset of U.S. inflation indexes described in Section \ref{fig:fhat_PCE}. The number of factors is such that $r_1=r_2=r=1$.
	\end{tablenotes}
\end{figure}

This section provides further information in relation to the factors estimated from the three large U.S. datasets of stock returns, macroeconomic time series and inflation indexes, respectively, as discussed in Sections \ref{section:Empirics_SR}, \ref{section:Empirics_macro} and \ref{section:Empirics_inflation}. These are shown in Figures \ref{fig:fhat_FF}, \ref{fig:fhat_FRED} and \ref{fig:fhat_PCE}, respectively.

\end{document}